%% file: caballero-thesis.tex
\documentclass[12pt,final,hyperref]{gatech-thesis}%
\usepackage[unofficial]{caballero-thesis}

\usepackage{framed}
\usepackage{rotating}
\usepackage{longtable}
\usepackage{array}
\usepackage{fix-cm}

\definecolor{red}{rgb}{1,0,0}
\definecolor{orange}{rgb}{1,0.5,0}
\definecolor{green}{rgb}{0.13,0.55,0.13}
\definecolor{purple}{rgb}{0.5,0,1}

\newcommand{\etal}{{\it et al.} }

\title{Evaluating and Extending a Novel Course Reform of Introductory Mechanics}  
\author{Marcos Daniel Caballero} 
\department{School of Physics}
\degree{Doctor of Philosophy}
\gradyear{2012} 
\firstreader{Prof. Michael Schatz}[School of Physics][Georgia Institute of Technology]
\secondreader{Prof. Andrew Zangwill}[School of Physics][Georgia Institute of Technology]
\thirdreader{Prof. Jennifer Curtis}[School of Physics][Georgia Institute of Technology]
\fourthreader{Prof. Richard Catrambone}[School of Psychology][Georgia Institute of Technology]
\fifthreader{Prof. Mark Guzdial}[College of Computing][Georgia Institute of Technology]

\hypersetup{
pdftitle={Evaluating and Extending a Novel Course Reform of Introductory Mechanics},
pdfauthor={Marcos D. Caballero}, 
pdfcreator={Marcos D. Caballero},
pdfsubject={Physics Education Research at an Engineering University}, 
pdfkeywords={physics, physics education, computation, modeling, VPython, FCI, mechanics, epistemology, Georgia Tech, NCSU},
bookmarks=true}

\submitdate{December 2011}
\settocstring{Contents}
\setlofstring{Figures}
\setlotstring{Tables}
\signaturepagetrue

\begin{document}
\bibliographystyle{is-unsrt}
\setchaptertocdepth{2}
\begin{preliminary}

\include{caballero-dedication}
\include{caballero-ack}
\contents
\include{caballero-summary}

\end{preliminary}

\include{01-intro}
\include{02-background}

\include{03-fci}

\include{04-comp}

\include{06-compass}

\include{07-conclusion}

\appendix
\include{appendix/vpapp}
\include{appendix/app-compass}
\include{appendix/app-stats}

\begin{postliminary}

\bibliography{caballero-thesis-bib}

\include{caballero-vita}

\end{postliminary}

\end{document}

%% file: caballero-dedication.tex
\begin{dedication}
  \begin{center}
For Tata and Juniper\\My favorite teachers
  \end{center}
\end{dedication}

%% file: caballero-ack.tex
\begin{acknowledgements}
I want to thank my advisor, Mike Schatz, for granting me the opportunity to perform research in physics education. 
I had became disenchanted with graduate school around the middle of my second year. 
The work was not difficult; I just found some of it unrewarding.
When I went to Mike with questions about my future, he gave me the chance to work in a field that I have come to truly enjoy. 
For that, I am extremely grateful.

I also want to thank all the friends that I have made while living in Atlanta. 
Without their ever present distractions, I might have finished earlier, but it certainly would not have been so much fun. 
I want to thank Ed Greco, Matt Kohlmyer and Alex Wiener for reading a number of drafts of different chapters of this thesis and giving helpful comments. 
I also want to thank the rest of the coffee crew, Daniel Borrero, Domenico Lippolis and Chris Malec for at least one hour of distraction a day.

I would not have been able to get here without my family. 
They always provided with me with love and support when I have needed it most.
They know that I wish I were closer, but I hope they also know that I am always thinking about them.

I certainly would have never finished this work if had not been for my loving wife, Jamie. 
She has been supportive and understanding, especially in the last few weeks that I have spent writing this beast.
I am very lucky to have her as my partner and my best friend.

Finally, I want to thank my two favorite teachers: Dan Caballero, my late grandfather and Juniper Jane Caballero, my daughter.
One helped me to become the person I hoped to be and the other helps me stay that way.
It it to them that this thesis is dedicated.

\end{acknowledgements}

%% file: caballero-summary.tex
\begin{summary}
The research presented in this thesis was motivated by the need to improve introductory physics courses.
Introductory physics courses are generally the first courses in which students learn to create models to solve complex problems.
However, many students taking introductory physics courses fail to acquire a command of the concepts, methods and tools presented in these courses.
The reforms proposed by this thesis focus on altering the content of introductory courses rather than content delivery methods as most reforms do.

This thesis explores how the performance on a widely used test of conceptual understanding in mechanics compares between students taking a course with updated and modified content and students taking a traditional course. 
Better performance by traditional students was found to stem from their additional practice on the types of items which appeared on the test.
The results of this work brought into question the role of the introductory physics course for non-majors.

One aspect of this new role is the teaching of new methods such as computation (the use of a computer to solve numerically, simulate and visualize physical problems).
This thesis explores the potential benefits for students who learn computation as part of physics course.
After students worked through a suite of computational homework problems, many were able to model a new physical situation with which they had no experience.

The failure of some students to model this new situation might have stemmed from their unfavorable attitudes towards learning computation. 
In this thesis, we present the development of a new tool for characterizing students' attitudes.
Preliminary measurements indicated significant differences between successful and unsuccessful students.

\end{summary}

%% file: 01-intro.tex
\chapter{Introduction}\label{chap:intro}

Each year more than 35\% of American college and university students enroll in a physics course \cite{sadler2001success}. Only a small fraction of these students ultimately complete a degree in physics; many are pursuing another science or engineering degree \cite{ies2006degrees}. The majority of students students who enroll in a college physics course take introductory calculus-based physics. These courses serve nearly 175,000 students every year \cite{mulvey2010focus}. However, many of these students fail to acquire effective understanding of concepts, principles, and methods from these introductory courses. The purpose of this thesis is to explore ways in which instruction in these introductory physics courses may be improved.

\input{intro/reform}

\input{intro/measurements}
\input{intro/rqs}

%% file: intro/reform.tex
\section{The Goals of Introductory Physics\label{sec:intro-imp}}

Introductory calculus-based physics courses are fundamental to the development of future scientists and engineers. These courses provide the foundational knowledge that students studying science or engineering will use in their advanced coursework and, ultimately, in their post-baccalaureate careers. Moreover, the methods, tools and thinking which students learn in introductory physics serves them well beyond their completion of the course.

Introductory physics courses are often the first in which students develop their abilities to solve problems. Students learn to translate physical descriptions to mathematical equations, represent physical phenomenon through figures and graphs, organize and carry out the solving of detailed problems and articulate the solutions of such problems in writing. The development of these skills is crucial to students' success in their future work.

Courses in introductory physics are also those in which budding scientists and engineers begin learning the tools of science. While taking these courses, students learn to set up and operate experimental equipment, acquire and manipulate data using software programs and, in some cases, learn to make models of physical phenomena using computers. Obtaining experience with these tools strengthens and diversifies students' problem solving skills beyond simply working with pencil and paper. 

Introductory physics courses can shape how students think about science, how they believe science is done and, perhaps most importantly, can influence if they continue to pursue science or engineering in the future. Students' attitudes toward learning physics, their beliefs about what it means to learn physics and their sentiments about the connection between physics to the ``real world'' can play a strong role in their performance in introductory physics courses. This performance can affect their decision to continue studying science or engineering. 

\section{\label{sec:intro-challenge}Instructional Challenges in Introductory Physics}

Instructors of introductory physics courses aim to provide students with sufficient opportunities to develop their problem solving skills with access to the tools of science and with information about how science is done. Each of these goals faces unique challenges.  

While introductory physics courses aim to develop students problem solving flexibility, many students fail to transition from naive ``equation hunters'' to expertly flexible problem solvers.  Research into this subject has shown that students retain physical misconceptions even after instruction \cite{hallouna1985csc} and are unable to solve even basic introductory problems after completing the course \cite{simon1978individual}. 

The core content of most introductory physics courses still focuses on 19$^{\mathrm{th}}$ century phenomena even though the modern world of science and engineering has progressed well into the 21$^{\mathrm{st}}$ century. Many modern scientists and engineers are exploring physics at the nano-scale, and yet, it is still rare for an introductory physics textbook to mention the atom, phenomenological models of solids or the interaction of electromagnetic fields with matter \cite{atomsajp}.

Furthermore, the problem solving tools taught in most introductory physics courses have not kept pace with the best professional practices of 21$^{\mathrm{st}}$ century science and engineering. For example, computation (the use of the computer to solve numerically, simulate or visualize a physical problem) has revolutionized scientific research and engineering practices. In modern science and engineering, computation is widely considered to be as important as theory and experiment \cite{siamstatementweb}. However, computation is virtually ignored in most introductory physics courses. By contrast, computers are used frquently in physics courses to handle administrative tasks (e.g., the delivery and grading of course assignments \cite{pritchardMasteringphysics,risleyWebAssign}), perform experiments (e.g., collecting and producing plots of data \cite{sokoloff95,sokoloff99}) and deliver content (e.g ``clicker'' questions \cite{keller2007research}).

Many introductory physics courses neglect student epistemology. A student's success in a  physics course might have more to do with her negative sentiments about the course and its content than her ability to learn \cite{elby2001helping}. Understanding such sentiments might help mitigate issues related to the number and diversity of students studying science and engineering \cite{adams2006new}.

Developing new instructional strategies that strengthen problem solving abilities, blend those skills with modern content and tools, and address issues raised by student epistemology is necessary to develop the next generation of scientists and engineers. 

\section{\label{sec:intro-reform}Approaches to Reforming Physics Instruction}

The approach taken by educational researchers to improve instruction in science mirrors the process used in scientific research. That is, efforts to improve education in the sciences generally follow the process of experimental design: development, testing, assessment and refinement \cite{wieman_clickers}. Furthermore, these efforts should be informed by results from cognitive science research \cite{reif2008applying}, the science that explores human learning.

Physics Education Research (PER) is a field of study which aims to address instructional challenges across a wide range of courses from the introductory to the graduate level. These challenges might include overcoming students' misconceptions about particular phenomena, making content relevant to the modern world or addressing students' own motivation to learn. Workers in PER have made some improvements to address these challenges. Generally, these improvements affected two broad areas of instruction: (1) the delivery of course materials and (2) the content of those materials.

Most work has focused on altering the delivery of content to help students overcome physical misconceptions. Traditionally, introductory physics courses have been taught in large lecture classrooms with passively delivered content that focused on derivations (i.e., traditional lecture). McDermott, Halloun and others found that students maintained physical misconceptions and had not developed strong qualitative reasoning skills even after instruction \cite{trowbridge1980investigation,trowbridge1981isu,clement1982spi,
mcdermott1984research,hallouna1985csc,mcdermott1999resource}. The shortcomings of the traditional lecture have been addressed by changes to content delivery methods that included a greater amount of qualitative content and more interaction with students \cite{tutorialswash, eandmtipers, mazurpeer}. Interactive student engagement with a focus on qualitative understanding has been shown to have a positive effect on student learning \cite{hake6000}. 

Some researchers have made fundamental changes to the content of the introductory physics course in an effort to make the course more relevant in the modern era \cite{mechajp,eandmajp}. As an example, which will be discussed in later in detail (Sec. \ref{sec:mandi}), Chabay and Sherwood altered the scope and sequence of topics in the introductory course in order to present modern content \cite{mandi1,mandi2} and introduce students to computational modeling \cite{computajp}, a tool used by most practicing scientists and engineers. Fundamental alterations of the introductory course are rarely implemented at large scale, hence, the impact of such changes on student learning is not yet well understood.

A number of educational researchers have worked to characterize students' attitudes and beliefs about learning physics. These epistemological studies have historically addressed the nature of scientific knowledge \cite{halloun1997views,redish1998student}, but, more recently, Adams et al. developed a survey meant to help instructors address issues related to students' motivation and interest as well as their thoughts about learning and knowledge \cite{adams2006new}. Results from surveys such as these might be helpful to instructors who choose to present topics that increase a students' personal interest in the course and their motivation to learn the material.

%% file: intro/measurements.tex
\section{Measuring the Effects of Reform\label{sec:intro-measure}}

When trying to understand the effects of an adopted reform in introductory physics, it is not only important to evaluate how students approach, reason about and perform on problems, but also to investigate how students think about the tasks, tools and practice of science. Characterizing students' attitudes provides insight into the type of effort they put forth in learning new material. To measure students' performance after adopting some reform, we require a set of instruments to probe students' abilities and compare those skills to students who took a course without any reform, a ``traditional'' course. A wide spectrum of tools exists; for example, we might use students' own coursework, a concept inventory, a detailed problem solving study or a survey instrument.
  
Coursework (e.g., exams, quizzes and projects) is normally used to evaluate what students have learned in a given course, however, performance on common tasks given to students in both a reformed and a ``traditional'' course can also be compared. Researchers often use common final exams to compare how students solve detailed problems \cite{dufresne2002effect}. Such comparisons are relatively straight-forward to perform but are generally limited to comparing students within a single institution. Others have used coursework to determine which reforms to a single course have most positively affected students' performance \cite{morote2009course}.

\begin{figure}[t]
\parbox{0.02\linewidth}{\hspace*{\linewidth}}\parbox{0.98\linewidth}{{\tt A stone dropped from the roof of a single story building to the surface of the earth:}}\\\\\\
\parbox{0.05\linewidth}{\hspace*{\linewidth}}
\parbox{0.90\linewidth}{{\tt (A) reaches a maximum speed quite soon after release and then falls at a constant speed thereafter.}\\
{\tt (B) speeds up as it falls because the gravitational attraction gets considerably stronger as the stone gets closer to the earth.}\\
{\tt (C) speeds up because of an almost constant force of gravity acting upon it.}\\
{\tt (D) falls because of the natural tendency of all objects to rest on the surface of the earth.}\\
{\tt (E) falls because of the combined effects of the force of gravity pushing it downward and the force of the air pushing it downward.}}\\
\caption{A sample item from a force and motion concept inventory. Figure reproduced from Hestenes.\label{fig:fci3}}
\end{figure}

A widely used tool to evaluate and compare performance across semesters \cite{pollock2009longitudinal}, courses \cite{kohlmyer2009tale} or institutions \cite{chasteen2009tapping} is a concept inventory; a instrument composed of a set of multiple-choice questions that is designed to probe performance on a particular topic or set of topics. Concept inventories are used for comparison more often than students' coursework because they are simpler to administer and provide a common baseline from which to measure performance. In mechanics, concept inventories exist to measure performance on qualitative force and motion concepts \cite{hestenes1992fci, thornton1998asl, nieminen2010force}, energy concepts \cite{thornton1998asl, ding2007energy}, the relationships between graphs of kinematic quantities \cite{beichner1994testing}, and ``expert-like'' approaches to solving problems \cite{hestenes1992mechanics,pawl2010mrt}. A sample item from a force and motion concept inventory \cite{hestenes1992fci} is shown in Fig. \ref{fig:fci3}.

\begin{figure}[t]
\begin{center}
\includegraphics[width=0.8\linewidth]{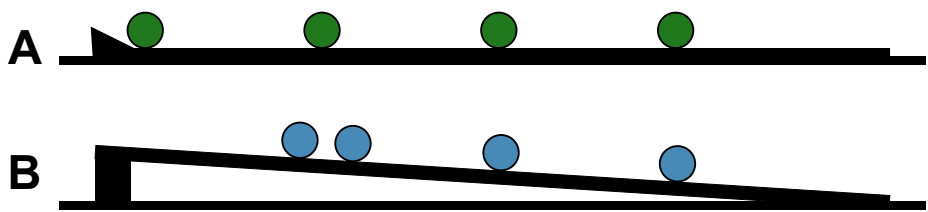}
\end{center}
\caption{[Color] - A question from detailed problem solving study about interpreting motion diagrams and graphing kinematic quantities. Students were presented with the stroboscopic image of a ball (A) rolling on level ground and another ball (B) rolling down a slight incline. Students were asked to draw graphs of the position and velocity of each ball versus time for each situation. The results from this study formed the basis for a section on kinematics developed for the Washington Tutorials in Physics, an instructional aid for introductory physics used by instructors at a large number of institutions. Figure reproduced from Rosenquist and McDermott.\label{fig:wip}}
\end{figure}

Detailed problem solving studies are equally useful to evaluate student performance on problems which they have not encountered in their coursework. These studies are typically designed to probe a particular concept in greater detail than is possible by using easy-to-administer instruments such as concept inventories. In these studies, students are asked by a researcher to verbalize their thinking while solving the problem. Figure \ref{fig:wip} reproduces a problem from a classic study by Rosenquist and McDermott on interpreting motion diagrams and graphing kinematic quantities \cite{rosenquist1987conceptual}. Studies such as these led to the development of some of the questions that appear on a number of concept inventories \cite{hestenes1992fci, thornton1998asl, ding2007energy} and problems that are available in some instructional aids \cite{tutorialswash,eandmtipers,mazurpeer}. Other detailed problem solving studies have been used to compare student populations \cite{bujak2011fci} and investigate students' problem solving flexibility \cite{kohlmyer_thesis}.

Surveys have generally been used by educational researchers to understand the role that students' epistemological beliefs play in their learning. When entering the course, many students have naive expectations about the nature of physics and how science, in general, is done. These surveys have been designed to probe the attitudes and views which experts expect students might acquire after completing the course \cite{halloun1997views, redish1998student, adams2006new}. It is believed that these views affect how students prepare for the course \cite{elby1999another} and can affect how much they learn \cite{perkins2005correlating}. Moreover, certain aspects of the course might alter students' epistemological beliefs \cite{redish2000discipline,elby2001helping,perkins2006towards}.



%% file: intro/rqs.tex
\section{Research Questions\label{sec:intro-rqs}}


The work presented in this thesis has been performed in the School of Physics at the Georgia Institute of Technology (Georgia Tech, GT). 
Georgia Tech offers two distinct large lecture introductory physics courses. Both courses are largely indistinguishable with respect to content delivery methods; all sections of both courses utilize reformed interactive engagement techniques such as ``clicker'' questions with similar intensity. 
These two courses are separated by their content. 
One course uses the textbook by Knight \cite{knight04} and follows the scope and sequence of topics that has remained largely unchanged in introductory physics courses for decades.
Furthermore, this ``traditional'' course does not introduce students to modern tools such as computation. 
The other course uses the {\it Matter \& Interactions} (M\&I) textbook by Chabay and Sherwood \cite{mandi1} which presents updated and reorganized content. 
M\&I differs from the traditional course in its emphasis on the generality of fundamental physical principles, the introduction of microscopic models of matter, its coherence in linking different domains of physics and its use of computation to predict the motion of sundry dynamical systems and compute and display visualizations of electric and magnetic fields due to charges  \cite{atomsajp, mechajp, computajp}. 
Georgia Tech has offered M\&I to one half of the total mechanics enrollment for a number of years.

With two introductory calculus-based mechanics courses offered concurrently, the opportunity to compare learning outcomes arises. One might ask: How do we quantify the effectiveness of this content reformed course, taken as a whole, with regard to student learning? What measurements can be made to compare students' performance on material that appears in both courses? What information can be gleaned about effective instructional strategies from these measurements?

Some of the content of M\&I overlaps with the content of the traditional course. However, a significant portion of M\&I content is absent from the traditional course. For example, students in the M\&I mechanics course learn to apply iterative methods to predict the motion of objects subject to non-constant forces both analytically and using a computer.  The value of learning this algorithmic approach to solve dynamics problems is still largely unknown.  What benefits do students obtain by learning this new material? How do we implement the teaching of this tool at a large scale? How do we evaluate the effects of learning this new material?

We have learned that what students think about learning plays a role in their success in physics courses. In a physics course in which students are learning computational modeling, we should attempt to understand what role student epistemology plays in learning this new tool. How can we measure what students think about learning computational modeling in an introductory physics course? Do our measurements suggest a role that motivation, interest or ability play in learning computation?

The rest of this thesis is presented as follows: 
We describe the outcomes of previous works in reforming physics courses, the history of computation in introductory physics and results from work in student epistemology in physics in Chapter \ref{chap:background} to set the background for the current work. 
The effectiveness of our reformed mechanics course (M\&I) was measured using student performance on a standard concept inventory (FCI). 
These results were compared to the performance by students from our traditional course.
In Chapter \ref{chap:fci}, we describe these measurements, discuss their implications and outline possible modifications to instruction in our reformed course.
In the reformed course, we have taught computation as part of the laboratory. 
We implemented a new instructional strategy to extend the teaching of computation beyond the laboratory. 
We discuss the benefits of teaching computation, the issues associated with implementation this new strategy in a large lecture setting and an evaluation of the learning outcomes from this strategy in Chapter \ref{chap:comp}.
This use of this new instructional strategy to teach computation has raised questions about the role of student epistemology in learning computation. 
In Chapter \ref{chap:compass}, we discuss the design of and the preliminary results from a new survey on students' epistemological beliefs about learning computation.  
We make concluding remarks and outline possible future research directions in Chapter \ref{chap:closing}.

%% file: 02-background.tex
\chapter{Background and Motivation}\label{chap:background}

Improvements of student learning in introductory physics courses have generally resulted from reforms of the delivery methods, namely, increasing active student engagement. 
Measurements of performance differences as a result of changes in content are still lacking.
To provide the context for comparative measures between courses with markedly different content (Ch. \ref{chap:fci}), we discuss previous efforts to reform introductory physics courses and their outcomes (Sec. \ref{sec:hist}).
The {\it Matter \& Interactions} course (Sec. \ref{sec:mandi}) is highlighted because it provides the foundation for the current work. 
{\it Matter \& Interactions} introduces computation as one significant modification to the content of the introductory physics course.
As it was not the first course to do so, we present a historical overview of the sundry attempts to introduce computation to introductory physics students (Sec. \ref{sec:history-comp}) in order to distinguish them from our approach (Ch. \ref{chap:comp}).
Students have expressed a lack of self-confidence and a considerable amount of anxiety when learning computation.
These sentiments are somewhat similar to what many students experience when they study science.
We present an overview of the work done to understand students' epistemology in science (Sec. \ref{sec:abs}) to provide the background for the development of a new instrument (Ch. \ref{chap:compass}) which helps characterize students' attitudes towards learning computation. 

\input{history/hist}

\input{history/mandi}

\input{history/comp-history}
\input{history/epistemology}

%% file: history/hist.tex
\section{Course Reform in Physics\label{sec:hist}}

Challenges to student learning in physics have generally been addressed by modifications to content delivery methods. Many institutions have adopted research-based instructional aids \cite{mazurpeer,tutorialswash} and some have developed introductory physics courses that focus on discovery and experiential learning \cite{laws1991workshop,beichner,etkina2006using,brewe2008modeling}. However, the core content of these courses is frequently little changed from the courses offered a century ago. 

Historically, reforms of physics instruction have focused on creating a more active environment for learning. These reforms stemmed from a series of novel problem studies in which students solved problems which they had not seen in class. 
Students who had completed a course in mechanics were invited to solve problems and view demonstrations related to kinematics \cite{trowbridge1980investigation, trowbridge1981isu, mcdermott1student}, dynamics \cite{clement1982spi,hallouna1985csc} and the work-energy theorem \cite{lawson1987student}. Afterwards, students were asked to explain their work or discuss the phenomenon that they had just observed. Researchers found that students were often unable to discuss the problem or phenomenon in a satisfactory way. In fact, many students had strong misconceptions and reverted to pre-instruction notions about the nature of the phenomenon \cite{hallouna1985csc}.

Students' challenges with conceptual understanding in mechanics led to the development of research-based instructional aids (e.g., University of Washington Tutorials in Physics \cite{tutorialswash} and Mazur's Peer Instruction \cite{mazurpeer}) that can be used in large lecture environments. These teaching tools are now used widely in introductory physics courses offered at high schools and colleges. In addition, a series of concept inventories were designed (e.g., the Force Concept Inventory (FCI) \cite{hestenes1992fci}, the Mechanics Baseline Test (MBT) \cite{hestenes1992mechanics} and the Force and Motion Conceptual Evaluation (FMCE) \cite{thornton1998asl}) using some of the questions and activities from previous studies. These assessments are typically given before and after instruction (pre-test and post-test, respectively) to compare student learning on qualitative concepts.

An active learning environment was shown to be instrumental in helping to build students' conceptual understanding of mechanics. Across a wide variety of instructional levels, mechanics students taught in an active learning environment using ``clicker'' questions (i.e., short, usually qualitative, questions posed to the class) performed significantly higher on a qualitative force and motion assessment (FCI) than those who took traditional lecture courses \cite{hake6000}.

The success of interactive instructional methods in mechanics courses drove content delivery improvements to be adopted in a broad spectrum of physics courses. Similar content delivery reforms have been made in introductory electromagnetism (E\&M) \cite{mcdermott1807research,shaffer1992research,bagno1997problem,deslauriers2011improved}. More recently, these reforms have percolated up to middle and upper-division courses: sophomore-level classical mechanics \cite{classmechreform}, junior-level E\&M \cite{pepper2010our,pollock-use,wallace2010upper}, senior-level quantum mechanics \cite{mckagan2006exploring,goldhaber2009transforming}, graduate courses \cite{carr2009graduate} as well as to other STEM (Science, Technology, Engineering and Mathematics) courses beyond physics \cite{keller2007research}. Preliminary results comparing the performance of students who took reformed upper-division courses to those who took traditionally taught courses suggest better performance by students in reformed courses \cite{chasteen2009tapping,upperdivreform}.

Some research-based courses have made sweeping changes to the learning environment. Some of the more well-known reconfigurations of the instructional environment are Workshop Physics \cite{laws1991workshop}, the Investigate Science Learning Environment (I.S.L.E.) \cite{etkina2006using}, the Arizona State modeling curriculum \cite{jackson2008modeling} and the Student-Centered Active Learning Environment for Undergraduate Programs (SCALE-UP) project \cite{beichner}. 

Workshop Physics is an instructional environment that is hands-on and experiment-driven. Students follow semi-scaffolded experiments to discover how physical principles operate. Workshop Physics students make heavy use of data collection techniques that would be necessary for practicing scientists and engage in  kinesthetic activities to obtain ``muscle memory'' of phenomena. Activities in the Workshop Physics course emphasize transferable skills of scientific inquiry \cite{laws1991calculus}.

I.S.L.E. is an experiment-driven instructional environment in which students engage in experimental design and execution to explore physical principles. I.S.L.E. students have demonstrated some ability to transfer experimental design skills (e.g., considering assumptions, evaluating the effect of uncertainties, etc.) between different experiments and between courses from different domains \cite{etkina2007spending,karelina2007design,ruibal2007physics}. 

The Arizona State modeling curriculum places a strong emphasis on the practice of constructing and applying conceptual models of physical phenomena. It makes extensive use of student discussion and reporting of findings. Students of the modeling curriculum use their own ability to construct knowledge through observation, inquiry and representation \cite{halloun1987modeling}.

Workshop Physics, I.S.L.E. and the modeling curriculum tend to follow a traditional sequence of topics: kinematics, constant force motion, and energy. Post 19$^{\mathrm{th}}$ century concepts such as phenomenological models of solids are absent. In addition, students' use of the modern tools of science is limited to collecting and displaying data using software programs  and interacting with simulations (visualizations of physical phenomena). These courses do not make use of modern problem solving tools such as computational modeling. 

The SCALE-UP project is slightly different from the others, because it is generally ``content neutral''; it attempts not only make courses hands-on but also social. SCALE-UP students collaborate in large groups (up to 9) to explore physical phenomena. SCALE-UP has been used by a number of traditional \cite{belcher2003improving} and content reformed \cite{gaffney2008scaling} physics courses. SCALE-UP has also been utilized by instructors in other domains such as calculus, chemistry and biology \cite{scaleupadopters}.

These reformed learning environments have transformed how physics is taught but the core content of most of them has remained unchanged. Fundamental changes to the core content of an introductory physics course are rare, often, because the adopted text is remains ``traditional''. Moreover, it is not yet understood how student learning is affected by courses which make use of drastically different texts. Three textbooks that have made significant alterations to the scope and sequence of topics in the introductory physics course are Moore's {\it Six Ideas That Shaped Physics}, Huggins's {\it Physics 2000} and Chabay and Sherwood's {\it Matter \& Interactions} (M\&I).  

{\it Six Ideas That Shaped Physics} \cite{mooresix} completely reorganizes the introductory course. Moore structures his texts around six ``grand ideas'' in physics: (1) conservation laws constrain interactions, (2) the laws of physics are universal, (3) the laws of physics are frame-independent, (4) electric and magnetic fields are unified, (5) particles behave like waves, (6) some processes are irreversible. Students learn about ideas 1-3 in their first semester and 4-6 in their second. {\it Six Ideas That Shaped Physics} has been positively reviewed and has been adopted, typically in small enrollment courses, by a number of institutions. Furthermore, post-instruction evaluations have so far been limited to course-instructor reviews \cite{moore2006post} rather than comparisons of learning outcomes with students taking other courses. 

{\it Physics 2000} \cite{physics2000} is example of a radical transformation of the introductory physics course. Students using {\it Physics 2000} follow a course that emphasizes expert-level problem solving sophistication (e.g., Lorentz transformations) and advanced experimental tools (e.g., Fourier analysis). No research has yet been performed to assess the effectiveness of using {\it Physics 2000} or compare it with other reformed courses. Furthermore, it is not clear if {\it Physics 2000} is widely used. 

{\it Matter \& Interactions} (M\&I) \cite{mandi1} is a reformed text that is being increasingly adopted for large lecture calculus-based introductory physics courses. As it plays an important role in the present work, we present a detailed overview of M\&I in the next section (Sec. \ref{sec:mandi}).

%% file: history/mandi.tex
\subsection{Matter \& Interactions\label{sec:mandi}}

{\it Matter \& Interactions} (M\&I) is an innovative and modern textbook \cite{mandi1,mandi2} which has fundamentally altered the core content and sequencing of topics in introductory physics to emphasize the ``modeling'' process. Modeling is the practice of creating and applying physical models of complex phenomenon. This practice includes the use and justification of simplifications and approximations that make such phenomena tractable either analytically or using computation. 
The M\&I mechanics course emphasizes the modeling process through its use of reductionism. M\&I students learn to construct models of complex systems and simplify analysis by starting from a few fundamental principles. The ``momentum principle'' relates a change in momentum ($\Delta \vec{p}$) experienced by a system in a short time ($\Delta t$) to the external net force ($\vec{F}$) applied to the system in that time, $\Delta \vec{p} = \vec{F} \Delta t$. The ``energy principle'' relates the change in the total energy ($\Delta E$) of a system to the work ($W$) that is done by or on the system's surroundings and the heat ($Q$) that the system exchanges with its surroundings, $\Delta E = W + Q$. The ``angular momentum principle'' relates the change in angular momentum ($\Delta \vec{L}$) of a system in a short time ($\Delta t$) to the net external torque ($\vec{\tau}$) applied to the system in that time, $\Delta \vec{L} = \vec{\tau} \Delta t$. 

M\&I's emphasis on the generality of these fundamental physical principles is well represented by its introduction of the iterative prediction of motion. 
Students learn to apply the impulse-momentum relationship iteratively over short time steps to predict motion of a variety of dynamical systems. 
This general applicability of the impulse-momentum relationship (i.e., Newton's Second Law) is further reinforced by the solving of dynamics problems using numerical computation. 
Students begin by constructing a physical model of the system and determining what information is applicable to the physical principles, namely, the momentum principle. 
Students then translate their physical model to a numerical model using VPython \cite{vpythonWebsite}, a module developed for the Python programming language. 
The relevant kinematic, dynamic and energetic information can be extracted from this model and displayed as graphs, if desired. 
An animated visualization gives students visual feedback on their model and the program may be refined. 
Students apply this technique to a number of different dynamical systems over the course of the semester. 

M\&I has been adopted by a variety of institutions (i.e., high schools, two-year colleges and universities) including a number of institutions that use M\&I in large-lecture sections \cite{mandiadopters}. Evaluations of student learning in M\&I have begun \cite{kohlmyer2009tale} and continue in the present work (Chs. \ref{chap:fci} \& \ref{chap:comp}).




%% file: history/comp-history.tex
\section{Computation in Physics Courses}\label{sec:history-comp}

Whether they are writing lab reports, solving online homework, reading a textbook or pursuing less scholarly activities, computer usage permeate every aspect of students' lives. However, it is rare for students to know how to use a computer to solve science and engineering problems. Teaching students computation can provide complementary instruction to regular analytic tasks and help students acquire the skills to be successful in 21$^{\mathrm{st}}$ century science and engineering.

\subsection{The Case for Teaching Computation\label{sec:comp-thecase}}

Several educational researchers discussed how computation can be beneficial to the development of 21$^{\mathrm{st}}$ century scientists and engineers. By using computation, complex problems become tractable at lower levels. This can be leveraged to engage students in work similar to professional scientists and engineers, which includes the modeling process \cite{redish1993student,schecker1993learning}. 
The use of a programming language itself constrains users to certain syntactic structures. 
Constructing programs requires students to contextualize the problem in order to produce a precise representation \cite{sherin1993dynaturtle}. 
More recently, model animation and the visualization of abstract quantities like momentum, angular momentum and field vectors in three dimensions have been cited as potential benefits \cite{computajp}.
The algorithmic approach for predicting the motion of physical systems, which computation affords, is quite general and is applicable to a broad number of complex problems.
It is possible that by learning computation, students might be able to solve problems related to physical systems with which they have had no exposure.

Redish and Wilson developed the Maryland University Project in Physics and Educational Technology to help resolve the differences between the activities in which their physics majors engaged and those in which professional physicists engage \cite{redish1993student}. 
Table \ref{tab:compareact} (reproduced from \cite{redish1993student}) summarizes some of these differences. In redesigning their course, Redish and Wilson aimed to develop their students' professional modeling skills; the skills that are necessary to solve the type of broad, open-ended problems that professional scientists and engineers encounter. Some skills were already addressed in their traditional course such as translating word problems to physics equations and algebraically solving a variety of equations. However, the development of other skills was nearly absent: making estimations and approximations, explaining and summarizing procedures and results and numerical modeling skills. 

\begin{table}[t]
\caption{Traditional course activities of introductory physics students are contrasted with activities of professional physicists. Course activities are characteristic of most introductory physics courses and stem from a focus on solving ``back-of-the-book'' style problems. Redish and Wilson aimed to address the mismatch between course preparation and professional practice by complementing their introductory courses with computation. Note that the final activity refers to ``using a computer for solving a science problem'' not casual usage of a computer (which is commonplace now). This table is reproduced from Redish and Wilson.\label{tab:compareact}}
\begin{center}
\begin{tabular}{| p{0.45\linewidth} | p{0.45\linewidth} |}\hline
{\bf Students} & {\bf Physicists}\\\hline
Solve narrow, pre-defined problems of no personal interest & Solve broad, open-ended and often self-discovered problems\\\hline
Work with laws presented by experts. Do not ``discover'' them on their own or learn why we believe them. Do not see them as hypotheses for testing. & Work with models to be tested and modified. Know that ``laws'' are constructs.\\\hline
Use analytic tools to get ``exact'' answers to inexact models. & Use analytic and numerical tools to get approximate answers to inexact models.\\\hline
Rarely use a computer. & Use a computer often.\\\hline
\end{tabular}
\end{center}
\end{table}

Niedderer and Schecker used STELLA, a commercially available graphical programming environment, with high school physics students to extend the discussion of physical principles to problems that could not be solved in analytic closed-form \cite{schecker1993learning}. Schecker opined that topics presented in most high school physics courses were chosen for mathematical convenience and that by using a computer to perform the more mathematically sophisticated operations, one could highlight physical relationships. He found that students' use of computational models could decrease their overuse of special case analytic solutions and increase qualitative analysis. 

DiSessa \etal used the Berkeley BOXER project (BOXER) in a high school physics course 
because, as a programming language, it required students to produce precise representations of the problem and contextualize abstractions \cite{sherin1993dynaturtle}. Precision is necessary when relating physical concepts to mathematical statements. But unlike hand-written work, feedback from imprecise program statements in a computational model is instantaneous. For example, the visualization of an imprecise program is completely wrong. 
Abstractions like the complex electric field patterns are contextualized using not only programmatic representations (e.g., iterative calculations to superpose source fields) but also through visualization (e.g., field vectors represented by arrows in space).
DiSessa \etal also noted that programming languages' constraint on syntax and causal realtionships
makes constructing computational models a useful mechanism for teaching physics \cite{disessa2001changing}.


Sherwood and co-workers developed VPython which aimed to help introductory physics students improve their conceptual understanding of the physical principles addressed in the course, produce a visualization of the problem that was not possible with static pictures and provide the tools to model complex real-world situations \cite{computajp}. The designers of VPython sought to leverage the development of high-level, object-oriented programming languages to make the construction of highly visual simulations accessible to introductory physics students \cite{scherer2000vpython}.

Any of these environments could help introductory students solve problems that would normally be intractable to them. 
For example, consider the mainstay of a typical introductory mechanics course, an object thrown into the air near the surface of the Earth with negligible air resistance. 
Students taking a typical introductory course would learn several equations 
to predict the motion that emphasizes kinematics, a way of describing the motion without explicitly connecting changes in the motion to forces (dynamics).
Hence, these students would be significantly challenged
if air resistance were not negligible. 
Moreover, these students might be inclined to use models or equations that are inappropriate for this new problem \cite{kohlmyer_thesis}.
By contrast, students who learn computation learn dynamics first. 
This initial introduction of Newton's Second Law facilitates the teaching of the iterative prediction of motion early on. 
For the computational student, predicting the motion of the object with air resistance requires a simple programming change to add air drag to their computational model. 

\subsection{Distinguishing Tools for Computational Instruction\label{sec:gen-approach}}

The tools used for computational instruction have a number of different characteristics: environment, programming language, limits on visualization, {\it et cetera}. Arguably, one of the most important characteristics is the nature of the environment, that is, whether or not students can gain access to the underlying model or modeling algorithm. Efforts to introduce computation in introductory physics courses have produced environments in which students have little or no access to the underlying model, a {\it closed} computational environment, and others that allow students to view and modify this model, an {\it open} computational environment. We further explicate closed and open environments below.

\paragraph{Closed Computational Environments} 
Some educational researchers \cite{perkins2006phet} have focused on creating closed computational environments. 
These are environments in which the user has little or no access to the underlying model or modeling algorithm (a ``black box'' environment). 
Closed computational environments are analogous to ``canned'' codes in scientific research. 
Users can set up and operate the canned programs but did not construct them.
User interaction in closed computational environments is often limited to setting or adjusting parameters and interactions with the mouse or keyboard.
The advantages of using closed computational environments are that they typically require no knowledge of computation to operate, run similarly on a variety platforms with little more than an Internet browser and produce highly visual simulations. 

\paragraph{Open Computational Environments}
Others \cite{redish1993student,schecker1994system,disessa1986boxer,computajp,esquembre2004easy} have created open environments in which students can construct computational models. 
Open computational environments are analogous to ``user-developed'' codes in scientific research. 
Students who learn to use one of these environments have the advantage of peering into the ``black box'' to view and alter the underlying algorithm on which the model depends. 
Moreover, students can learn to develop their own computational models that solve new problems (Ch. \ref{chap:comp}).  
It is possible for students to interact with open environments as if they were closed; users can be restricted (formally or informally) from viewing or altering the underlying model of any simulation developed in an open environment.

All historical attempts to introduce computation used open environments because the level of sophistication to program interactive closed environments exceeded technological resources at the time. 
Most of these environments had limited visualization capabilities as well; output was limited to graphs \cite{redish1993student} or, perhaps, two dimensional line animations \cite{schecker1994system,disessa1986boxer}. 
More modern implementations of computation take advantage of high-level, object-oriented programming languages, cheap and available hardware and efficient video cards not only making possible the development of closed environments \cite{perkins2006phet} but also the rendering of three dimensional graphical environments \cite{computajp,esquembre2004easy}. 

It is true that to use an open computational environment, students must devote additional time and cognitive effort to learning syntax and procedures of the language the environment supports. Depending on the environment that is used, students might devote more time and cognitive effort to the details of constructing a working simulation (e.g., message handling, drawing graphics, garbage collection) than to modeling the physics behind it. It is, therefore, important to consider students' experience (or lack thereof) with computation when choosing a computational environment.

\subsection{Samples of Computational Environments Used in Introductory Physics \label{sec:comp-imp-hist}}

Constructing computational models can be beneficial to students learning physics, not only by engaging them in the modeling process but also by reaping some of the other benefits listed in Sec. \ref{sec:comp-thecase}. 
Several notable attempts have been made since the development of small and inexpensive microcomputers with visual displays. 
Each has tried to engage students constructively in the modeling process.
Some attempts to use open computational environments were limited to small classes. The program statements written by students in these attempts were minimal, if written by students at all \cite{redish1993student,schecker1993learning,disessa1986boxer}. 
Others have been successful with using a closed environment to introduce computation to students in a variety of class environments including large lecture sections \cite{perkins2006phet}. 
One open environment developed for teaching computation has been effectively used like a closed one in a variety of settings \cite{esquembre2004easy}. 
There has been one successful attempt at teaching computation to introductory physics students in a large lecture setting which uses an open environment \cite{computajp}. 

We present a short chronological discussion of these attempts. 
When possible, we have highlighted their use to solve a problem that is not analytically tractable for introductory students in a traditional setting, i.e., the prediction of motion of an object subjected to turbulent ($F \sim v^2$) air drag in one or two dimensions.


\subsubsection{Maryland University Project in Physics and Educational Technology}

\begin{figure}
	\centering
		\includegraphics[width=0.90\linewidth]{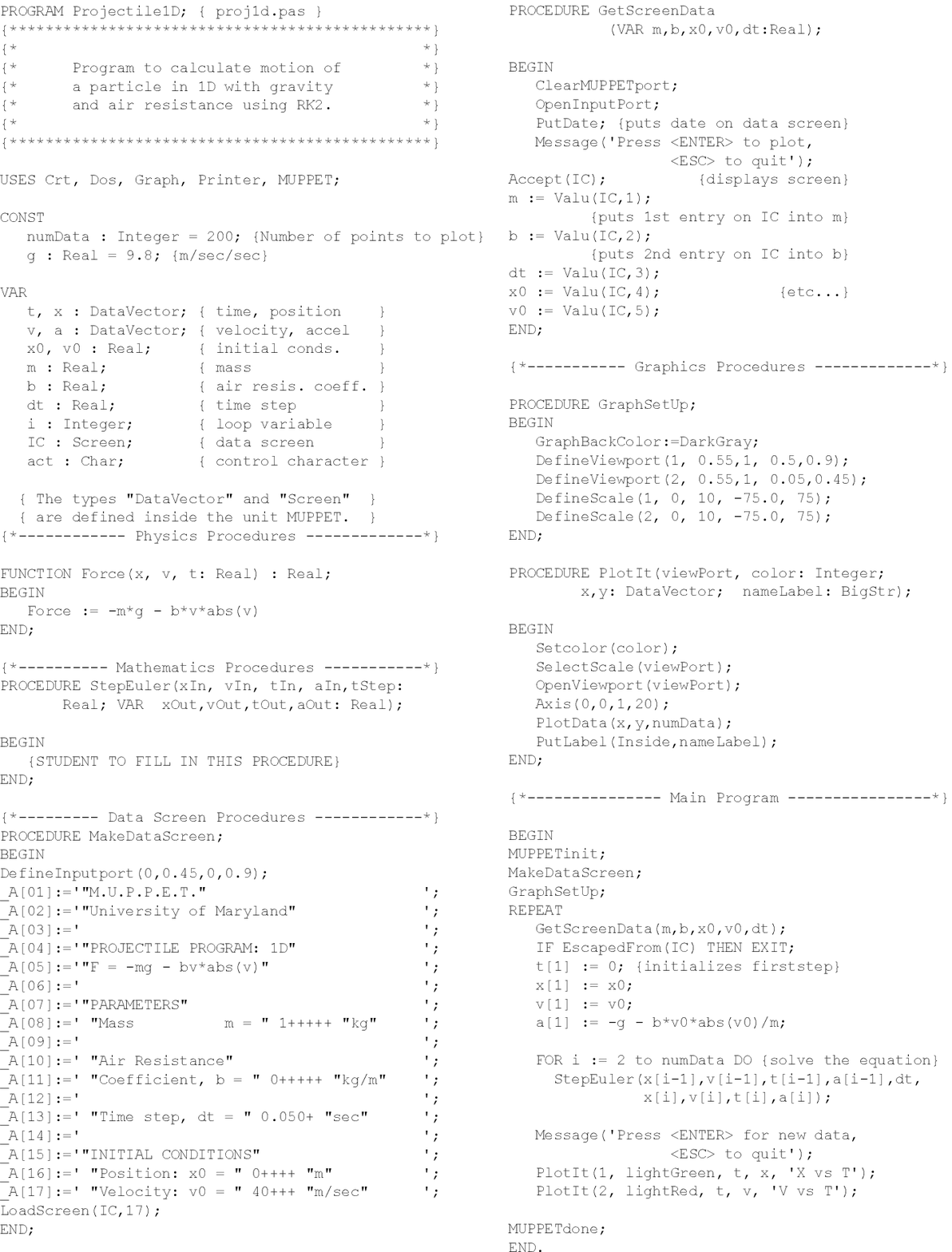}
	\caption{A M.U.P.P.E.T. program that models the one dimensional motion of an object moving under the influence of the gravitational force subjected to turbulent drag in one dimension. The Turbo Pascal code that generates each piece of the visualization can be seen clearly (\texttt{MakeDataScreen}, \texttt{GraphSetUp}, \texttt{PlotIt}). Students filled in the section of code that performed the numerical integration (i.e., \texttt{StepEuler}). Once compiled students worked with the user interface shown in Fig. \ref{fig:muppet2}, entering initial conditions and parameter values.}\label{fig:muppet}
\end{figure}

The Maryland University Project in Physics and Educational Technology (M.U.P.P.E.T.) was designed as part of a complete rewrite of the sequence of introductory physics courses taken by freshman physics majors \cite{redishmuppet}. Between 1986 and 1989, these courses served around 25 students at a time. M.U.P.P.E.T. provided a platform to broaden the scope of problems which were addressed in the introductory physics sequence. Over the course of the sequence, students solved a variety of problems (e.g., turbulent air drag in two dimensions, large amplitude pendulum and billiard dynamics) in class and on their homework. Not all students were novice programmers; two-thirds of these students had some familiarity with a programming language. 

\begin{figure}[t]
	\centering
		\includegraphics[scale=.6]{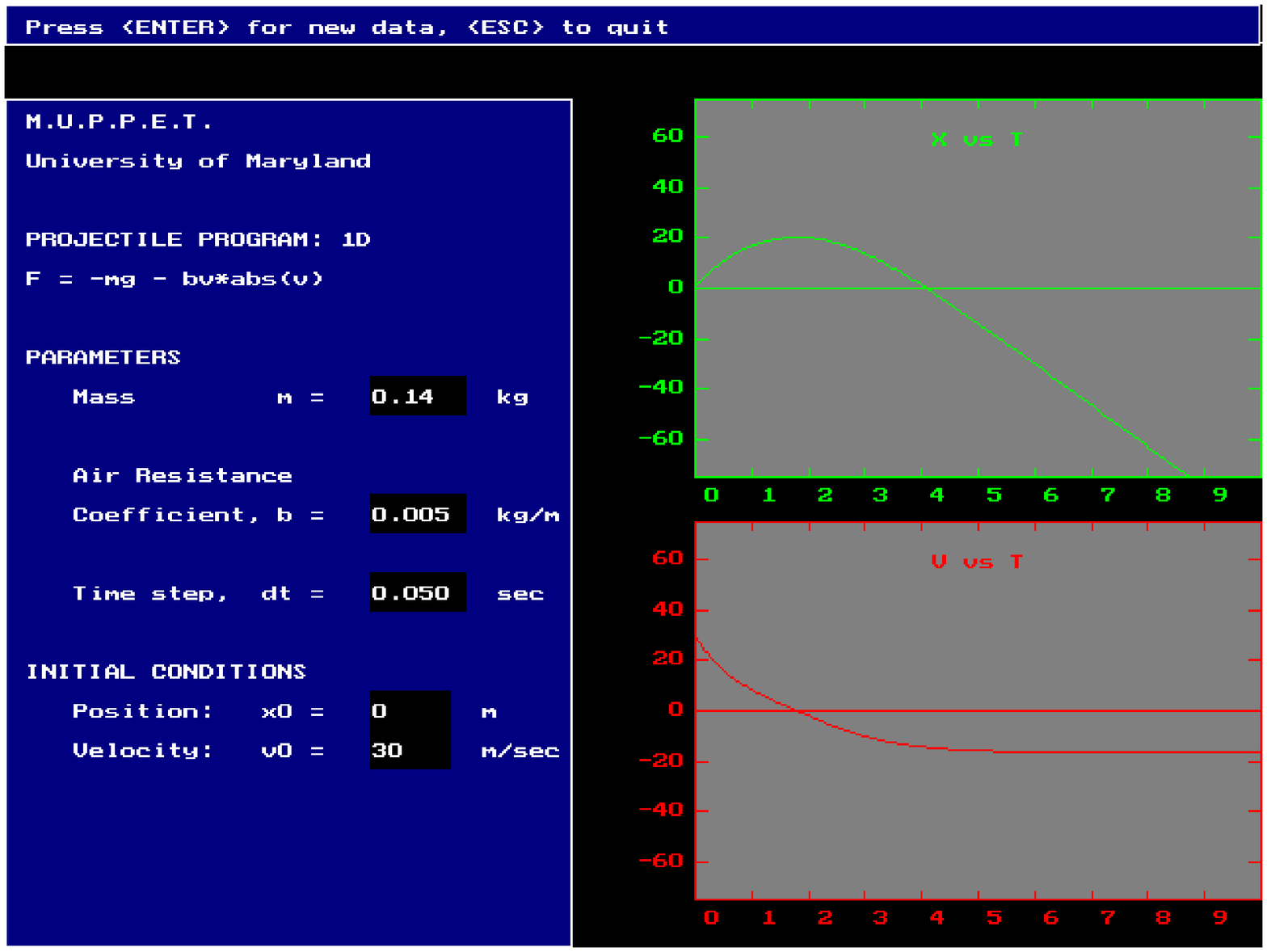}
	\caption{[Color] - A compiled M.U.P.P.E.T. program (Fig. \ref{fig:muppet}) produces this user interface with which students interact when modeling the motion of an object falling with air resistance. On the left, students view the model (i.e., the forces acting on the particle), the relevant parameter values (e.g., mass and drag coefficient) and the initial conditions. The program produces plots of the object's position and velocity versus time on the right.}\label{fig:muppet2}
\end{figure}

A  M.U.P.P.E.T. program that models the one dimensional motion of a particle near the surface of the Earth subject to turbulent drag is shown in Fig. \ref{fig:muppet}. M.U.P.P.E.T. programs were written using Turbo Pascal because it was believed that by using M.U.P.P.E.T. students might learn skills that were transferable to other professional languages (C or FORTRAN) which they would use in their future careers \cite{redish1993student}.  This program handles variable creation and assignment (\texttt{CONST} and \texttt{VAR} sections of Fig. \ref{fig:muppet}), the physical model and numerical integration (\texttt{Physics Procedures} and \texttt{Mathematics Procedures} section of Fig. \ref{fig:muppet}), and the screen and graphics procedures (\texttt{Data Screen Procedures} and \texttt{Graphics Procedures} section of Fig. \ref{fig:muppet}). Most of this program was constructed by experts; students were only responsible for filling in the algorithm needed to model the system (i.e., the Euler step procedure, \texttt{StepEuler}). After the program was compiled, the resulting simulation was seeded with initial conditions and parameter values. The solution was plotted (Fig. \ref{fig:muppet2}). Due to hardware and software limitations at the time, output was limited to graphs and tables.

Redish and Wilson report that M.U.P.P.E.T. was used successfully in introductory physics and sophomore level classical mechanics courses \cite{redishmuppet}. Classical mechanics students selected complex open-ended problems to explore analytically and using computation. A large fraction of students were reported to have produced ``valuable and interesting'' projects which might have helped them in future research with faculty. However, little is said about the challenges that students experienced with learning and using M.U.P.P.E.T., and nothing is mentioned about how M.U.P.P.E.T. helped develop students' conceptual understanding. It is possible that the success of M.U.P.P.E.T. might be attributed to its use by physics majors in small class sizes who had some programming experience. Most introductory students were less computationally experienced than these physics majors. Future efforts would aim to reduce the cognitive effort devoted to learning traditional programming syntax and semantics. 

\subsubsection{STELLA}

\begin{figure}[t]
	\centering
		\includegraphics[width=0.60\linewidth]{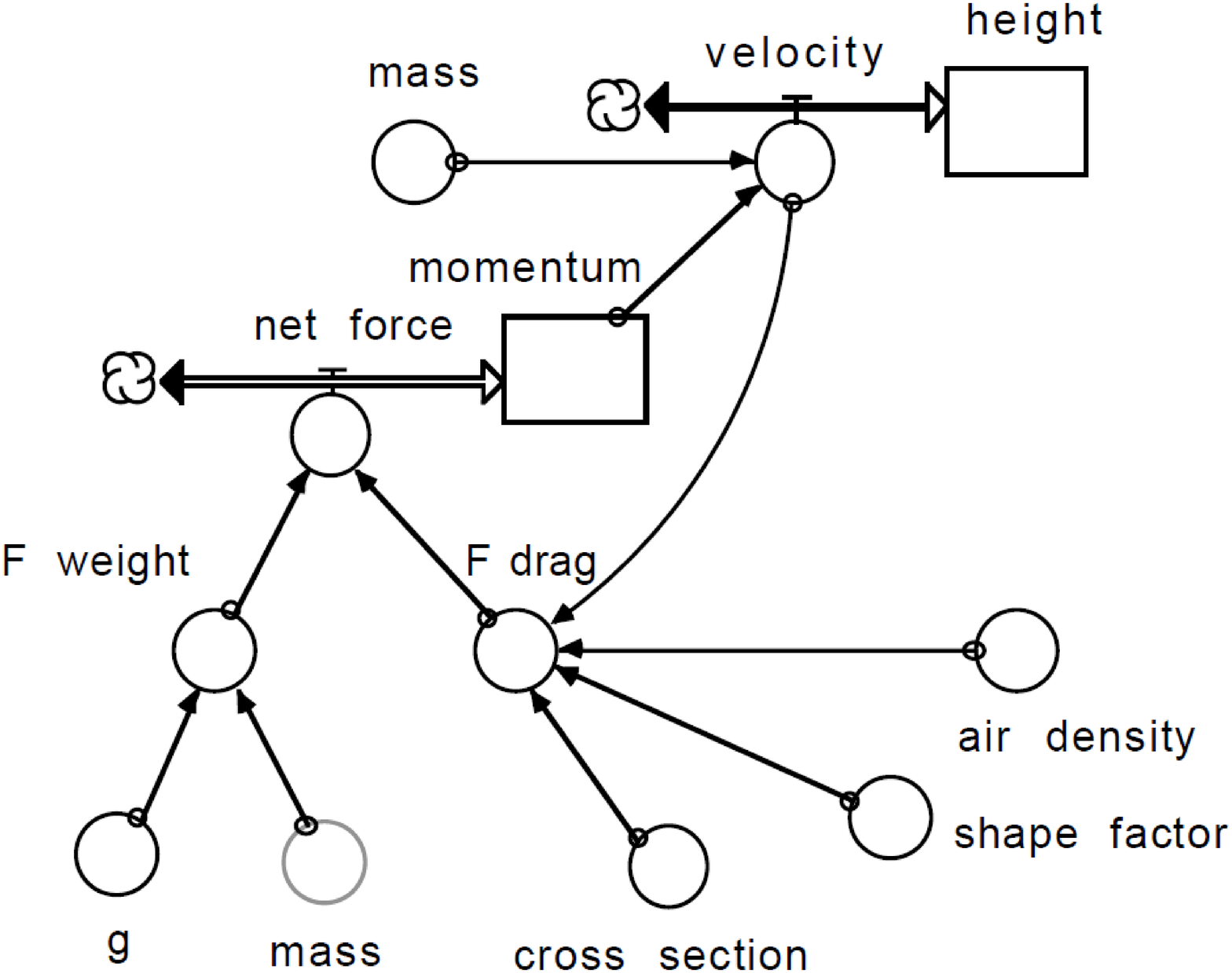}
		\includegraphics[width=0.65\linewidth, clip, trim=0mm 110mm 10mm 0mm]{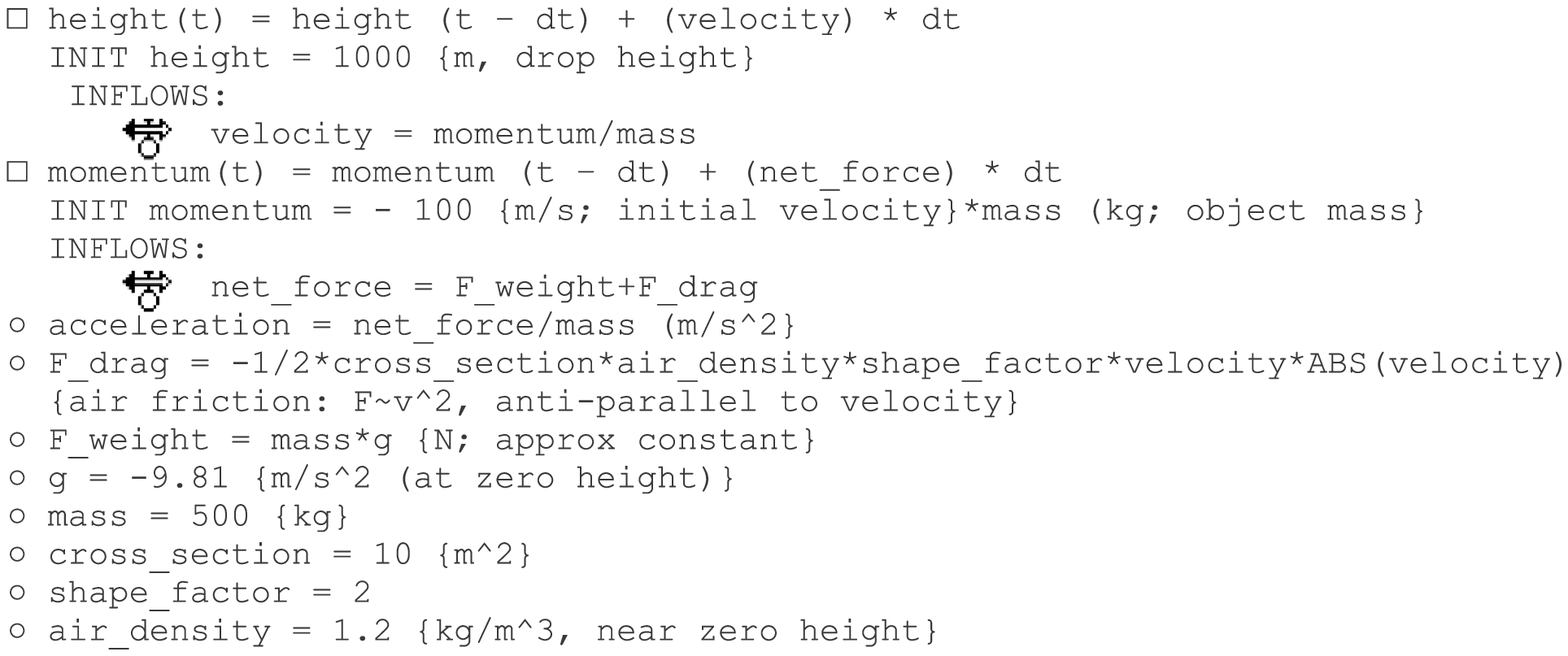}
	\caption{A graphical STELLA program that models the motion of an object falling with turbulent air drag. STELLA programs were connected (upper diagram) using structural elements that represent constants (open circles); functional relationships (arrows, curved or straight); rates of change (valves with arrows) and quantities that accumulate over time (boxes). Once compiled, the input equations became available for modifying initial conditions (lower diagram). This figure was reproduced from Schecker.}\label{fig:stella}
\end{figure}

Students who learned to use STELLA to model physical systems were not burdened by the additional cognitive load of learning a programming language \cite{niedderer1991role}. 
STELLA's low-level of sophistication and intuitive graphical programming environment made it accessible to high school students. In the late 1980's, Niedderer and Schecker taught a class of eleventh grade German physics students with no programming experience to construct STELLA programs that modeled kinematics and dynamics problems as well as work-energy problems. The programs were constructed in small groups or as part of a large class activities. 

\begin{figure}[t]
	\centering
		\includegraphics[width=0.60\linewidth, clip, trim=10mm 0mm 0mm 0mm]{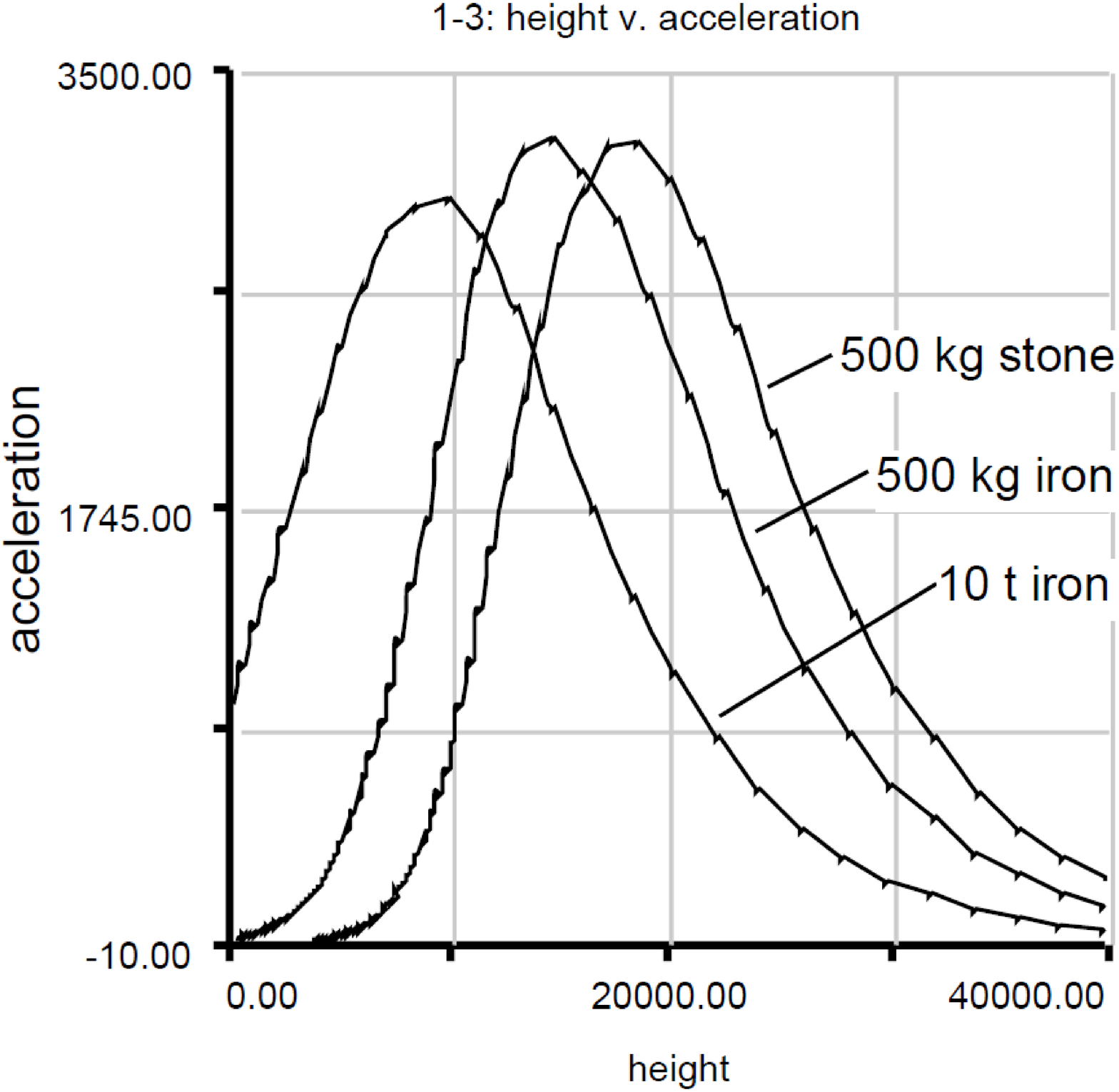}
	\caption{The visual output of the STELLA program (graph of acceleration vs. height fallen) illustrated in Fig. \ref{fig:stella}  is shown for three sets of initial conditions. After the program was compiled the user might input any number of initial scenarios and view output in the form of graphs. Due to hardware and software limitations, visual output was limited to plots and two dimensional animations. This figure was reproduced from Schecker.}\label{fig:stella2}
\end{figure}

A STELLA program that models the motion of a particle falling to the Earth experiencing turbulent air drag is shown in Fig. \ref{fig:stella} (reproduced from \cite{schecker1994system}). Students interacted with STELLA using a graphical programming environment; programs were ``written'' by making connections between various structural elements. These structural elements represented constants like {\tt mass} and {\tt g} (open circles in Fig. \ref{fig:stella}), functional relationships like the product of {\tt mass} and {\tt g} gives {\tt F weight} (arrows, curved or straight in Fig. \ref{fig:stella}), rates of change like {\tt net force} (valves with arrows in Fig. \ref{fig:stella}) and quantities that accumulate over time like {\tt momentum} (boxes in Fig. \ref{fig:stella}). After compiling this program, a set of input equations were made available with which students could define initial conditions and parameter values (lower diagram in Fig. \ref{fig:stella}). Visual output was generated after running the model using these inputs. Output was limited to graphs and some animation (Fig. \ref{fig:stella2}).

Schecker believed that a graphical environment (Fig. \ref{fig:stella}) that emphasized functional relationships, rates of changes and methods of accumulation made the basic structural features of the dynamics more explicit than a laundry list of formulas. By emphasizing mechanism, that is, how state variables change, it was thought that this tool could be used to develop students' conceptual understanding and help students to connect key concepts in physics at a younger age. After two or three introductory examples, Schecker noted that students were able to work with the software or, at least, contribute to class discussion using the model. Furthermore, he found that class discussion shifted from ``back-of-the-book'' problems to open-ended, inquiry-based problems which increased student-student and student-instructor interaction \cite{schecker1994system}. 

\subsubsection{The Berkeley BOXER Project}

\begin{figure}[t]
	\centering
		\includegraphics[width=0.50\linewidth]{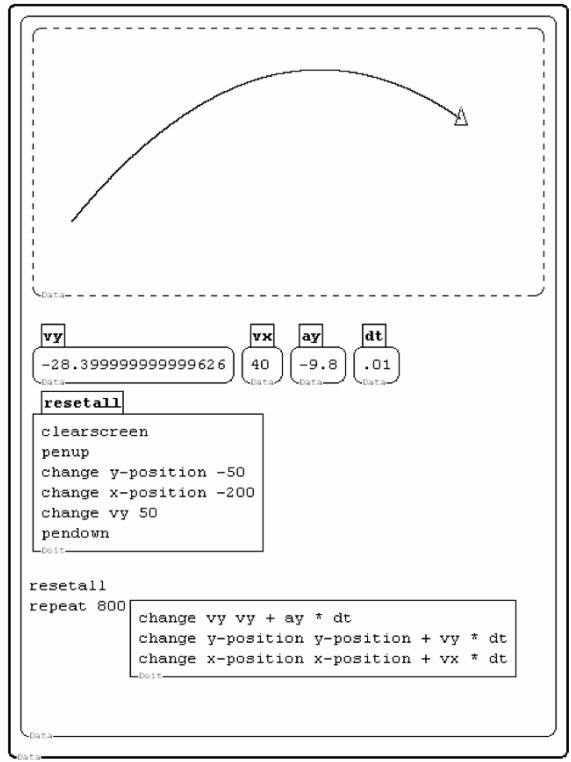}
	\caption{A BOXER program that models the motion of an object under the influence of the gravitational force in two dimensions. In lower half of the figure, the user inputs the initial conditions and parameters (i.e., within the ``Data'' boxes), sets up the model and drawing (i.e., within the upper and lower ``Doit'' boxes respectively. The visualization is a two dimensional drawing in the upper half of the figure (i.e., in the upper ``Data'' box). This figure was reproduced from DiSessa \etal.}\label{fig:boxer}
\end{figure}

The Berkeley BOXER project (BOXER), an outgrowth of the widely used Logo language, was designed to make programming more accessible. Developers claimed that BOXER was not a programming environment but a ``reconstructable medium'' \cite{disessa1986boxer}. The BOXER group experimented in the early 1990's with using BOXER to teach introductory mechanics to eight private high school physics students with no programming experience \cite{sherin1993dynaturtle}. Over a ten week period, students learned the syntax and structure of the BOXER language, programmed some of their own simple models and, eventually, modeled the motion of thrust-driven space craft. The space craft model was developed through group discussion using a program template which was filled in on the blackboard. After the program was completed to the students' satisfaction, the instructor typed the program into the computer exactly as it appeared on the blackboard. Students then experimented with the model. 

A BOXER program that models the motion of an object under the influence of the gravitational force in two dimensions is shown in Fig. \ref{fig:boxer}. BOXER crossed the line between text-based and graphical programming environments. While users wrote procedures and subroutines line-by-line, much of BOXER's organizational structure was graphical. The upper half of this BOXER program (``Data box'') displays a graph of the x-y position of the object in ``virtual'' time. Additional data boxes contain the parameters (e.g., $a_y$) and initial conditions (e.g., $v_y$). Two ``Doit'' boxes control the drawing of the graph and the numerical integration routine. The absolute position of these boxes has no bearing on the program. If subroutines were necessary, then the relative position of boxes (i.e., which box is contained by which other box) becomes important \cite{disessa1991overview}.

The use of BOXER in introductory physics facilitated student inquiry of physics principles. Furthermore, students were able to successfully confront and contextualize abstractions and reason through issues related to precision to build a working program. DiSessa \etal noted challenges with this approach that included the overhead of learning BOXER syntax and students' manipulation of programming statements without understanding. 





\subsubsection{Physics Education Technology Simulations}

\begin{figure}[t]
	\centering
		\includegraphics[width=0.60\linewidth]{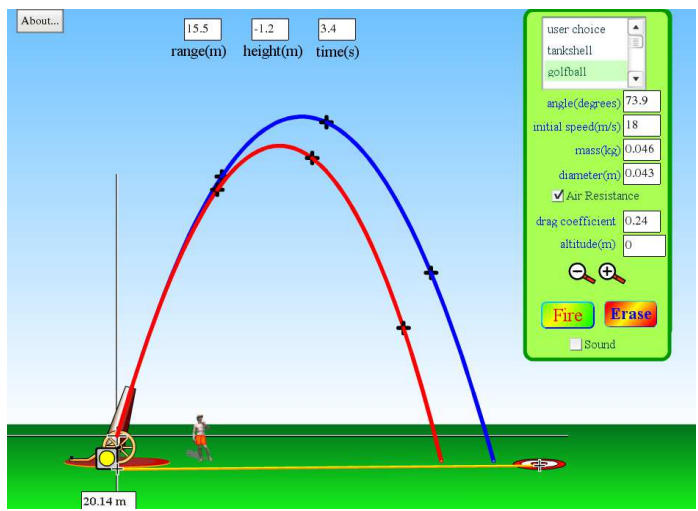}
	\caption{[Color] - A PhET that models the motion of an object under the influence of the gravitational force, which could be subjected to turbulent air drag, in two dimensions. Users may select from a variety of projectiles and set a number of parameters including the projectile's mass and initial speed, the launch angle and the drag coefficient. This simulation is highly visual, including a mobile target and tape measure, but students have no access to the underlying model.}\label{fig:phet}
\end{figure}

The University of Colorado at Boulder's Physics Education Technology (PhET) group has developed a suite of JAVA and Adobe Flash simulations which cover a wide variety of physical phenomenon and are used at all levels of physics and physical science \cite{phetwebsite}. Each PhET was carefully studied in focus groups and through student interviews to enhance functionality and maximize student engagement \cite{perkins2006phet,wieman2006powerful}. PhETs can be used to promote engaged exploration into science in ways similar to professional scientists \cite{podolefsky2010factors}. 

Fig. \ref{fig:phet} shows a PhET that models the motion of an object near the surface of the Earth with (red curve) or without (blue curve) the influence of turbulent ($F \sim v^2$) air drag. Students may select from a variety of projectiles, input parameters, and initial conditions. The simulation includes a target which students may try to strike with the projectile and a tape measure to determine the range of the projectile. This simulation is a highly visual, research tested tool, but students have no access to the underlying model. In fact, a description of the PhET's model of air resistance was not indicated in the PhET; it appeared in supporting documentation.

Because PhETs are a closed computational environment, they do not provide the type of educational support that is possible with open computational environments; see Redish \etal \cite{redish1993student} or Schecker \cite{schecker1993learning,schecker1994system}. PhETs, like other closed environments, also fail to take advantage of the useful demands that a programming language imposes, namely, students' precision with the relationships between variables and their contextualization of abstractions like physical equations; see DiSessa \etal \cite{sherin1993dynaturtle}. However, PhETs are accessible to students at nearly any all levels of instruction, require next to no training to operate and, used appropriately, have been shown to be effective at promoting students' conceptual understanding \cite{finkelstein2006high}. 

\subsubsection{Easy Java Simulations}

Esquembre developed Easy Java Simulations (EJS) to provide a platform for creating JAVA simulations of physical phenomena \cite{esquembre2004easy}. EJS was intended for programmers and novices alike to easily prototype, test and distribute their own simulations \cite{christian2007modeling}. Fully constructed EJS programs have been used at a variety of levels including upper-division quantum mechanics \cite{belloni2007osp}. EJS has made authoring high quality simulations so straight-forward that Esquembre and others have proposed teaching upper-divison science majors to construct simulations using EJS \cite{esquembre2007integrate,ejsupper}.

\begin{figure}[t]
	\centering
		\includegraphics[scale=0.4]{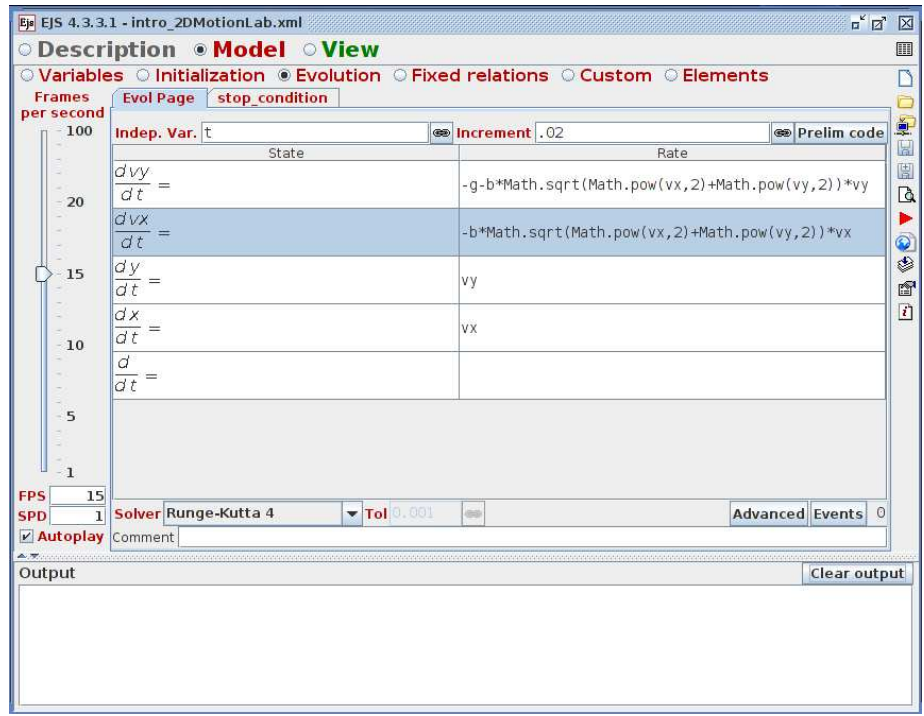}
	\caption{[Color] - The {\it Model} view of an Easy Java Simulation (EJS) that models the motion of an object moving near the surface of the Earth subject to turbulent air drag. The {\it Evolution} tab is shown to highlight how EJS handles modeling the dynamics. Users simply type the ODEs that govern the dynamics into the cells under the {\it Evolution} tab and then select one the integrators. In this case, we have selected 4\textsuperscript{th} order Runge-Kutta. Once compiled, the program produces the output shown in Fig. \ref{fig:ejsoutput}.}\label{fig:ejsmodel}
\end{figure}

\begin{figure}[t]
	\centering
		\includegraphics[scale=0.6]{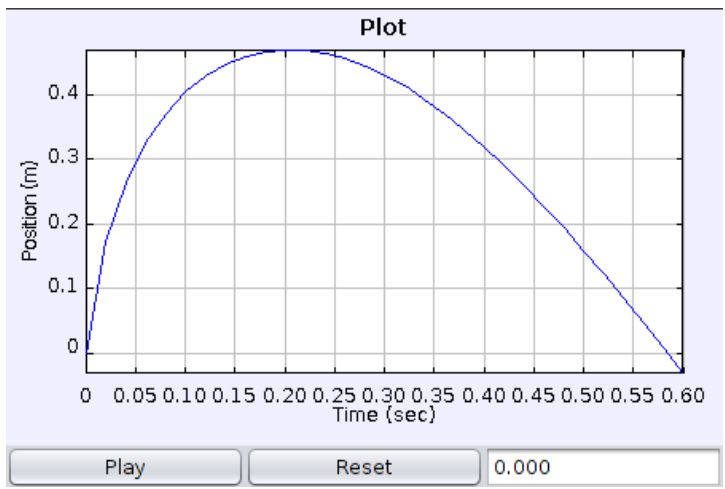}
	\caption{[Color] - The graphical output from the compiled Easy Java Simulation shown in Fig. \ref{fig:ejsmodel}. Here the vertical position of the object is plotted as a function of time. Other plots may be selected in by altering the EJS {\it View} view. In addition to plotting a variety of quantities, it is also possible to use EJS to produce an animation of the system as it evolves.}\label{fig:ejsoutput}
\end{figure}

An EJS program that models the motion of particle near the surface of the Earth with the influence of turbulent air drag appears in Fig. \ref{fig:ejsmodel}. Users have three views in an EJS program. The {\it Description} view contains an {\it HTML} document about the program. The {\it Model} view contains all the necessary variables and functions for modeling the evolution of the system. In Fig. \ref{fig:ejsmodel}, we have highlighted the {\it Evolution} tab within the {\it Model} view because it contains, in a set of cells, the ordinary differential equations (ODEs) that govern the dynamics of the system. Other {\it Evolution} tabs contain the variables, initialized quantities and the relationships between quantities that are not governed directly by the ODEs. The {\it View} view describes the visualization of the EJS program. After simulation is compiled, it can be run and it produces the output shown in Fig. \ref{fig:ejsoutput}.

It is possible to use Easy Java Simulations to engage each stage of the modeling process (Sec. \ref{sec:mandi}) and reap each of the benefits of learning computational modeling (Sec. \ref{sec:comp-thecase}). However, it appears that EJS have been largely used as a closed computational environment to create ``black box'' simulations (similar to PhETs). Students might interact with the program without peaking at the underlying model \cite{christian2007modeling}. However, all the features of the physical model are available (the {\it Model} view in Fig. \ref{fig:ejsmodel}); EJS is an open environment. Usage that would be congruent with previous implementations of open environments with introductory students \cite{redish1993student,schecker1994system,sherin1993dynaturtle}  appears to be limited to upper-division students \cite{esquembre2007integrate,ejsupper}. 




\subsubsection{VPython}

In the mid 1980's, significant efforts were made by Sherwood and co-workers to develop the cT programming language, an outgrowth of the TUTOR language written for the PLATO computer-based education system \cite{sherwood1993ct}. 
Initially, cT was an open computational environment designed for instructors who had little or no programming experience but wanted to construct physics simulations for use in their classes. In 1997, Chabay and Sherwood taught a small class of introductory physics students at Carnegie Mellon University, most of whom had never written a program before, a subset of cT's capabilities.  Students were able to create simulations of physical systems \cite{ctWebsite}. The cT language was simple and intuitive but still lacked three dimensional graphics.

VPython succeeded cT as a simpler open environment based on a professional language in Python. Furthermore, it extended the visual experience by including full three dimensional graphics. VPython has been used to teach computation to large sections of introductory physics students with no background in computer modeling or programming for a number of years \cite{computajp}. It contains features that make writing programs to simulate physical systems straight-forward \cite{vpythonWebsite}.

\begin{figure}
\begin{center}
\includegraphics[width=0.80\linewidth, clip, trim=2mm 0mm 2mm 0mm]{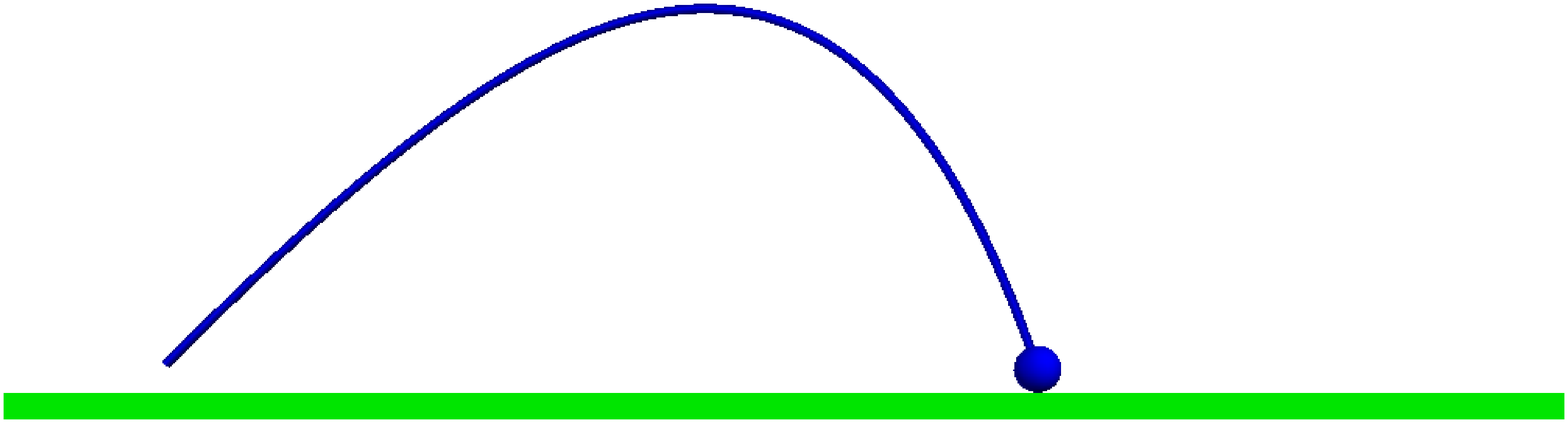}
\includegraphics[width=0.80\linewidth, clip, trim=2mm 25mm 2mm 0mm]{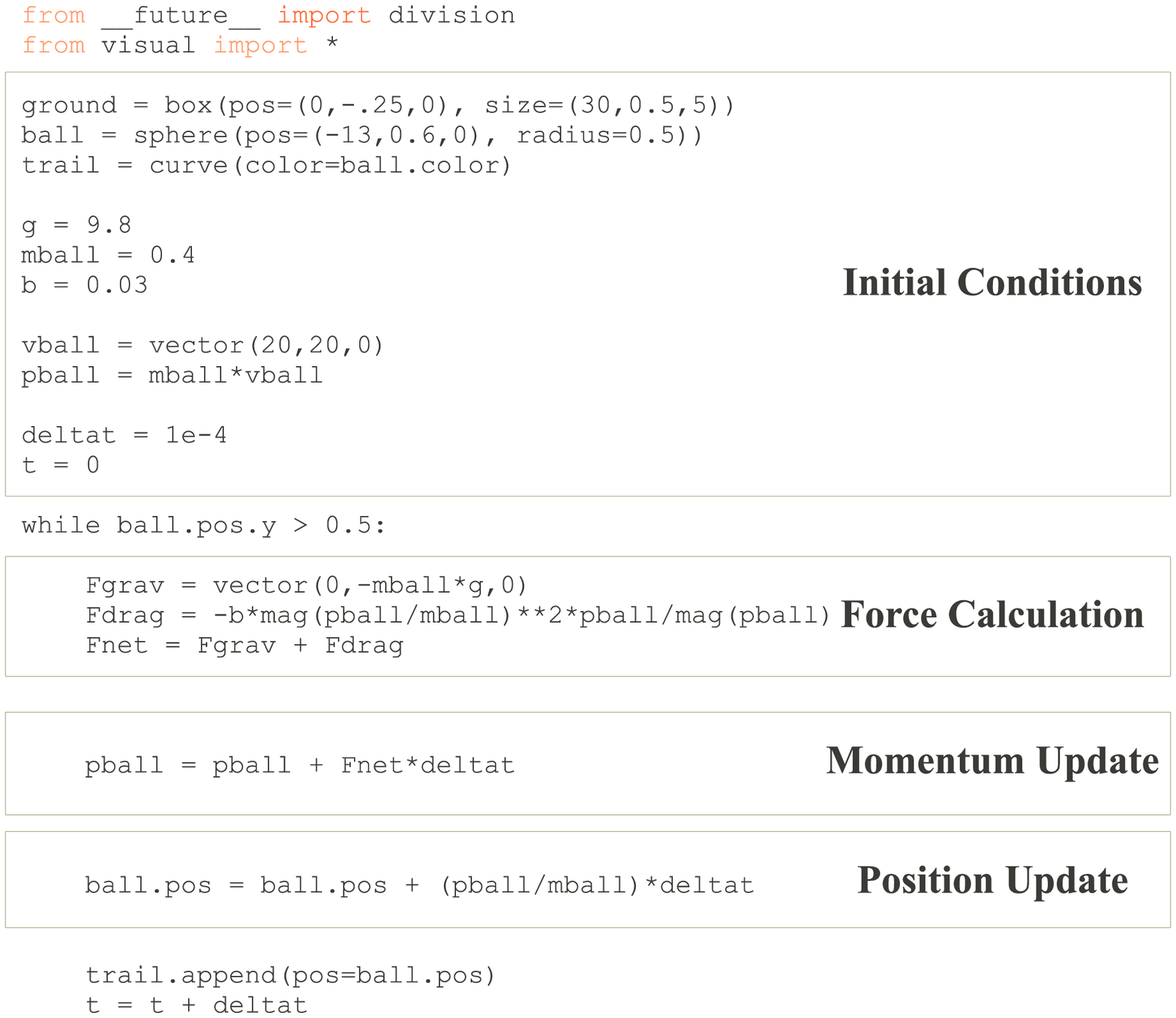}
\end{center}
\caption{[Color] - A VPython program and visualization that models the motion of a ball moving near the surface of the Earth subject to turbulent air drag. Objects and initial conditions are defined outside the calculation (\texttt{while}) loop. Over the course of the semester, students write several VPython programs, but the activities focus on translating the physical model to VPython code using the appropriate syntax and on updating the object's momentum and position.}\label{fig:vpsamplecode}
\end{figure} 

A VPython program that models the motion of a ball moving near the surface of the Earth subject to turbulent air drag is shown in Fig. \ref{fig:vpsamplecode}. The program creates three objects (the ground, a ball and a curve which follows the ball); assigns a few initial conditions and parameters (the ball's position and momentum and the mass of the ball); and integrates numerically. VPython is open computational environment; students write all the program statements necessary to model the physical system (e.g., creating objects, assigning variables and numerical calculations).  The additional details of simulation construction (e.g., drawing graphics, creating windows, mouse interactions) are handled by VPython and are invisible to the students. They can focus on the core details of the physical (and computational) model. The visualization produced by the interpreted VPython code appears at the top of Fig. \ref{fig:vpsamplecode}. VPython supports full three dimensional graphics and animation without the additional burden to students of learning object-oriented programming \cite{scherer2000vpython}.

If the activities are framed appropriately, constructing computational models using VPython engages each stage of the modeling process (Sec. \ref{sec:mandi}) and reaps each of the benefits of learning computational modeling (Sec. \ref{sec:comp-thecase}) . Students can engage in the modeling process by writing programs to explore broad, open-ended problems in a highly visual open environment. Moreover, by learning to use a professional language, it is possible that students will develop transferable skills that could be used in their future work \cite{redish1993student}. We note that it is possible, as with any open computational environment, to use VPython as a closed environment in which students do not access the underlying model. In this case, the added benefits to learning computational modeling are not realized.


With the introduction of changes to courses as significant as learning computation, it becomes important to consider what students think about this new learning this new tool. In particular, it is important to understand what role student epistemology might play in their learning of and success with computation.

%% file: history/epistemology.tex
\section{Student Epistemology}\label{sec:abs}

Students entering university-level physics courses are not ``blank slates'' onto which we can impress what it means to learn and understand physics. 
Their experiences with the natural world shaped their views far before they entered a classroom.
Students tend to have views about physics that are divergent from those that instructors expect \cite{redish1998student,adams2006new}.  
Many view physics as a set of disconnected ideas and a collection of formulas.
In fact, they have a diverse spectrum of views. 
More fluent students might understand, as experts do, that physics is a science with overarching themes and 
interconnected concepts \cite{hammer1994epistemological}. Epistemology is important because the views that students hold affect their success in their science courses \cite{halloun1997views,perkins2005correlating}. These attitudes influence how students prepare for the course \cite{elby2001helping}. 
Student epistemology is a diverse subject with researchers working to understand and measure the role of students' beliefs about knowledge and learning.

Hammer \cite{hammer1994epistemological} proposed a framework aimed at understanding student epistemology in physics. His framework contained three dimensions: {\it Independence},
{\it Coherence} and {\it Concepts}. 
Hammer's {\it Independence} dimension focused on students' beliefs about learning physics -- whether students thought that learning is simply receiving information or that learning involved the active practice of constructing their own understanding.
His {\it Coherence} dimension was aimed at students' beliefs about the structure of knowledge in physics -- whether students considered physics to be a coherent system of knowledge or, naively, believed physics is a collection of facts. 
The {\it Concepts} dimension considered students' beliefs about the content of physics knowledge -- whether students thought content knowledge in physics is simply knowing which formulas to use or, more expertly, that physics knowledge includes knowing how physical concepts underlie these formulas. 

Hammer interviewed a number of students over the course of the semester to refine his framework. 
Using a variety of interview activities, he discovered consistent patterns in each student's beliefs across a number of topics in mechanics. 
Moreover, Hammer found that the students' beliefs (whether novice or expert) suggested whether or not they retained physical misconceptions or conceptual misunderstandings. 
He found that students with more fragmented knowledge, those who felt physics was a collection of facts, were more likely to think that conceptual understanding was unnecessary and that they should not work to create their own understanding \cite{hammer1989two}. 
Hammer's data, though rich, was limited to a small number of students. Several survey instruments have been developed to compare student and expert epistemology in the sciences at much larger scales. 

Halloun's Views About Sciences Survey (VASS) \cite{halloun1997views} provided one of the first assessments of students' views about the nature of science and what it means to learn science. 
VASS is a 33 item survey with a contrasting alternatives design. VASS statements asked students to select how often they perform one action compared to another (i.e., only vs. never, mostly vs. rarely, equally, etc.). 
VASS's complicated design was meant to deal with validity and reliability issues. However, its contrasting alternatives design made completing the survey a difficult task. Furthermore, validity issues were not probed.

Initially, VASS was given to $\sim$300 college students and $\sim$2500 high school students before instruction in their introductory physics courses. VASS classified students by the number of responses that they selected which demonstrated an {\it expert} view. 
The number of {\it expert} responses was used to place students in one of four profiles. These profiles, in decreasing order of VASS performance, were {\it expert}, {\it high transitional}, {\it low transitional}, and {\it folk}.

A majority of both college and high school students achieved low VASS scores and were classified as having {\it folk} or {\it low transitional} views.
Between the college and high school populations, there was no significant difference in the fraction of students that appeared in each of the four profiles.
Because many college students had taken physics in high school, Halloun concluded that instruction had no practical effect on VASS profiles. 
However, within groups (i.e., college and high school), students' VASS scores correlated weakly with level of instruction, course grade, and gain on the Force Concept Inventory \cite{hestenes1992fci}. 
Halloun argued that students with a greater interest in physics, that is, those who took higher (college and high school) level courses and who earned better grades, 
were more likely to achieve higher VASS scores and were classified as having {\it high transitional} or {\it expert} views.

Developed by Redish, Saul and Steinberg at nearly the same time as VASS, the Maryland Physics Expectations (MPEX) survey \cite{redish1998student} explored how students' 
attitudes and beliefs about physics changed as a result of instruction. The MPEX is a 34 item Likert scale (agree-disagree) survey. 
The designers of the MPEX drew upon previous work \cite{hammer1989two} to construct a survey which probed not only student epistemology but students' expectations about the 
physics course as well.

The framework for the MPEX was based Hammer's original work \cite{hammer1994epistemological} but included three additional dimensions: {\it Reality link}, {\it Math link} and {\it Effort}. Redish {\it et al.}'s {\it Reality link} dimension focused on how students perceive that physics is related to the real world -- whether students think 
that physics is related to their experiences outside the classroom or not. Their {\it Math link} dimension considered the role of mathematics in learning physics -- whether 
students think about mathematical formalism as simply a way to compute numbers or if mathematics is a way of representing physical phenomena.
The {\it Effort} dimension explores what kinds of activities students think are necessary to make sense out of physics -- whether they think carefully and evaluate what they 
are doing or not.

The MPEX's statements were carefully crafted to reflect these dimensions and its wording was validated through interviews with students. 
Redish {\it et al.} placed statements into one or more of six subsets (i.e., MPEX categories) that they believed reflected the dimensions the MPEX was meant to probe. 
``Expert'' responses were selected {\it a priori}. These responses were validated through a calibration of the survey. The MPEX was given to several groups of respondents ranging from engineering students to high school teachers to college faculty. 
The average fraction of overall alignment with these {\it a priori} responses (i.e., the percentage of favorable responses) positively correlated with experience. 
Positive correlation with experience was also observed on each of the six subsets of statements.

To measure how students' views changed, the MPEX was given to $\sim$1500 students at six institutions both before and after instruction in introductory physics. 
At all institutions, students' percentage of favorable responses decreased after instruction. 
On each of the MPEX categories, students' responses became less expert-like at all institutions save one. 
This school used an innovative hands-on curriculum \cite{laws1991workshop}, and students' expectations increased on all dimensions except {\it Effort}. 

The largest drop in students' MPEX expectations occurred on statements in the {\it Effort} dimension. 
Redish {\it et al.} suggested that this might stem from a mismatch in the amount of effort students initially expected to 
exert in order learn the material well and the amount they actually put forth to do ``well'' (i.e., earning good marks) in the course. 
They posited that such courses might reward students who had lower expectations, students who preferred memorizing to constructing their own understanding and evaluating their progress. 
Redish {\it et al.} warned against neglecting epistemology in course design; doing so might unnecessarily drive away students who would excel in science 
if given a stronger support structure \cite{redish1998student}.

Adams, Perkins and others developed the Colorado Learning Attitudes about Science Survey (CLASS) \cite{adams2006new} to improve upon VASS and MPEX. 
The CLASS is a 43 item Likert scale (agree-disagree) survey. 
Statements on the CLASS were subjected to interview studies to ensure abstractions (that appeared on VASS and MPEX statements) like ``domain'' and ``concept'' were removed, wording was clarified and students' 
interpretations of statements were predictable. These additional considerations made CLASS suitable for use in lower levels of instruction \cite{otero2008attitudinal} and easily modified to other domains \cite{adams2008modifying}.

Adams {\it et al.} used a pragmatic design approach to determine the underlying framework of CLASS. A reduced-basis principal component analysis informed by {\it a priori} categorization ensured that scores on 
subsets of statements (i.e., categorical percent favorable) were statistically robust \cite{adams2005design}. Eight robust categories emerged: {\it Real World Connection}, {\it Personal Interest}, {\it Sense Making/Effort}, {\it Conceptual Connections}, {\it Applied Conceptual Understanding}, {\it Problem Solving General}, {\it Problem Solving Confidence} and {\it Problem Solving Sophistication}. Students' overall and category scores on CLASS tended to decrease or remain the same after instruction. 
This effect was not surprising given similar results were observed on the MPEX. However, Adams {\it et.al.} noted that students CLASS scores were positively affected by instructors who specifically addressed his or her beliefs about the nature and learning of physics in their classes \cite{adams2006new}.

Demographic factors were found to be important when interpreting results from CLASS. Older students tended to earn higher marks in the {\it Real World Connections} and {\it Personal Interest} categories while younger students did so on the {\it Problem Solving} categories. Women generally responded less expert-like than men to statements in the {\it Real World Connections}, {\it Personal Interest}, {\it Problem Solving Confidence} and {\it Problem Solving Sophistication} categories but responded more expert-like to statements in the {\it Sense Making/Effort} category. An effect of major was clearly evident in the {\it Personal Interest} category \cite{adams2006new}.

Perhaps more notable than demographic influences was the correlation of CLASS scores with conceptual understanding as measured by the Force and Motion Conceptual Evaluation (FMCE) \cite{thornton1998asl}. FMCE scores correlated strongly with overall CLASS performance as well as scores on the {\it Conceptual Understanding} category \cite{perkins2005correlating}.

In lab studies, students' epistemological expertise was found to correlate with academic performance in math and science even after controlling for factors such as socioeconomic status \cite{schommer1990effects,schommer1993epistemological}. In classroom environments, directly addressing students' epistemology has been shown to have an effect on students' performance, specifically, on their conceptual understanding. Elby \cite{elby2001helping} developed a high school physics curriculum that addressed students' attitudes and beliefs through epistemological lessons. Students were asked to confront and resolve differences between their intuition and results from problem solving on their classwork and labs. Class discussion helped students refine their intuition and highlight epistemological insights. Elby also assigned homework questions that asked students to reflect on the nature of knowledge in physics and the acquisition of that knowledge.

Having students expend this extra effort to develop their epistemological beliefs produced large overall and subset gains on the MPEX. Furthermore, students in Elby's classes achieved high gains on the Force Concept Inventory \cite{hestenes1992fci}. According to Elby, these results demonstrated that a curriculum that directly addressed the nature of knowledge helped to develop better learners. However, the additional time devoted to address students' attitudes and beliefs limited the content that could be covered in these courses; Elby had to leave out quite a few topics from the traditional course \cite{elby2001helping}.

Student epistemology appears to play an important role in how students learn physics. When making curricular changes that require students to learn additional tools (i.e., computation), it is important to consider how students think about learning this new tool. An instrument similar to VASS, MPEX, and CLASS has not yet been developed for students who are learning new tools in science courses such as computation. Surveys about computer science \cite{wiebe14computer,heersink2010measuring} are too domain specific and surveys about computer usage \cite{bear1987attitudes, eastman2004elderly} are inappropriate for such purposes. In designing a new epistemological survey about computation in science (Ch. \ref{chap:compass}), much can be gleaned from previous work in the sciences \cite{redish1998student,adams2006new}.

%% file: 03-fci.tex
\chapter{Evidence of curricular effects on performance in mechanics}\label{chap:fci}

This chapter presents the performance by over 5000 students in introductory calculus-based mechanics courses on a standard concept inventory. Results from two courses using different textbooks were compared: a course using a traditional text \cite{knight04} and a reformed course using the Matter \& Interactions (M\&I) mechanics textbook \cite{mandi1}. The effectiveness of the M\&I course is quantified using student performance on the Force Concept Inventory (FCI). Comparative measures find that M\&I students are less prepared than traditional students to solve the types of problems appearing on the FCI even though students of M\&I tend to solve more sophisticated problems on their homework. Further exploration of these results suggest that performance differences were due to an instructional mismatch. Students in the traditional course solved significantly more problems like those that appear on the FCI than M\&I students. This work raises questions about how the context of learning and how the role of practice within that context can improve performance in a particular domain. We comment on how performance improvements, within the framework of the M\&I course, on the types of problems that appear on the FCI might be made. We also discuss the importance of the broader goals related to students' success on novel problems.

\newpage

\input{fci/intro2}

\input{fci/mechatGT}

\input{fci/global2}
\input{fci/item2}

\input{fci/origins2}
\input{fci/closing2}

%% file: fci/intro2.tex
\section{\label{sec:fciintro}Introduction}

Many students taking introductory physics courses fail to acquire effective understanding of concepts, principles, and methods from these courses. 
Rates of failure and withdrawal from these courses are often high and research into this subject has shown that students' misconceptions in physics persist after instruction \cite{hallouna1985csc}.
To address shortcomings in introductory physics courses, researchers have developed and tested modifications to content delivery methods (pedagogy) designed to improve student learning \cite{tutorialswash, eandmtipers, mazurpeer}. 
There have also been efforts to improve student learning by modifying the content of introductory physics courses; one prominent example is Matter and Interactions (M\&I) \cite{mandi1}.
M\&I revises the learning progression of the first semester introductory mechanics course by reorganizing and augmenting the traditional sequence of topics.  
M\&I differs from the typical traditional course where the early emphasis is on kinematics before introducing dynamics;  M\&I places relatively little emphasis on kinematics, as such.   

Because such major changes in introductory physics courses are rare, the impact of such changes on student learning is not well understood. 
Given the differences between the M\&I course and the traditional course, the question arises as to how well M\&I students might fair on standard assessments of mechanics knowledge which have long been used to assess performance in the traditional curriculum.  
The current ``gold standard'' of assessment is the Force Concept Inventory (FCI), a widely used instrument for measuring and comparing performance in introductory physics courses \cite{hestenes1992fci}.  
The FCI has been used to assess, at Georgia Tech, students' understanding of force and motion concepts in M\&I \cite{mandi1} and a traditionally sequenced course \cite{knight04}.

The questions appearing on the FCI probe performance on a subdomain of the mechanics curriculum (force and motion) and do so using multiple-choice conceptual questions. 
In designing the FCI, the authors prepared questions aimed at drawing out common misconceptions and naive notions about the nature of force and motion. 
To review the questions that appear on the FCI, the reader is directed to \cite{hestenes1992fci}. 
The nuances of interpreting student performance on the FCI have been well-documented \cite{huffman1995dfc, hestenes1995ifc, heller1995ifc, steinberg1997pmc, rebello2004eds}.  
Hereafter, we refer to the content of and concepts covered by the FCI as {\it FCI force and motion concepts}. 
We distinguish between FCI force and motion concepts and broader force and motion concepts (e.g., quantitative mechanics problems) presented in both courses. 

The description of our study is presented below as follows: In Sec. \ref{sec:narrative}, we describe the organizational structure of the Georgia Tech mechanics courses.  Sec. \ref{sec:summary} summarizes the results of the in-class testing. In Sec. \ref{sec:item_analysis}, we present an analysis of FCI performance by individual item and concept. Sec. \ref{sec:origins} examines possible reasons for performance differences observed in Secs. \ref{sec:summary} and \ref{sec:item_analysis}.  In Sec. \ref{sec:discussion}, we provide more insight into the performance differences, make concluding remarks, and outline possible future research directions.

%% file: fci/mechatGT.tex
\section{\label{sec:narrative}Introductory Mechanics at Georgia Tech}

The typical introductory mechanics course at Georgia Tech is taught with three one-hour lectures per week in large lecture sections (150 to 250 students per section) and three hours per week in small group (20 student) laboratories and/or recitations. In the traditional (TRAD) course, each student attends a two-hour laboratory and, in a separate room, a one-hour recitation each week. In the M\&I course, each student meets once per week in a single  room for a single three-hour session involving both lab activities (for approximately 2 hours on average) and separate recitation activities (for approximately 1 hour on average). The student population of the mechanics course (both traditional and M\&I) consists of approximately 
85\% engineering majors and 
15\% science (including computer science) majors.

Table \ref{tab:summary_data} summarizes the FCI test results for individual sections. In most traditional (T6-T22) and all M\&I sections, $N_O$ students in each section took the FCI during the last week of class at the completion of the course. In all of the traditional sections and in the majority of M\&I sections (M2-M6), $N_I$ students in each section took the FCI at the beginning of the course during the first week of class. For a given section, $N_I$ is approximately equal to the number of students enrolled in that section. $N_O$ is usually smaller than $N_I$, sometimes substantially so (e.g., T12, T13 and T20). M\&I students took both the pre- and post-test during their required laboratory section. Students of the traditional course typically took the pre-test during the first lecture or lab section. Traditional students were asked to attend an optional section during their evening testing period to take the post-test. Students become busy with other coursework near the end of the semester, hence fewer traditional students attended this optional evening section. 
In each section, only those $N_m$ students who took the FCI both on entering and on completion of the course are considered for the purposes of computing any type of gain (Sec. \ref{sec:summary}). The FCI was administered using the same time limit (30 minutes) for both traditional and M\&I students. M\&I students were given no incentives for taking the FCI; they were asked to take the exam seriously and told that the score on the FCI would not affect their grade in the course. Traditional students taking the FCI were given bonus credit worth up to a maximum of 0.5\% of their final course score, depending in part on their performance on the FCI. This incentive difference between the two courses has no bearing on the performance differences we observe in our data (see Sec. \ref{sec:discussion}).

%% file: fci/global2.tex
\section{\label{sec:summary}Summary of Results from In-class Testing}

\begin{table}
\caption{Georgia Tech FCI test results are shown for twenty-two traditional sections (T1-T22) and six Matter \& Interactions sections (M1-M6). Different lecturers are distinguished by a unique letter in column L. 
The average FCI score $I$\% for $N_I$ students entering the course are indicated.
In those sections where data are available, the average FCI score $O$\% for $N_O$ students completing the course is shown for all sections.
$N_m$ is the number of students in a given section who took the FCI both at the beginning and at the end of their mechanics course.}
\begin{center}
\begin{tabular}{|l||c|c|c||c|c|c|}\hline
{\bf ID} & {\bf L} & {\bf I\%} & {\bf N$_{\mathrm{I}}$} & {\bf O\%} &  {\bf N$_{\mathrm{O}}$} & {\bf N$_{\mathrm{m}}$}\\ \hline
T1 & A & 49.95$\pm$3.05 & 194 & N/A & N/A &  N/A\\\hline
T2 & A & 52.13$\pm$2.80 & 208 & N/A & N/A &  N/A\\\hline
T3 & B & 51.76$\pm$2.88 & 207 & N/A & N/A &  N/A\\\hline
T4 & B & 51.39$\pm$2.91 & 196 &  N/A & N/A &  N/A\\\hline
T5 & C & 46.39$\pm$2.69 & 205 & N/A & N/A &  N/A\\\hline
T6 & D & 45.83$\pm$3.53 & 139 & 70.13$\pm$3.60 & 103 & 97\\\hline
T7 & C & 47.27$\pm$2.86 & 182 & 64.01$\pm$3.05 & 158 & 139\\\hline
T8 & C & 42.03$\pm$2.55 & 194 & 61.26$\pm$3.14 & 140 & 133\\\hline
T9 & A & 52.16$\pm$2.99 & 182 & 73.44$\pm$2.97 & 127 & 122\\\hline
T10 & A & 48.12$\pm$2.72 & 188 & 73.97$\pm$2.92 & 116 & 113\\\hline
T11 & B & 49.82$\pm$2.88 & 182 & 75.35$\pm$3.48 & 104 & 98\\\hline
T12 & B & 49.58$\pm$3.43 & 168 & 72.04$\pm$4.06 & 93 & 88\\\hline
T13 & E & 52.81$\pm$3.25 & 141 & 77.20$\pm$3.38 & 88 & 84\\\hline
T14 & E & 40.36$\pm$2.65 & 183 & 67.33$\pm$3.53 & 140 & 132\\\hline
T15 & F & 46.39$\pm$3.05 & 180 & 69.59$\pm$3.36 & 131 & 120\\\hline
T16 & F & 40.74$\pm$2.84 & 194 & 65.22$\pm$3.60 & 115 & 108\\\hline
T17 & E & 48.02$\pm$3.17 & 160 & 71.82$\pm$3.57 & 121 & 109\\\hline
T18 & A & 50.19$\pm$3.05 & 175 & 74.05$\pm$3.44 & 107 & 105\\\hline
T19 & A & 53.49$\pm$3.37 & 174 & 72.10$\pm$3.52 & 103 & 94\\\hline
T20 & E & 53.36$\pm$3.27 & 143 & 78.52$\pm$3.68 & 97 & 89\\\hline
T21 & B & 49.43$\pm$3.00 & 180 & 75.79$\pm$3.12 & 121 & 115\\\hline
T22 & B & 51.48$\pm$3.09 & 182 & 79.92$\pm$2.81 & 119 & 116\\\hline
M1 & G & N/A & N/A & 35.71$\pm$5.62 & 28 & N/A\\\hline
M2 & H & 54.12$\pm$3.86 & 127 & 64.68$\pm$4.16 & 116 & 111\\\hline
M3 & G & 45.01$\pm$3.11 & 145 & 56.49$\pm$3.38 & 148 & 133\\\hline
M4 & H & 45.57$\pm$3.51 & 143 & 62.27$\pm$3.37 & 141 & 128\\\hline
M5 & I & 45.35$\pm$3.61 & 134 & 62.70$\pm$3.44 & 132 & 110\\\hline
M6 & J & 44.83$\pm$2.50 & 214 & 54.15$\pm$3.06 & 196 & 180\\\hline
\end{tabular}
\end{center}
 \label{tab:summary_data}
\end{table}

\begin{figure}[t]
	\centering
		\includegraphics[width=0.70\linewidth]{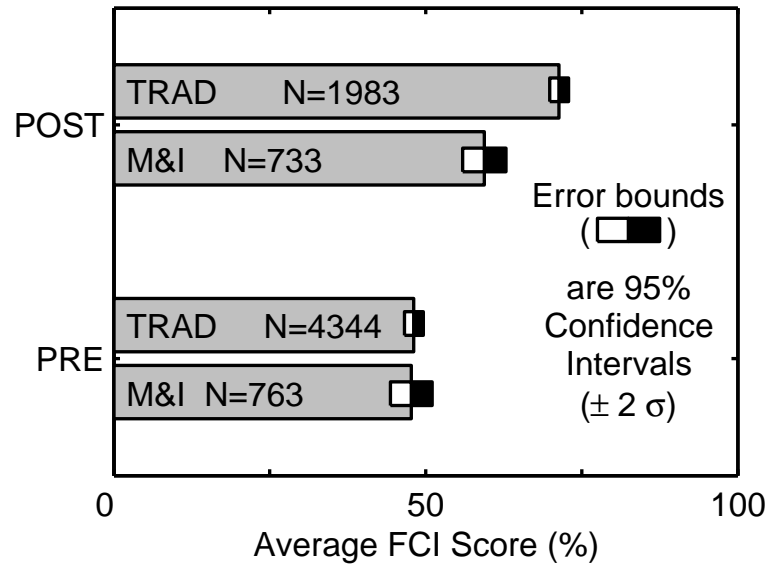}
	\caption{Average pre- and post-instruction FCI scores at Georgia Tech. The average FCI post-test scores are shown for students who have completed a one-semester mechanics course with either the traditional (TRAD) or Matter \& Interactions (M\&I) curriculum. Additionally, the average FCI pre-test score are shown for students before instruction in either the TRAD or M\&I course. The number of students ($N$) tested for each course is indicated in the figure. The error bounds represent the 95\% confidence intervals (estimated from the t-statistic) on the estimate of the average score.}\label{fig:gt_summary}
\end{figure}

\begin{figure}[t]
	\centering
		\includegraphics[width=0.70\linewidth]{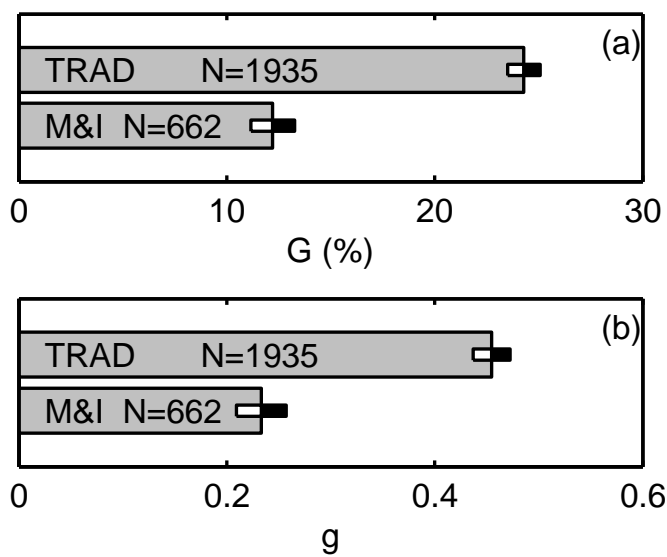}
	\caption{Gain in understanding of mechanics as measured by the FCI. The increase in student understanding
resulting from a one-semester traditional (TRAD) or Matter \& Interactions (M\&I) course is measured using (a) the average raw gain $G$ and (b) the average normalized gain $g$. The average gains in FCI post-test scores are shown for students who have completed a one-semester mechanics course with either the traditional (TRAD) or Matter \& Interactions (M\&I) course. Only students with matched scores were used for this figure (see Table \ref{tab:summary_data}). The error bounds represent the 95\% confidence intervals (estimated from the t-statistic) on the estimate of (a) the raw gain and (b) the normalized gain.}\label{fig:gains}
\end{figure}

\begin{figure}[t]
	\centering
		\includegraphics[width=0.70\linewidth]{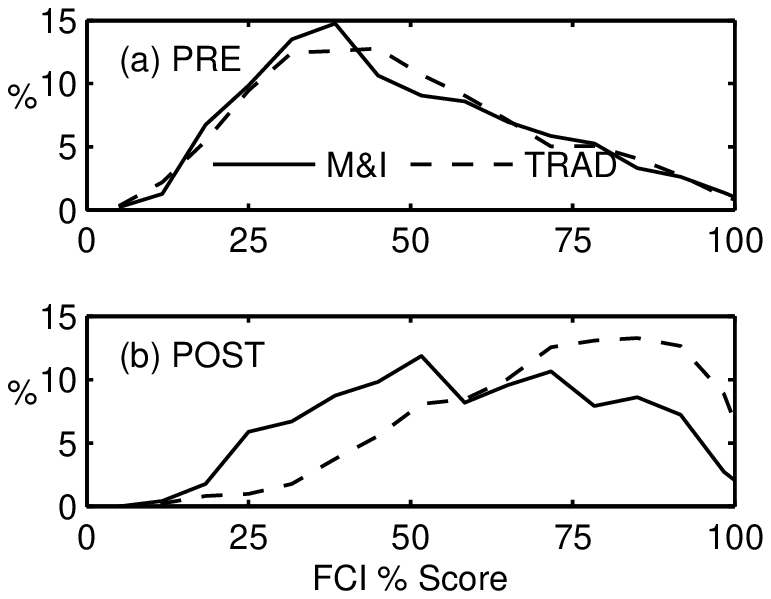}
	\caption{FCI score distributions by course. (a) The distribution of FCI test scores for students before completing a mechanics course with either a traditional (dashed line) or M\&I course (solid line) are shown for data from GT. (b) The percentage of students with a given FCI test score is plotted for students who have completed a mechanics course with either a traditional (dashed line) or M\&I course (solid line) at GT. The total number of students tested in each course is the same as in Fig. \ref{fig:gt_summary}. The plots are constructed from binned data with bin widths equal to approximately 6.7\% of the maximum possible FCI score (100\%). }
\label{fig:gt_prepost}
\end{figure}

The FCI pre-test scores for Matter \& Interactions (M\&I) and traditional students did not differ significantly (mean FCI score, 48.9\% for TRAD vs. 47.4\% for M\&I). 
By contrast, on the FCI  post-test, traditional students significantly outperformed M\&I students (mean FCI score, 71.3\% for TRAD vs. 59.3\% for M\&I). In Fig. \ref{fig:gt_summary}, these mean scores have been reported with 95\% confidence intervals estimated from the t-statistic for each distribution \cite{zhou_biostats}.
A common measure of the change in performance from pre-test to post-test \cite{hake6000} is the average percentage gain, $G = (O - I)*100\%$, where $I$ is the average fractional FCI score for students entering a mechanics course, and $O$ is the average end-of-course fractional FCI score. We also report an average normalized gain $g$, where $g = (O-I)/(1 - I)$, and where $(1 - I)$ represents the maximum possible fractional gain that could be obtained by a class of students with an average incoming fractional FCI score of $I$. For the gains reported in Fig. \ref{fig:gains}, 95\% confidence intervals have been estimated from the t-statistic for the distributions of $G$ and $g$. The data are shown for $N_m$ students (Table \ref{tab:summary_data}).



FCI pre-test score distributions were found to be statistically indistinguishable between the two courses, which is evident from Fig. \ref{fig:gt_prepost}(a). By contrast, distributions of post-test FCI scores were dissimilar; the traditional distribution was shifted towards higher scores (Fig. \ref{fig:gt_prepost}(b)). This is consistent with the finding that the mean score achieved by traditional students were higher than their M\&I peers on the post-test (Fig. \ref{fig:gt_summary}). 
Because the distributions of FCI pre- and post-test scores were non-normal, the similarity of the distributions was compared using a rank-sum test \cite{conover_nonpara,nonparabook}. 

An examination of measures of student performance entering each course suggests that the incoming student population of both courses were identical. 
We obtained and examined students' grade point averages (GPA) upon entering the mechanics course (2.93 for TRAD vs 2.97 for M\&I), SAT Reasoning Test (SAT) scores (1336 for TRAD vs 1339 for M\&I), and the grades earned in the mechanics course (2.47 for TRAD vs 2.46 for M\&I); we found no significant difference in the distributions of any of these metrics using a rank-sum test. 


Mean scores differed between one or more sections within a given course as measured by a Kruskal-Wallis test \cite{nonparabook}. Given this section effect, we compared the three lowest performing traditional sections (T7, T8, \& T16) to the three highest performing M\&I sections (M2, M4, \& M5) to determine if this section effect enhanced the overall observed differences in the normalized gains. Post-test FCI scores were statistically indistinguishable between these subsets (65.7\% for TRAD vs 63.2\% for M\&I) when compared using a rank-sum test. However, traditional students in these sections had significantly lower pre-test FCI scores (43.3\% for TRAD vs 48.2\% for M\&I). Hence, students in these lower performing traditional sections achieved significantly higher normalized gains (0.43 for TRAD vs 0.21 for M\&I). We also compared the FCI post-test scores achieved by the three traditional sections with lowest normalized gains (T14, T18, \& T22) to the M\&I sections with the highest normalized gains (M3, M4, \& M5). Pre-test FCI scores were significantly higher for the M\&I subset (44.0\% for TRAD vs 48.5\% for M\&I) while post-test scores were higher for the traditional subset (66.3\% for TRAD vs 63.7\% for M\&I). Thus normalized gains achieved by traditional students in this subset were higher (0.40 for TRAD vs 0.22 for M\&I). 

%% file: fci/item2.tex
\section{\label{sec:item_analysis}Item Analysis of the FCI}




\begin{figure}[t]
	\centering
		\includegraphics[width=0.70\linewidth, clip, trim = 0mm 15mm 26mm 15mm]{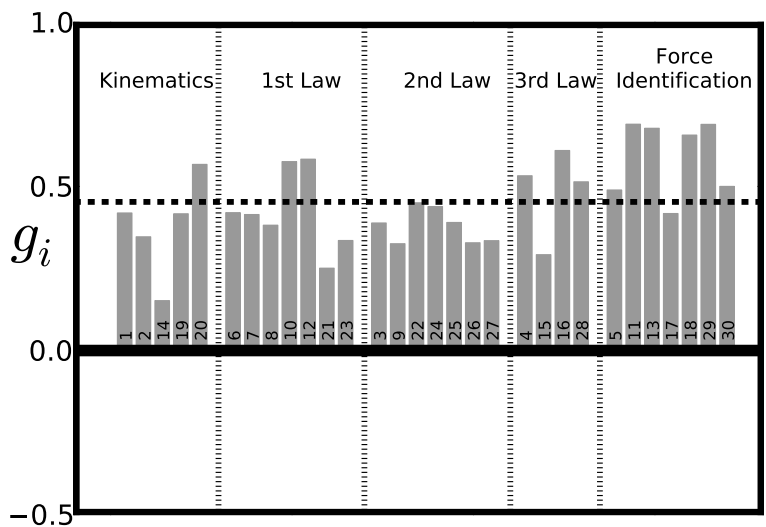}
	\caption{The normalized item gain ($g_i$) achieved by traditional students is shown for each question on the FCI. Positive (negative) $g_i$ indicates better (worse) performance on the post-test. The numerical labels indicate the corresponding question number in order of appearance on the FCI. The items are grouped together into one of five concepts: Kinematics,  Newton's first law, Newton's second law, Newton's third law, and Force Identification. The horizontal line (dash) illustrates the value of $\bar{g}$, the average item gain.}\label{fig:pca-trad}
\end{figure}

\begin{figure}[t]
	\centering
		\includegraphics[width=0.70\linewidth, clip, trim = 0mm 15mm 26mm 15]{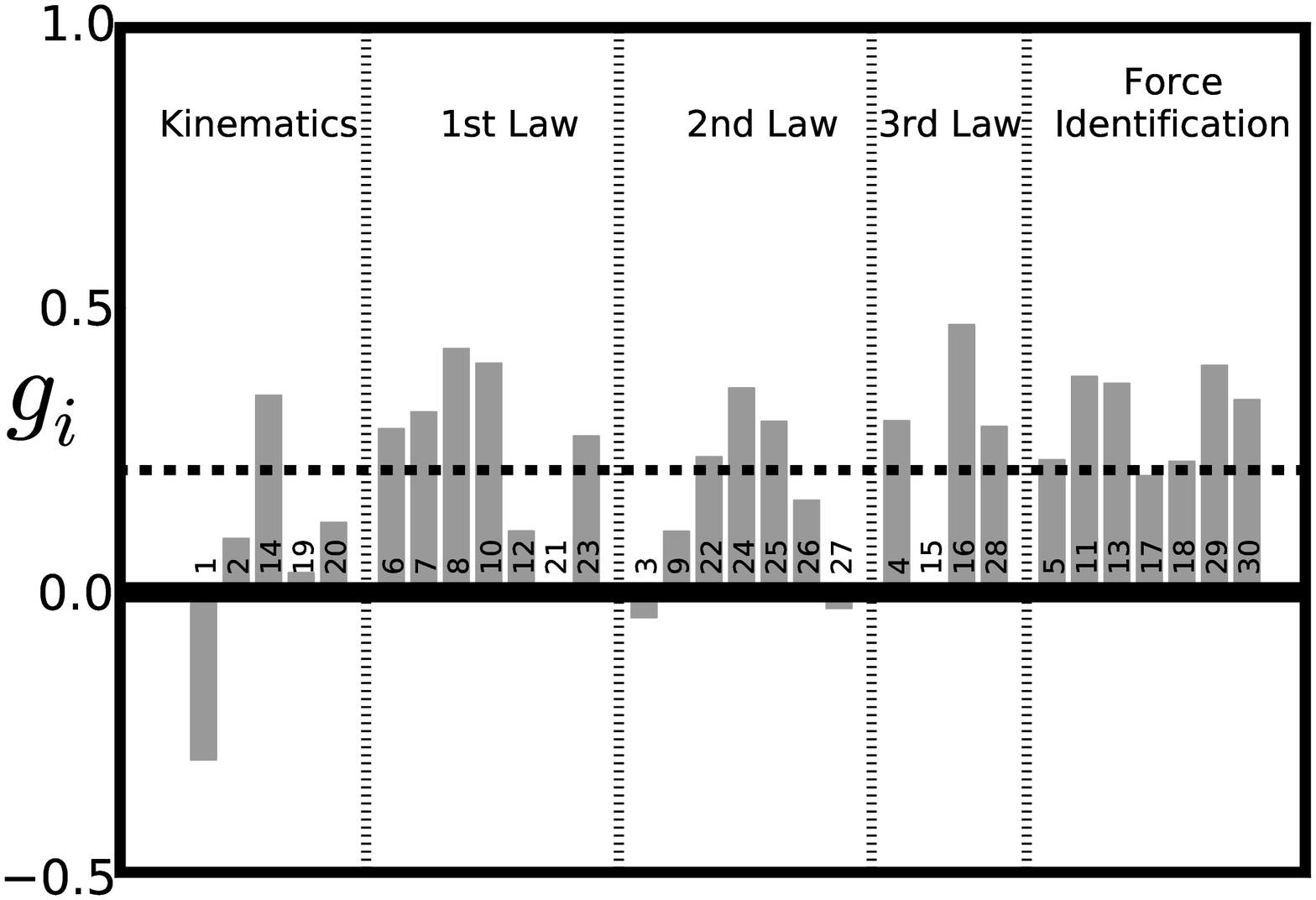}
	\caption{The normalized item gain ($g_i$) achieved by M\&I students is shown for each question on the FCI. Positive (negative) $g_i$ indicates better (worse) performance on the post-test. The numerical labels indicate the corresponding question number in order of appearance on the FCI. The items are grouped together into one of five concepts: Kinematics,  Newton's first law, Newton's second law, Newton's third law, and Force Identification. The horizontal line (dash) illustrates the value of $\bar{g}$, the average item gain.}\label{fig:pca-mi}
\end{figure}


\begin{figure}[t]
	\centering
		\includegraphics[width=0.70\linewidth, clip, trim = 0mm 15mm 26mm 15mm]{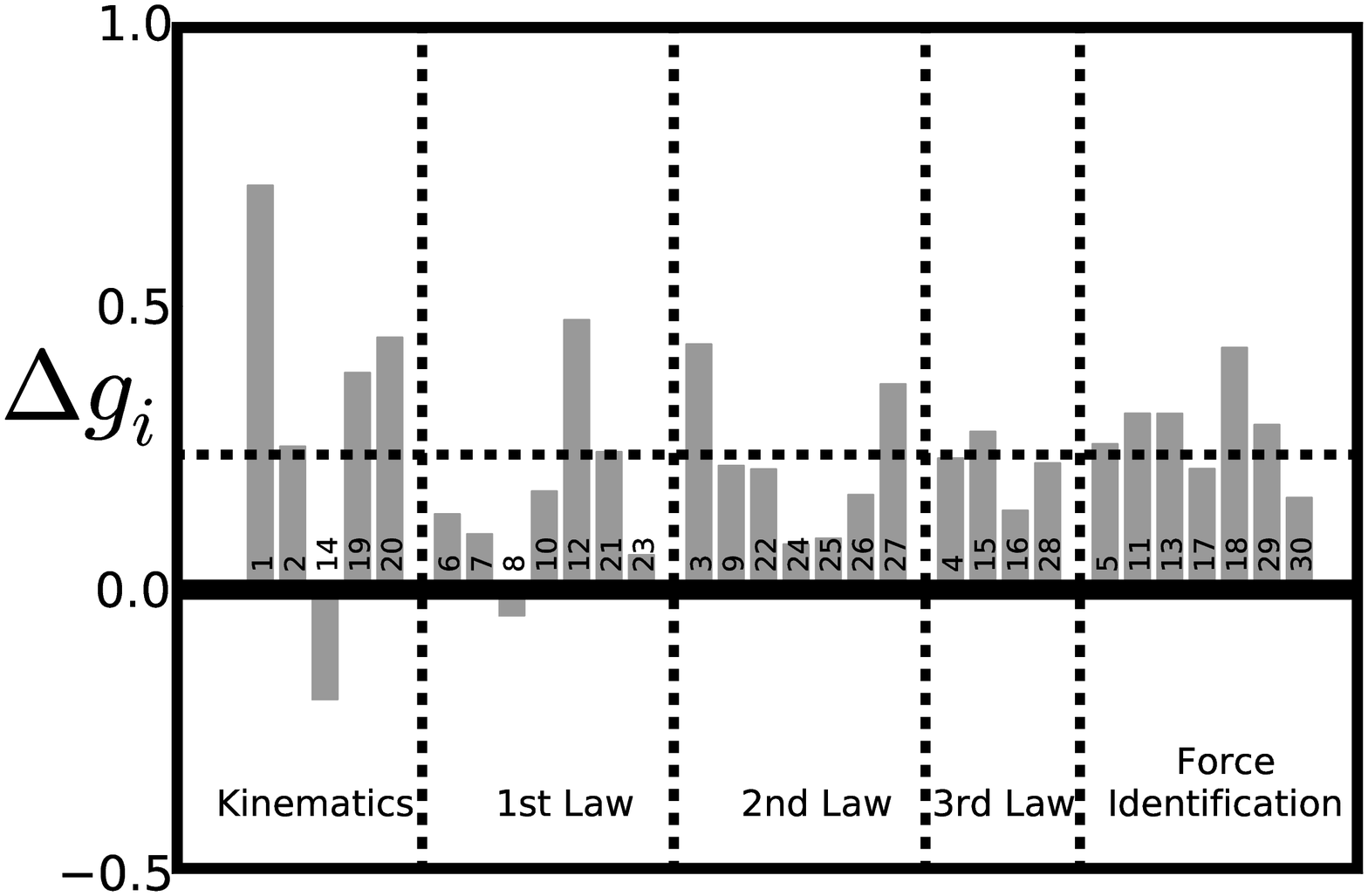}
	\caption{Difference in performance for individual FCI items and mechanics concepts. The difference in performance ${\Delta g_{i}}$ between traditional and M\&I students is shown for each question on the FCI. Positive (negative) ${\Delta g_{i}}$ indicates superior performance by traditional (M\&I) students on individual questions. The numerical labels indicate the corresponding question number in order of appearance on the FCI. The items are grouped together into one of five concepts: Kinematics,  Newton's first law, Newton's second law, Newton's third law, and Force Identification. The horizontal line (dash) illustrates the value of $\overline{\Delta{g}}$ the mean difference in the item gains between courses.}\label{fig:pcanew}
\end{figure}

Student performance on individual questions or groups of questions was used to determine on which FCI force and motion concepts students in the traditional course outperformed M\&I students. Questions on the FCI were sorted into concept categories using Hestenes' original conceptual dimensions \cite{hestenes1992fci}, but we required that each question be placed in only one category. In our work only five concept categories were used: Kinematics, Newton's 1st Law, Newton's 2nd Law, Newton's 3rd Law, and Force Identification. The first four of these categories were identical to Hestenes' dimensions and Force Identification was a renamed category which contained questions from Hestenes' Kinds of Forces dimension. In Figs. \ref{fig:pca-trad} and \ref{fig:pca-mi} the items that comprise each category are listed. Note that this was an \emph{a priori} categorization based on our judgment of the concepts covered by the items; it is not the result of internal correlations or factor analysis based on student data.


We used the normalized gain in performance on a per question basis to quantify item performance. We define an item gain, ${g}_{i} =  (f_{post,i}-f_{pre,i})/(1-f_{pre,i})$ where $f_{pre,i}$ and $f_{post,i}$ are the fraction of students responding correctly to the $i^{th}$ item on the pre- and post-test respectively. This measure normalizes the gain in performance on a single item by the largest possible gain given the students' pre-test performance on that item; ${g}_{i}$ is essentially the Hake gain for a single item. The sign of $g_{i}$ is important; a positive (negative) $g_{i}$ corresponds to an item on which students' performance was better (worse) on the post-test. To discern which questions have large item gains, we  compare $g_{i}$ for each question to the mean item gain, $\bar{g} = \Sigma_i^N (g_i/N)$ where $N$ is the number of items on the FCI.

The plots of $g_i$ for each course (Fig. \ref{fig:pca-trad} for TRAD, Fig. \ref{fig:pca-mi} for M\&I) provide a visual representation on which items and concepts students in each course achieved large gains. Students in the M\&I course had positive gains for 27 of the 30 items. By contrast, traditional students had positive gains for all items. On 18 of the questions, M\&I students achieved item gains higher than $\bar{g} = 0.21$. Traditional students achieved item gains higher than $\bar{g} = 0.45$ on 12 questions. Looking across FCI force and motion concepts, M\&I students achieved their highest item gains (compared to $\bar{g}$) on 1st Law and Force Identification questions, while traditional students did so for 3rd Law and Force Identification questions.

It is important to note that the average item gain, $\bar{g}$, is different from the usual Hake gain, $g$. The Hake gain overweights (underweights) questions where students initially performed worse (better) than their average performance. By contrast, all questions that yield the same relative improvement (regardless of initial performance) are given equal weight in the computation of $\bar{g}$

To illustrate the differences between courses more succinctly, we computed difference in normalized item gains between the two courses. We define the difference in normalized item gain, $\Delta g_{i} = {g}_{i}^{T} - {g}_{i}^{M}$ where ${g}_{i}^{T}$ and ${g}_{i}^{M}$ are the normalized gain for the $i^{th}$ item achieved by traditional and M\&I students respectively. The sign of $\Delta g_{i}$ is important; a positive (negative) $\Delta g_{i}$ corresponds to an item on which the traditional students achieved a higher (lower) gain than M\&I students. We discovered on which questions students' item gains in each course differed the most by comparing the $\Delta g_{i}$ of each item to the mean difference in the item gains between courses, $\overline{\Delta{g}} = \Sigma_i^N (\Delta g_i/N)$.

The plot of ${\Delta g_{i}}$ illustrates better performance by traditional students across all concepts on the FCI (Fig. \ref{fig:pcanew}). We observed that $\Delta{g_{i}}$ is positive for almost all questions, and 45\% of the questions had values of $\Delta{g_{i}}$ greater than $\overline{\Delta g} = 0.238$. The grouping of the FCI questions by category permits one to visualize which concepts contributed most 
strongly to the difference in performance. For example, the difference in performance on the Force Identification concept was striking, where 5 of the 7 questions in this category had $\Delta g_i > \overline{\Delta g}$.

The grouping of the questions by concept helps one to determine on which concepts differences in item gains were greatest. We computed the difference in the average concept gain, $\overline{\Delta {g}_c} = \sum_{i \epsilon c} (\Delta g_{i}/N_c)$ where $N_t$ is the number of items covering concept $c$. Concepts with higher $\overline{\Delta {g}_c}$ were those on which traditional students achieved higher normalized gains than M\&I students. The Kinematics and Force Identification concepts had the highest values of $\overline{\Delta {g}_c}$ (shown in Table \ref{tab:deltag}). By contrast, we found $\overline{\Delta {g}_c}$ for Newton's 1st Law which was well below $\overline{\Delta {g}}$. The remaining two concepts had values of $\overline{\Delta {g}_c}$ slightly below $\overline{\Delta {g}}$.

\begin{table}[t]
\caption{The average difference in item gains between courses are computed for the items in each FCI force and motion concept, $\overline{\Delta {g}_c}$. Each $\overline{\Delta {g}_c}$ is positive, indicating better average item gains for traditional students across all FCI force and motion concepts. Concepts with higher $\overline{\Delta {g}_c}$ are those for which traditional students achieve higher normalized gains than M\&I students. Traditional students achieve the highest values of $\overline{\Delta {g}_c}$ on the Kinematics and Force Identification concepts and lowest on Newton's 1st Law concept.}
\begin{center}
\begin{tabular}{|l|c|}\hline
{\bf FCI force and motion concepts \hspace*{50pt}} & {${\overline{\Delta {g}_c}}$}\\\hline
Kinematics & 0.32\\ 
Newton's 1st Law & 0.16\\
Newton's 2nd Law & 0.22\\
Newton's 3rd Law & 0.22\\
Force Identification & 0.28\\\hline
\end{tabular}
\end{center}
\label{tab:deltag}
\end{table}

%% file: fci/origins2.tex
\section{\label{sec:origins}Origins of the Performance Differences}



We turn now to the examination of factors that might lead to higher FCI post-test scores by traditional students, including grade incentives, differences in pedagogy, and differences in instruction (e.g., homework and lecture topics).  




The incentive given to traditional students to take the FCI was too small to account for the marked differences in performance indicated in Figs. \ref{fig:gt_summary}, \ref{fig:gains}, and \ref{fig:gt_prepost}(b).  As mentioned earlier (Sec. \ref{sec:narrative}), traditional students were provided with an incentive to take the FCI while M\&I students received no incentive. In principle, sufficiently large incentives can impact FCI outcomes \cite{ding4etc}.  To check for this incentive effect, we offered similar incentives (i.e., a maximum of 0.5\% bonus to overall course grade) to both traditional and M\&I students who took the FCI post-test at Georgia Tech in the Fall of 2009. During this term, we found the performance differences for M\&I and traditional students were similar to those reported in this thesis.
FCI data from Fall 2009 was not included in this paper because instructional changes had been made to the M\&I course; M\&I sections M1-M5 had similar homework exercises, lectures, and laboratories.



The performance differences cannot be attributed to differences in pedagogy. It is well-known that using interactive engagement (i.e., ``clicker'' questions, ConceptTests, Peer Instruction, etc.) can improve students' conceptual understanding in introductory and advanced courses \cite{hake6000,crouch2001peer,pollock-use}. However, all sections (both traditional and M\&I) were largely indistinguishable with respect to interactive engagement: all sections used similar methods (``clicker'' questions) with similar intensity (3-6 ``clicker'' questions per lecture period). 






We examined whether differences in coursework (homework) could be connected to performance differences on the FCI. We categorized the 575 traditional homework questions and the 756 M\&I homework questions. Questions were placed into one or more categories depending on the topical nature of the problem and the principles needed to answer the question. Categories included the five FCI force and motion concepts discussed in Sec. \ref{sec:item_analysis} as well as several other concepts which do not appear on the FCI (e.g., Angular Momentum). The {\it Kinematics category} included questions about the relationships between position, velocity, and acceleration that did not refer to the underlying dynamical interactions that cause changes in these quantities. Questions in the {\it Newton's 1st Law category} included qualitative questions which discussed the direction of motion and its relationship to applied forces. The {\it Newton's 2nd Law category} included questions with a heavy emphasis on contact forces and resolving unknown forces, but excluded open-ended questions in which the prediction of future motion is the goal (e.g., using iterative methods to predict the motion of an object). Questions in the {\it Newton's 3rd Law category} included questions in which Newton's 3rd law was treated as an isolated law, that is, where there was no reference to the underlying reciprocity of long range electric interactions which causes it. Generally, it was applied to contact forces and gravitational interactions. The {\it Force Identification category} included questions in which the direction and relative strength of forces acting on a body or set of bodies were represented by diagrams (i.e., force-body diagram).  
The aforementioned categories represent those concepts that are covered extensively in the first half of a traditional physics course and were heavily represented on the FCI. 

The difference in the relative fraction of homework questions covering FCI force and motion concepts between the courses (Table \ref{tab:hwcat}) correlated with the overall performance differences observed in Figs. \ref{fig:gt_summary} - \ref{fig:gt_prepost}. Furthermore, the differences in the relative fractions of homework questions corresponding to individual FCI concepts were consistent with the results from our item analysis (Fig. \ref{fig:pcanew}). The relative fraction of homework questions was computed by first categorizing questions, then counting the number of questions covering the concepts of interest and dividing by the total number of homework questions given in a course. The relative fraction of homework questions covering FCI force and motion concepts differed by more than a factor of 2 in favor of the traditional course (0.57 for TRAD vs 0.26 for M\&I). On individual FCI concepts, we found a lower relative fraction of homework questions in the M\&I course compared to the traditional course on four of the five concepts: Kinematics (0.26 for TRAD vs 0.10 for M\&I), Newton's 2nd Law (0.25 for TRAD vs 0.15 for M\&I), Newton's 3rd Law (0.04 for TRAD vs $<$0.01 for M\&I), and Force Identification (0.11 for TRAD vs 0.01 for M\&I). On most FCI questions about these concepts traditional students outperformed M\&I students (Sec. \ref{sec:item_analysis} \& Fig. \ref{fig:pcanew}). We found that the relative fraction of Newton's 1st Law questions were similar ($<$0.01 for both). This signature was also observed in our item analysis (Fig. \ref{fig:pcanew}); the Newton's 1st Law FCI concept had the smallest $\overline{\Delta {g}_c}$ (Sec.\ref{sec:item_analysis}).

\begin{table}[t]
\caption{An estimate of the fraction of homework questions covering a particular FCI concept in the two mechanics courses is compared. Subtopics for these homework questions were not mutually exclusive. The relative fraction of homework questions covering FCI force and motion concepts and some individual FCI concepts (i.e., Kinematics, Newton's 2nd Law, Newton's 3rd Law, and Force Identification) is greater in the traditional course. This is consistent traditional students' superior overall performance (Figs. \ref{fig:gt_summary}, \ref{fig:gains}, \ref{fig:gt_prepost}) and their better performance on particular FCI concepts (Fig. \ref{fig:pcanew}). }
\begin{center}
\begin{tabular}{|l|c|c|}\hline
{\bf Est. Fraction of HW Questions \hfill} & {\bf M\&I} & {\bf TRAD}\\\hline
FCI force and motion concepts & 0.26 & 0.57 \\\hline
\multicolumn{3}{|l|}{\bf HW Subtopics (not exclusive)} \\\hline
Kinematics & 0.10 & 0.26\\ 
Newton's 1st Law & $<$0.01 & $<$0.01\\
Newton's 2nd Law & 0.15 & 0.25\\
Newton's 3rd Law & $<$0.01 & 0.04\\
Force Identification & 0.01 & 0.11\\\hline
\end{tabular}
\end{center}
\label{tab:hwcat}
\end{table}

The difference in the relative fraction of force and motion lectures/readings between the courses (Table \ref{tab:readcat}) was consistent with the overall performance differences observed in Figs. \ref{fig:gt_summary}, \ref{fig:gains}, \& \ref{fig:gt_prepost}(b). However, the differences in the relative fractions of lectures/readings corresponding to individual FCI concepts did not completely correlate with the results from our item analysis (Fig. \ref{fig:pcanew}). Lecture and reading topics were examined and categorized for each course using the same categories as our homework question analysis.
The relative fraction of lectures/readings which cover FCI force and motion concepts was greater by nearly a factor of 2 for the traditional course (0.44 for TRAD vs 0.26 for M\&I). This result is consistent with the difference in the relative fraction of homework questions (Table \ref{tab:hwcat}). However, the differences in the relative fraction of lectures/readings which cover individual FCI concepts were mixed. The relative fractions for three of five concepts were greater for the traditional course: Kinematics (0.21 for TRAD vs 0.07 for M\&I), Newton's 3rd Law (0.03 for TRAD vs 0.01 for M\&I), and Force Identification (0.11 for TRAD vs 0.06 for M\&I). But on two concepts, the relative fractions of lectures/readings were roughly similar: Newton's 1st Law (0.01 for TRAD vs 0.02 for M\&I) and Newton's 2nd Law (0.09 for TRAD vs 0.08 for M\&I).

\begin{table}[t]
\caption{An estimate of the fraction of lecture/reading topics in the two mechanics courses is compared. Subtopics for these lectures/readings were not mutually exclusive. The relative fraction of lectures/readings in the traditional course is greater for the Kinematics, Newton's 3rd Law, and Force Identification topics which is consistent with their superior performance in those concepts on the FCI. However, on Newton's 1st and 2nd Laws, the relative fraction of lectures/readings are roughly similar.}
\begin{center}
\begin{tabular}{|l|c|c|}\hline
{\bf Est. Fraction of Lecture Topics \hfill} & {\bf M\&I} & {\bf TRAD}\\\hline
FCI force and motion concepts & 0.26 & 0.44 \\\hline
\multicolumn{3}{|l|}{\bf Lecture Subtopics (not exclusive)} \\\hline
Kinematics & 0.07 & 0.21\\ 
Newton's 1st Law & 0.02 & 0.01\\
Newton's 2nd Law & 0.09 & 0.08\\
Newton's 3rd Law & 0.01 & 0.03\\
Force Identification & 0.06 & 0.11\\\hline
\end{tabular}
\end{center}
\label{tab:readcat}
\end{table}

%% file: fci/closing2.tex
\section{\label{sec:discussion}Closing Remarks}

We have found that students who completed an introductory mechanics course which employs the Matter \& Interactions course earned lower post-test FCI scores than students who took a traditionally sequenced course. The differences in performance were significant and were supported by the number of students involved in the measurement. We demonstrated that these differences cannot be explained by differences in the incoming population of students between the courses (i.e., SAT scores, GPA, etc.). The overall performance differences between the courses on the post-test correlated with instruction within each course. The relative fraction of FCI force and motion concepts that appeared on students' homework and in their lectures was roughly twice as large for the traditional course (Tables \ref{tab:hwcat} \& \ref{tab:readcat}). We observed this signature in the differences of the means and distributions of FCI scores (Fig. \ref{fig:gt_summary}, \ref{fig:gains}, \& \ref{fig:gt_prepost}(b)) as well as the average item gain, $\bar{g}$. The average item gain for traditional students was roughly twice as large when compared to M\&I students (Sec. \ref{sec:item_analysis}).

Furthermore, we found that traditional students outperformed M\&I students across all subtopics on the FCI (Fig. \ref{fig:pcanew}) and that these differences correlated with instruction on individual FCI force and motion concepts that appeared on students' homework (Table \ref{tab:hwcat}). In terms of decreasing average topical gain, $\overline{\Delta g_c}$, students of the traditional course outperformed M\&I students on: Kinematics, Force Identification, Newton's 2nd Law, Newton's 3rd Law, and Newton's 1st Law (Sec. \ref{sec:item_analysis}). The difference in the relative fraction of homework questions covering FCI force and motion concepts followed a similar order with Kinematics and Force Identification having the largest difference, Newton's 2nd and 3rd Laws next, and Newton's 1st Law last. Each course had, roughly, the same relative fraction of homework questions covering Newton's 1st Law.

The relatively poor performance of M\&I students on the FCI might appear surprising given the sophistication of some of the mechanics problems addressed in the M\&I course, for example, planetary motion, ball-and-spring models of solids, multi-particle systems, etc. From a physicist's perspective, M\&I students should be able to successfully solve the sorts of problems appearing on the FCI; yet, apparently they were unable to extend (i.e., transfer) what they had learned, for example, in the context of the momentum principle, to questions on the FCI.  Two inter-related factors are operating here: first, the context of learning; and second, the role of practice within that context. In general, students, especially at the introductory level in physics, are sufficiently challenged to learn what they have to learn and tend not be very successful in generalizing their skills to novel situations with which they have had little practice \cite{gick1983schema,ross1990generalizing}. 

We believe that the differences in instruction, how much and how long students learn about particular mechanics concepts, had a direct effect on their performance on the FCI. The relative fraction of homework questions and lecture topics covering FCI force and motion concepts provides a connection to the time students' devoted to learning particular concepts and the depth to which concepts are covered in their respective courses (i.e., time-on-task). It is well-accepted that increased time-on-task will generally improve learning gains on the topics for which more time is devoted \cite{dweck1986motivational,fisher1998differential}. While an accurate measure of student time-on-task requires interviewing individual students, our results suggest that students of the traditional course devoted more time to learning FCI force and motion concepts than students of M\&I. 

As we have shown, traditional students had much greater practice in the sorts of problems the FCI presents and their relative performance shows the importance of that practice. It is possible that additional exposure to FCI force and motion concepts would improve M\&I students' performance on the FCI. However, making changes to the course in this manner requires instructors to reflect on the learning goals for their course. The M\&I course was not designed to improve performance on the FCI. As mentioned previously, the M\&I course includes significant changes to the content of the introductory course, not just pedagogy, and the goals of its content might not align with those of the traditional course. 
For example, traditional students work extensively with kinematics equations and constant force motion -- a staple of the FCI. Students of M\&I learn about motion from a dynamical perspective, that is, they use the net force to derive the equations of motion or predict the motion iteratively. 
The amount of time in a semester is finite and including additional practice on FCI force and motion concepts might require the instructor to leave out other M\&I topics (e.g., elementary statistical mechanics) and/or tools (i.e., computation). These changes would be informed by a single measurement of a subset of mechanics concepts and problem types.


One might not want to use the results from an FCI post-test as the sole measurement to inform where improvements to a physics course should be made. We have recently completed a think-aloud study which demonstrates better performance by M\&I students compared to traditional students when solving mechanics problems informed by the M\&I course \cite{bujak2011fci}. 
Traditional students in this study were found to be unable to express the reasoning behind their correct responses to the FCI.
M\&I students worked from fundamental principles to solve these problems.
While students might always perform best on problems similar to ones they studied in class or solved for homework, the goal is to help them achieve a good level of success on novel problems. To compare students' problem solving abilities between courses comprehensively additional metrics are needed, including measures of performance on other topics in mechanics (e.g., energy and angular momentum), complex problems, and non-traditional problems.  The net sum of all these measurements would provide a more complete picture of the nuanced differences between these two courses. These detailed comparisons would inform where improvements to both courses could be made to help introductory physics students to become flexible problem solvers.

%% file: 04-comp.tex
\chapter{Implementing and Assessing Computation in Introductory Mechanics}\label{chap:comp}

In this chapter, we present the design and implementation of computational homework problems for students taking the reformed introductory mechanics course \cite{mandi1}. 
These problems were designed to engage students in the modeling process by exploring the generality and utility of certain physical principles \cite{redish1993student,schecker1994system},  provide a platform for students to contextualize problems into novel tasks \cite{sherin1993dynaturtle} and develop students' abilities to use a new problem solving tool in computation \cite{computajp}. 
Over three different semesters, nearly 1400 students taking this course solved this suite of problems. 
Their proficiency was evaluated in a proctored environment using a computational problem which they had not solved before, a novel problem. 
The majority of students (60.4\%) successfully completed the evaluation. 
Analysis of erroneous student-submitted programs indicated that a small set of student errors explained why most programs failed. 
Errors indicated that students would benefit from additional exposure to computation that focused on qualitative analysis rather than rigorous training. 
This work raises questions about instructional design, knowledge transfer and student epistemology.
We also discuss the broader implications of teaching computational modeling in STEM courses.

\newpage

\input{vpython/intro}
\input{vpython/deploy}
\input{vpython/hw}

\input{vpython/eval}

\input{vpython/cluster}

\input{vpython/discussion}

\input{vpython/conclusion}

%% file: vpython/intro.tex
\section{\label{sec:vpintro}Introduction}

Computation (the use of the computer to solve numerically, simulate or visualize a physical problem) has revolutionized scientific research and engineering practice. 
In science and engineering, computation is considered to be as important as theory and experiment \cite{siamstatementweb}. 
Systems that are too difficult to solve in closed-form are probed using computation; experiments that are impossible to perform in a lab are studied numerically \cite{fenton2005modeling,gonzalez2009black}.
Yet, in sharp contrast, most introductory courses fail to introduce students to computation's problem solving powers. 

Using computation in introductory physics courses has several potential benefits. 
Students can engage in the modeling process to make complex problems tractable. This use of computation can be leveraged to explore the generality and utility of physical principles. 
In a way, students are participating in work that is more representative of what they will do as professional scientists and engineers \cite{macdonald1988muppet,redish1993student,schecker1993learning,schecker1994system} .
When constructing simulations, students are constrained by the programming language to the certain syntactic structures. 
Hence, they must learn to contextualize problems in a way that produces a precise representation of the physical model \cite{disessa1986boxer,sherin1993dynaturtle}. 
Arguably, one of computation's key strengths lies in its utility in visualizing and animating solutions to problems. 
These visualizations can improve students' conceptual understanding of physics \cite{finkelstein2006high}.

We have used computation in a large enrollment introductory calculus-based mechanics course at the Georgia Institute of Technology to develop students' modeling and numerical analysis skills. 
We have built upon previous attempts to introduce computation in introductory physics laboratories \cite{computajp,beichner2010labs} by extending its usage to other aspects of students' coursework.
In particular, we have taught students to construct models that predict the motion of physical systems using the VPython programming environment \cite{vpythonWebsite}. 
We describe the design and implementation of homework problems to develop students' computational modeling skills in a high enrollment foundational physics course (Sec. \ref{sec:vpdeploy}). 
We also provide the first evaluation and explication of students' skills when they attempt individually to solve a novel computational problem in a proctored environment (Secs. \ref{sec:eval}--\ref{sec:cluster}). 
We discuss implications for instructional design, considerations regarding student epistemology and the assessment of knowledge transfer as well as the broader implications of teaching computation to introductory physics students (Sec. \ref{sec:closing}).

%% file: vpython/deploy.tex
\section{Approaches to implementing computation}\label{sec:comp-approaches}

Since the development of inexpensive modern microcomputers with visual displays, there have been a number of attempts to introduce computation into physics courses. We review these attempts by decomposing them along two dimensions ({\it size of intended population} and {\it openness of the environment}) to illustrate how our approach fits with previous work. 

Some have worked closely with a small number of students to develop computational models in an {\it open computational environment}. 
Historical examples include the Maryland University Project in Physics and Educational Technology, \cite{macdonald1988muppet,redish1993student} STELLA \cite{schecker1993learning,
schecker1994system} and the Berkeley BOXER project \cite{disessa1986boxer,sherin1993dynaturtle}.
Open computational environments are analogous to ``user-developed'' codes in scientific research. 
Students who learn to use an open environment have the advantage of viewing and altering the underlying algorithm on which the computational model depends. 
Moreover, students might learn to develop their own models that solve new problems.
It is true, however, that students must devote time and cognitive effort to learning the syntax and procedures of the programming language that the open environment supports. 
Students might spend more time and cognitive effort to the details of constructing a working simulation (e.g., message handling, drawing graphics, garbage collection) rather than to developing the physical model behind it. 
It is, therefore, important to consider students' experience (or lack thereof) with computation when choosing an open computational environment. 

Others have developed {\it closed computational environments} for use at a variety of instructional levels. 
These environments have been deployed in a number of settings ranging from a few students to large lecture sections.
Examples of closed environments include Physlets \cite{christian1998developing} and the University of Colorado's Physics Educational Technology simulations \cite{perkins2006phet, wieman2006powerful}.
Closed computational environments are analogous to ``canned'' codes in scientific research. 
Students can set up and operate the program but do not construct it; nor do they have access to the underlying model or modeling algorithm (``black box'' environment). 
User interaction in closed computational environments is often limited to setting or adjusting parameters. 
Closed computational environments are useful because they typically require no programming knowledge to operate, run similarly on a variety of platforms with little more than an Internet browser and produce highly visual simulations. 

It is possible for computational models created in any open environment to be used as if they were developed for a closed one. 
Users can be restricted (formally or informally) from viewing or altering the underlying model. 
Models developed using Easy Java Simulations \cite{esquembre2004easy} (EJS) have been used in a closed manner at a variety scales and instructional levels \cite{christian2007modeling,belloni2007osp}.
However, all the features of the physical and computational model in an EJS simulation are available as it is an open environment. 
Furthermore, EJS has made authoring high quality simulations accessible to students with some (but not much) programming experience. Some have proposed teaching upper-divison science majors to develop computational models using EJS \cite{esquembre2007integrate}.

VPython, \cite{vpythonWebsite} an open computational environment, has been used to teach introductory physics students to create computational models of physical phenomena \cite{computajp}.
Typically, students write all the program statements necessary to model the physical system (e.g., creating objects, assigning variables and numerical calculations). 
The additional details of model construction (e.g., drawing graphics, creating windows, mouse interactions) are handled by VPython and are invisible to the students. 
VPython supports full three dimensional graphics and animation without the additional burden to students of learning object-oriented programming \cite{scherer2000vpython}. Given its roots in the Python programming language, VPython can be a powerful foundation for students to start to learn the tools of their science or engineering trade.
Moreover, VPython is an open-source, freely available environment that is accessible to users of all major computing platforms.



The Matter \& Interactions (M\&I) textbook \cite{mandi1} introduces computational modeling as an integral part of the introductory physics course. 
Many of the accompanying laboratory activities are written with VPython in mind and a number of lecture demonstrations are VPython programs.
In the traditional implementation of M\&I, the practice of constructing computational models is limited to the laboratory. 
In a typical lab, students work in small groups to complete a computational activity by following a guided handout. 
They pause periodically to check their work with other groups or their teaching assistant (TA).
Students' computational modeling skills are evaluated by solving fill-in-the-blank test questions in which they must write in the VPython program statements missing from a computational model.

Our approach to teaching computation uses an open environment (in VPython) and builds on our experience with M\&I to extend the computational experience beyond the laboratory.
We chose to use an open environment to teach computation in order to provide students with the opportunity to look inside the computational ``black-box'' and alter or construct the model.
Furthermore, we aimed to teach students how to develop solutions to non-analytic problems.
We chose VPython (e.g., instead of Java, C or Matlab) because it has a number of helpful features for novice programmers, can be used to construct high-quality three-dimensional simulations easily and is freely available to our students.
VPython is also conveniently coupled to M\&I allowing us to leverage our years of experience with teaching M\&I.
While our implementation builds on our M\&I experience, it is not limited to it.
We describe our implementation philosophy in the next section.

%% file: vpython/hw.tex
\section{Design and Implementation of Computational Homework}\label{sec:vpdeploy}

We aimed to develop an instructional strategy that helps computation permeate the course but does not require that students have previous programming experience. 
Furthermore, this implementation had to be easily deployable across large lecture sections; the setting in which most introductory calculus-based courses are taught.
Our philosophy was that students should learn computation by altering their own lab-developed programs to solve slightly modified problems. 
This design philosophy was informed by what research scientists do quite often; they write a program to solve a problem and then alter that program to solve a different problem that is of interest to them.
We envisioned developing computational activities that would start with guided inquiry and exploration in the laboratory followed by independent practice on homework. 
Students would work with TAs in the laboratory to develop a program that solves a problem. 
Students would then use that program individually to solve a different problem on their homework by making any modifications that were necessary.


The class of problems that becomes available to students who have learned computation is large and diverse; we chose to focus our efforts on teaching students to apply Newton's second law iteratively to predict motion.
Students taking a typical introductory mechanics course would learn several equations to predict the motion that emphasizes kinematics, a way of describing the motion without explicitly connecting changes in the motion to forces (dynamics).
These kinematic formulas are quite limited; students can only apply them to problems in which the forces are constant.
This can confuse students when they are presented with a situation where such formulas do not apply \cite{kohlmyer_thesis}.
Furthermore, the special case of constant force motion is usually the capstone of motion prediction in an introductory mechanics course. We acknowledge that some courses might teach students to determine the velocity as a function of time for a falling object subjected to linear ($F \sim v$) or turbulent ($F \sim v^2$) air drag in one dimension analytically, but this hardly demonstrates the full predictive power of Newton's second law. Furthermore, such problems are in the upper-range of tractability for introductory students and it is not typical that such problems are carried through to the prediction of motion. 
By contrast, computation allows instructors to start first and foremost with Newton's second law and emphasize its full predictive power. 
Students can numerically model the motion of a system as long as they are able to develop a physical model of the interactions and express it in the computational environment. 
The numerical integration technique used to predict motion is a simple algorithm.


\begin{figure}
\includegraphics[width=\linewidth]{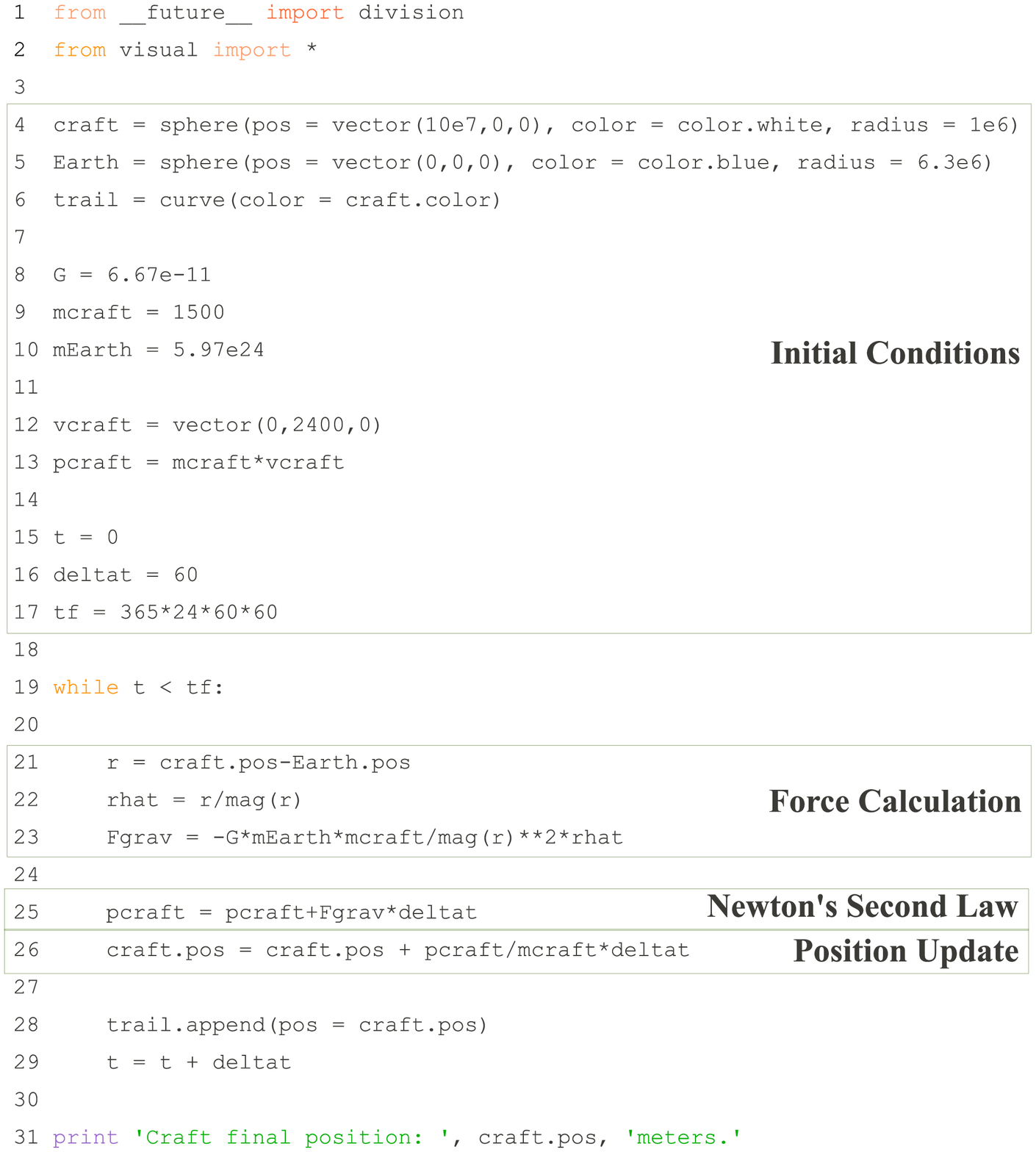}
\caption{[Color] - Under the guidance of their TAs, students wrote the VPython program above in the laboratory. This program modeled the motion of a craft (size exaggerated for visualization) orbiting the Earth over the course of one ``virtual'' year. To construct this model, students must create the objects and assign their positions and sizes (lines 4--6), identify and assign the other given values and relevant initial conditions (lines 8-10, 12--13 and 15--17), calculate the net force acting on the object of interest appropriately (lines 21--23) and update the momentum and position of this object in each time step (lines 25--26).}\label{fig:hwsamplecode}
\end{figure}

As a concrete example of our design, we show a mid-semester laboratory activity and homework problem in which students modeled the gravitational interaction between two bodies. 
In this example, students develop a VPython program that models the motion of a craft as it orbits the Earth (Fig. \ref{fig:hwsamplecode}).
Students later make a number of modifications to this program to solve a new problem on their homework.
This example is useful because it illustrates not only the level of sophistication we expect of students but it also illustrates the types of alterations that students are asked to make on their homework.

In groups of three, students wrote a program in the laboratory to model the motion of low-mass craft as it orbits the Earth (Fig. \ref{fig:hwsamplecode}). 
In VPython, they created the objects (lines 4--6), assigned the constants and initial conditions (lines 8-10, 12--13 and 15--17) and set up the numerical integration loop (lines 19--29). 
The program statements in this loop included those that calculated the net force (lines 21--23) and updated the momentum (using Newton's second law) and position of the craft (lines 25--26). 
When developing their physical model, students discussed that while the Earth experiences the same (magnitude) force as the craft, the change in the Earth's velocity due to this force is negligible.
Hence, students did not model the motion of the Earth in their VPython program.
When writing this program in the lab, students could seek help from TAs at any time.
The accuracy of the students' completed code was checked by their TAs. 
After completing the lab, students had written a VPython program that modeled the motion of the craft moving around the Earth for any arbitrary amount of time.  

In the week following the lab, students solved a computational homework problem in which they used the computational model that they had written in lab to solve a problem that differed only slightly from the lab problem.
Students were asked to alter their initial conditions to predict the position and velocity of the craft after some integration time.
To solve this problem successfully, students had to identify and make changes to their initial conditions (lines 4, 6, 9--10 \& 12) and integration time (line 17).
In addition, students had to add an additional print statement (after line 31) to print the final velocity of the craft.

Computational homework problems were deployed using the WebAssign course management system, which facilitated the weekly grading of students' solutions. 
To create the homework problem, we numerically integrated several hundred initial conditions and stored the solutions, including final quantitative and qualitative results. 
Each student was assigned a random set of initial conditions corresponding to a particular set of results. 
Randomization ensured that each student received a unique realization with high probability. 
Students used their assigned initial conditions and wrote additional statements to answer the questions posed in the problem. 
Students entered numeric answers into answer blanks and selected check-boxes to answer to qualitative questions.
On these weekly assignments, only students' final results were graded; their code was uploaded for verification purposes, but not graded.
Grading programs for structure and syntax at this large scale requires additional work by TAs who are already charged with a number of other teaching and grading tasks.
Computational homework problems were generally completed in the week that followed the associated laboratory activity.

To facilitate student success and help them learn to debug their programs, each assignment contained a {\it test case} -- an initial problem was posed for which the solution (i.e., the results from the numerical integration) was given.
When writing or altering any computer program, making programming errors (bugs) is possible.
Learning to debug programs is part of learning how to develop computational models.
This test case ensured that a student's program worked properly and helped to instill confidence in students who might otherwise have been uncomfortable writing VPython programs without the help of their group members or TAs.
After a student checked her program against the {\it test case}, she completed the {\it grading case}, a problem without a given solution. 

In keeping with our overall design philosophy, most homework problems that students solved had similar designs as the aforementioned example. 
In particular, students built a computational model in the laboratory and independently used that fully functioning model to solve a modified problem on their homework. 
On the first four homework assignments, of which the previous example is the fourth, students made few modifications to their programs; altering their initial conditions and adding a new print statement.
In the next several labs, students learned to model more complicated systems (e.g., three body gravitational problem, spring dynamics with drag) while learning new algorithms such as decomposing the net force vectors into radial and tangential components.
Students also learned to represent these force components as arrows in VPython.
On the homework problems associated with these labs, students still used their lab-developed programs to solve new problems by changing initial conditions and representing new quantities with arrows but also made some of their earlier programs more sophisticated.

The last two homework problems which students solved were not related to the laboratory; we intended to emphasize the utility of learning to predict motion using Newton's second law. 
To solve these problems, students wrote all the statements missing from a partially completed code to predict the motion of two interacting objects. 
These were interactions which students had not seen before (e.g., the anharmonic potential and Lennard-Jones interaction).
In these problems, we omitted the appropriate initial conditions and the statements that numerically integrated the equations of motion.
Students had to contextualize the word problem into a programming task and produce a precise representation of the problem in the VPython programming environment.
With regard to programming tasks, students had to do no more than identify and assign variables and implement the motion prediction algorithm for these two problems.
A similarly designed problem was used as an evaluative assignment and is discussed in detail in Sec. \ref{sec:eval}.

%% file: vpython/eval.tex
\section{Evaluating computational modeling skills}\label{sec:eval}

Students performed as well on computational modeling homework problems as they did on their analytic homework; we found no statistical difference in students' performance using a rank-sum test (Analytic 84.6\% vs Computational 85.8\%) \cite{conover_nonpara}. 
However, this result did not indicate what fraction of students solved these computational homework problems without assistance (e.g., textbook, notes, study partners, etc.). 
Randomizing the initial conditions for each student's realization ensured that students' solutions differed with high probability. This might appear to handle issues related to cheating, but working programs could be distributed easily from student to student by email.
We note that the distribution of students' programs might not be deleterious; students who receive these programs must still read and interpret the program statements to enter in their initial conditions, make changes to the force law or print additional quantities.
This is a more complex interaction than simply plugging numbers into a algebraic solution that they discovered online.
In a sense, students who work with shared code are using a ``closed'' computational environment.

Nevertheless, we wanted to measure how effective students were at individually solving computational problems.
We delivered a proctored laboratory assignment during the last lab of three different semesters to evaluate students' computational skills on an individual basis. 
Students received a partially completed program that created two objects (one low-mass and one high-mass), initialized some constants and defined the numerical integration loop structure.
We aimed to evaluate students' engagement of the modeling process by contextualizing a physics problem into programming task. 
Furthermore, certain programming skills were being assessed, namely, students' abilities to identify and assign variables and implement the numerical integration algorithm. 
The assignment was delivered using WebAssign in a timed mode (30 minutes), and TAs were not permitted to help students debug their programs.
A timed assignment opens with a pop-up dialog box that informs the student of the time limit. 
After the student acknowledges the limit (by clicking the ``OK'' button) the full assignment opens with a countdown clock in the upper corner of the browser window.
When the time runs out, the answers that have been selected or entered are automatically submitted and the student is locked out of the assignment.
The format of the assignment was identical to students' final two homework problems; students were given a test case to check their solution before solving the grading case. 

For this assignment, students modeled the motion of the low-mass object as it interacted with the high-mass object through a central force. 
The nature of the force (attractive or repulsive) and its distance dependence ($r^n$) were randomized on a per student basis. 
We also randomized some of the variable names in the partially completed program  to hinder copying. 
After adding and modifying the necessary program statements, students ran their program and reported the final location and velocity of the low-mass object.
During the assignment, students did not receive feedback from the WebAssign system about the correctness of their solution, but they were given three attempts to enter their answers.
Similar to students' online homework, only the final numerical answer was graded.

\begin{table}[t]
\caption{As part of a final proctored lab assignment, students completed a partially constructed program that modeled the motion of an object under the influence of a central force. The partially written program defined the objects, some constants and the numerical integration loop structure. Delivered initial conditions, the sign ($\pm$) and distance dependence ($r^n$) of the force and object names were randomized on a per student basis. Slightly modified versions (Ver.) of this assignment were given at the end of three different semesters. Modifications were made to streamline delivery (Version 1 to Version 2), minimize transcription errors and improve presentation (Version 2 to Version 3). Students' performance on Version 1 was likely inflated because some students were allowed to work the problem on two separate occasions. \label{tab:eval}}
\begin{center}
\begin{tabular}{c|cc|c}
\multicolumn{1}{c}{\bf Ver.} & \multicolumn{1}{c}{\bf Correct} & \multicolumn{1}{c}{\bf Incorrect} & \multicolumn{1}{c}{\bf \% Correct}\\\hline\hline
1 & 303 & 168 & 64.3\\\hline
2 & 201 & 193 & 51.0\\\hline
3 & 316 & 176 & 64.2\\\hline\hline
\multicolumn{1}{c}{Overall} & \multicolumn{1}{c}{820} & \multicolumn{1}{c}{537} & \multicolumn{1}{c}{60.4}\\
\end{tabular}
\end{center}
\end{table}

Performance varied from semester to semester (Table \ref{tab:eval}) because the assignment was modified slightly between each semester in order to streamline delivery (Version 1 to Version 2), reduce transcription errors and improve presentation (Version 2 to Version 3). 
In the first semester, students were permitted to attempt Version 1 of the assignment twice due to a logistical issue with the initial administration of the assignment. The majority of students (64.3\%) were able to model the grading case successfully on the second administration of the assignment. 
Students' performance on Version 1 was likely inflated because some students were able to work the problem twice. \footnote{The first administration of this assignment was during a regular hour exam. Roughly, 40\% of the students modeled the motion correctly. However, students used their own laptop computers which created several logistical challenges.} 
Students solved Version 2 only once, and student performance dropped.
A number of students were confused by the randomized exponent on the units of one of their initial conditions (Sec.  \ref{sec:errorfreq}).  
About half of the students (51.0\%) were able to model the grading case successfully. 
Students were more successful on Version 3 of the assignment; 64.2\% modeled the grading case correctly.

Overall, roughly 40\% of the students were unable to model the grading case. 
To determine exactly what challenges they faced while completing this assignment, we reviewed the program of each student who failed to model the grading case. 
Through a $\sim$60\% bonus on the proctored assignment, we encouraged all students to upload their programs to the WebAssign system. 
We limited our review to the programs submitted for Versions 2 and 3 of the assignment.

%% file: vpython/cluster.tex
\begin{table}[t]
\caption{Incorrectly written programs were subjected to an analysis using a set of codes developed from common student mistakes. The codes focused on three procedural areas: {\it using the correct given values} (IC), {\it implementing the force calculation} (FC) and {\it updating with the Newton's second law} (SL). We reviewed each of the incorrectly written student programs for each of the features listed below. These codes are explained in detail in Appendix \ref{sec:vpcodes}.}\label{tab:vpcodes}
\begin{center}
\begin{tabular}{ p{0.10\linewidth} p{0.85\linewidth} }\hline
\multicolumn{2}{p{0.95\linewidth}}{\bf Using the correct given values (IC)} \\
IC1 & Used all correct given values from grading case\\
IC2 & Used all correct given values from test case\\
IC3 & Used the correct integration time from either the grading case or test case\\
IC4 & Used mixed initial conditions\\
IC5 & Exponent confusion with $k$ (interaction constant)\\\hline

\multicolumn{2}{p{0.95\linewidth}}{\bf Implementing the force calculation (FC)} \\
FC1 & Force calculation was correct\\
FC2 & Force calculation was incorrect but the calculation procedure was evident\\
FC3 & Attempted to raise separation vector to a power\\
FC4 & Direction of the force was reversed\\
FC5 & Other force direction confusion\\\hline

\multicolumn{2}{p{0.95\linewidth}}{\bf Updating with Newton's second law (SL)} \\
SL1 & Newton's second law (N2) was correct\\
SL2 & Incorrect N2 but in an update form\\
SL3 & Incorrect N2 attempted update with scalar force\\
SL4 & Created new variable for $\vec{p}_f$\\\hline

\multicolumn{2}{p{0.95\linewidth}}{\bf Other (O)} \\
O1 & Attempted to update (force/momentum/position) for the massive particle\\
O2 & Did not attempt the problem\\\hline

\end{tabular}
\end{center}
\end{table}

\section{Systematically unfolding students' errors}\label{sec:vpcat}

Students must perform several tasks to successfully write and execute the program for the proctored assignment. 
Students must interpret the problem statement; that is, they must contextualize a word problem into a programming task. 
They must review the partially completed program and identify the variables to update. 
Students need to apply their knowledge of predicting motion using VPython to the problem. 
They must identify that the force is non-constant and then write the appropriate programming statements to calculate the vector force. 
Students need to then complete the motion prediction routine by writing a statement to update the momentum of the low-mass object. 

Using an iterative-design approach, we developed a set of binary (affirmative/negative) codes to check which tasks students performed correctly and which errors they made. 
An initial review of students' uploaded programs yielded the mistakes that were made most often. 
These common mistakes formed the basis for the codes. The codes were developed empirically and several iterations were made before they were finalized. 
Two raters tested the codes by coding a single section of student submitted programs ($N = 45$). 
The raters resolved their differences which further explicated the codes and then recoded the section. 
The final codes (Table \ref{tab:vpcodes}) were used by both raters independently to code the remaining sections ($N = 324$). 
The final codes had high inter-rater reliability; both raters agreed on 91\% of the codes. 

We classified the codes into one of three procedural areas: {\it using the correct given values} (IC), {\it implementing the force calculation} (FC) and {\it updating with the Newton's second law} (SL). 
These areas were congruent with the broad range of difficulties which students exhibited through their erroneous programs. 
Each code is explained in greater detail in Appendix \ref{sec:vpcodes}.

Determining where students encountered difficulties with these tasks might help explain how students learn this algorithmic approach to use Newton's second law to predict motion.  
Because we reviewed students' programs after they were written, we are unable to comment directly on students' challenges with contextualizing the problem. 
Our work was limited to analyzing students' procedural efforts (i.e., identifying variables and implementing the numerical integration algorithm). 
However, some information about students' thoughts and actions could be inferred from this analysis.

%% file: vpython/discussion.tex
\section{Frequency of errors in students' programs\label{sec:errorfreq}}

We measured the frequency of students' errors within each category (IC, FC and MP) by mapping binary patterns extracted from our coding scheme to common student mistakes.
The number of possible binary patterns that we could observe in our data ranged from nine for MP to seventeen for FC with 13 possible for IC. 
Not all the codes within a given category are independent, hence, the number of possible binary patterns is much less than $2^n$.
Within a given category, we found that a large percentage of students could be characterized by just a few error patterns (between four and seven). 

The errors we observed were not necessarily unique to computational problems. 
The most notable errors involved calculating forces or updating the momentum. 
Most of these errors appeared to be physics errors reminiscent of those made on pencil and paper problems.
Many of them could have been mitigated by qualitative analysis. 
Some errors were unique to computational models and the iterative description of motion because they could produce a program that ran but did not model the system appropriately. 
Still others (e.g., replacing initial conditions) appeared to be simple careless mistakes, but, when investigated, highlighted the fragility of students' knowledge.

\subsection{Initial Condition Errors}

\begin{table}[t]
\caption{Only seven of the fourteen distinct code patterns for the IC category (Table \ref{tab:vpcodes}) were populated by more than 3\% of the students. The patterns (ICx) are given by affirmatives (Y) and negatives (blank) in the code columns (IC\#). The percentage of students with each pattern is indicated by the last column (\%). These 7 patterns accounted for 88.8\% of students with erroneous programs.}
\begin{center}
\begin{tabular}{c|ccccc|c}
\multicolumn{7}{c}{\bf Initial Condition Codes} \\
\multicolumn{1}{c}{\bf Pattern} & {\bf IC1} & {\bf IC2} & {\bf IC3} & {\bf IC4} & \multicolumn{1}{c}{\bf IC5} & \multicolumn{1}{c}{\bf \%}\\\hline\hline
{\bf ICa} & & Y & Y & & & 27.6\\\hline
{\bf ICb} & & & Y & Y & & 16.0\\\hline
{\bf ICc} & Y & & Y & & & 14.4\\\hline
{\bf ICd} & & & Y & & Y & 13.8\\\hline
{\bf ICe} & & & & Y & & 7.9\\\hline
{\bf ICf} & & & Y & & & 5.2\\\hline
{\bf ICg} & Y & & & & & 3.8\\
\end{tabular}\label{tab:ic}
\end{center}
\end{table}

Students had to identify and update a total of eight given values: the interaction constant ($k$), the ``interaction strength'' ($n$), the mass of the less massive particle, the position and velocity of both particles and the integration time.
Most students with incorrect programs (88.8\%) fell into one of seven IC patterns (Table \ref{tab:ic}). 
Students in ICa (27.6\%) identified and correctly replaced all the initial conditions with those from the test case (IC2), including the integration time (IC3). 
Those in ICb (16.0\%) mixed up the initial conditions (IC4), but used the correct integration time (IC3). 
Students in ICc (14.4\%) identified and correctly replaced all the initial conditions with those from the grading case (IC1), including the integration time (IC3). 
Students who appeared in ICd (13.8\%) were confused by the exponent on the units of the interaction constant (IC5), but used the correct integration time (IC3). 
Students in ICe (7.9\%) used a variety of initial conditions and given values (IC4). 
Those students in ICf (5.2\%) and ICg (3.8\%) used incorrect initial conditions (IC3) or the wrong integration time (IC1), respectively.
Most students might have simply forgotten to update one or more of the initial conditions from either the default case or the test case (ICb, ICe, ICf and ICg). 
A small fraction of students with mixed initial conditions had values from all three cases. 

Students in ICa were most likely stuck on the test case because they had trouble with another aspect of the problem. 
These students were unable to obtain the solutions provided in the test case and kept working on it.
It is possible a number of these students ran out of time while trying to debug their programs.

It is difficult to say definitively if students with mixed initial conditions (ICb and ICe) were unable to identify the appropriate values, as we reviewed students' programs only after they were submitted. 
It is possible that these students were just careless when making changes, but they might have been unable to identify and update these quantities. 
Some students could have been in the process of updating these quantities when they ran out of time and uploaded their programs.

Identifying and updating variables in a program is not a trivial task for students. 
In fact, their challenges with updating variables highlights the fragility of their computational knowledge.
As an example, consider the students who confused the exponent on the length unit of the interaction constant ($k$) for the exponent in scientific notation of $k$ when they defined it in their programs (ICd). 
The distance dependence of the central force was randomized, and hence the units of the interaction constant ($k$) were dependent on a student's realization. 
In Version 2 of the assignment, the exponent on the length unit of $k$ was colored red (WebAssign's default behavior for random values). 
A student in ICd would read $k = 0.1$ Nm$^3$ to mean $k = 100$ rather than $k = 0.1$ {\it Newton times meters cubed}. 
In Version 3 of the assignment, we changed the exponent's text color to black like the rest of the non-random text. 
The overall frequency of this mistake dropped from 30.5\% to 9.1\%.

\subsection{Force Calculation Errors}

Students were given the magnitude of the force as an equation ($F=kr^n$) and told that their (attractive or repulsive) force acted along the line that connected the two objects. 
In solving this problem, students had to correctly calculate the magnitude of the central force and identify the unit vector ($\hat{r}$) and sign ($\pm$) for their own realization. 
Almost all students (98.8\%) appeared with one of five FC patterns (Table \ref{tab:fc}). 
Students in FCa (23.9\%) implemented the force calculation algorithm correctly (FC2), but reversed the direction of the net force (FC4). 
Those in FCb (22.2\%) performed the force calculation correctly (FC1). 
Students in FCc (15.7\%) implemented the procedure correctly (FC2) but were likely to include a force irrelevant to the problem (i.e., gravitational or electric interactions) or compute only the magnitude of the net force. 
Students who appeared in FCd (14.6\%) attempted to raise the separation vector to a power. 
Students in FCe (14.0\%) showed no evidence of an appropriate force calculation procedure; the procedure was either completely incorrect (e.g., used the differential form of the Impulse-momentum theorem) or was calculated outside the numerical integration loop (i.e., a constant force). 
Those students in FCf (8.4\%) had an appropriate force calculation procedure (FC2) but invented a unit vector for the net force (FC5).

\begin{table}[t]
\caption{Only six of the nine distinct code patterns for the FC category (Table \ref{tab:vpcodes}) were populated by more than 3\% of the students. The patterns (FCx) are given by affirmatives (Y) and negatives (blank) in the code columns (FC\#). The percentage of students with each pattern is indicated by the last column (\%). These 6 patterns accounted for 98.8\% of students with erroneous programs.}
\begin{center}
\begin{tabular}{c|ccccc|c}
\multicolumn{7}{c}{\bf Force Calculation Codes} \\
\multicolumn{1}{c}{\bf Pattern} & {\bf FC1} & {\bf FC2} & {\bf FC3} & {\bf FC4} & \multicolumn{1}{c}{\bf FC5} & \multicolumn{1}{c}{\bf \%}\\\hline\hline
{\bf FCa} & & Y & & Y & & 23.9\\\hline
{\bf FCb} & Y & & & & & 22.2\\\hline
{\bf FCc} & & Y & & & & 15.7\\\hline
{\bf FCd} & & Y & Y & & & 14.6\\\hline
{\bf FCe} & & & & & & 14.0\\\hline
{\bf FCf} & & Y & & & Y & 8.4\\
\end{tabular}\label{tab:fc}
\end{center}
\end{table}

The difficulties that students' faced when numerically computing the net force could stem from a weak grasp of the concept of vectors. 
Students in FCa made directional mistakes (e.g., changing the sign of one of lines 21--23 in Fig. \ref{fig:hwsamplecode}) that could have been easily identified and rectified by drawing a sketch of the situation, a problem-solving strategy that is practiced in the laboratory. 
Those who raised the separation vector to a power (FCd) likely transcribed the central force equation (replacing $r$ by $\vec{r}$) without thinking that this operation was mathematically impossible ($\vec{F} \sim k(\vec{r})^n$ vs. $\vec{F} \sim k|\vec{r}|^n\hat{r}$). 
We have found that students attempt a similar operation on pencil and paper problems; raising components of a vector to a power (e.g., $(\vec{r})^n = \langle r_x^n, r_y^n, r_z^n \rangle$). 
However, in the pencil and paper case, students are not immediately directed to their mistake as they are in a programming environment.
VPython raised an exception error when this operation was attempted. 
These students appeared to be unable to parse this error into any useful information. 
Students who make this type of error might be helped by additional exposure to translating force equations to precise programmatic representations \cite{sherin1993dynaturtle}.
Some students invented a unit vector (FCf) for the net force. 
This was most likely because they had computed a scalar force and tried to add a scalar impulse to the vector momentum. 
VPython raised a different exception error if an attempt to add a scalar to a vector was made. 
These students were able to parse this error, but resolved it incorrectly.

Other students (FCc) might have incorrectly contextualized the problem by including an irrelevant force (i.e., gravitational or electric interactions). 
The problem clearly stated that the two objects were far from all other objects. 
It did not explicitly state to neglect the gravitational interaction between the objects. 
However, the gravitational interaction could be safely neglected for the range of masses and distances we had chosen. 
Furthermore, nothing about the charge of the objects was mentioned in the problem statement. 
It is surprising that students included these interactions in their models. 
One possible explanation for the inclusion of these interactions is that students had memorized how to solve the gravitational and Coulomb problems because these problems had appeared on their homework several times and on an exam. 
They might have panicked and simply wrote all possible forces they could remember.

A number of students (FCe) did not employ the force calculation algorithm at all. 
Some of these students computed the net force (e.g., lines 21--23 in Fig. \ref{fig:hwsamplecode}) outside the numerical integration loop (e.g., before line 19 in Fig. \ref{fig:hwsamplecode}). 
In this case, the net force was effectively constant and therefore only correct at $t=0$ . 
A program with correct syntax will run regardless of the physical implications.
This error is unique to computational problems in which motion is predicted iteratively. 
Students in introductory physics rarely use Newton's second law to predict motion due to non-constant forces.
Other students who fell into FCe wrote ``creative'' program statements. 
Students in this group manipulated some quantities in the loop but did not perform any physically relevant calculations. 
The number of students with ``creative'' program statements was relatively small.

\subsection{Newton's Second Law Errors}

Students had to write a program statement similar to line 25 in Fig. \ref{fig:hwsamplecode} to properly update the momentum using Newton's second law. 
Most students demonstrated no difficulty in remembering the formula for the momentum update but some met challenges with making that description precise \cite{sherin1993dynaturtle}.
Nearly all students (95.7\%) fell into one of four SL patterns (Table \ref{tab:mp}). 
Most students appeared in SLa (69.7\%) because they wrote the momentum update correctly (SL1). 
A much smaller number of students fell into SLb (13.2\%) and attempted to update the vector momentum with a scalar force. 
Students in SLc (7.9\%) were unable to write Newton's second law in any form that updated (all codes negative). 
A small fraction (SLd, 4.9\%) wrote Newton's second law in an iterative form, but did so incorrectly (SL2). 

\begin{table}[t]
\caption{Only four of the nine distinct code patterns for the SL category (Table \ref{tab:vpcodes}) were populated by more than 3\% of the students. The patterns (SLx) are given by affirmatives (Y) and negatives (blank) in the code columns (SL\#). The percentage of students with each pattern is indicated by the last column (\%). These 4 patterns accounted for 95.7\% of students with erroneous programs.}
\begin{center}
\begin{tabular}{c|cccc|c}
\multicolumn{6}{c}{\bf Second Law Codes} \\
\multicolumn{1}{c}{\bf Pattern} & {\bf SL1} & {\bf SL2} & {\bf SL3} & \multicolumn{1}{c}{\bf SL4} & \multicolumn{1}{c}{\bf \%}\\\hline\hline
{\bf SLa} & Y & & & & 69.7\\\hline
{\bf SLb} & & Y & Y & & 13.2\\\hline
{\bf SLc} & & & & & 7.9\\\hline
{\bf SLd} & & Y & & & 4.9\\
\end{tabular}\label{tab:mp}
\end{center}
\end{table}

Students who attempted to update the momentum with a scalar force (SLb) might still face difficulties with understanding vectors. 
The momentum update is presented as a vector equation ($\vec{p}_f = \vec{p}_i + \vec{F} \Delta t$). 
These students might be unable to unpack that representation into a precise programmatic description, but it was more likely that they calculated a scalar force (FCc) and then simply wrote the correct (vector) second law syntax.
VPython raised an exception error if an attempt to add a vector to a scalar was made. 
The students appeared unable to parse this error into any useful information.

Students who were unable to write Newton's second law in any form that updated  (SLc) might have experienced difficulties with converting the second law formula into a precise and useful programmatic representation.
Students in this category either wrote Newton's second law in a non-update form (e.g. writing {\tt deltap = Fnet*deltat} or {\tt pf - pi = Fnet*deltat} as line 25 in Fig. \ref{fig:hwsamplecode}) or wrote a number of program statements that manipulated quantities but performed no useful calculations. 
In either case, these students could benefit from the precision required by a programming language \cite{sherin1993dynaturtle}. By forcing them to accurately represent Newton's second law in their programs, they might begin to distinguish between the utility and applicability of its various algebraic forms.

Students who wrote Newton's second law in form that updated incorrectly (SLd) either remembered the formula for the second law incorrectly or made a typo.
These students would generally leave off the time step in the momentum update (e.g., {\tt p = p + F}) or divide by it ({\tt p = p + F/deltat}). 
Dividing by the time step is a particularly egregious error because it was quite small. Hence, the impulse added in this case would be large. 
Students who made this error were unable to assess the state of the visualization (the particle flew off to ``infinity'') to debug this error.


\section{Common Error Patterns in Students' Programs}\label{sec:cluster}

The patterns within individual categories (IC, FC and SL) indicated the frequency of common mistakes students made when solving the proctored assignment, but a single student could make one or more of these mistakes. 
Evaluating a student's complete solution requires an analysis using all the codes (Table \ref{tab:vpcodes}).
In principle, the codes we developed could have up to $\sim4300$ possible error patterns using all sixteen codes.
In fact, the intersections of code categories indicated that the number of distinct errors made by students across all categories was relatively small; we found only 111 distinct binary patterns.
It is possible to relate these unique patterns in a manner that suggests dominant  common errors.

Cluster analysis, a technique borrowed from data mining, is particularly well suited for this application because it characterizes patterns in complex data sets \cite{tan2006introduction, clustereveritt}.
This technique has been used previously to classify students' responses to questions about acceleration and velocity in two dimensions \cite{springuel2007applying}.
It was used here to determine the major features in students' incorrect programs which were responsible for their failure. 

We applied the cluster analysis technique to the data generated from our set of binary codes. 
We used the Jaccard metric \cite{jaccard1901study} to measure inter-cluster distances and linked clusters using their average separation \cite{sokal1975statistical}.
We tested several other metrics (e.g., Hamming, city block, etc.). 
The Jaccard metric was chosen because it neglects negative code pairs. 
Both the Hamming and city block metrics produced similar pairings at low levels, but higher order clusters were difficult to interpret.
We used average linkage to avoid the effects of ``chaining'' that appeared when nearest \cite{sneath1957application} and because useful clusters were more difficult to distinguish when farthest \cite{sorenson1948method} neighbor linkage was used. Additional information on cluster analysis is available in Appendix \ref{sec:clusteranalysis}.

Thirty clusters with inter-cluster distances below 0.5 were reviewed in detail. This cutoff was selected to minimize the number of unique clusters while still rendering clusters with useful interpretations. 
Most students (86.5\%) appeared in seven of the thirty clusters (Table \ref{tab:clus}). 
These clusters had very few students ($<$1\%) with affirmatives in the ``Other'' category. 
Codes O1 and O2 were dropped from Table \ref{tab:clus} for this reason. 
Each of the other 23 clusters were populated by less than 3\% ($N \approx 10$) of the students, and the bottom 18 clusters had less than 1\% ($N \approx 3$) each.
Each of the dominant clusters demonstrated a unique challenge that students faced while solving the proctored assignment (Table \ref{tab:clus}). 

\begin{sidewaystable}[t]
\caption{Only seven of the thirty clusters with an inter-cluster distance of less than 0.3 were populated by more than 3\% of the students. The bottom 18 clusters were populated by less than 1\% of students each. These seven clusters accounted for 86.5\% of students. The percentage of affirmatives for each code (Table \ref{tab:vpcodes}) within any given cluster (A-G) is given to the nearest whole percentage. Codes with affirmative percentages greater than 60\% are bolded. These clusters had very few students ($<1$\%) with any affirmatives in the `Other'' category, hence the results from this category are not reported. The percentage of students in each cluster is indicated in the last column (\%). 
}
\begin{center}
\begin{tabular}{c|ccccc|ccccc|cccc|c}
\multicolumn{1}{l}{\hspace*{1pt}} & \multicolumn{5}{c}{\bf Initial Conditions} & \multicolumn{5}{c}{\bf Force Calculation} & \multicolumn{4}{c}{\bf Second Law} & \multicolumn{1}{l}{\hspace*{1pt}}\\
\multicolumn{1}{c}{\bf Cluster} & {\bf IC1} & {\bf IC2} & {\bf IC3} & {\bf IC4} & \multicolumn{1}{c}{\bf IC5} & {\bf FC1} & {\bf FC2} & {\bf FC3} & {\bf FC4} & \multicolumn{1}{c}{\bf FC5} & {\bf SL1} & {\bf SL2} & {\bf SL3} & \multicolumn{1}{c}{\bf SL4} & \multicolumn{1}{c}{\bf \%}\\\hline\hline
A & 00 & {\bf 68} & {\bf 93} & 18 & 15 & 00 & {\bf 100} & 22 & {\bf 66} & 09 & {\bf 95} & 00 & 00 & 01 & 23.8\\\hline
B & 21 & 01 & {\bf 86} & 37 & 41 & {\bf 88} & 00 & 00 & 00 & 00 & {\bf 97} & 00 & 00 & 00 & 19.8\\\hline
C & 04 & 33 & {\bf 76} & 31 & 22 & 00 & {\bf 94} & 00 & 08 & 00 & 00 & {\bf 98} & {\bf 98} & 08 & 13.3\\\hline
D & {\bf 98} & 00 & {\bf 85} & 00 & 00 & 00 & {\bf 85} & 18 & 50 & 00 & {\bf 98} & 00 & 00 & 00 & 10.8\\\hline
E & 00 & 00 & 57 & {\bf 75} & 36 & 00 & {\bf 100} & {\bf 79} & 00 & 04 & {\bf 89} & 00 & 00 & 00 & 7.6\\\hline
F & 00 & {\bf 100} & {\bf 96} & 00 & 00 & {\bf 65} & 00 & 00 & 00 & 00 & {\bf 73} & 19 & 00 & 04 & 7.1\\\hline
G & 27 & 00 & {\bf 93} & 53 & 07 & 00 & {\bf 100} & 00 & 07 & {\bf 100} & {\bf 93} & 00 & 00 & 00 & 4.1\\
\end{tabular}\label{tab:clus}
\end{center}
\end{sidewaystable}

Students in cluster A (23.8\%) tended to remain stuck on the test case (ICa) due to an error in their force calculation. Reversing the direction of the force (FCa) was the most common mistake, followed by raising the separation vector to a power (FCd). Most students in this cluster had no trouble expressing Newton's second law (SLa).
These students worked diligently to solve the test case but were unable to do so. As a result, they did not proceed to the grading case.

Cluster B (19.8\%) contained students who made mistakes while replacing the given values and initial conditions (any IC code except ICa). 
Some of these students worked with the grading case (ICc and ICg).
Others might have been working with either case and had mixed conditions (ICb and ICe) or simply incorrect ones (ICf).
Still others might have incorrectly assigned the exponent on the units of $k$ to the value of $k$ (ICd).
At any rate, most students in this cluster were able to construct a working albeit incorrect program. 
Given their unfamiliarity with general central force interactions, these students might have believed their solutions were correct. 
In fact, it is possible that students who were working with the grading case (ICc and ICg) had solved the test case correctly and simply made a typo.

Students in cluster C worked with either the grading or test case and might have made a number of mistakes with their initial conditions (any IC code except ICa).
The dominant error in cluster C were students who computed the magnitude of the net force (FCc) and attempted to update the vector momentum with this scalar force (SLb). 
This mathematically impossible operation would have raised a VPython error. Students in this cluster were unable to parse this error into any useful information.

Cluster D (10.8\%), like cluster A, was populated by students who tended to make errors in the force calculation (FCa and FCd), but students in Cluster D worked with the grading case (ICc). 
The most common error in Cluster D was reversing the direction of the net force (FCa) followed by raising the separation vector to a power (FCd). 
Again, like cluster A, most students met no challenges when updating the momentum using Newton's second law (SLb). 
These students might have started working with the test case, but we think it is more likely that they jumped right into working with the grading case because the dominant error appears in their force calculations.

Students in cluster E (7.6\%) tended to raise the separation vector to a power (FCd) and have mixed initial conditions (ICb, ICd, ICe and ICf). 
These students generally had no difficulty with writing Newton's second law correctly (SLa). The dominant error for students in cluster E was raising the separation vector to a power (FCd). This mathematically impossible operation would have raised a VPython error. Students in this cluster were unable to parse this error into any useful information.

Cluster F (7.1\%) contained students who worked solely with the test case (ICa) and either had no issue with their force calculation (FCb) or had no evident force calculation procedure (FCe). 
Most of these students had no difficulty updating the momentum using Newton's second law (SLa).
Students in cluster F were able to construct a program which ran without raising any VPython errors. 
Students who had no issue with their solution likely completed test case but simply ran out of time before turning to the grading case. 
Students with no evident procedure generally computed the net force outside the numerical calculation loop, essentially making this force constant in time. 
Given students' unfamiliarity with general central force interactions, it would not be surprising if students who treated the central force outside the loop believed their solutions were correct.

Students in cluster G (4.1\%) all invented an incorrect unit vector for the force rather than using $\hat{r}$ (FCf) regardless of the case with which they worked (ICc,  ICb and ICf).
These students generally had no difficulty updating the momentum using Newton's second law (SLa). 
Most likely, these students computed the magnitude of the force, similar to students in cluster C, but were able to parse the resulting VPython error. 
Students in cluster G corrected their mistake by assigning some unit vector to the force before the momentum was updated.

%% file: vpython/conclusion.tex
\section{Closing remarks}\label{sec:closing}

In large introductory physics courses, students can develop the skills necessary to predict the motion of different physical systems. 
After a solving a suite of computational homework problems, most students ($\sim 60$\%) were able to model the motion of a novel problem successfully.
Students had no previous experience with the physical system on this evaluation. Students transferred the algorithmic approach used to solve other problems to this problem.
In our work, we discovered that most students who were unsuccessful encountered challenges when calculating the net force acting on the object in the motion prediction algorithm (Clusters A and C through G in Table \ref{tab:clus}).
By contrast, there were fewer students whose primary challenge was identifying and assigning variables (Cluster B in Table \ref{tab:clus}). We acknowledge that we have limited the development of our students' computational skill set to contextualizing a word problem into a programming task, identifying and updating input variables and applying a motion prediction algorithm. We believe that further development of our homework problems and other novel deployments could broaden the scope of the skills students develop.

Procedural errors such as those we have documented (Secs. \ref{sec:errorfreq} \& \ref{sec:cluster}) could be corrected through additional materials aimed at addressing each error in turn. 
However, the results from this work indicate that instructional efforts should be focused not only on correcting procedural mistakes but also on developing students' qualitative habits of mind. 
Training students to write programs to predict motion might help them to be successful in a highly structured environment, but they would be better served by learning the practice of debugging. 
Here, debugging includes identifying syntax errors, of which we found few, and, more importantly, performing the type of qualitative analysis that is typically taught for solving analytic problems. 
Students who could synthesize their analytic and computational skills would be better prepared to solve the open-ended problems they will face in their future work. 

Developing the materials to teach these skills requires an evaluation of how students contextualize computational problems. 
We do not claim to understand this presently, although we have been able to glean some suggestive information based on students' errors. 
Some students' weak grasp of vectors is responsible for their inability to model the motion in the evaluative assignment.
Others exhibit fragility with respect to identifying and assigning variables.
Investigating what students think about when solving computational problems requires structured student interviews (i.e., a think-aloud study). 
In the future, we plan to perform such a study to not only characterize students' abilities to contextualize but also to elucidate the mechanism for some of the errors we reported in Secs. \ref{sec:errorfreq} \& \ref{sec:cluster}.
We believe that such a study will demonstrate a number of students correcting their errors by working the problem out loud in the absence of a timed and stressful environment. 
However, students who  are unable to solve the  problem presented in this study will most likely lack the skills to systematically debug their own programs.

Research into skill development in math and science has shown a strong correlations with student epistemology \cite{schommer1990effects,schommer1993epistemological}.
Epistemology is important because the views that students hold affect how they learn \cite{elby2001helping} and, utimately, how successful they are in their science courses \cite{halloun1997views,perkins2005correlating}.
It is therefore crucial that we understand students' sentiments about learning a new tool such as computation. 
Our students expressed anxiety and demonstrated a lack of self confidence, even with their additional exposure to computational problems. 
We are developing an attitudinal survey aimed at exploring these and other beliefs in detail.
Students who learn to use computational modeling and are confident in their abilities will be better prepared to solve challenging problems.

It is unlikely that computation will make students better adept at solving analytical problems.
The mathematical skills needed to solve analytical problems are directly addressed in most computational work. 
However, students who learn computation could develop their qualitative reasoning skills through the debugging of programs and their learning to interpret the results of their program.
These predictions might be tested by comparing the solutions to analytic and qualitative problems between students who received computational instruction and those who do not.
It is important that only the instruction in computation differ between the two populations.
The results from this study could inform a balanced instructional approach which develops analytic, qualitative and computational skills and thereby better prepares students to approach and investigate challenging problems.

We have not claimed to have assessed a transfer of computational knowledge. 
We designed a set of problems (Sec. \ref{sec:vpdeploy}) that students solved over the course of the semester with an eye towards a final assessment of their skills using a novel problem. 
This problem (Sec. \ref{sec:eval}) was similar to some of the homework problems students had solved previously. 
It focused on key skills that we desired students to acquire: contextualizing a problem, identifying and assigning variables in a program and carrying out the motion prediction algorithm.
An evaluation of transfer would require that students apply these computational skills to a different domain (e.g., electromagnetism) or a different task (e.g., open-ended inquiry). 
A study in which students apply these algorithmic approaches in a open-ended setting would likely find that students could model the selected system, but that students would face challenges with interpreting the results.
Demonstrating transfer of computational knowledge is a necessary step in developing students into flexible problem solvers for the 21$^{\mathrm{st}}$ century.



%% file: 06-compass.tex
\chapter{Towards Characterizing Student Epistemology in Computation}\label{chap:compass}

This chapter presents development of a new tool for characterizing how students think about learning computation.
In developing this instrument, we were able to leverage much from the work of others in the physics \cite{redish1998student,adams2006new} and computer science \cite{wiebe14computer} educational domains.
In short, we have composed a valid and statistically robust survey for characterizing students  with regard to a number of epistemology features: 
their reasons for and interest in learning computation, 
the efforts they put forth to learn computation,
their confidence with using computation and 
a self-evaluation of their aptitude with computation.
Such an instrument is useful not only as a research tool but as a guide for STEM instructors using computation in their courses.
Student epistemology and their performance in science are interrelated \cite{schommer1990effects,schommer1993epistemological} and instructional design efforts which leverage student epistemology not only affect how students learn but how well they learn \cite{elby2001helping}.
We present the development and validation of this survey as well as preliminary measurements from three different populations.
Results from these measurements paint an interesting picture of what students think about learning computation and offer suggestions for teaching computation (similar to those proposed in Ch. \ref{chap:comp}) such as additional practice with the modeling process and more experiential learning  earlier.
We discuss issues related to the survey's robustness with other populations, its reliability and the validity of its use before instruction.

\newpage

\input{compass/01-intro}
\input{compass/02-design}
\input{compass/02a-scoring}

\input{compass/03-results}

\input{compass/03a-diffpop}
\input{compass/03b-compeval}

\input{compass/04-concluding}

%% file: compass/01-intro.tex
\section{\label{sec:compassintro}Introduction}


Students' ideas about learning can affect how well they learn. 
In controlled studies, students' epistemological expertise correlated with their performance in math and science even after controlling for factors such as socioeconomic status \cite{schommer1990effects,schommer1993epistemological}.
In classroom environments, directly addressing students' epistemology
has been shown to have a positive effect on their performance, specifically, on students' conceptual understanding in physics \cite{elby2001helping}. 
A number of possible factors have been proposed to explain this effect including students' beliefs about acquiring knowledge, the relevance of course activities to their own lives and students' motivation and effort to learn new material \cite{hammer1994epistemological,hammer1989two}. 

Several instruments have been created to characterize students' epistemology in physics courses \cite{halloun1997views,redish1998student,adams2006new}. 
These instruments compare students' responses to experts' on the same set of statements about the nature and learning of science.
Results from each instrument have indicated that a significant rift exists between how students and how experts think about science. 
It also appears that students' background and performance in science has a measurable effect on their alignment with expert sentiments \cite{halloun1998interpreting,redish1998student,perkins2005correlating}.
These instruments have been used to compare changes in student epistemology between different content delivery reforms \cite{redish1998student,elby2001helping} and science disciplines \cite{perkins2007chemistry}. 
They are used at a variety of instructional levels \cite{otero2008attitudinal} and some have been adapted for use in other STEM (Science, Technology, Engineering and Mathematics) courses \cite{adams2008modifying}.

One STEM area which has been largely ignored in all this is numerical computation.
While observations of students using computation indicate they experience considerable anxiety and a lack of confidence, formal knowledge of students' attitudes towards computation is lacking.
However, no instrument yet exists for characterizing student epistemology with regard to learning computation.
Surveys about computer science \cite{wiebe14computer,heersink2010measuring} are too domain specific and surveys about computer usage \cite{bear1987attitudes,eastman2004elderly} are inappropriate for such purposes. 


To meet this need, we have designed and validated the Computational Modeling in Physics Attitudinal Student Survey (COMPASS), a new instrument that addresses students' attitudes towards learning computation. 
In creating this survey, guiding design principles (Sec. \ref{sec:compass-design}) were leveraged from previous work in physics \cite{adams2006new} and computer science education \cite{wiebe14computer}. 
After developing an initial draft of the COMPASS, discussions with both students and experts were essential to ensure the validity and clarity of the statements on the COMPASS (Sec. \ref{sec:compass-valid}).
The COMPASS was administered (before and after instruction) to introductory physics students taking a reformed calculus-based physics course in which computation is introduced \cite{mandi1,mandi2}. 
Students' alignment with experts on each statement were compared and used to compute overall scores (Sec. \ref{sec:compass-scoring}). 
Overall pre- and post-instruction COMPASS scores were observed to be influenced by student demographics (Sec. \ref{sec:compass-gt-results}).
Students' responses on a number of statements were strongly correlated with each other which gave rise to the construction of statistically robust dimensions (Sec. \ref{sec:compass-statdim}). 
Differences in students scores on each of these dimensions suggest not only demographic influences, but also influences of self-identification and academic preparation (Secs. \ref{sec:compass-pop} \& \ref{sec:compass-comp-sig}).
The COMPASS is still under development, but these preliminary measurements suggest its utility to researchers and instructors alike (Sec. \ref{sec:compass-applied}).
Furthermore, several avenues for future testing and measurement, including usage with a variety of populations and in controlled studies, are possible (Sec. \ref{sec:compass-closing}).

%% file: compass/02-design.tex
\section{Guiding Principles \label{sec:compass-design}}

In developing the COMPASS, a number of design principles which have been expressed previously by the creators of the Maryland Physics Expectations Survey (MPEX) \cite{redish1998student} and the Colorado Learning Attitudes about Science Survey (CLASS) \cite{adams2006new} were followed. The survey must be valid, the wording clear and concise and the format simple to use and score. It should also be reliable and informative. 

\paragraph{Validity} 
The COMPASS should be valid for the domain of interest; in particular, statements should be identified with the use of computation in science and not with other computer related domains such as computer science \cite{wiebe14computer,heersink2010measuring} or the casual usage of computers \cite{bear1987attitudes,eastman2004elderly}. 
Issues related to domain validity are typically controlled by discussion with domain experts. 
However, this form of validity does not ensure item validity; it is possible for students to interpret statements differently from experts.
Hence, interviews with students are essential to secure item validity. 
Both students and experts must interpret statements similarly to make any valid conclusions about the results. 
In Sec \ref{sec:compass-valid}, the results of both domain and item validity for the COMPASS are discussed in detail.

\paragraph{Clarity}
Wording of COMPASS statements should be clear and concise to facilitate usage at different instructional levels. 
Furthermore, statements should be constructed such that adapting the COMPASS to other domains (e.g., computational biology, engineering) is relatively straightforward.
Designers of surveys in science invest a significant amount of time casting and revising the wording of their surveys to ensure their accessibility.
The domain of the COMPASS makes this goal somewhat more challenging; introductory STEM students typically have little familiarity with the vocabulary of computation. 
Issues related to the selection and wording of COMPASS statements are discussed in Sec \ref{sec:compass-valid}.

\paragraph{Simple and Familiar Format}
The design of the COMPASS should simple enough to facilitate an automated administration. 
Moreover, students should be familiar with its format so that they have little trouble completing it. 
While a wealth of knowledge might be gained from using open-ended surveys, there are significant administrative challenges with collecting and scoring such instruments.
Some surveys which are simple to administer have used complicated designs \cite{halloun1997views}. 
These designs are unfamiliar to students and unnecessarily limit the types of items that can appear on the survey.
Others have adopted the more typical Likert scale (agree/disagree), a much simpler design with which most students are familiar \cite{redish1998student,adams2006new}. 
Additional details about format and scoring this design are discussed in Sec. \ref{sec:compass-scoring}.

\paragraph{Reliability}
Results from the COMPASS should be consistent for similar populations. Unreliable instruments are useless because their outcomes are not based on population or treatment differences, but simply on artifacts of a poorly designed survey. 
The reliability of attitudinal surveys is generally checked through a modified test-retest scenario; the consistency of results from students with similar backgrounds taking the same course in different semesters are compared \cite{adams2006new}. 
The latest version of the COMPASS has been administered in only one semester.
However, results from the COMPASS as it relates to different sections of the same course and students with similar backgrounds taking the same course are discussed in Sec. \ref{sec:compass-scoring}. We also discuss results from different student populations in Sec. \ref{sec:compass-pop}.

\paragraph{Interrelatedness of Statements}
Individual statements on the COMPASS are likely to be interrelated in some fashion. 
Selecting which statements are interrelated (i.e., dimensions of the instrument) has typically followed two distinct paths: {\it a priori selection} and {\it emergent categorization}. 
Selecting categories a priori provides the simplicity of interpreting the categories \cite{redish1998student}.
Emergent categorization is statistically robust and can help to uncover connections between concepts which respondents make that designers did not \cite{rennie1987scale}.
A pragmatic approach to categorizing statements on the COMPASS that balances the benefits of a priori selection with the statistical robustness of emergent categorization was borrowed from the design of the CLASS \cite{adams2006new} and is discussed in Sec. \ref{sec:compass-statdim}.

\paragraph{Informative for Instructional Design}
Results from the COMPASS should be useful for guiding improvements to computational instruction. 
Epistemology has been shown to affect student learning  \cite{schommer1990effects,schommer1993epistemological}.
Moreover, when students' attitudes about learning are addressed directly, student performance improves \cite{elby2001helping}.
Hence, some subset of COMPASS statements should guide instructional reforms.
In \ref{sec:compass-applied}, possible instructional improvements based on preliminary results are discussed.

\section{Survey Design and Validation \label{sec:compass-valid}}

The Computational Modeling in Physics Attitudinal Student Survey (COMPASS) is a 36-item, five-point Likert scale (strongly agree to strongly disagree) survey that was designed to be used in courses that teach computation alongside science. 
This survey was intended to be used in science courses in which computation is introduced, for example, introductory physics courses in which computation is taught \cite{mandi1,mandi2}, interdisciplinary courses in science and computer science \cite{sedgewick2007introduction}, courses in which knowledge about computation is assumed (e.g., nonlinear science) \cite{cross2009pattern,ChaosBook} or upper-division and graduate level computational physics courses.

The COMPASS was designed to address several themes that instruction in computation is meant to communicate: 
(1) Computation is a useful tool for solving science and engineering problems.
(2) Computation is an algorithmic process, but new methods can arise. 
(3) Computational models have limitations and users must be aware of them. 
(4) Anyone can learn computation and use it effectively; but, they must be motivated and put forth the effort to do so. 
(5) Memorizing the syntax and structure of computational models is insufficient for developing solutions to new problems. 
(6) Visualizing the solution of a problem through the use of a computational model can add another dimension to one's understanding of the problem.

In designing the COMPASS, the considerable efforts made by others in the physics and computer science education communities were drawn upon.
Several attitudinal surveys in physics and computer science were reviewed to find statements that were, at least tangentially, congruent with the themes of the COMPASS.
Some of the initial COMPASS statements were altered statements from the CLASS \cite{adams2006new}.
Changes made to CLASS statements varied from simple word replacements (e.g., replacing ``physics'' with ``computer modeling'') to complete rewrites which altogether changed the intent of the statement. For example, the CLASS statement, ``When I solve a physics problem, I explicitly think about which physics ideas apply to the problem.'', was rewritten for the COMPASS to read, ``When I solve a computer modeling problem, I explicitly think about the limitations of my model.''
Others statements were borrowed from a survey on the utility and challenges of computer programming \cite{wiebe14computer}. 
All statements taken from this survey were rewritten to remove references to ``programming'', ``programming languages'' and ``programs'' as these are somewhat domain specific words.
A small number of original statements were written to cover all the themes  adequately. 
For example, the statement, ``When I solve a problem using a computer, I have a better understanding of the solution than if I solve it with pencil and paper.'', addresses, in part, the additional understanding to a problem that might be gained from solving it computationally. 
A complete list of the latest version of COMPASS statements appears in Appendix \ref{sec:compass-statements}.

Regardless of their origin, each statement was subjected to discussions of validity with a total of twenty four faculty members and graduate students who were familiar with computation in science. 
Out of the twenty four experts, eleven were graduate students and thirteen were faculty; nineteen were male and five were female.
These experts completed the survey and provided feedback about the intent and wording of each statement.
Several experts provided detailed comments about each statement and the survey overall.
Experts generally agreed that, overall, the statements probed the domain of computation adequately but some raised a few important issues about the use of certain words.

Some experts raised serious questions about the clarity of ``computer model'' and ``computer modeling'' as some might intend to give this survey to students taking introductory science courses.
Experts pointed out that some students might not think about constructing a computational model when a statement referred to ``using a computer model.''
This distinction between simply interacting with a computer model and building one was crucial to understand our intent of certain statements.
As one expert noted, 
\begin{quote}
I am concerned that you don't make a distinction up front between building a computer model and using a computer model (i.e., running a simulation of someone else's model)...[f]or example, [one of the original COMPASS statements reads,] ``To learn how the computer model works, it is important to understand all the statements.''
\end{quote}
His comments were echoed by another,
\begin{quote}
I am concerned that the concept of ``computer modeling'' is not clearly defined...[w]hat does ``computer modeling'' mean here?  One might assume that students are simply using pre-existing computer models rather than writing their own code.
\end{quote}
Both suggested that we define ``computer modeling'' with the interpretation that we intended students to use at the beginning of the survey.

Other experts raised minor issues with switching the operational definition of the word ``model'' between statements.
``Model'' used in some statements meant the physical model of system (e.g., ``After I solve a problem using a computer model, I feel that I understand how the model works''), and in other statements, it meant the computer model (e.g., ``When I solve a computer modeling problem, I explicitly think about the limitations of my model''). 
These experts were concerned that students might unable to distinguish between the two usages.
 
In order to better understand the concerns of experts from the students' perspective, five introductory physics students (three male and two female) were invited to take the survey while being interviewed.
These students were selected at random from a pool of volunteers enrolled in an introductory electromagnetism course that teaches computation as part of the laboratory \cite{mandi2}. 
Two students had taken an introductory mechanics course in which computation was used in the lab \cite{mandi1} and to solve homework problems (Ch. \ref{chap:comp}). 
The other three took a mechanics course without computation \cite{knight04}.
Students were interviewed after they had completed two computational lab activities.

Discussions with these students helped to clarify issues related to the use of the word ``computer modeling'' and validate our operational definitions of the word ``model''.
At the beginning of the interview session, students were asked to define ``computer modeling'' in their own words. 
Each interviewee defined it similarly and quite narrowly as ``constructing computer programs using some sort of programming language.''
The fact that all interviewees identified ``constructing programs'' as a key feature of computer modeling might not be surprising given their exposure to computation in the lab.
This raises questions about the validity of using this survey before instruction with students who have had no computational experience.
However, based on the language used by interviewees, introductory students' definition of ``computer modeling'' appears colloquial; they do not distinguish between ``programming'' and ``computer modeling''. 
Hence, defining the word ``computer modeling'' on the COMPASS was not necessary. 
In the interview, students were questioned directly about the usage of ``model'' in certain statements. 
Their operational definitions for ``model'' changed appropriately depending on the statement, hence, the wording of these statements was not altered.

To make any valid conclusions about the results from the COMPASS, students must interpret all statements in the same way that experts would.
During the interviews, students were asked to briefly discuss the intent of each statement. 
Their interpretations were generally congruent with those of experts and consistent with each other.  
It is possible that this might be attributed to our refactoring of a number statements from previously developed and validated surveys \cite{adams2006new,wiebe14computer}.

To help make the COMPASS easily modifiable for use in other domains (e.g., computational biology, engineering, etc.), we used the word ``physics'' in the statements sparingly. 
In fact, only three statements use the word ``physics'' (i.e., items 16, 21 and 25 in Appendix \ref{sec:compass-statements}).
In each of these statements, the name of another domain might simply be inserted in lieu of ``physics'' without completely rewriting the statement.
Such uses are ``physics problems'' and ``physics ideas''.

%% file: compass/02a-scoring.tex
\section{Scoring the Survey\label{sec:compass-scoring}}

A wealth of information might be gleaned from an instrument like the COMPASS. 
In order to make this information more manageable yet still useful, a data reduction technique that has been used previously was employed \cite{redish1998student,adams2006new}.
The level of complexity of our data was reduced by compacting students' responses from a five-point Likert scale (strongly agree to strongly disagree) to a three-point Likert scale (agree to disagree). 
The alignment of students' responses with experts' were then compared to obtain two scores, the percentage of responses aligned with experts' and the percentage anti-aligned. 
This methodology mitigates issues of validity with respect to a students' conviction about a particular statement.

The Likert scale attempts to measure not only direction (positive or negative responses) but also magnitude (conviction). 
However, students' interpretations of magnitude are often not consistent. 
A student who selects ``agree'' over ``strongly agree'' might do so for a number of reasons.
That same student might select ``strongly agree'' over ``agree'' on the same statement when asked at a later time.
This raises issues about the consistency of the magnitude of a student's response when interpreting data from a large population.  
Others \cite{halloun2001student,lee2002cultural} have found that polarized responses, the predilection to select responses from the ends of the spectrum, reflect ``liberalism'' within a population. 
The differences in the distribution of responses make comparison between populations challenging without coarse graining. 

On the COMPASS, 
students were presented with a five-point Likert scale but for scoring purposes their responses were collapsed to a three point scale.
\begin{center}
{ $\:\:\:\:\:\underbrace{\mathrm{Strongly\:Agree - Agree}}_{\mathrm{Generally\:Agree}} \mathrm{-} \mathrm{Neutral} - \underbrace{\mathrm{Disagree - Strongly\: Disagree}}_{\mathrm{Generally\:Disagree}}$}
\end{center}
By reducing the level of complexity, issues related to response validity were mitigated by measuring only direction; measuring magnitude was sacrificed.
Furthermore, data interpretation was simplified by considering only if students generally agree, disagree or feel neutral about individual statements.
Even though coarse grained scoring was performed, it was important to maintain the five-point scale. 
Others \cite{adams2006new} have found, as we did in our student interviews, that students are more likely to select neutral responses if presented with a three-point scale.

To present compactly the differences that exist between expert and novice epistemology regarding computation, the alignment of students' responses with experts was compared. 
Students received two overall scores on the COMPASS, the percentage with which their responses agreed with experts on scored statements (``percent favorable'') and the percentage with which they disagreed (``percent unfavorable''). Formulae for computing these quantities are shown in Eqn. \ref{eqn:pf}.

\begin{subequations}\label{eqn:pf}
\begin{equation}
\%\:\mathrm{Favorable} = \frac{\#\:\mathrm{Align\:with\:Expert\:Opinion}}{\#\:\mathrm{Scored\:Statements}} \times 100
\end{equation}
\begin{equation}
\%\:\mathrm{Unfavorable} = \frac{\#\:\mathrm{Opposite\:of\:Expert\:Opinion}}{\#\:\mathrm{Scored\:Statements}} \times 100
\end{equation}
\end{subequations}

The responses of the twenty four experts who provided feedback on COMPASS statements (Sec. \ref{sec:compass-valid}) formed the pool of responses (``expert opinion'') against which students' responses were compared. 
Expert responses to individual statements were reviewed to determine if experts generally agreed, disagreed or were inconsistent (neutral) on the statement.
Expert responses to statements for which we had selected the expert opinion {\it a priori} as ``agree'' or ``disagree'' were quite consistent.
On these 30 statements, more than 75\% of the experts agreed with our judgments.
The rest generally selected neutral responses on these statements. 
For the six statements on which we thought experts would have varying opinions, 
expert responses were inconsistent or neutral.
Only the 30 COMPASS statements with non-neutral expert opinions are scored.
While the ``expert neutral'' statements were not scored, students' responses to these six statements might still be informative (Sec. \ref{sec:compass-applied}). 
If a student skips a statement, that statement is not calculated into their score. 
The overall mean COMPASS scores are computed from the average of all students' scores. 
Students who skip more than 20\% of scored statements (more than 6) are dropped from the mean score calculation. 

%% file: compass/03-results.tex
\section{Results from Mechanics Students at Georgia Tech \label{sec:compass-gt-results}}

The COMPASS has been administered to students before and after instruction in both mechanics and electromagnetism courses at Georgia Tech and to students taking mechanics courses at NCSU. 
In each of these courses, students learned to create computational models as part of the laboratory component of the courses \cite{mandi1,mandi2} using the VPython programming environment \cite{vpythonWebsite}. 
Students in the Georgia Tech mechanics courses also solved weekly computational homework sets (Ch. \ref{chap:comp}). 
Delivery of the COMPASS was facilitated by the students' homework system \cite{risleyWebAssign}.
The version of the COMPASS (v2.3) for which we have checked validity (Sec. \ref{sec:compass-valid}) has been given in only one semester thus far.
The results from Georgia Tech mechanics students are reported in this section.
These results are later compared to those from students who took the Georgia Tech electromagnetism course (Sec. \ref{sec:compass-em}) and mechanics students at NCSU (\ref{sec:compass-ncsu}).

513 students took one of four sections of mechanics taught by three different instructors. 
Certain students could register for a small honors section (N = 37) that met with a larger section taught by one of the instructors. 
Students were encouraged to take the COMPASS through a small credit, a bonus of 0.025\% to their overall average. 
The responses of students who did not take the survey seriously were dropped from the data set.
A statement told students to select ``agree'' to preserve their answers; responses from students who selected any other option were dropped. 
Less than 10\% of Georgia Tech mechanics students who took the survey responded to this statement incorrectly. 
As a result of this filtering process, we were left with 480 mechanics students who took the pre-instruction COMPASS, 354 who took the post-instruction COMPASS and 316 who took both. 
The results of non-honors students are discussed before contrasting these results with honors students (Sec. \ref{sec:compass-dimscores}).

\begin{figure}[t]
  \begin{center}
  \subfigure[COMPASS distribution before instruction]{\label{fig:compass_pre}\includegraphics[clip,trim=30mm 5mm 30mm 5mm, width=0.45\linewidth]{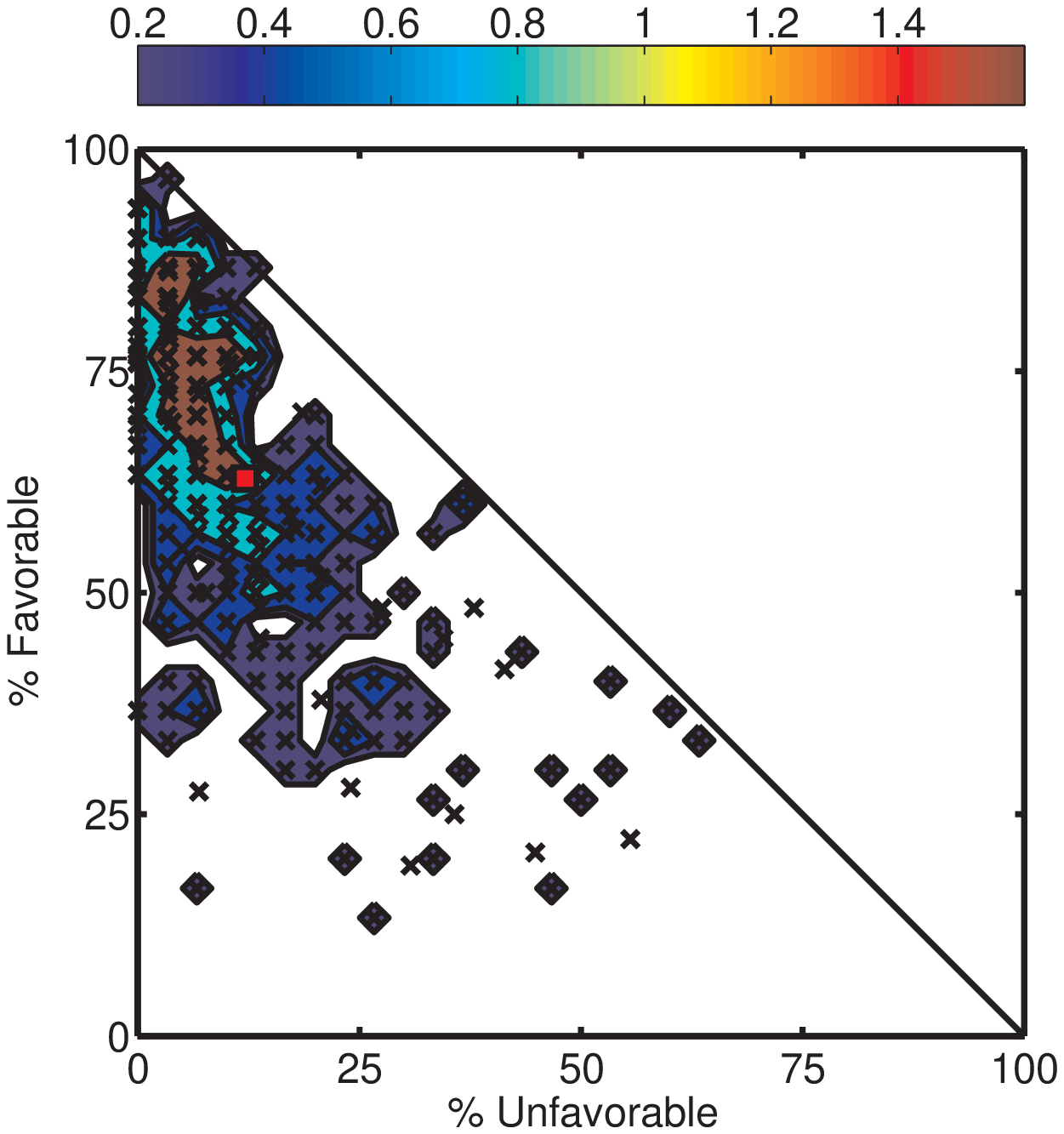}}                
  \subfigure[COMPASS distribution after instruction]{\label{fig:compass_post}\includegraphics[clip,trim=30mm 5mm 30mm 5mm,width=0.45\linewidth]{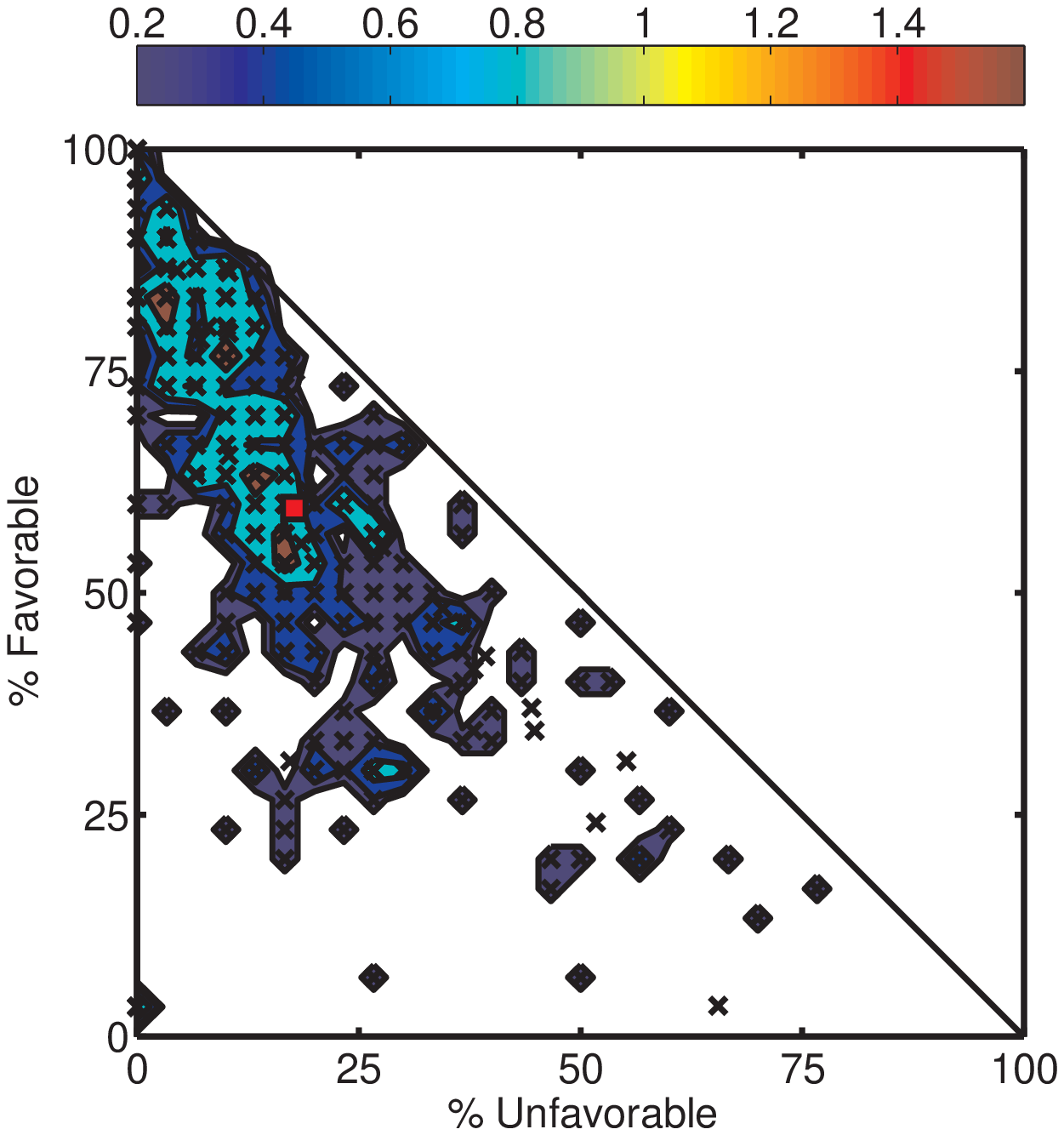}}
\end{center}
  \caption{[Color] - Non-honors students' ($N$ = 316) percentage of favorable and unfavorable responses to COMPASS statements given (a) before and (b) after instruction in a introductory calculus-based mechanics course which uses computer modeling homework (Ch. \ref{chap:comp}) are plotted (black x's). The distribution of responses in both figures is highlighted using a colored contour map of the percentage of students lying at each `x'. The mean percentages for both pre- and post-instruction COMPASS results are shown by a bold red square.}
  \label{fig:compass_prepost}
\end{figure}

The mean favorable score for non-honors mechanics students decreased from pre- to post-instruction while the mean unfavorable score increased from pre- to post; students responded less expert-like on the post-test. 
Students earned a mean favorable score of 62.8\% and an unfavorable score of 12.2\% on the pre-instruction COMPASS. 
On the post-instruction COMPASS, these scores shifted to 59.6\% and 17.7\%, respectively. 
The shifts in these scores were statistically significant.
The characteristic drop in students' expert-like responses after instruction on instruments such as the COMPASS is well-documented \cite{redish1998student,adams2006new}. This drop might stem from a number of issues including students who have a more idealistic view of computation upon entering the course, students who regard computation highly in principle but not in practice, {\it et cetera}. Mechanisms that underlie the drop in COMPASS scores could be investigated through interviews with students throughout the semester or by delivering the COMPASS throughout the semester to obtain a temporal profile.

Plotting students' percentage of unfavorable and favorable responses on a two dimensional grid permits one to visualize the distribution of students' COMPASS scores (Fig. \ref{fig:compass_prepost}). 
On this grid, the horizontal axis represents the percentage of unfavorable responses and the vertical, the percentage of favorable responses. 
A point in this space in then given by the pair of scores (e.g., $\langle$\% unfavorable, \% favorable $\rangle$). 
The plot is bounded by the negatively sloped line from the point $\langle$0, 100$\rangle$ to the point $\langle$100, 0$\rangle$ corresponding to all favorable or all unfavorable responses respectively.
A student's score might lie anywhere to the left and below this line. 
For any given student, either the vertical or horizontal distance to this line is her percentage of neutral responses. 
In Fig. \ref{fig:compass_pre}, the mean overall pre-instruction COMPASS score (bold red square) for non-honors students who took both the pre- and post-test (mean score, $\langle$12.2, 62.8$\rangle$) is plotted along with individual student scores (black x's). 
The density of student scores has been highlighted using a colored contour map. 
The approximate percentage of students lying at a single point in a given region is represented by a color from blue (0.2\%, N $\approx$ 0.6)  to red (1.6\%, N $\approx$ 5). 
Students' post-instruction COMPASS scores appear in Fig. \ref{fig:compass_post}. 
The mean overall post-instruction COMPASS score (bold red square) for non-honors students (mean score, $\langle$17.7, 59.6$\rangle$) is shown along with individual student scores (black x's). 
Again, the density of student scores is shown using a colored contour map. 

Typically, only the mean scores are shown in a plot like Fig. \ref{fig:compass_prepost} \cite{redish1998student,adams2006new} because comparisons are made between a number of different populations.
Doing so for a single population obscured interesting features of the distribution of student scores. 
In Fig. \ref{fig:compass_pre}, there was a large peak in scores (represented by the red color) in the region near $\langle$3, 77$\rangle$ while the mean score was closer to $\langle$12, 63$\rangle$. 
A broad and sparse tail below favorable scores of 50\% shifted the mean to lower favorable and higher unfavorable scores. 
Apparently, a large number of students had near expert-like responses on the COMPASS before instruction.
The minor shift down and to the right of the post-instruction mean score was the result of a larger rearrangement of the distribution of scores. 
Two peaks appeared on the post-test (Fig. \ref{fig:compass_post}); a more favorable peak centered near $\langle$10, 76$\rangle$ and a less favorable one centered near $\langle$16, 53$\rangle$. 
It appeared that after instruction individual students are subdivided into those whose responses remained roughly the same and others who selected less favorable responses.
The shift in the distribution of scores is highlighted in Fig \ref{fig:compass-mech-shift} in Appendix \ref{sec:app-add-figs}.

\subsection{Possible Influences of Students' Backgrounds on the COMPASS \label{sec:compass-inf}}

COMPASS scores might be distributed differently for students with different backgrounds;
the effect of background and performance in science appears on other epistemological assessments \cite{redish1998student,adams2006new}. 
An analysis of the variance (ANOVA) was performed to determine what elements of a student's background might affect her scores on the COMPASS. 
An ANOVA is a statistically robust method of making simultaneous comparisons of mean scores for different groups that are classified by a number of ``independent variables'' \cite{keppel1991design}. 
The independent variables in an ANOVA are possible influences (e.g., major, GPA, etc.) on the dependent variable for which the ANOVA is performed (e.g., COMPASS score).
Variations in the dependent variable can demonstrate that mean scores for some groups of (one or more) independent variables are statistically different. 
If the means of the dependent variable differ significantly between groups characterized by a single independent variable, 
this independent variable is a ``main effect''.
Hence, the mean of the dependent variable between one or more groups of that independent variable are statistically different. 
If there exist significant differences in the means of the dependent variable between groups characterized by combinations of (more than one) independent variables, these variables are said to be ``confounded''.
In a sense, the confounding of one or more variables in this way is because the variables were not inherently independent in the first place.
For our ANOVA, students' overall incoming grade point averages (GPAs), the grade they earned in the course, their classification and the college in which their declared major is offered were the independent variables.

Only a student's choice of college was observed as a main effect on her overall pre-instruction COMPASS scores.
For the ANOVA of pre-instruction COMPASS scores, 
students' course grades were not used because they had not yet completed the course.
Students majoring in computing (e.g., computer science and computational media) tended to have highest pre-instruction COMPASS scores followed by engineering students. 
Science (e.g., biology, chemistry, mathematics, etc.) students tended to earn the lowest pre-instruction scores. 
Students from other colleges (architecture, liberal arts and management) made up less than 5\% of the population and together earned roughly the same scores as engineering students. 
The effect of other independent variables (classification and GPA) was not significant nor were there any confounded variables.
The highly favorable scores achieved by computing students are not surprising given their strong interest in computer science. 

\begin{figure}[ht]
  \begin{center}
  \subfigure[Shift in COMPASS scores (Overall GPA)]{\label{fig:bygpa}\includegraphics[width=0.45\linewidth]{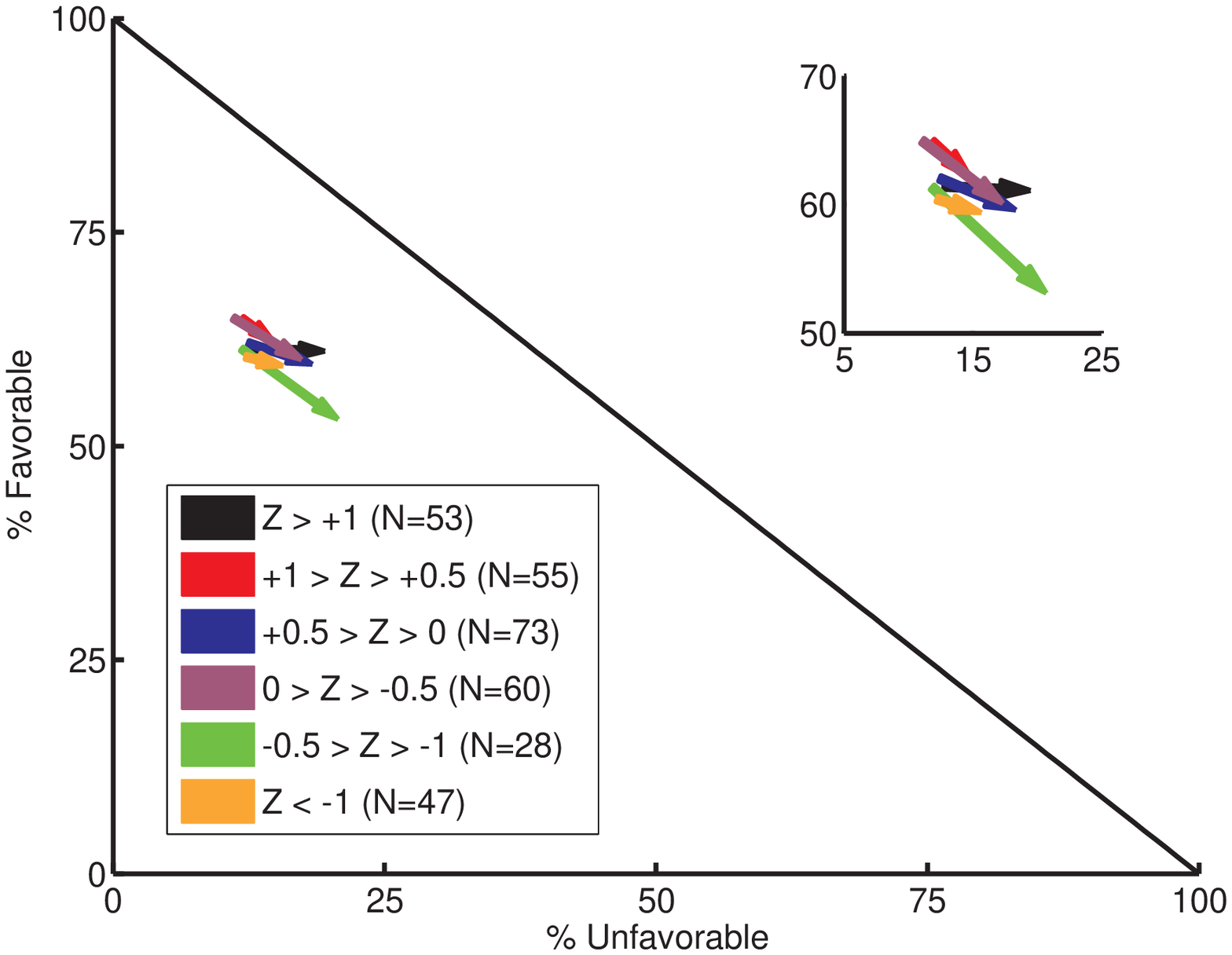}}                
  \subfigure[Shift in COMPASS scores (Course Grade)]{\label{fig:bygrade}\includegraphics[width=0.45\linewidth]{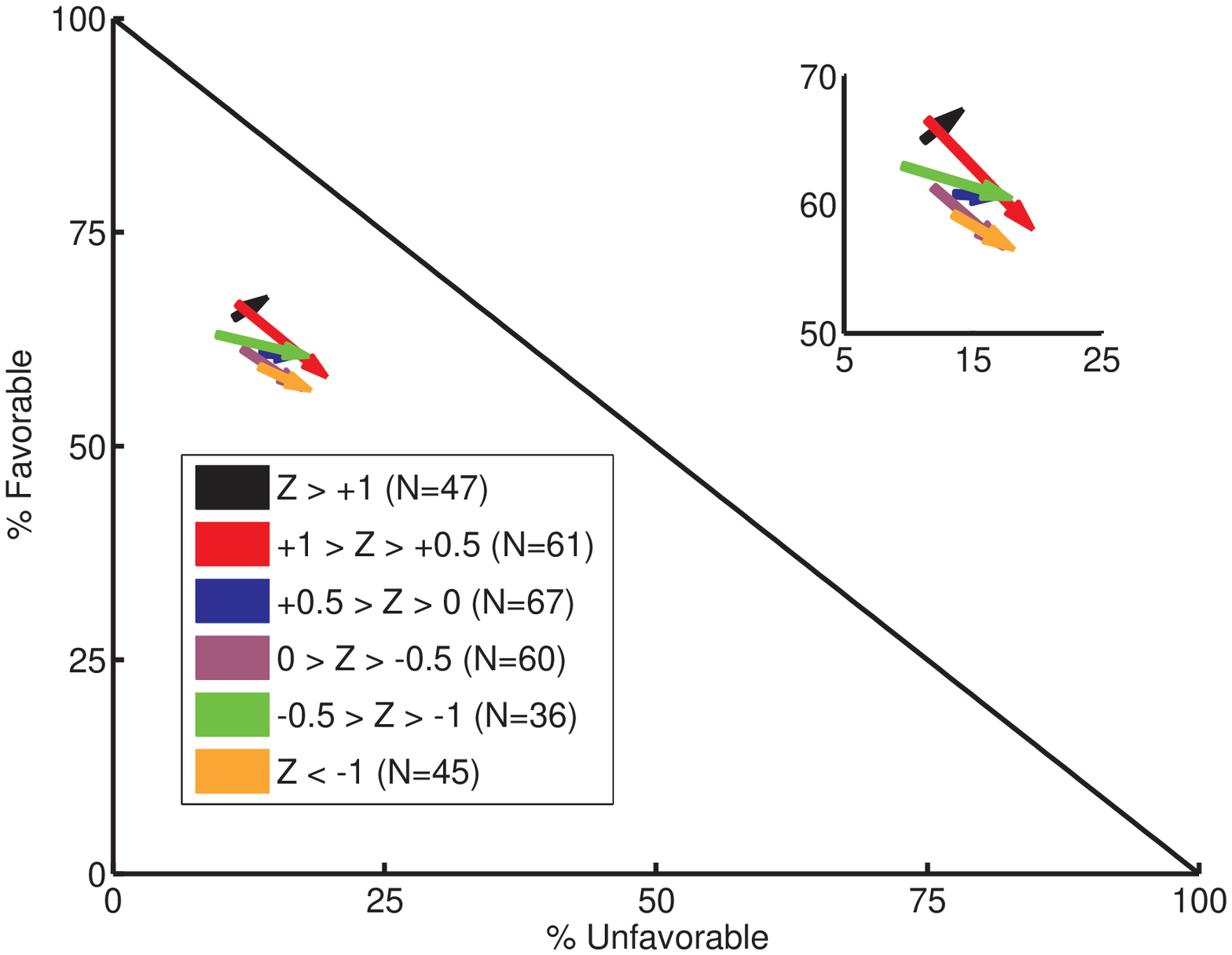}}\\
  \subfigure[Shift in COMPASS scores (Classification)]{\label{fig:byclass}\includegraphics[width=0.45\linewidth]{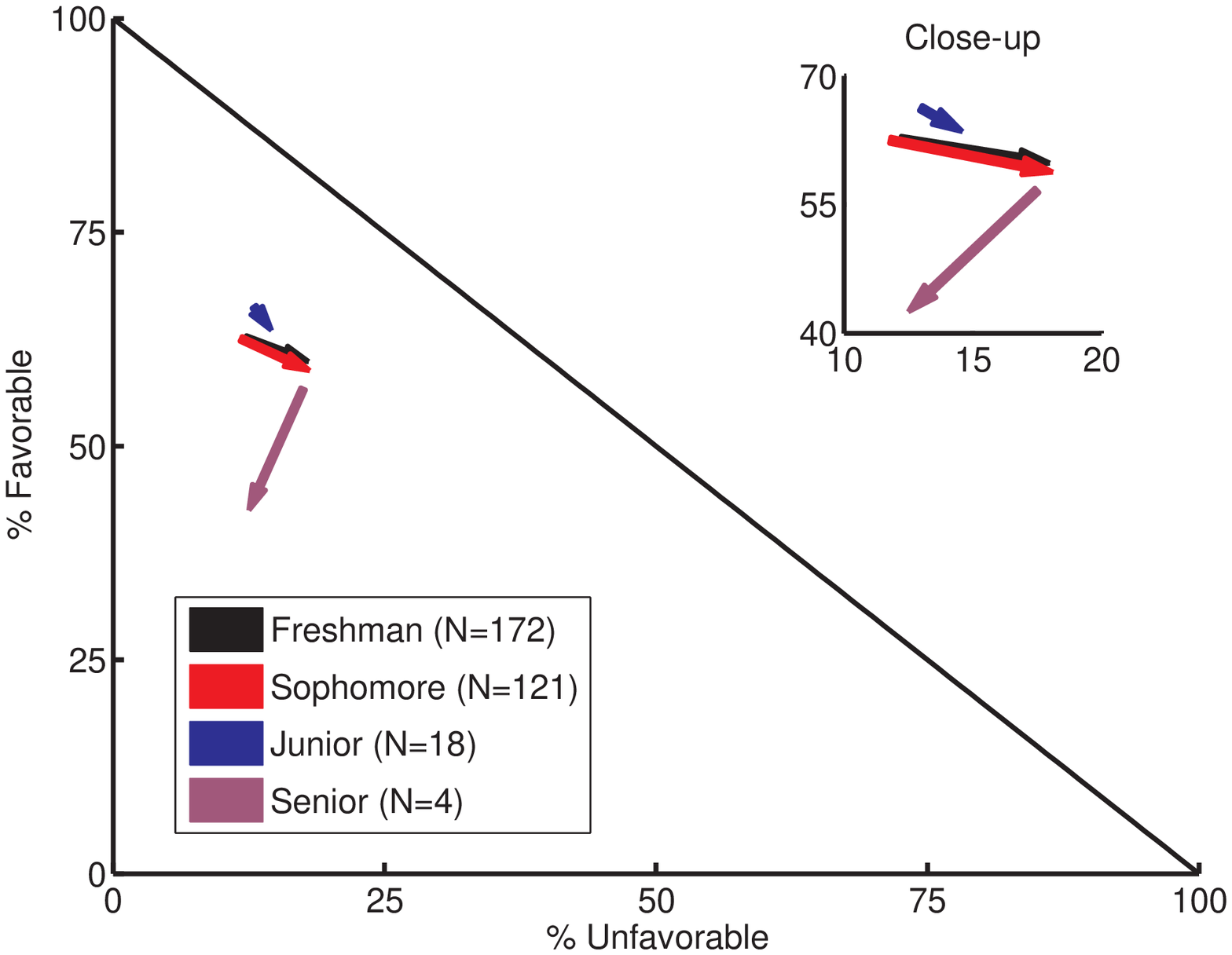}}
    \subfigure[Shift in COMPASS scores (College)]{\label{fig:bycollege}\includegraphics[width=0.45\linewidth]{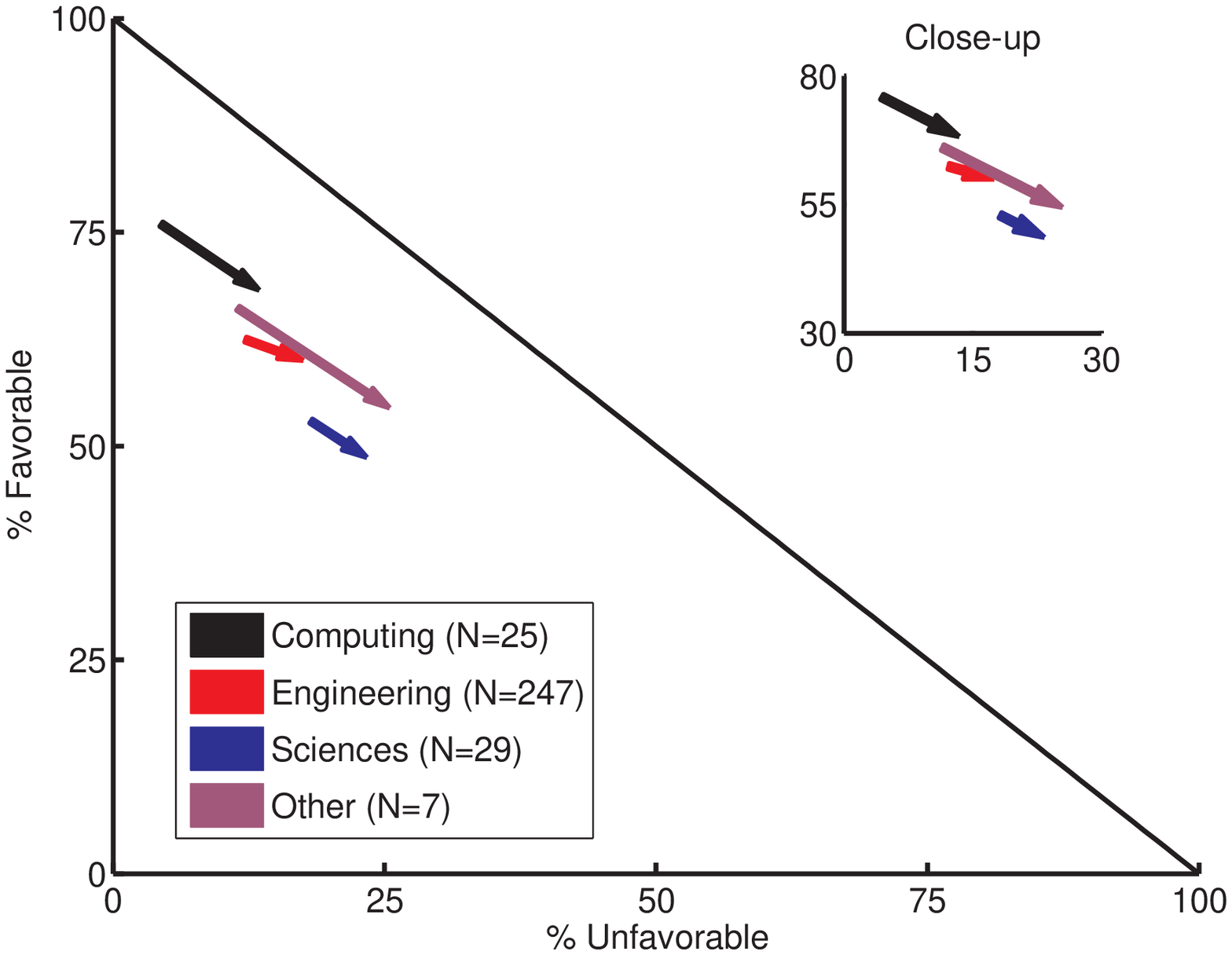}}
    \end{center}
  \caption{[Color] - The shift in the mechanics students' mean COMPASS scores are shown. Colored arrows indicate the magnitude and direction of the shift from pre- to post-instruction. Mean scores are shown for students based on: (a) their Z-scored (standard deviations from the mean) overall GPA, (b) their Z-scored (standard deviations from the mean) grade in the mechanics course, (c) their classification and (d) the college of their declared major college. Architecture, liberal arts and management majors (Other) are included for completeness but these students represented less than 5\% of the total population.}
  \label{fig:compass_dem}
\end{figure}

While students' pre-instruction COMPASS scores were the most significant effect on their post-instruction COMPASS performance, students' choice of major was also a main effect. 
We added students' final course grade and pre-instruction COMPASS scores as additional independent variables for the post-instruction ANOVA.
The trend by major of overall COMPASS post-instruction performance was similar to pre-instruction performance; computing students followed by engineering students followed by students in the sciences. 
Grouped together, students from other colleges earned similar post-instruction scores to engineering majors.
The effect of other independent variables (classification, GPA and course grade) was not significant nor were there any confounded variables.



In Fig. \ref{fig:compass_dem}, we have visually represented pre- and post-instruction scores as well as the shift in scores for students who took both the pre and post-instruction COMPASS. 
In these diagrams, students have been grouped by incoming GPA (Fig. \ref{fig:bygpa}), course grade (Fig. \ref{fig:bygrade}), classification (Fig. \ref{fig:byclass}) and college (Fig. \ref{fig:bycollege}). 
The tail of the arrows in these diagrams represent a group of students' mean pre-instruction COMPASS score; post-instruction mean scores are located at the arrow's tip. 
The arrow representation helps to visualize the shift in the mean towards or away from expert-like sentiments as well as demonstrate the change of neutral responses. 

An arrow pointed upward or to the left indicates an increase in expert-like sentiments. 
An arrow pointed to the right or downward indicates a decrease in expert-like sentiments. 
Neutral responses that shifted to favorable point upwards and unfavorable responses that became neutral point to the left. 
Neutral responses the shifted to unfavorable point to the right and favorable responses that shifted neutral point downwards.
An arrow parallel to the negatively sloped boundary indicates that the percentage of neutral responses remained unchanged, but the overall responses became less favorable (down and to the right) or more favorable (up and to the left).
Arrows perpendicular to this line indicates that responses became more (towards line) or less (away from line) polarized.

Most subgroups of students shifted from more favorable to less favorable responses by converting both neutral responses to unfavorable and favorable responses to neutral. 
Students could make the shift from favorable to unfavorables by flipping their response but this effect accounted for less than 3\% of all changes.
The vast majority of arrows shown in Figs. \ref{fig:bygpa} -- \ref{fig:bycollege} are roughly parallel to the boundary and point downward and to the right. 
Two arrows indicated quite different effects for two subgroups of students. 
Students who performed one standard deviation above the mean in the mechanics course (black arrow in Fig. \ref{fig:bygrade}) appeared to have slightly more polarized responses after instruction. 
In fact, these high performing students' pre- and post-instruction scores are statistically indistinguishable; these students maintained their sentiments.  
Senior students (purple arrow in Fig. \ref{fig:bycollege}) had far less favorable responses on the post-instruction survey. 
This might be an artifact of the small number of seniors in this second semester freshman level mechanics course.

Overall COMPASS scores indicated that students have less expert-like sentiments about computation after instruction. 
Some statements appeared to contribute more to the shift than others. 
For example, 74\% students responses shifted unfavorable on the post-test for the statement, "I do not spend more than 30 minutes stuck on a computer-modeling problem before giving up or seeking help from someone else."
On other statements, students sentiments became more favorable.
Favorable responses to the statement, "A significant problem in learning computer modeling is being able to memorize all the information I need to know.", increased by more than 30\% on the post-instruction COMPASS.
The shift in percentage of favorable responses for each scored statement from pre to post are summarized by Fig. \ref{fig:statementshifts} in Appendix \ref{sec:app-add-figs}.
It appeared that a number of statements on the COMPASS might be interrelated in some fashion.

\section{Searching for Robust Dimensions\label{sec:compass-statdim}}


The COMPASS might have underlying dimensions  that provide additional information about the mismatch between student and expert sentiments and changes in these sentiments after instruction in computation.
To uncover these dimensions, a pragmatic approach that balances the utility of selecting dimensions a priori and the statistical robustness of emergent categorization was taken. 
The pre-instruction responses of Georgia Tech mechanics students to all statements were investigated for correlations and groups of highly correlated statements informed the formation of several working categories. 
These categories were refined using the tools of emergent categorization, forming a new set of clearly interpretable yet statistically robust dimensions. This technique was borrowed from the designers of the CLASS \cite{adams2006new}.

\begin{figure}[t]
\begin{center}
\includegraphics[clip, trim = 15mm 0mm 25mm 1mm, width=0.95\linewidth]{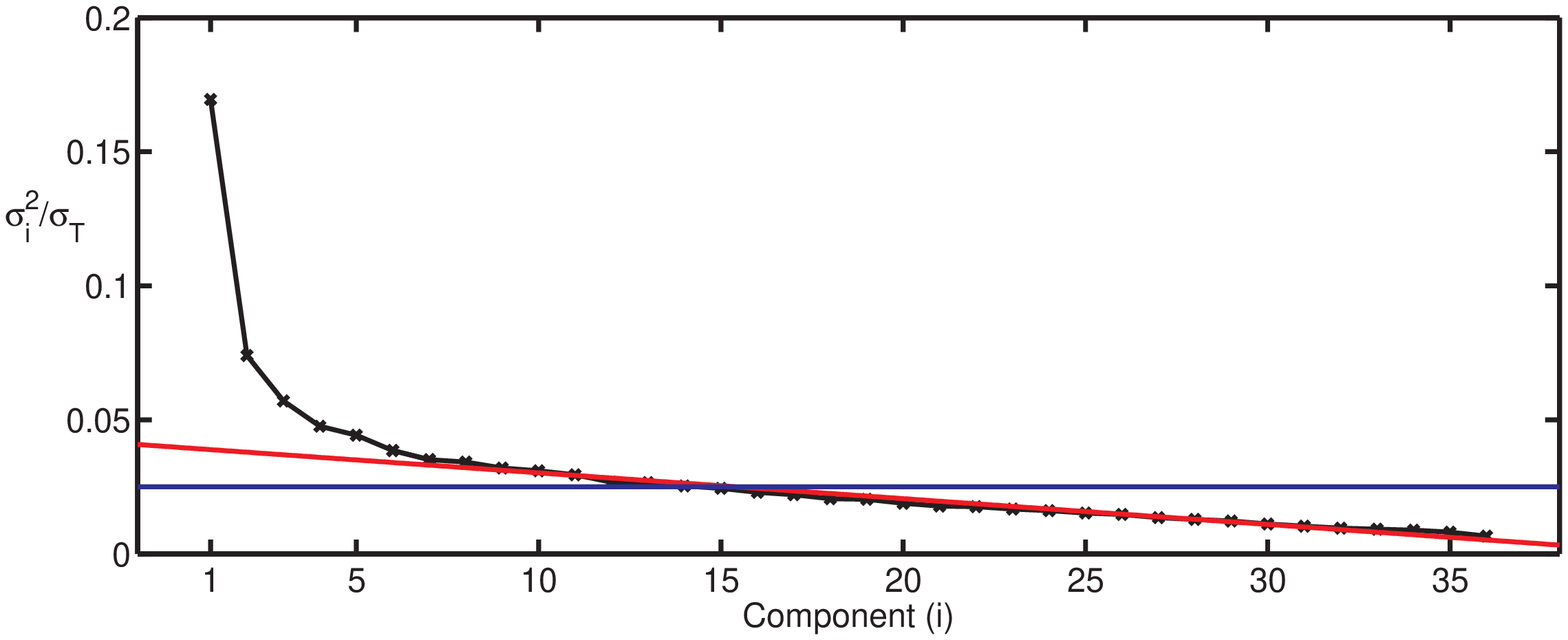}
\end{center}
\caption{[Color] - The fraction of the overall variance (eigenvalue) from an exploratory principal component analysis are plotted as a function of the extracted component (eigenvector). The shape of the diagram (scree plot) indicates there are at least six significant eigenvectors to consider. The first six eigenvectors account for more than 40\% of the total variance. The red line indicates a linear fit to the eigenvalues of the last 30 components (the scree). A good linear fit here indicates the rest of the components could be neglected. For this linear fit, the fraction of the variance accounted for by the fit is 99\% ($R^2$=0.992). An alternative method indicates that up to fourteen eigenvectors could be considered. The blue line indicates the level above which an eigenvector accounts for more than the average fractional variance.\label{fig:compass-pca}}
\end{figure}

An exploratory principle component analysis (PCA) can help determine the number of different dimensions of the COMPASS.
Principal component analysis is mathematical technique by which a set of measurements or observations that might be correlated are orthogonally transformed into a set of uncorrelated variables, ``principal components'' or eigenvectors.
These eigenvectors tend to describe some unique features about the data set.
PCA can be performed on the mean-centered data set (singular value decomposition) or on the correlation matrix of the data set (eigenvalue decomposition).
The extraction of the eigenvectors by the transformation is done in the order of the eigenvectors' eigenvalues, the amount of the variance in the original data set for which each eigenvector accounts.
That is, the first eigenvector has the largest eigenvalue, and hence, accounts for the most variance in the data set. 
The next accounts for the second most and so on.
The number of significant eigenvectors in a principle component analysis is usually much less than the total number of variables in the original data set.

The Georgia Tech mechanics students' pre-instruction responses were used in a PCA to determine the number of useful working categories.
We can use the fraction of the total variance (eigenvalue) accounted for by each of the components (eigenvector) to determine a range for the number of working categorizes.
In Fig. \ref{fig:compass-pca}, we plot the eigenvalues determined by the PCA in order of decreasing eigenvalue (fraction of the variance). 
Such a plot is a called a ``scree plot''. 
Some argue that the significant eigenvectors are those that account for more than the mean variance \cite{gorsuch1983factor}. 
In Fig. \ref{fig:compass-pca}, this value (one over the number of components) is indicated using a blue line. 
The fourteen eigenvectors with eigenvalues above this line account for roughly 67\% of the total variance. 
Others think that this technique overestimates the number of components to extract.
They recommend using the characteristic linear drop-off of eigenvalues (the near-linear slope after the scree) \cite{child2006essentials}. 
A linear fit to the lowest lying eigenvectors was performed repeatedly (adding the next eigenvector to the fit each time) until the coefficient of determination for the fit ($R^2$) was maximized in the region before the plot diverges. 
The linear fit (red line in Fig. \ref{fig:compass-pca}) indicated that as few as six eigenvectors were significant. 
These six eigenvectors accounted for approximately 42\% of the overall variance. 
Arguably, the number of significant eigenvectors and, hence, useful working categories should lie between six and fourteen.

The statements that were correlated strongly with each of the top fourteen eigenvectors were reviewed to form the working categories. 
Only the first six eigenvectors showed any coherence between their most highly correlated statements. 
Statements with neutral expert opinions were not among the highly correlated.
From the highly correlated statements, six working categories were deduced: (1) Perceived Ability, (2) Perceived Utility, (3) Real-World Connections, (4) Sophistication, (5) Personal Interest and (6) Learning.  
These categories were constructed to be useful to instructors (Sec. \ref{sec:compass-design}).
As an example, the statements that should appear in the Perceived Ability category as those for which students evaluated their own performance or skill in order to select a response.
Instructors might find scores on this subset to be useful feedback about students' confidence with using computation.
These six categories were also congruent with our original themes that computational instruction was meant to communicate (Sec. \ref{sec:compass-valid}). 
The aforementioned example is also related to the computational theme that all students can learn to use computation effectively.
A detailed discussion of each of these working categories is deferred to Appendix \ref{sec:compass-six}.

Two raters placed COMPASS statements into one or more of six working categories suggested by the PCA. 
Each performed his initial categorization individually without discussion with the other. 
Using a two-scale rating system, raters distinguished between statements that were representative of the category and those that could be. 
The raters discussed the results of their categorizations and any conflicts were resolved.
In the final version of the working categories, each category contained between three and eleven statements.

While each of the initial categories was practical, they were not necessarily statistically robust. 
Two qualities may be used determine robustness of a working category:
(1) statements in the category should be well correlated with all other statements
and
(2) categories should be well represented by a single eigenvector in a reduced basis PCA.
Four quantities were used to characterize these two qualities: the average absolute value of the linear correlation matrix for all the statements in the category ($\bar{r}$), the average absolute value of the linear correlation coefficients between each statement in the category and the first eigenvector produced from a reduced basis PCA ($\bar{l}$), the difference between the fraction of the variance attributed to the first and second eigenvectors and the average fractional drop between subsequent eigenvectors (normalized by number of statements, $|\Delta E|/N$) and the fraction of the variance accounted for by this linear fit to the scree ($R^2$).

\begin{table}[t]
\caption{Each of the six working categories were subjected to a reduced basis principal component analysis. The outcome of that analysis (Column rPCAi) suggested that two categories might be robust dimensions (PL), three were quite weak (WL) and one category might have multiple dimensions (ML). Statements were systematically added or removed from the categories and a new reduced basis PCA performed. The outcome of those results (Column rPCAf) revealed robust dimensions. Some dimensions contained roughly the same statements as the working categories (BQ). Others were formed from dissections of weak or multidimensional categories (NF). \label{tab:compass-dim}}
\begin{center}
\begin{tabular}{llll}
\multicolumn{1}{l}{\bf Working Category} & \multicolumn{1}{l}{\bf rPCAi} & \multicolumn{1}{l}{\bf Robust Dimension(s)} & \multicolumn{1}{l}{\bf rPCAf}\\\hline\hline
Perceived Ability & PL & Perceived Ability & BQ\\\hline
Perceived Utility & PL & Perceived Utility & BQ\\\hline
Real-World Connections & WL & Real-World Connections & BQ\\\hline
Sophistication & ML & Sense-making & NF\\
 &  & Expert Behaviors & NF\\
 &  & Avoiding Novice Behaviors & NF\\\hline
Personal Interest & WL & Personal Interest & BQ\\\hline
Learning & WL & Avoiding Rote & NF\\\hline\hline
\multicolumn{2}{p{0.375\linewidth}}{PL -- Properly loaded} & \multicolumn{2}{p{0.375\linewidth}}{ML -- Multiply loaded}\\
\multicolumn{2}{p{0.375\linewidth}}{WL -- Weakly loaded} & \multicolumn{2}{p{0.375\linewidth}}{NF -- Newly formed}\\
\multicolumn{4}{p{0.75\linewidth}}{BQ -- Better quality achieved by adding or removing statements}
\end{tabular}
\end{center}
\end{table}
Students' responses to statements in the working categories were subjected to a reduced basis PCA; categories with well correlated statements and represented by a single eigenvector were statistically robust. 
Such categories were considered underlying dimensions of the COMPASS.
The outcome of this analysis suggested that two categories might be robust dimensions (``properly loaded''), three were quite weak (``weakly loaded'') and one category might have multiple dimensions (``multiply loaded'').
Two of the six categories (Perceived Ability and Perceived Utility) were found to be strong single dimensions, but with slight modifications (adding or removing a statement) became stronger. 
The statements that were added were those that the raters had selectively removed in their earlier categorization. 
Three categories (Real-World Connections, Personal Interest and Learning) were weakly loaded.
Generally, this was because not all the statements in these categories were well correlated with each other and, hence, their principal eigenvector. 
Statements were removed or added to strengthen weak categories but still maintain an interpretation similar to the original working category. 
In the Learning category, a number of statements were removed which changed the meaning of the category altogether.
One category (Sophistication) appeared to have several strong eigenvectors.
This category was dissected and, ultimately, it was split into three robust dimensions.
Table \ref{tab:compass-dim} summarizes the results of the reduced basis PCA on the six working categories and shows from which categories the robust dimensions were formed.
In Table \ref{tab:compass-corr}, the values used to quantify the robustness of each dimension ($\bar{r}$, $\bar{l}$, $|\Delta E|/N$ and $R^2$) are reported.


\begin{table}[t]
\caption{For each of the eight robust COMPASS dimensions, we report average linear correlation component between all the statements ($\bar{r}$), the average linear correlation component between all the statements and the first eigenvector for the subset ($\bar{l}$), the difference between the fraction of the variance attributed to the first to second eigenvectors minus the average fractional drop between subsequent eigenvectors normalized by number of statements in the subset ($|\Delta E|/N$) and the fraction of the variance accounted for by a linear fit to the scree ($R^2$), the nearly linear drop off in variance attributed to the rest of the eigenvectors. \label{tab:compass-corr}}
\begin{center}
\begin{tabular}{l| c | c | c | c }
\multicolumn{1}{l}{{\bf COMPASS Dimension}} & \multicolumn{1}{c}{$\bar{\mathbf{r}}$} & \multicolumn{1}{c}{$\bar{\mathbf{l}}$} & \multicolumn{1}{c}{$|\Delta \mathbf{E}|/\mathbf{N}$} & \multicolumn{1}{c}{$\mathbf{R}^2$}\\\hline\hline
Perceived Ability & 0.26 & 0.30 & 0.16 & 0.98\\
Perceived Utility & 0.27 & 0.35 & 0.14 & 0.96\\
Real-World Connections & 0.35 & 0.43 & 0.11 & 0.99\\
Sense-making & 0.32 & 0.44 & 0.12 & 0.97\\
Expert Behaviors & 0.33 & 0.40 & 0.16 & 0.98\\
Avoiding Novice Behaviors & 0.34 & 0.41 & 0.16 & 0.96\\
Personal Interest & 0.39 & 0.44 & 0.18 & 0.99\\
Avoiding Rote & 0.35 & 0.50 & 0.10 & 0.97\\\hline\hline
\end{tabular}
\end{center}
\end{table}

Using this methodology, a total of eight practical and statistically robust dimensions were uncovered: (1) Perceived Ability, (2) Perceived Utility, (3) Real-World Connections, (4) Sense-making, (5) Expert Behaviors, (6) Avoiding Novice Behaviors, (7) Personal Interest and (8) Avoiding Rote. 
As the titles of the dimensions suggest, these dimensions characterize students' feeling towards learning and using computation as well as a self-evaluation of their abilities. 
Students' scores on each of these dimensions are useful to understand their confidence with using computation (Perceived Ability), their reasons for and interests in learning computation (Perceived Utility,  Real-World Connections and Personal Interests), how they characterize the efforts they put forth when learning computation (Sense-making and Avoiding Rote) and a self-evaluation of their own aptitude with computation (Expert Behaviors and Avoiding Novice Behaviors).
This purpose is quite different from the intent of other surveys in science whose objectives include what it means to have acquired knowledge in science and how that knowledge organized \cite{halloun1997views,redish1998student}.
A more detailed discussion of each dimension is reserved for Appendix \ref{sec:compass-eight} and the scree plots for each of the dimensions  (the visual representation of their robustness) appear as Fig. \ref{fig:compass-comp} in Appendix \ref{sec:app-add-figs}.

\begin{table*}[htp]
\begin{center}
\caption{Pre- and post-instruction COMPASS scores are reported for non-honors students ($N$ = 316) who took an introductory mechanics course. 
Scores are reported with a 95\% confidence interval estimated from the $t$-statistic in parentheses. 
Overall COMPASS scores for non-honors mechanics students were less favorable. 
Favorable post-instruction scores decreased on most dimensions but remained the same within error on Perceived Ability, Expert Behaviors and Avoiding Rote.
Unfavorable post-instruction scores increased on all dimensions except for Avoiding Rote which remained the same within error. 
\label{tab:compass-mech}}
\begin{tabular}{l|cc|cc}
\multicolumn{1}{c}{}& \multicolumn{2}{c}{{\bf PRE}} & \multicolumn{2}{c}{{\bf POST}}\\
\multicolumn{1}{c}{{\bf Dimension}}& \multicolumn{1}{c}{{\bf Favorable}} & \multicolumn{1}{c}{{\bf Unfavorable}} & \multicolumn{1}{c}{{\bf Favorable}} & \multicolumn{1}{c}{{\bf Unfavorable}}\\\hline\hline
Overall & 63 (2) & 12 (1) & 59 (2) & 18 (2) \\
Perceived Ability & 57 (2) & 14 (2) & 57 (3) & 19 (2) \\
Perceived Utility & 59 (2) & 13 (1) & 52 (3) & 22 (2) \\
Real-World Connections & 77 (2) & 8 (1) & 69 (3) & 13 (2) \\
Sense-making & 71 (2) & 8 (1) & 57 (3) & 16 (2)\\
Expert Behaviors & 53 (2) & 16 (2) & 55 (3) & 23 (2) \\
Avoiding Novice Behaviors & 67 (2) & 12 (2) & 61 (3) & 21 (2)\\
Personal Interest & 64 (3) & 12 (2) & 57 (3) & 21 (3) \\
Avoiding Rote & 57 (3) & 18 (2) & 58 (3) & 19 (2) \\\hline\hline
\end{tabular}
\end{center}
\end{table*}

\subsection{Measurements across Dimensions\label{sec:compass-dimscores}}

\begin{figure}[t]
  \begin{center}
  \includegraphics[clip,trim=0mm 5mm 0mm 5mm, width=0.60\linewidth]{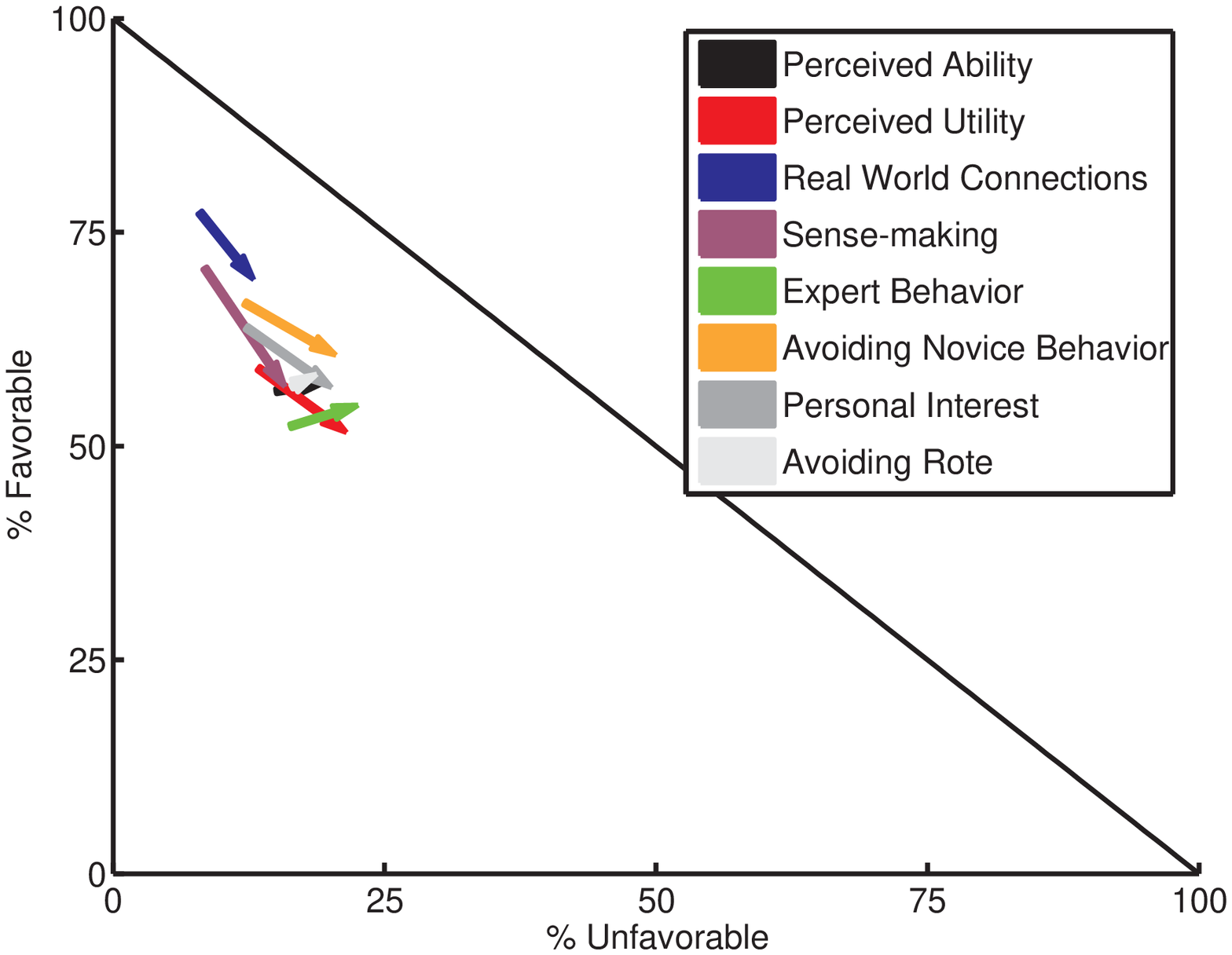}
\end{center}
  \caption{[Color] - The shift of students' percentage of favorable and unfavorable to subsets of COMPASS statements is shown for non-honors students taking an introductory calculus-based mechanics course which uses computer modeling homework. 
Students tend to shift away from expert opinion on all dimensions though three were not statistically significant (Perceived Ability, Expert Behavior and Avoiding Rote). 
Only scores from students who took both the pre- and post-instruction COMPASS were used.}
  \label{fig:compass_prepost_cat}
\end{figure}

Overall scores by non-honors mechanics students were less favorable after instruction; however, scores on COMPASS dimensions might improve, fall or remain the same.
Favorable scores on dimensions for which students evaluated their own confidence with using computation (Perceived Ability) or reported performing expert-like actions (Expert Behaviors) remained steady after instruction. 
Based on their favorable scores, students appear to prefer less effort devoted to making sense of problems (Sense-making) after instruction, but they reported the same effort devoted to rote memorization (Avoiding Rote).
Their interest in computation appeared to drop after instruction (Perceived Utility, Real-World Connections and Personal Interest).
These results are summarized in Table \ref{tab:compass-mech}.
We can also visualize changes to scores on COMPASS dimensions using arrow diagrams to follow the shift in the mean (Fig. \ref{fig:compass_prepost_cat}).

%% file: compass/03a-diffpop.tex
\section{Performance by Different Populations \label{sec:compass-pop}}

We have so far limited discussion to results from non-honors students who took an introductory calculus-based mechanics course at Georgia Tech. 
The COMPASS was also given to honors introductory calculus-based mechanics students, non-honors introductory calculus-based electromagnetism (E\&M) students at Georgia Tech and introductory calculus-based mechanics students at North Carolina State University (NCSU).
All courses taught computation as part of the laboratory activities \cite{beichner2010labs}; students did not solve computational homework problems (Ch. \ref{chap:comp}) either in the E\&M course or in the course taught at NCSU.
The populations highlight differences due to self-identification (honors mechanics), instruction (E\&M) and entirely different student populations (NCSU).

\subsection{Honors Mechanics Students at Georgia Tech \label{sec:compass-honors}}

The COMPASS was administered to a small number of honors students who were enrolled in a separate introductory mechanics section.
After the filtering process, 36 honors students took the pre-instruction COMPASS, 21 took both the pre- and post-test.
This section met with the same lecture section as a much larger non-honors section. 
Honors students received identical instruction in computation and solved the same computational homework sets (Ch. \ref{chap:comp}) as their non-honors classmates.
In fact, coursework, including exams, across all the mechanics sections was identical.  
Yet, honors students had more favorable responses to the COMPASS after instruction than their non-honors classmates.

\begin{figure}[t]
  \begin{center}
  \includegraphics[clip,trim=0mm 5mm 0mm 5mm,width=0.60\linewidth]{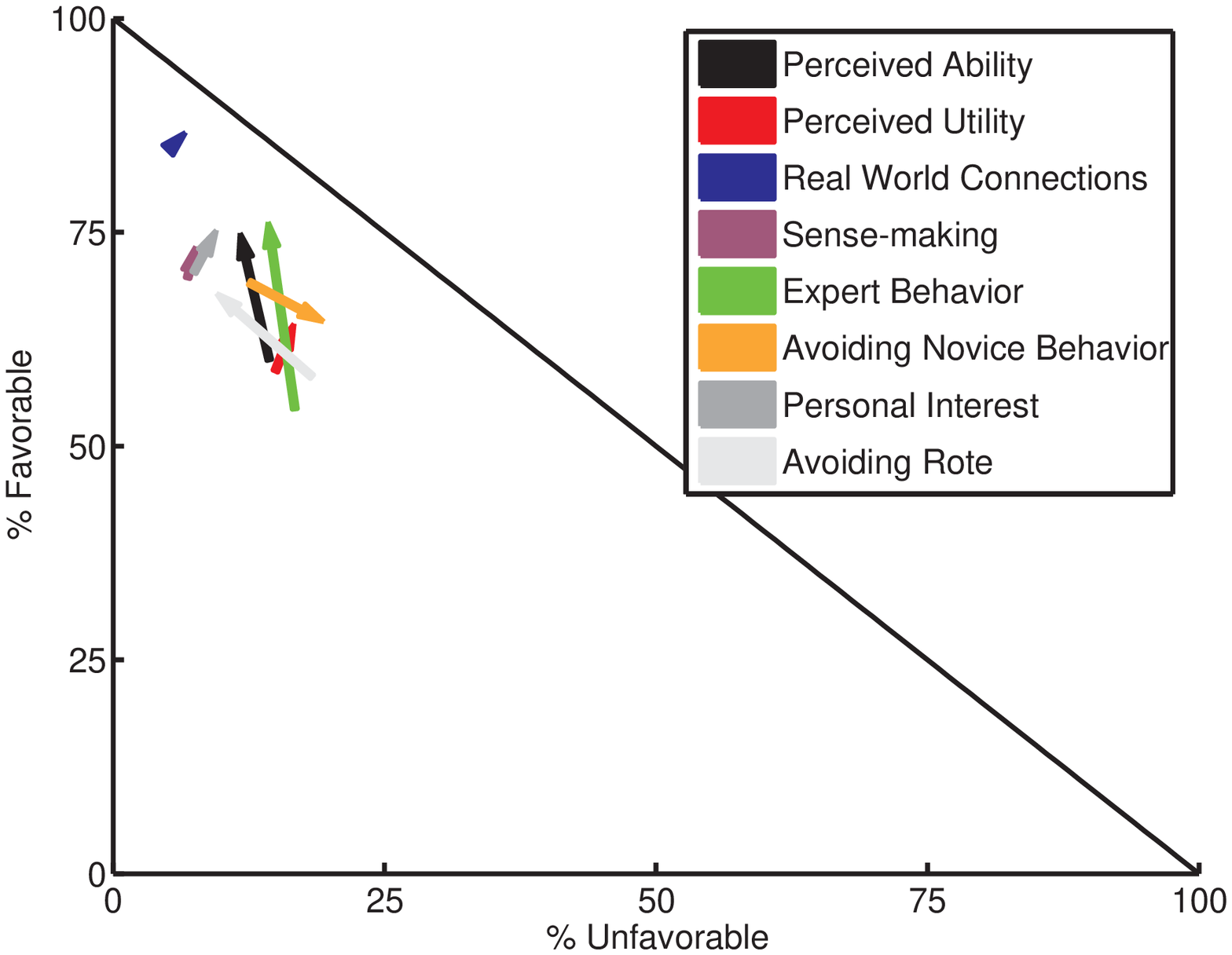}
\end{center}
  \caption{[Color] - The shift of students' percentage of favorable and unfavorable to subsets of COMPASS statements is shown for honors students taking an introductory calculus-based mechanics course which uses computer modeling homework. 
Honors students maintained their scores on all dimensions though on two dimensions were significantly more favorable (Perceived Ability and Expert Behaviors). 
Only scores from students who took both the pre- and post-instruction COMPASS were used.}
  \label{fig:compass_prepost_cat_h}
\end{figure}

COMPASS scores both overall and on individual dimensions appear to be affected by self-identification. 
Honors students achieved pre-instruction percent favorable scores overall and on each dimension were statistically indistinguishable between non-honors and honors students.
After instruction, their post-instruction scores were more favorable than their non-honors classmates on all dimensions except Avoiding Novice Behaviors and Avoiding Rote, which were identical to non-honors students.
After instruction, honors students appeared to exhibit more confidence (Perceived Ability) and report to perform more expert-like behaviors (Expert Behaviors). 
However, this does not mean that honors students abandoned novice like behaviors; scores on the Avoiding Novice Behaviors dimension remained the same.
Honors students maintained their level of interest (Perceived Utility, Real-World Connections and Personal Interest) and the effort they will put forth to learn computation by making sense of problems or memorizing by rote (Sense-making and Avoiding Rote).
These shifts are summarized in Fig. \ref{fig:compass_prepost_cat_h}

Such differences between non-honors and honors students did not arise as a result of differences between instructors or student demographics. 
Honors students were taught in the same section by the same instructor as some non-honors students. 
A comparison by instructor, using a Kruskall-Wallis test \cite{nonparabook}, indicated no difference in the mean percent favorable and unfavorable scores among non-honors students in the three sections. 
We have observed that some demographic factors influence COMPASS scores (Fig. \ref{fig:compass_dem}).
However, honors students had incoming GPAs and outgoing course grades that were statistically indistinguishable from non-honors students.
Furthermore, we observed no association between choice of major or classification and honors status using a contingency table analysis \cite{numericalrecipes,bevington}.
The honors section was composed of essentially the same (academically) student population as other sections. 
It is possible that the more favorable responses by honor students on the post-instruction COMPASS might stem from their personal identification as honor students, additional research experience or, perhaps, some other experience outside the classroom.
Interviews with honors and non-honors students should be carried out to determine the source of the differences in epistemology.

\subsection{Electromagnetism Students at Georgia Tech \label{sec:compass-em}}

We administered the COMPASS to two sections of noon-honors introductory E\&M students at Georgia Tech taught by two different instructors to a total of 364 students. 
E\&M students' experience with computation was limited to the laboratory; they solved no computational homework problems (Ch. \ref{chap:comp}).
The percentage of faithful E\&M respondents was much lower than mechanics students; $\sim$20\% of students did not read the statements carefully.
After the filtering process, we were left with 293 students who took the pre-instruction COMPASS, 241  who took the post-instruction and 238 who responded to both.

The COMPASS is a valid instrument for Georgia Tech E\&M students; most COMPASS dimensions remained robust for this population.
A reduced basis PCA of E\&M students pre-instruction COMPASS responses showed that most of COMPASS dimensions described in Sec. \ref{sec:compass-statdim} were still robust.
The values of the metrics used to measure robustness ($\bar{r}$, $\bar{l}$, $\Delta |E|/N$ and $R^2$) were somewhat different from those presented in Table \ref{tab:compass-corr}, but still indicated robustness for six dimensions.
The Personal Interest and Avoiding Rote dimensions appeared less robust than they did for mechanics students. 
Statements in these dimensions could be revisited and, perhaps, reworded and retested. 
However, as we note in Sec. \ref{sec:compass-ncsu}, it is more important to collect responses from students with varying backgrounds before constructing dimensions.
Scree plots that summarize these results are reserved for Appendix \ref{sec:app-add-figs} (Fig. \ref{fig:compass-comp-em}).

\begin{figure}[t]
  \begin{center}
  \subfigure[COMPASS distribution before instruction]{\label{fig:em_pre}\includegraphics[clip,trim=30mm 5mm 30mm 5mm, width=0.45\linewidth]{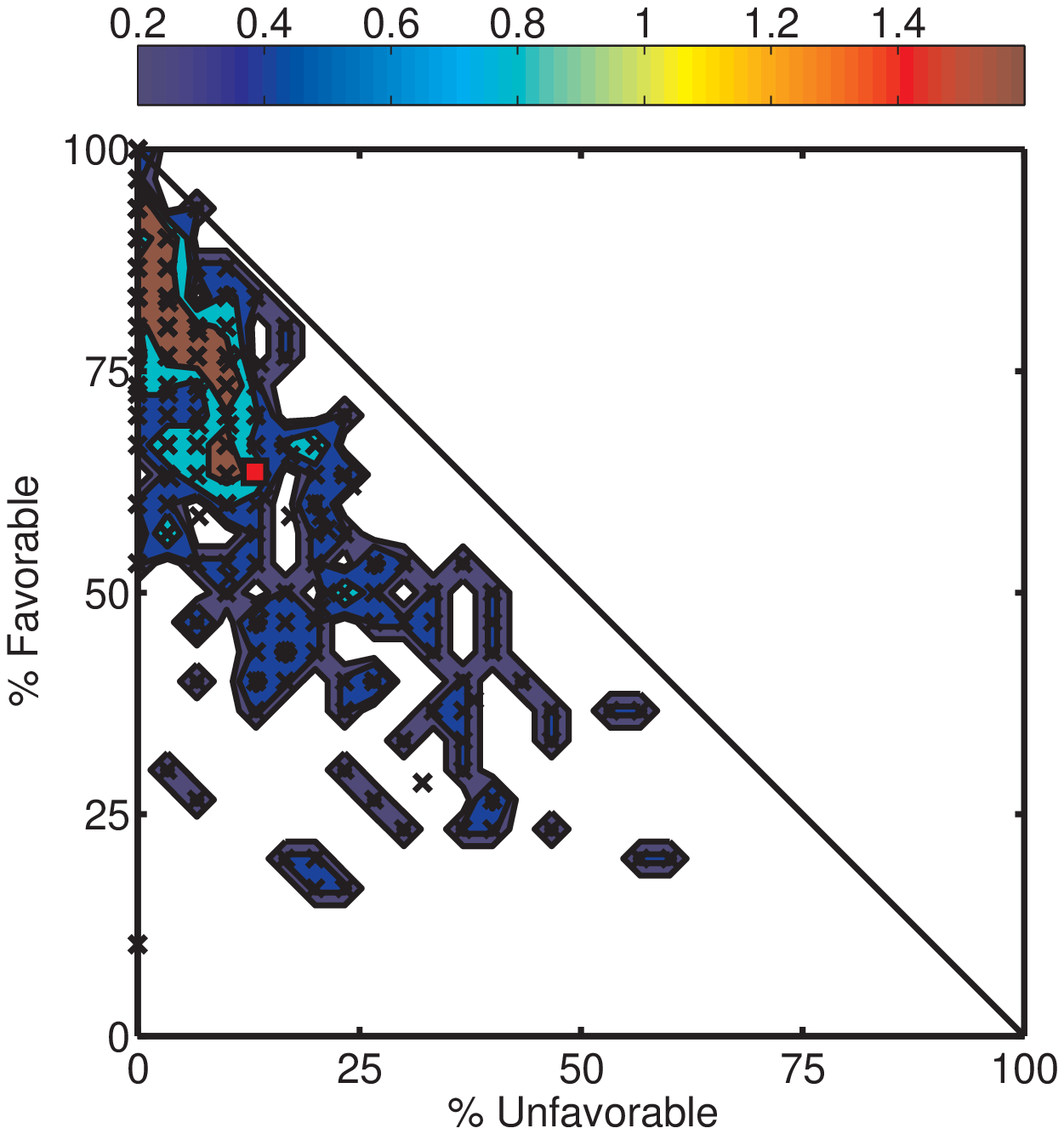}}                
  \subfigure[COMPASS distribution after instruction]{\label{fig:em_post}\includegraphics[clip,trim=30mm 5mm 30mm 5mm,width=0.45\linewidth]{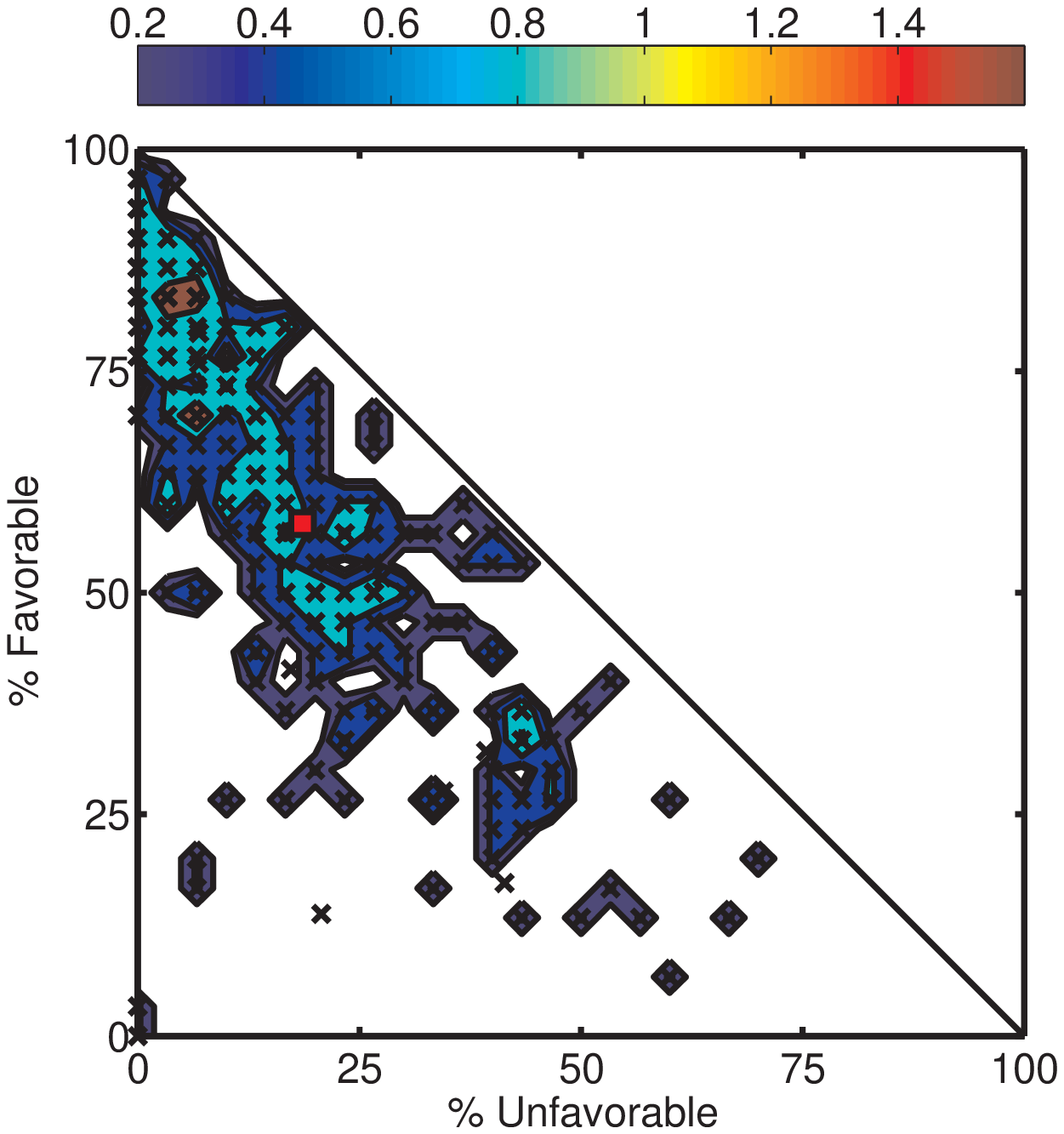}}
\end{center}
  \caption{[Color] - Students' ($N$ = 238) percentage of favorable and unfavorable responses to COMPASS statements given (a) before and (b) after instruction in a introductory calculus-based electromagnetism course at Georgia Tech are plotted (black x's). The distribution of responses in both figures is highlighted using a colored contour map of the percentage of students lying at each `x'. The mean percentages for both pre- and post-instruction COMPASS results are shown by a bold red square.}
  \label{fig:em_prepost}
\end{figure}


It appears that the differences in instruction, namely, mechanics students' computational homework problems (Ch. \ref{chap:comp}), might have a small negative effect on some COMPASS dimensional scores and no effect on others. 
E\&M students mean overall COMPASS scores were statistically indistinguishable from Georgia Tech non-honors mechanics students. 
E\&M students earned pre-instruction favorable and unfavorable COMPASS scores of 63.6\% and 13.2\% (red bold square, Fig. \ref{fig:em_pre}).
After instruction, mean COMPASS scores were less expert-like (red bold square, Fig. \ref{fig:em_post}), 57.8\% and 18.6\% for percentage favorable and unfavorable respectively.
The shift of E\&M students' mean scores was significant.
E\&M students' distribution of COMPASS scores appear quite similar their mechanics colleagues before and after instruction (Fig. \ref{fig:em_prepost}).
Both had a number of students with expert-like responses before instruction, peaked around $\langle$15, 76$\rangle$ for E\&M students, and a long sparse tail of students with more novice-like responses (Fig. \ref{fig:em_pre}).
On the post-instruction COMPASS, E\&M students have appeared to split into two groups as did mechanics students. For E\&M students, one peaked around $\langle$10, 80$\rangle$ and the other was closer to the mean, $\langle$16, 60$\rangle$.
The shift in percentage of favorable responses for each scored statement from pre to post are summarized by Fig. \ref{fig:emshift} in Appendix \ref{sec:app-add-figs}.

Dimensional scores for E\&M students generally became less favorable.
Post-instruction favorable scores on three dimensions (Perceived Ability, Expert Behaviors and Avoiding Rote) remained at their pre-instruction levels.
One the other five dimensions, E\&M students' achieved lower favorable scores.
Post-instruction unfavorable scores were higher across all dimensions expect Perceived Ability, Real-World Connections and Avoiding Rote which remained the same after instruction.
E\&M students' performance is summarized by Table \ref{tab:compass-em}.

Most E\&M students' dimensional scores were statistically similar to those achieved by Georgia Tech mechanics students both before and after instruction.
However, Georgia Tech mechanics students had a significantly larger shift than E\&M students on the unfavorable scores for three dimensions (Perceived Ability, Personal Interest and Expert Behaviors).
The absence of a positive effect due to mechanics students additional instruction in computation might not be surprising given considering the nature of that instruction; mechanics students' computational homework is prescriptive rather than experiential.
Generally speaking, improved favorable scores on instruments such as the COMPASS have been affected by content delivery methods such as facilitating experiential learning \cite{laws1991workshop,redish1998student}.


\begin{table*}[t]
\begin{center}
\caption{Pre- and post-instruction COMPASS scores are reported for non-honors students ($N$ = 238) who took an introductory electromagnetism (E\&M) course at Georgia Tech. 
Scores are reported with a 95\% confidence interval estimated from the $t$-statistic in parentheses. 
Overall COMPASS scores for E\&M students were less favorable. 
Favorable post-instruction scores decreased on most dimensions but remained the same within error on Perceived Ability, Expert Behaviors and Avoiding Rote. 
Unfavorable post-instruction scores increased on most dimensions but remained the same with error on Perceived Ability, Real-World Connections and Avoiding Rote. 
\label{tab:compass-em}}
\begin{tabular}{l|cc|cc}
\multicolumn{1}{c}{}& \multicolumn{2}{c}{{\bf PRE}} & \multicolumn{2}{c}{{\bf POST}}\\
\multicolumn{1}{c}{{\bf Dimension}}& \multicolumn{1}{c}{{\bf Favorable}} & \multicolumn{1}{c}{{\bf Unfavorable}} & \multicolumn{1}{c}{{\bf Favorable}} & \multicolumn{1}{c}{{\bf Unfavorable}}\\\hline\hline
Overall & 64 (2) & 13 (1) & 58 (3) & 19 (2) \\
Perceived Ability & 55 (3) & 18 (2) & 55 (3) & 21 (2) \\
Perceived Utility & 57 (3) & 16 (2) & 50 (3) & 22 (3) \\
Real-World Connections & 74 (3) & 9 (2) & 66 (4) & 13 (3) \\
Sense-making & 69 (3) & 10 (2)& 54 (4) & 17 (3) \\
Expert Behaviors & 51 (3) & 21 (3) & 53 (4) & 26 (3) \\
Avoiding Novice Behaviors & 69 (3) & 12 (2) & 60 (3) & 21 (3) \\
Personal Interest & 61 (4) & 16 (3) & 53 (4) & 22 (3)\\
Avoiding Rote & 61 (3) & 15 (2) & 59 (4) & 17 (3) \\\hline\hline
\end{tabular}
\end{center}
\end{table*}

While students' performance between E\&M and mechanics students is fairly similar, the influences on E\&M students' COMPASS scores are different from their mechanics classmates.
This is likely because the population of students taking sophomore-level E\&M is somewhat different.
Nearly all students at Georgia Tech are required to take introductory calculus-based mechanics. 
By contrast, introductory E\&M is only required by some engineering departments; although, many non-science students choose to take it as their second required science credit. 
Furthermore, introductory E\&M is typically a sophomore-level course. 
Freshman students taking this course have typically tested out of introductory mechanics and are usually strong students.
A contingency table analysis (Appendix \ref{sec:ctables}) confirms that the distributions of E\&M students by major and classification are statistically different from mechanics course.
The overall GPA of students taking the E\&M is statistically indistinguishable from mechanics students.
However, the mean outgoing course grade for E\&M is lower.

An ANOVA of pre-instruction COMPASS performance by E\&M students found scores were strongly dependent on a student's choice of major and, to a lesser extent, on her overall GPA. 
This is different from mechanics students in which major was the only main effect (Sec. \ref{sec:compass-inf}).
Students taking E\&M might not have lower GPAs, but their GPAs are more indicative of their performance at Georgia Tech. 
Many students taking introductory mechanics courses are freshman with a significant amount of Advanced Placement credit which is calculated into their GPA.
COMPASS performance after instruction was found to depend strongly on pre-instruction scores, but course grade, major, GPA and classification, in that order, were also significant main effects. 
Classification and GPA are confounded variables indicating that GPA and classification are interrelated. 
Indeed, we find that freshman taking E\&M tend to have higher GPAs than other students and seniors tend to have lower ones.
Hence, course grade and college are the two main demographic effects on post-instruction COMPASS performance.
Students who perform better in the course and those in majoring in computing tend to have higher post-instruction scores. 
Those who perform poorly and those majoring in the sciences tend to have lower post-instruction scores.
This is not very different from mechanics students; the trend of students' performance by major was quite similar (Fig. \ref{fig:bycollege}) and students who performed extremely well in the mechanics courses appeared to maintain their pre-instruction COMPASS scores (Fig. \ref{fig:bygrade}). 
These results are summarized by arrow diagrams (Fig. \ref{fig:compass_dem-em}) in Appendix \ref{sec:app-add-figs}.


\subsection{Mechanics Students at North Carolina State University  \label{sec:compass-ncsu}}

At North Carolina State University (NCSU), a large enrollment engineering university, the COMPASS was administered to two sections of introductory calculus-based mechanics sections taught by two different instructors to a total of 243 students. 
NCSU mechanics students experience with computation was limited to the laboratory.
After filtering roughly $20$\% of the respondents out, we were left with 198 students who took the pre-instruction COMPASS, 180 who took the post-instruction and 164 who took both.

The COMPASS appears to still be a valid overall instrument for comparing NCSU students' thoughts on computation.
However, NCSU students selected neutral responses more often than Georgia Tech students on almost all statements.
On some dimensions with a significant mismatch in the number of neutral responses between NCSU and Georgia Tech students (e.g., Personal Interest, Expert Behaviors and Avoiding Rote), the dimensions were weakly loaded (Fig. \ref{fig:compass-comp-ncsu} in Appendix \ref{sec:app-add-figs}).
The weak loading of certain COMPASS dimensions might bring into question their robustness.
However, this predilection to select neutral responses is likely a reflection of the differences in the conservatism of the two populations of students \cite{halloun2001student,lee2002cultural}.
Arguably, the best methodology for uncovering these dimensions is using data collected from a variety of students with differing backgrounds and experiences.
A check of the robustness of dimensions using all available data (Fig. \ref{fig:compass-comp-all}) found that robustness was generally preserved. 

\begin{figure}[t]
  \begin{center}
  \subfigure[COMPASS distribution before instruction]{\label{fig:ncsu_pre}\includegraphics[clip,trim=30mm 5mm 30mm 5mm, width=0.45\linewidth]{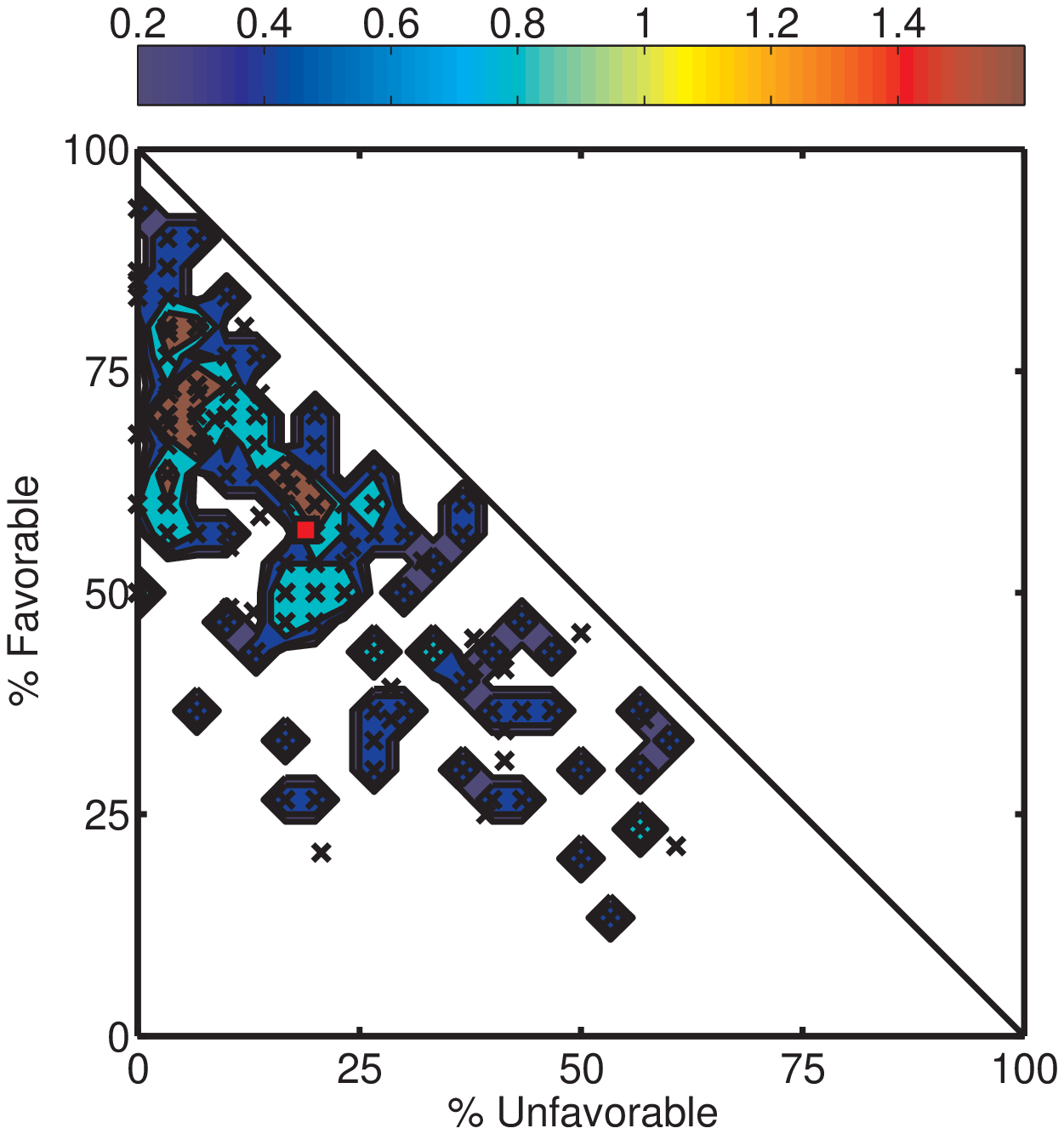}}                
  \subfigure[COMPASS distribution after instruction]{\label{fig:ncsu_post}\includegraphics[clip,trim=30mm 5mm 30mm 5mm,width=0.45\linewidth]{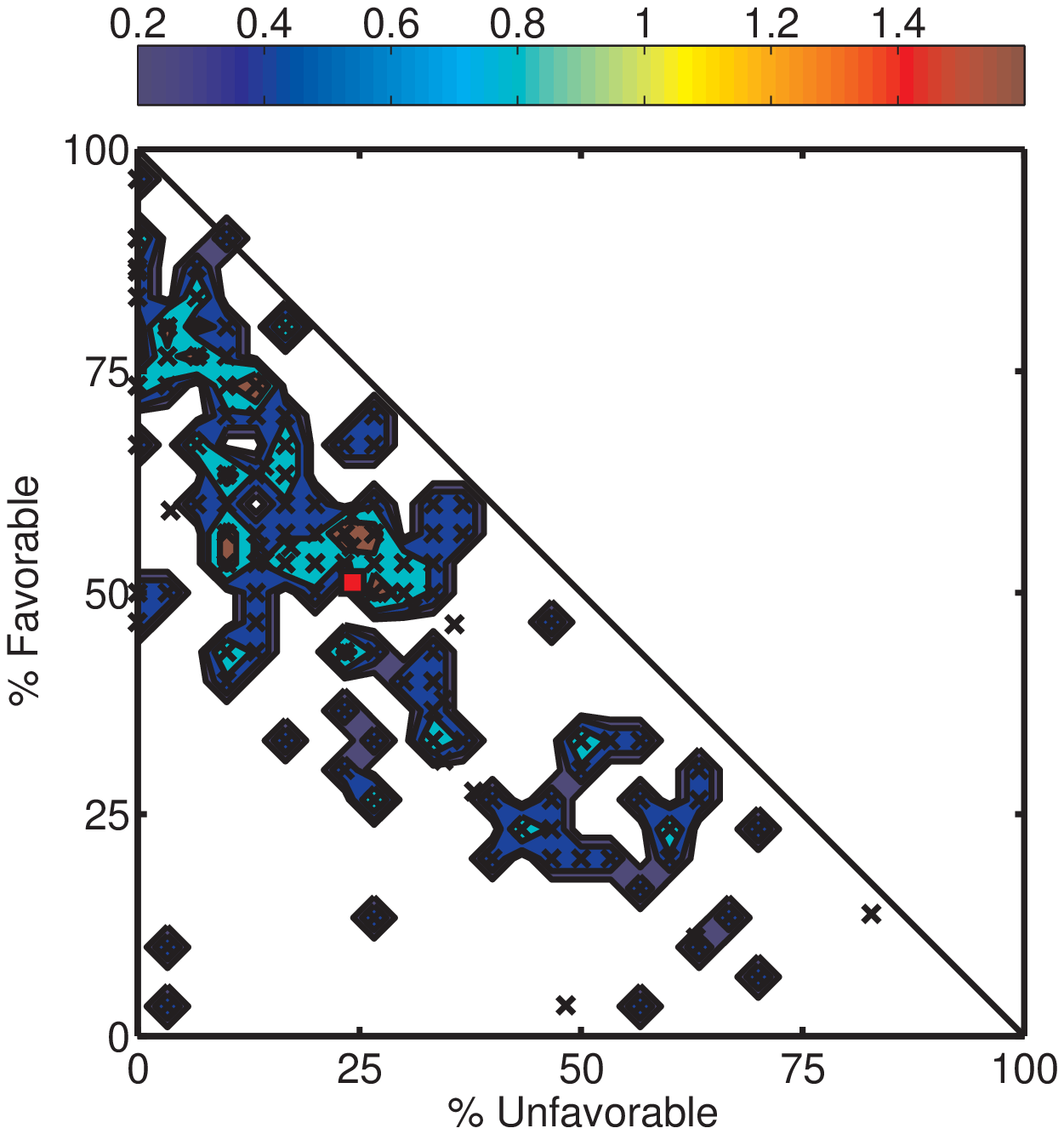}}
\end{center}
  \caption{[Color] - Students' ($N$ = 164) percentage of favorable and unfavorable responses to COMPASS statements given (a) before and (b) after  instruction in a introductory calculus-based mechanics course at NCSU are plotted (black x's). The distribution of responses in both figures is highlighted using a colored contour map of the percentage of students lying at each `x'. The mean percentages for both pre- and post-instruction COMPASS results are shown by a bold red square.}
  \label{fig:ncsu_prepost}
\end{figure}


NCSU mechanics students' mean overall COMPASS scores were less favorable than non-honors mechanics students at Georgia Tech, possibly because of academic differences.
NCSU students earned pre-instruction favorable and unfavorable COMPASS scores of 
57.1\% and 18.9\% (bold red square in Fig. \ref{fig:ncsu_pre}). 
After instruction, mean COMPASS scores were significantly less expert-like (bold red square, Fig. \ref{fig:ncsu_post}), 51.1\% and 24.5\% for percentage favorable and unfavorable respectively.
The distribution of NCSU scores appear to be more sparsely populated than those of their Georgia Tech colleagues both before and after instruction.
This might be an effect of fewer students responding to the survey or, perhaps, less diverse responses. 
Additional data must be collected to investigate this observation.
The small perpendicular shift of the distribution away from the boundary line in Figs. \ref{fig:ncsu_pre} \& \ref{fig:ncsu_post} indicated that NCSU mechanics students' responses tended to be less polarized than responses from Georgia Tech students.
However, some structural elements of the shift in post-instruction distributions observed for both Georgia Tech mechanics students (Fig. \ref{fig:compass_post}) and E\&M students (Fig. \ref{fig:em_post}) appeared for NCSU students. 
Two peaks of students appear in Fig. \ref{fig:ncsu_pre} near $\langle$6, 73$\rangle$ and $\langle$17, 63$\rangle$.
The more favorable of these peaks is similar to those observed in Georgia Tech, but the other appears to stem from NCSU students' responding with more neutrals than Georgia Tech students. 
After instruction, three peaks are visible in Fig. \ref{fig:ncsu_post}; a favorable one (near $\langle$13, 73$\rangle$) and two less favorable ones.
One of the less favorable peaks (near $\langle$20, 56$\rangle$) is similar to those observed in Georgia Tech post-instruction data.
The other (near $\langle$10, 56$\rangle$) has significantly more neutral responses.
The shift in percentage of favorable responses for each scored statement from pre to post are summarized by Fig. \ref{fig:ncsushift} in Appendix \ref{sec:app-add-figs}.
NCSU students' dimensional scores tended to be significantly less favorable than Georgia Tech mechanics students both before and after instruction.
On all dimensions expect Perceived Utility, NCSU students have less expert-like responses than Georgia Tech students. 
However, students from NCSU have more expert-like responses to statements concerning the utility of computation both on the pre- and post-instruction COMPASS. 
Their performance is summarized by Table \ref{tab:compass-ncsu}.

The less favorable performance by NCSU students might stem from the differences in academic preparation between students at Georgia Tech and those at NCSU. 
Favorable scores on epistemological instruments for science scale with preparation in STEM courses \cite{redish1998student,adams2006new}.
In a recent study of E\&M students \cite{kohlmyer2009tale}, the mean SAT reasoning test score of NCSU students was near 1240 while Georgia Tech students' mean SAT score was closer to 1340.
While not a direct measure of differences in their preparation in STEM courses, these results are suggestive.
Moreover, the differences in COMPASS performance are likely not a result of instructor or content delivery methods. 
Both mechanics sections were taught by instructors with as much or more experience teaching this course than Georgia Tech mechanics instructors and both used interactive engagement techniques in their classes \cite{kohlmyer_chat}.
These differences are also not a results of differences in student demographics; for example, fewer engineering majors at NCSU.
The mechanics course at NCSU predominantly serves freshman and sophomore engineering majors and to a lesser extent science (including computer science) majors.
A contingency table analysis (Appendix \ref{sec:ctables}) confirms that both the distribution of majors and the classifications of students in the NCSU mechanics course are statistically indistinguishable from Georgia Tech mechanics courses.
The overall GPA of students taking mechanics at NCSU was unavailable.

\begin{table*}[t]
\begin{center}
\caption{Pre- and post-instruction COMPASS scores are reported for students ($N$ = 168) who took an introductory mechanics course at NCSU. 
Scores are reported with a 95\% confidence interval estimated from the $t$-statistic in parentheses. 
Overall COMPASS scores for NCSU mechanics students were less favorable. 
Favorable post-instruction scores were lower for Avoiding Novice Behavior. 
and unfavorable scores were lower for Perceived Ability, Sense-making, Avoiding Novice Behavior.
All other dimensional scores remained the same within error.
\label{tab:compass-ncsu}}
\begin{tabular}{l|cc|cc}
\multicolumn{1}{c}{}& \multicolumn{2}{c}{{\bf PRE}} & \multicolumn{2}{c}{{\bf POST}}\\
\multicolumn{1}{c}{{\bf Dimension}}& \multicolumn{1}{c}{{\bf Favorable}} & \multicolumn{1}{c}{{\bf Unfavorable}} & \multicolumn{1}{c}{{\bf Favorable}} & \multicolumn{1}{c}{{\bf Unfavorable}}\\\hline\hline
Overall & 57 (2) & 19 (2) & 51 (3) & 25 (2) \\
Perceived Ability & 45 (3) & 14 (2) & 44(3) & 18 (2)\\
Perceived Utility & 65 (3) & 10 (2) & 62 (4) & 11 (2) \\
Real-World Connections & 57 (4) & 7 (2) & 52 (4) & 10 (3)\\
Sense-making & 37 (3) & 10 (2) & 36 (3) & 13 (2)  \\
Expert Behaviors & 39 (3) & 14 (2) & 41 (3) & 14 (2)  \\
Avoiding Novice Behaviors & 60 (3) & 20 (3) & 43 (4) & 28 (4) \\
Personal Interest & 47 (3) & 8 (2) & 45 (3) & 9 (2)\\
Avoiding Rote & 47 (3) & 7 (2) & 45 (4) & 8 (2)\\\hline\hline
\end{tabular}
\end{center}
\end{table*}

While mechanics students' scores at NCSU were less favorable than their Georgia Tech colleagues, the influences on those scores were identical.
Pre-instruction COMPASS scores at NCSU were found to depend strongly on a students' choice of major using an ANOVA. 
This result was similar to Georgia Tech mechanics students in which major was the only main effect (Sec. \ref{sec:compass-inf}).
However, computing students at NCSU have the least favorable responses; students in engineering and the sciences perform equally well and respond more favorably the computing students on the pre-test.
Post-instruction performance is most closely tied to pre-instruction performance; but choice of major also plays a role.
These influences are quite similar to those for Georgia Tech mechanics students.
An insignificant influence on post-instruction COMPASS scores is students' performance in the course as it was for Georgia Tech's mechanics students, but NCSU students who earn the highest scores in the course are likely to maintain their overall COMPASS score just as Georgia Tech students.
These results are summarized by arrow diagrams (Fig. \ref{fig:compass_dem-ncsu}) in Appendix \ref{sec:app-add-figs}.


%% file: compass/03b-compeval.tex
\section{Epistemological Signatures in Computational Modeling Performance \label{sec:compass-comp-sig}}

Student epistemology and performance in science are interrelated \cite{schommer1990effects,schommer1993epistemological} and such effects are measurable on attitudinal instruments for science \cite{perkins2005correlating}.
While differences in students' COMPASS scores were not observed between non-honors mechanics and E\&M students at Georgia Tech, we did find differences between non-honors mechanics students who successfully completed a final proctored evaluation assignment (Sec. \ref{sec:eval}) and those who were unable to do so. Students who successfully solved the assignment had more favorable overall pre- and post-instruction COMPASS scores than their unsuccessful classmates.
Furthermore, successful students had more favorable responses on nearly all dimensions both pre and post. Scores on the Avoiding Rote dimension on the pre-test and Sense-making dimension on the post-test were indistinguishable between successful and unsuccessful students.

\begin{table*}[t]
\begin{center}
\caption{Pre- and post-instruction COMPASS percentage favorable scores are reported for non-honors students who completed a final computational evaluation (Sec. \ref{sec:eval}). 
Students were divided into two groups, those that completed the assignment successfully (Passed, $N$ = 210) and those who did not (Failed, $N$ = 129).
Scores are reported with a 95\% confidence interval estimated from the $t$-statistic in parentheses. 
Students who passed the evaluation earned more favorable overall scores on both the pre- and post-test than students who did not pass.
Passing students also had more favorable scores on nearly all dimensions on both tests. For ``Avoiding Rote'' on the pre-test and ``Sense-making'' on the post-test, scores were indistinguishable.
\label{tab:compass-compeval}}
\begin{tabular}{l|cc|cc}
\multicolumn{1}{c}{}& \multicolumn{2}{c}{{\bf Passed}} & \multicolumn{2}{c}{{\bf Failed}}\\
\multicolumn{1}{c}{{\bf Dimension}}& \multicolumn{1}{c}{{\bf PRE}} & \multicolumn{1}{c}{{\bf POST}} & \multicolumn{1}{c}{{\bf PRE}} & \multicolumn{1}{c}{{\bf POST}}\\\hline\hline
Overall & 66 (2) & 61 (3) & 58 (3) & 55 (4) \\
Perceived Ability & 60 (3) & 59 (3) & 50 (4) & 54 (5) \\
Perceived Utility & 62 (3) & 55 (4) & 54 (4) & 44 (4) \\
Real-World Connections & 79 (3) & 72 (4) & 74 (4) & 64 (5) \\
Sense-making & 77 (3) & 58 (4) & 67 (4) & 54 (6) \\
Expert Behaviors & 56 (3) & 57 (4) & 47 (4) & 51 (5) \\
Avoiding Novice Behaviors & 69 (3) & 63 (3) & 64 (4) & 56 (4) \\
Personal Interest & 66 (3) & 60 (4) & 61 (4) & 49 (6)\\
Avoiding Rote & 59 (4) & 61 (4) & 54 (5) & 53 (5) \\\hline\hline
\end{tabular}
\end{center}
\end{table*}

Students who had more expert-like attitudes were more likely to solve the proctored assignment, likely, because they prepared for the assignment differently from unsuccessful students.
As mentioned in Sec. \ref{sec:eval}, students who solved computational homework problems successfully might attempt them individually, work with others or share solutions. 
However, on the proctored assignment, students must faithfully execute this problem without any outside assistance.
Students must prepare for solving the problem on the proctored assignment by practicing problems like it on their homework.
Differences in students' COMPASS scores, dissected in this manner, begin to unfold the role that students' attitudes towards computation might play in their learning to use computation.

%% file: compass/04-concluding.tex
\section{Possible Applications \label{sec:compass-applied}}

With these preliminary results, we have raised a number of issues including how self-identification, instruction and academic preparation might affect what students think about learning and using computation. 
Such questions might be answered in controlled or classroom studies in which the COMPASS is used as the common instrument for researchers to assess how students learn computation and what about learning computation they value.
This work might be foundational; for example, making attempts to understand how self-identification affects computational skills and contrasting that with reported attitudes.
The COMPASS, as a research tool, might also be used in more practical studies; carefully unfolding what elements of instruction or student background in computation affect scores overall and on different dimensions.
Furthermore, some might be interested in contrasting attitudes of students who learn computation in introductory courses with more advanced (but still novice) computational students.

Use of the COMPASS is not limited to research; instructors might use the COMPASS in their classes to help provide more customized learning opportunities for their students.
In a subject like computation, many introductory students have no practical experience with it.
However, it is likely that students have already formed attitudes about learning it.
Instructors teaching computation could use the results from a pre-instruction COMPASS to identify students with low interest or, largely, novice-like attitudes.
The instructor might provide additional support or a more active learning environment for these students.
In addition, COMPASS could also identify students with more expert-like attitudes.
After initial instruction, the students could be challenged with sophisticated tasks and problems.

Instructors might use post-instruction COMPASS results to identify areas to improve content or content delivery methods.
For example, instructors might address issues related to the connection of computation with the real scientific world by changing the focus of activities to more practical examples or framing the activity in terms of a design task.
Students following a design task learn to build and explore computational models in a manner that is different from following prescribed activities.
While completing design tasks, students perform spend more time making sense of the problems they have posed for themselves. 
Moreover, students' notions about their abilities  would be challenged as memorization of program statements is insufficient for solving these types of problems.

\section{Concluding Remarks \label{sec:compass-closing}}

Before any such applications are sought, the COMPASS must be tested in a number of different populations, its reliability determined and the robustness of its dimensions fully understood. 
The previous discussion has been based entirely on one sample, albeit from three different courses at two institutions with slightly dissimilar populations. 
We have not provided a measure of reliability that is typical of such instruments.
We have also raised questions about the robustness of dimensions of the COMPASS in different populations.

The COMPASS appears to be valid in a number of courses and its wording clear, but additional testing is needed. 
Our conclusions about item validity were based on discussions with a few students taking a second semester course with computation (Sec. \ref{sec:compass-valid}).
Item validity, particularly for use before instruction, depends on students with no computational experience interpreting statements and wording in the same manner as experts.
Furthermore, it is not clear if the usage of COMPASS in courses in which computation is taught using closed computational environments such as PhETs \cite{perkins2006phet} is appropriate. 

We have not demonstrated the reliability of the survey to the extent that is needed to call it ``reliable''.
Typically, reliability measurements involve a test-retest scenario.
However, we have shown that demographic influences on COMPASS scores are reasonable and consistent among several different populations (Secs. \ref{sec:compass-gt-results} \& \ref{sec:compass-pop}). 
Furthermore, we have also shown how that performance might be affected by academic preparation and a slightly more conservative population (Sec. \ref{sec:compass-ncsu}).

The robustness of COMPASS dimensions should be revisited after more data have been collected. 
The selection of COMPASS dimensions was done using a PCA on data from a single population in a single semester (Sec. \ref{sec:compass-statdim}). 
The robustness of some dimensions became questionable when reliability was checked using the responses by students from other populations (Secs. \ref{sec:compass-em} \& \ref{sec:compass-ncsu}).
Such issues need to be explored more fully after additional data have been collected.
The broad robustness of dimensions is key for comparing how students from different populations think about different aspects of learning computation.

While a number of different avenues for improving COMPASS are possible, its utility is clear from the preliminary measurements that have been made.
The COMPASS can provide additional information (e.g., about anxiety, confidence and sense-making) to explain students' performance with using computation.
The COMPASS adds another dimension of understanding, beyond simply performance, for researchers and instructors looking to improve instruction for the next generation of scientists and engineers.


%% file: 07-conclusion.tex
\chapter{Concluding Remarks}\label{chap:closing}

\section{Summary}

The purpose of the work presented in this thesis has been to extend the understanding of how novel content reforms to introductory physics courses affect student learning. In this work, we presented three complementary studies: a comparison of student performance by students in a reformed and a traditional course on well-known concept inventory (FCI), the development and evaluation of students' computational knowledge in a large lecture setting and the assessment of reform students' attitudes about learning computation.

With regard to course comparison, we discussed the context of learning; how the choice of content, even within the same domain, can affect what students learn. 
Students taking a traditional course outperformed reform students on the FCI, a measure of conceptual knowledge in mechanics. 
We found that the practice which traditional students received was more congruent with the items that appeared on the inventory.
However, this inventory only represents a small slice of mechanics, and only one class of problems.
We raised questions about the role of introductory physics for non-majors. 
Specifically, should a physics course for engineering students focus on particular topics in physics (i.e., kinematics, constant force motion) or should we aim to introduce other ideas, methods and tools (i.e., computation)?

As a reexamination of that role, we highlighted a particular choice of content, computation, in the reform course to examine the possible benefits of teaching new methods for solving problems.
We presented the first large scale implementation of teaching computation to introductory physics students using computational modeling homework and the first evaluation of students' computational skills in this setting.
The majority of students were able to apply general numerical problem solving methods to new problems with some success; however, a number of students were unsuccessful.

Success in science is related closely to how students prepare themselves during the course, how much effort they put forth to learning new material and what type of effort. 
As suggested by work in physics education, we examined the role that students' attitudes towards learning computation might play in their learning.
We developed the first tool to assess students' attitudes about learning computation and, in doing so, discovered students' have a wide range of attitudes about computation both before and after instruction.
These attitudes tended to be less expert-like after instruction, but this can be affected by self-identification and academic preparation.
Furthermore, performance on the final proctored and time assignment correlated with expert-like attitudes.

\section{Future Research Directions}

Two courses that as markedly different as those we discussed could be compared on a number of different dimensions (qualitative energy problems, quantitative problem solving, etc.). 
However, such comparisons will tend to favor the class of students that have had practice that is more closely aligned with the evaluation as we have observed in our work.

Future work should focus on developing best practices for using new tools (computation) to develop students' qualitative understanding in physics, their ability to create physical and computational models of new systems and strengthen their general problem solving skills. 
Students taking these courses are likely to become the next generation of scientists and engineers. 
As such, they will need a grasp of these tools as success in their professional lives will be increasingly defined by using new tools such as computation. 
To this end, we propose three avenues for future research: engaging students in the modeling process early, attempting to understand where students encounter difficulties with using computation and how students attitudes influence their abilities to learn and use computation.

Computation can help students engage in the modeling process, the skills necessary to become practicing scientists and engineers.
Engaging students in this process early in their academic careers is crucial for their success in later coursework and their professional work.
However, computation is absent from most of these courses, especially from those that are conceptual or algebra-based. 
The work in this thesis has presented the use of computation in calculus-based physics courses for engineering majors, but it is worth considering how computation might be introduced at lower levels. 
We are presently working to teach computation to students taking a conceptual high school physics course. 
The challenge in doing this has been reducing the level of detailed syntax needed to build highly visual computational models while preserving the syntax that engages the modeling process (i.e., the initial conditions, force calculation and momentum update). 
To this end, we have developed and tested a new Python module \cite{physutil} that reduces the need for complicated programming statements. 
Preliminary results from our pilot high school section have been positive. Students are capable of creating computational models of one and two dimensional motion with constant and non-constant (i.e., turbulent drag) forces. 
Future work in this area will be focused on embedding computation  within the Arizona State modeling curriculum \cite{halloun1987modeling,jackson2008modeling}.
The curriculum emphasizes the use of physical models (e.g., constant velocity model) in a variety of representations (e.g., graphical, analytic, etc.). 
We aim to introduce the prediction of motion using computation as another representation. 
Through these efforts, we aim to introduce young students to the practice of science and strengthen their problem solving toolbox for their future work.

To aid in the efforts of teaching computation at all levels, studies to understand why students have trouble using computation and if successful students can transfer computational knowledge to new domains and tasks should be pursued in parallel.
We have postulated why the errors we observed in programs written for the proctored assignment appeared.
However, think-aloud studies in which students solve such problems using a computer are necessary to put these errors in context. 
Moreover, such studies might be used to test if students who have learned computation can apply these algorithms to a new task in the same way that research scientists do with their own programs.

The influence of students' attitudes towards learning computation appear, in part, to affect their success when using it.
Additional work is needed to complete our preliminary work on the COMPASS. 
While the COMPASS has been validated for use at engineering schools, its reliability has yet to be determined.
No traditional (test-retest) measurements of reliability have been made, but some are planned for future semesters.
Moreover, responses to the COMPASS have come solely from students at two engineering universities.
In the near future, we plan to adminster the COMPASS to introductory physics students at other large enrollment engineering universities (e.g., Purdue University) to compare to Georgia Tech and NCSU students.
The robustness of COMPASS dimensions should be revisited after more data have been collected.
In the longer term, interviews with students at lower academic levels should be conducted as such interviews are central to securing item validity across all instructional levels.
Item validity of the COMPASS, particularly for use before computational instruction, depends on students with no computational experience interpreting statements and wording in the same manner as experts.

\section{Final Remarks}

It is the goal of many reforms in physics education to develop students into flexible problem solvers while exploring the practice of science. 
Yet, the development of generalizable problem solving skills are relatively absent from most courses. 
Teaching new content such as computation alongside physics can provide support develop the modeling process while also introducing students to powerful tools for solving problems. 
By learning computation, students learn the tools for doing science while developing a qualitative understanding of physical systems, exploring the generality of physics principles and learning broadly applicable problem solving methods. 
The acquisition of these skills are necessary to develop 21$^{\mathrm{st}}$ century scientists and engineers.

%% file: appendix/vpapp.tex
\appendix
\chapter{More details on the evaluation codes} \label{sec:vpcodes}

The codes shown in Table \ref{tab:vpcodes} were developed empirically. The procedure followed an iterative-design approach. We reviewed student work for common errors and devised a rough coding scheme. We then tested the scheme on a new set of student submitted programs. The scheme was refined and re-tested. This iterative procedure was repeated several times until we captured the majority of students' mistakes. Each code is explained in detail below.

\section{Using the correct given values (IC) Codes}

We reviewed the variables in each student's program. The default values had to be updated with the values given in the problem statement in the partially completed program. We present the codes used to categorize each student's program with respect to identifying and updating the appropriate initial conditions for their realization.

\paragraph*{IC1 -- Student used all the correct given values from grading case.}
A student must replace the values of all the variables (mass, position, and velocity, interaction constant $k$ and the exponent in the force law $n$ in $F=kr^n$) with those given in the {\it grading case}. This code excluded the integration time. It was intended that the larger mass object was to remain at its location. This was made explicit in the problem statement; the initial position $\langle 5,4,0 \rangle$ m and velocity $\langle 0,0,0 \rangle$ m/s of the larger mass of object were given in the problem statement, even though these same values appeared in the partially written program.

\paragraph*{IC2 -- Student used all the correct given values from test case.}
A student must replace the values of all the variables (mass, position, and velocity, interaction constant $k$ and the exponent in the force law $n$ in $F=kr^n$) with those given in the {\it test case}. This code excluded the integration time. It was intended that the larger mass object was to remain at its location (See IC1). 

\paragraph*{IC3 -- Student used the correct integration time from either the grading case or test case.}
A student must replace the default integration time (1 s) with the values given in the case with which they intended to work (grading or test). A student who mixed initial conditions was given an affirmative on this code if the majority of their initial conditions were from the same case as the integration time.

\paragraph*{IC4 -- Student used mixed initial conditions.}
A student who used some but not all of the initial conditions from any of the cases (default, test, or grading) was given an affirmative on this code. This code excluded the integration time.

\paragraph*{IC5 -- Students confused the exponents on the units the exponent of $k$ (interaction constant).}
Many students incorrectly thought the exponent on the length unit of the interaction constant was the scientific notation exponent for the interaction constant itself. For example, a student thought $k = 0.1$ Nm$^3$ meant $k = 100$ rather than $k = 0.1$ {\it Newton times meters cubed}.

\section{Implementing the force calculation (FC) Codes}

We reviewed how the students employed the force calculation algorithm in each of the programs written for the proctored assignment. The partially written program given to the students left out all statements related to the force calculation. Students were required to fill in this procedure using the appropriate VPython syntax. We present the codes used to categorize each student's program with respect to computing the vector force acting on the low-mass object.

\paragraph*{FC1 --The force calculation was correct.}
A student must compute the separation vector, its magnitude, its unit vector, the magnitude of the force and the vector force correctly. Each of these steps may be combined as long as the final result computes the vector force acting on the less massive particle at each instant. These steps must all appear in the numerical integration loop.

\paragraph*{FC2 -- The force calculation was incorrect, but the calculation procedure was evident.}
In the numerical integration loop, the student must perform a position vector subtraction, a calculation of the force magnitude and some attempt at combining magnitude with unit vector (any unit vector was acceptable). If a student treated the problem using components and had some force which is a vector, it was coded as evident. If any part of the calculation was performed outside the loop, it was coded as {\it not} evident.

\paragraph*{FC3 -- The student attempted to raise the separation vector ($\vec{r}$) to a power.}
Students who raised the separation vector to a power generated a VPython exception error:\\
{\tt unsupported operand type(s) for ** or pow(): 'vector' and 'int'}.\\
This error told them that VPython cannot raise a vector to a power, as it is a mathematically impossible operation.

\paragraph*{FC4 -- The direction of the force was reversed.}
Students had to assign the correct unit vector and sign to the force depending on whether their force was attractive or repulsive. This code was not used if the student calculated the force as a magnitude only, raised $\vec{r}$ to a power, or invented a unit vector (e.g., $\langle 1,0,0 \rangle$). Visual feedback (i.e., the lower mass particle flying off to infinity) indicated a simple sign mistake.

\paragraph*{FC5 -- Student had some other force direction confusion.}
Some students used vectors other than $\vec{r}$ or $-\vec{r}$ to compute $\vec{F}$. Other students computed the force as a magnitude and then multiplied it by an ``invented'' unit vector (e.g., $\langle 1,0,0 \rangle$, $\hat{p}$). Both of these errors were given an affirmative for this code.

\section{Updating with the Newton's second law (SL) Codes}

We reviewed how the students employed the momentum update in each of the programs written for the proctored assignment. The partially written program given to the students left out the one line of code necessary to update the momentum. Students were required to fill in this line using the appropriate VPython syntax. We present the codes used to categorize each student's program with respect to updating the momentum of the low-mass particle.

\paragraph*{SL1 -- Newton's second law was correct.}
Correct Newton's second law meant that it was ``correct as a physics principle'' and also that it appeared ``in the update form''. This meant that {\tt pfinal = pinitial + Fnet*deltat} alone in a loop did not fall under ``correct Newton's second law''. It is an incorrect update form.

\paragraph*{SL2 -- Newton's second law was incorrect but in form that updates.}
Newton's second law updates the momentum, but not necessarily correctly. (e.g., \texttt{p = p + Fnet}, \texttt{p = p + Fnet/dt}, \texttt{pf = p + Fnet}, etc. )

\paragraph*{SL3 -- Newton's second law was incorrect and the student attempted to update it with a scalar force.}
Some students computed the magnitude of the force acting on the particle and then used this magnitude to update the momentum. Students who did this raised a VPython exception error:\\
{\tt unsupported operand type(s) for +: 'vector' and 'int'}\\
This might have lead some to invent unit vectors in the momentum update, for example,\\{\tt p = p + vector(Fmag,0,0)*dt} and {\tt p = p + Fmag*vector(1/sqrt{2},1/sqrt{2},0)*dt}.

\paragraph*{SL4 -- Student created a new variable for $\vec{p}_f$.}
In computational modeling, the equal sign in a update line means ``add and replace''. Some students used a new symbol for the final momentum (e.g. {\tt pfinal}) and then replaced the momentum in the next step (e.g. {\tt p = pfinal}). Others only did the former, that is, they did not replace the momentum with its updated value.

\section {Other Codes}

Two common errors were not included in the above codes because they do not reflect errors in the procedure of modeling the motion of the low-mass particle. We present two miscellaneous codes which were common enough to consider relevant.

\paragraph*{O1 -- Student attempted to update force, momentum or position for the massive particle.}
The massive particle was intended to remain in place.

\paragraph*{O2 -- Student did not attempt the problem.}
Some students uploaded plain text files to the receive bonus credit for uploading their code. We assumed they did not attempt solving the problem.

%% file: appendix/app-compass.tex
\chapter{Additional details about COMPASS dimensions}\label{sec:compass-dim}

In this appendix, we provide full descriptions of working categories and emergent dimensions alluded to in Ch. \ref{chap:compass}.

\section{The Initial Working Categories\label{sec:compass-six}}

The descriptions listed below were used by the rater to select which statements should be classified in which category. The descriptions are short and somewhat broad. This was by design to give the raters some flexibility in interpretation before discussing the results. Furthermore, some descriptions have additional commentary about splitting categories. This was not done until after the check of category robustness described in Sec. \ref{sec:compass-statdim}.

\paragraph{Perceived Ability}
Statements placed in this category probe how comfortable students are with using computational models. It is possible that there might be overlap with statements in the Personal Interest category, we should check for this. Statements in which students are asked directly about their skill with using computational models are of particular interest.

\paragraph{Perceived Utility}
Statements placed in this category probe how useful students think computation is. This could include how well computational models describe phenomena or how they see computation fitting in to their future work. Statements in which students are asked directly about the utility of computation are of particular interest.

\paragraph{Real-World Connections}
Statements selected for this category offer explanations of the utility of computation. In particular, statements should ask students about its use in the professional world. Of particular interest are statements in which students are asked specifically about how computation connects to real-world phenomena.

\paragraph{Sophistication}
Statements selected for this category contrast students with expert with respect to usage of computation. Statements should ask students about how they use computation and when it is appropriate to do so. Statements in which students are asked directly about their usage of computation outside of class or their behaviors when using computation are of particular interest.

\paragraph{Personal Interest}
Statements placed in this category are concerned with students' personal feelings towards not only computation but also learning computation. The two ideas might be split up later. Statements in which students must judge whether they have a vested interest in learning computation are noteworthy.

\paragraph{Learning}
Statements placed in this category probe how students learn computation. Of particular interest are the particular actions they take to learn how to write and develop computational modeling programs.

\section{The Robust Dimensions\label{sec:compass-eight}}

After checking the robustness of the original six categories, a number of changes were introduced (Sec. \ref{sec:compass-statdim}). 
Ultimately, statements appeared to have one or more of eight statistically robust dimensions. 
After reviewing the statements in each dimension, the following descriptions were written as the to described the common thread which appeared to link the statements together. 
Some (e.g., Perceived Ability) were obvious because they were quite similar to the original categories. 
Others (e.g, Avoiding Rote) had changed enough that the original descriptions only weakly applied.

\paragraph{Perceived Ability}
Statements with this dimension are those on which students will make an assessment of their own skills. 
This dimension is concerned with how confident students feel about using or learning computation.
Statements that may be included could ask students to make an assessment of others ability to use or learn computation.
Favorable responses to these statements indicate that students are confident in their abilities to learn or use computation.

\paragraph{Perceived Utility}
Statements with this dimension are those on which students evaluate the utility of learning computation for their future work or the utility of computation itself for helping the to understand science.
This dimension does not make a distinction between these two different aspects of utility; robustness is sacrificed if this is done. 
However, it is relatively clear from reading individual statements which statements imply which aspect.
Favorable responses to these statements indicate that students believe their is some utility to learning or using computation.

\paragraph{Real-World Connections}
Statements with this dimension connect students' use of computation to their future career or the use computation in the ``Real World'' of science and engineering. 
This dimension does not make a distinction between these two aspects as robustness is sacrificed when statements are extracted. 
However, it is clear which statements probe which aspect.
Favorable responses to these statements indicate that students see how computation is crucially connected to science and engineering practice.

\paragraph{Sense-making}
Statements with this dimension describe the effort which students put forth to make sense of a computational model or the physical model that it describes. 
In particular, many of the statements asks students to evaluate how important to them that they understand how computational models are constructed and how the physical model connects to it.
Favorable responses to these statements indicate that it is important to students to understand how computational models are constructed from physical descriptions.

\paragraph{Expert Behaviors}
Statements with this dimension contrast what experts do when using or developing computational models to what students might do.
This dimension probes if students perform different behaviors which experts use to construct computational models of physical systems.
Favorable responses to these statements indicate that students have adopted expert-like behaviors when using computation.

\paragraph{Avoiding Novice Behaviors}
Statements with this dimension also contrast what novices think or do when using or developing computational models to experts thoughts and actions.
This dimension does not make the distinction between thoughts and actions; doing so sacrifices robustness.
Favorable responses to these statements indicate that students think about computation or use computation in a more expert-like manner by avoiding novice like thoughts or actions.

\paragraph{Personal Interest}
Statements with dimension probe students own interest for learning computation.
This dimension has a variety of different statements which either ask directly about students interest in computation or do so somewhat tangentially.
Favorable responses to these statements indicate that students have some personal interest in learning computation.

\paragraph{Avoiding Rote}
Statements with this dimension describe how students learn computation.
In particular, most of these statements asked students if it is sufficient to simply memorize details about computation to learn it.
Favorable responses to these statements indicate that students avoid learning by rote and that they might be trying to construct their own understanding of computation.

\chapter{Computational Modeling Attitudinal Student Survey (v2.3)}\label{sec:compass-statements}

In this appendix, we provide all the statements from the most recent version of the COMPASS, along with the dimensions into which they were classified.

\begin{center}
\begin{longtable}{|p{0.45\linewidth}|c|c|c|c|c|c|c|c|c|}\hline

& 
\begin{sideways}{Perceived Ability}\end{sideways} &
\begin{sideways}{Perceived Utility}\end{sideways} & 
\begin{sideways}{Real-World Connections}\end{sideways} & 
\begin{sideways}{Sense-making}\end{sideways} & 
\begin{sideways}{Expert Behaviors}\end{sideways} & 
\begin{sideways}{Avoiding Novice Behaviors\hspace*{1mm}}\end{sideways} & 
\begin{sideways}{Personal Interest}\end{sideways} & 
\begin{sideways}{Avoiding Rote}\end{sideways} & 
\begin{sideways}{Not Scored}\end{sideways}\\\hline

\endhead

{(1) A significant problem in learning computer modeling is being able to memorize all the information I need to know.} & & & & & & & & X & \\\hline

{(2) When using a computer to solve a problem, I try to decide what would be a reasonable value for the answer.} & & & & & X & & & & \\\hline

{(3) It is useful for me to solve lots of computer modeling problems when learning computer modeling.} & & & & & & & & & X \\\hline

{(4) After I solve a problem using a computer model, I feel that I understand how the model works.} & X & & & & X & & & & \\\hline

{(5) I find that I can use a computer model that I've written to solve a related problem.} & X & & & & X & & & & \\\hline

{(6) There is usually only one correct approach to solving a problem using a computer.} & & X & & & & X & & & \\\hline

{(7) I am not satisfied until I understand how my working computer model connects to a real world situation.} & X & & X & X & & & X & & \\\hline

{(8) I cannot learn computer modeling if the teacher does not explain things well in class. }& & & & & & & & & X \\\hline

{(9) I do not expect computer modeling to help my understanding of the ideas; it is just for doing calculations. }& & X & & & & X & & & \\\hline

{(10) If I get stuck on a computer modeling problem my first try, I usually try to figure out a different way that works.} & X & & & X & & & & & \\\hline

{(11) Nearly everyone is capable of using a computer to solve problems if they work at it.} & X & & & & & & & & \\\hline

{(12) To understand how to use a computer to solve a problem I discuss it with friends and other students.} & & & & & & & & & X \\\hline

(13) I do not spend more than 30 minutes stuck on a computer-modeling problem before giving up or seeking help from someone else. & & & & & & X & & & \\\hline

(14) If I want to apply a computer modeling method used for solving one problem to another problem, the problems must involve very similar situations. & & X & & & & X & & & \\\hline

(15) In doing a computer modeling problem, if my calculation gives a result very different from what I'd expect, I'd trust the calculation rather than going back through the problem. & & & & & &  X & & & \\\hline

(16) It is important for me to understand how to express physics concepts in a computer model. & & & & X & & & & & \\\hline

(17) I enjoy solving computer modeling problems. & & & & & & & X & & \\\hline

(18) To learn how to solve problems with a computer, I only need to see and to memorize examples that are solved using a computer. & & & & & & & & X & \\\hline

(19) Spending a lot of time understanding how computer modeling methods work is a waste of time. & & & & & & & X & X & \\\hline

(20) I find carefully analyzing only a few problems in detail is a good way for me to learn computer modeling. & & & & & & & & & X \\\hline

(21) I can usually figure out a way to solve physics problems. & X & & & & & & & & \\\hline

(22) If I have trouble solving a problem with pencil and paper, I will try using a computer. & X & X & & & X & & & & \\\hline

(23) Computer models have little relation to the real world. & & & X & & & X & & & \\\hline

(24) Reasoning skills used to understand a computer model could be helpful to me in my everyday life. & & X & & & & & & & \\\hline

(25) When I solve a computer modeling problem, I explicitly think about which physics ideas apply to the problem. & X & & & & X & & & & \\\hline

(26) When I solve a computer modeling problem, I explicitly think about the limitations of my model.  & X & & & & X & & & & \\\hline

(XX) We use this statement to discard the survey of people who are not reading the questions. Please select agree-option D (not strongly agree) for this question to preserve your answers. & \multicolumn{9}{c|}{Filter Statement -- Not Scored} \\\hline

(27) If I get stuck on a computer modeling problem, there is no chance I'll figure it out on my own. & X & & & & & & & & \\\hline

(28) When studying computer modeling, I relate the important information to what I already know rather than just memorizing it the way it is presented. & & & & & & & & X & \\\hline

(29) I would rather have someone give me the solution to a difficult computer modeling problem than to have to work it out for myself. & & & & & & & & & X \\\hline

(30) I expect to have little use for solving problems using a computer when I get out of school. & & X & X & & & & X & & \\\hline

(31) I'll need to solve problems using a computer for my future work. & & X & X & & & & X & & \\\hline

(32) When my computer model does not work immediately, I stick with it until I have the solution. & X & & & X & & & & & \\\hline

(33) When I solve a problem using a computer, I have a better understanding of the solution than if I solve it with pencil and paper. & & X & & & & & & & \\\hline

(34) Computer models are useful for solving science and engineering problems. & & & X & & & & & & \\\hline

(35) Watching a computer model helps me understand the solution to a problem. & & X & & X & & & & & \\\hline

(36) The results of the computer model are more important than the computer modeling method. & & & & & & & & & X \\\hline
\end{longtable}
\end{center}

\chapter{Additional Figures for COMPASS data}\label{sec:app-add-figs}

Some figures were kept from the main text in Ch. \ref{chap:compass} because they were too large or simply provided another way of visualizing data that was already presented in tables. Those figures are provided in this appendix.

\begin{figure}[t]
\centering
 \includegraphics[clip,trim=30mm 5mm 30mm 5mm,width=0.60\linewidth]{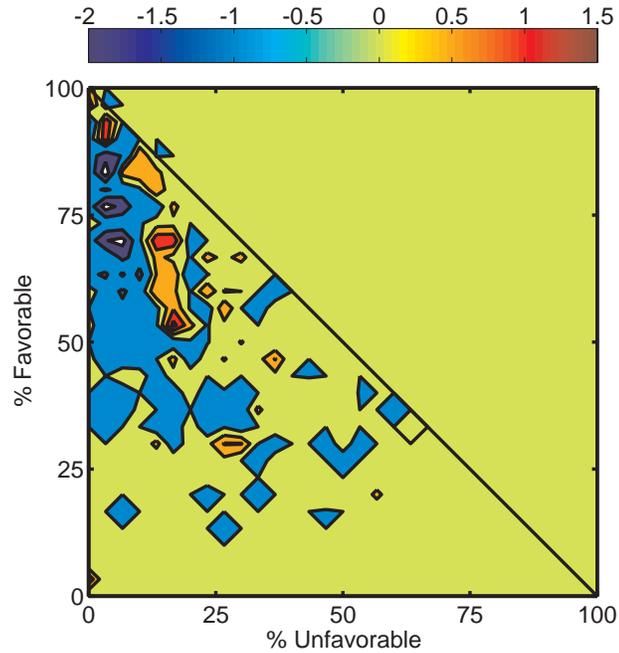}
\caption{[Color] - The distribution of the percent change (from pre to post) in COMPASS scores for non-honors mechanics students at Georgia Tech is shown. Students tended to shift to less favorable scores on the post-test (red block near center). Although, some students moved to much more favorable scores (red island in upper left corner). \label{fig:compass-mech-shift}}
\end{figure}

\begin{figure}[t]
\centering
 \includegraphics[clip,trim=30mm 5mm 30mm 5mm,width=0.90\linewidth]{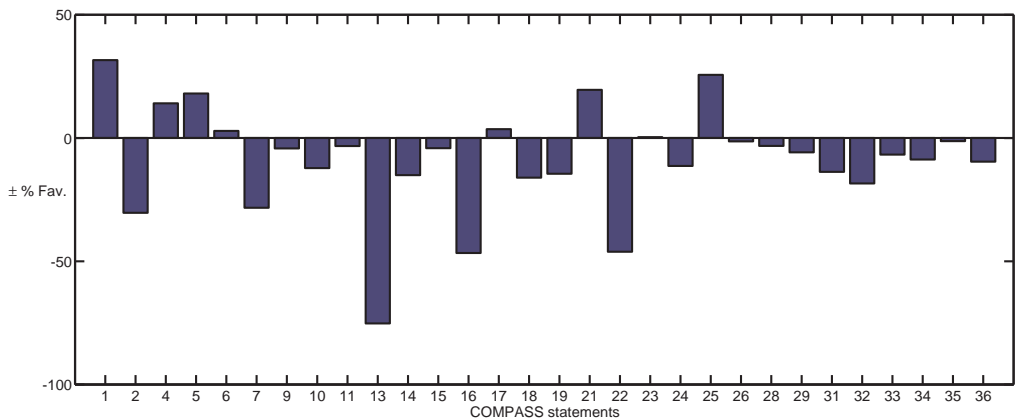}
\caption{[Color] - The percentage of non-honors mechanics students who shifted the responses to more (+) or less (-) favorable on the COMPASS post-test  \label{fig:statementshifts}}
\end{figure}

\begin{figure}[t]
\centering
 \includegraphics[clip,trim=30mm 5mm 30mm 5mm,width=0.60\linewidth]{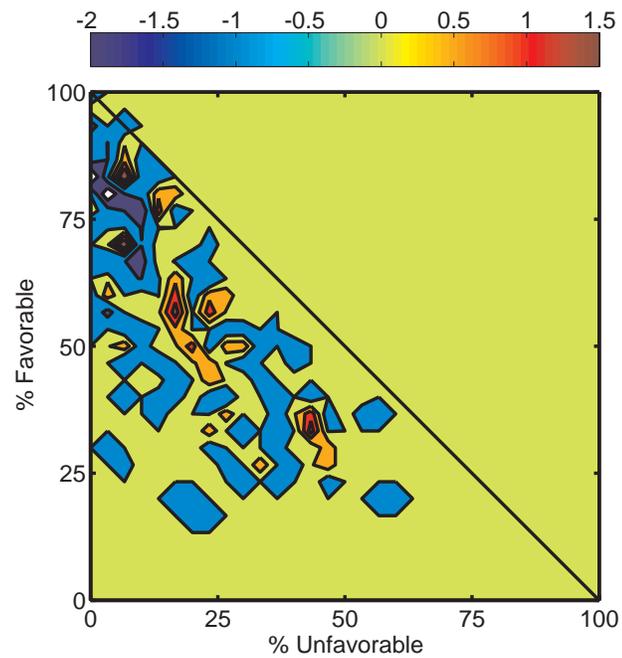}
\caption{[Color] - The distribution of the percent change (from pre to post) in COMPASS scores for non-honors E\&M students at Georgia Tech is shown. Students tended to shift to less favorable scores on the post-test (red block near center). Although, some students moved to much more favorable scores (red island in upper left corner) and far less (red island near lower right corner). \label{fig:emshift}}
\end{figure} 

\begin{figure}[t]
\centering
 \includegraphics[clip,trim=30mm 5mm 30mm 5mm,width=0.60\linewidth]{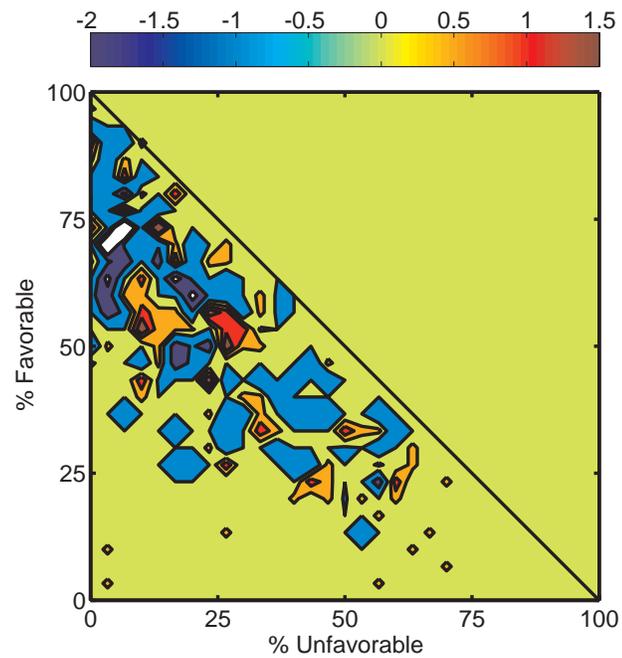}
\caption{[Color] - The distribution of the percent change (from pre to post) in COMPASS scores for mechanics students at NCSU is shown. Students tended to shift to less favorable and more neutral scores on the post-test (red block near center). Although, some students moved to much more favorable scores (red island in upper left corner) and far less (red islands near lower right corner). \label{fig:ncsushift}}
\end{figure} 

%
%

\begin{figure}[t]
\begin{center}
\includegraphics[clip, trim = 15mm 20mm 15mm 21mm, width=0.95\linewidth]{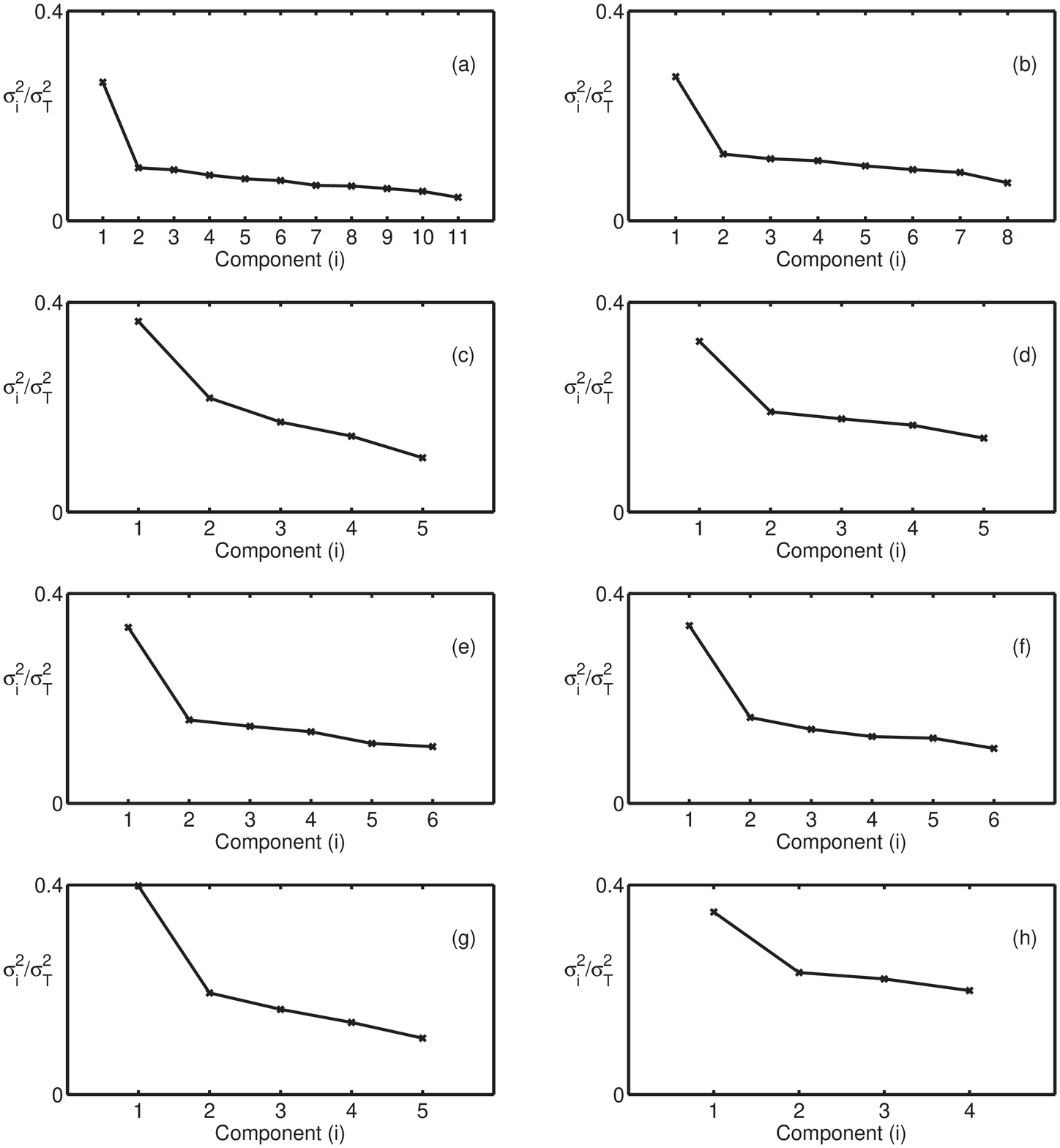}
\end{center}
\caption{The eigenvalues for each component are plotted for the eight subsets of statements selected as robust dimensions of the COMPASS. For this plot, Georgia Tech mechanics pre-instruction data was used. The first eigenvalue for each represents the fraction of the total variance that can be attributed to the first eigenvector (i.e., the first component). In each subset, the linear drop-off of the variance that can be attributed the other eigenvectors (i.e., the scree) in the subset is representative of a strong single factor (i.e., a single dimension). We have titled the dimensions (a) Perceived Ability, (b) Perceived Utility, (c) Real-world Connections, (d) Sense-making, (e) Expert behaviors, (f) Novice behaviors, (g) Personal Interest and (h) Avoiding Rote. \label{fig:compass-comp}}
\end{figure}

\begin{figure}[t]
\begin{center}
\includegraphics[clip, trim = 15mm 20mm 15mm 21mm, width=0.95\linewidth]{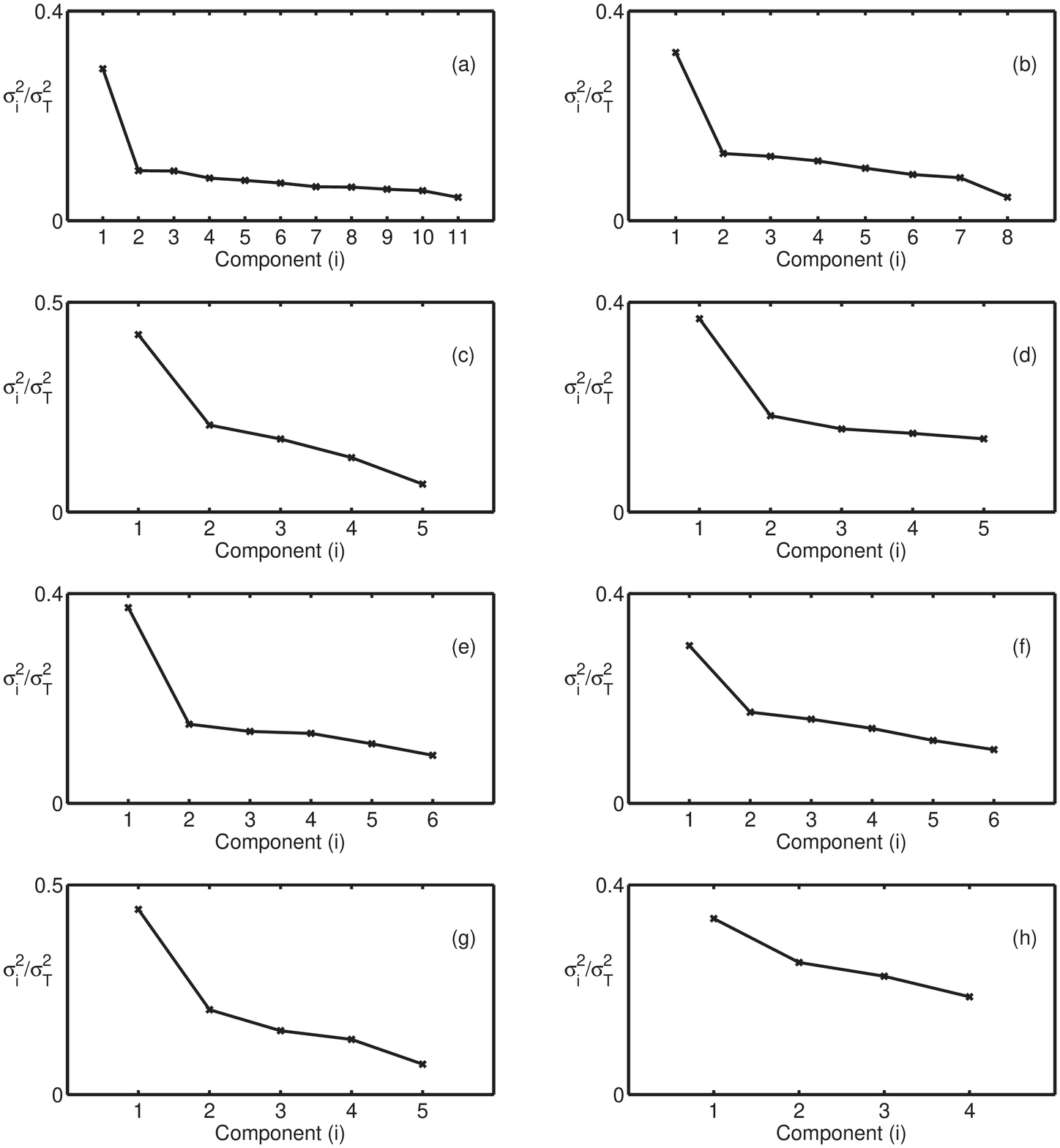}
\end{center}
\caption{The eigenvalues for each component are plotted for the eight subsets of statements selected as robust dimensions of the COMPASS. For this plot, Georgia Tech mechanics post-instruction data was used. The first eigenvalue for each represents the fraction of the total variance that can be attributed to the first eigenvector (i.e., the first component). In each subset, the linear drop-off of the variance that can be attributed the other eigenvectors (i.e., the scree) in the subset is representative of a strong single factor (i.e., a single dimension). We have titled the dimensions (a) Perceived Ability, (b) Perceived Utility, (c) Real-world Connections, (d) Sense-making, (e) Expert behaviors, (f) Novice behaviors, (g) Personal Interest and (h) Avoiding Rote. \label{fig:compass-comp-post}}
\end{figure}

\begin{figure}[t]
\begin{center}
\includegraphics[clip, trim = 15mm 20mm 15mm 21mm, width=0.95\linewidth]{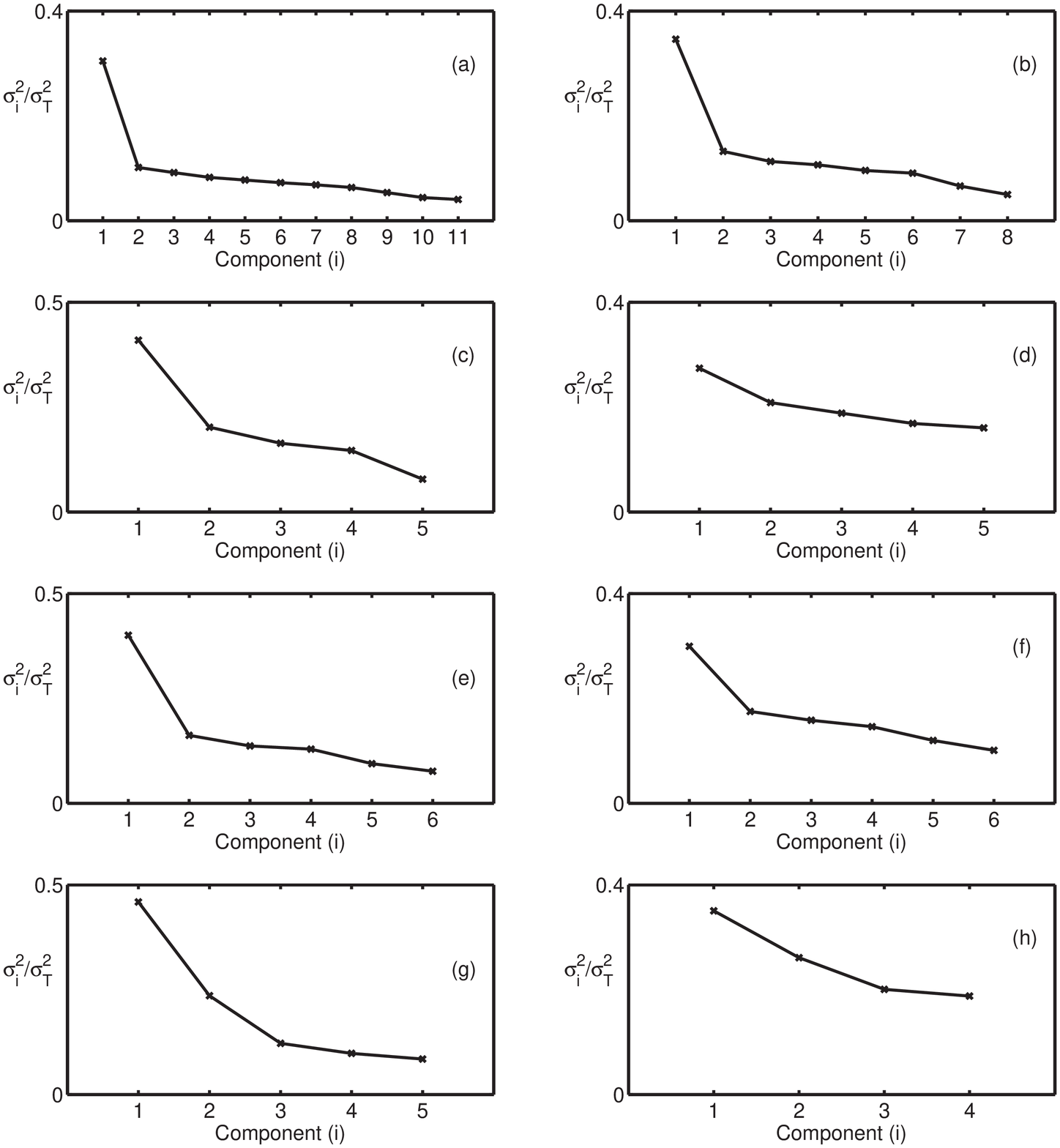}
\end{center}
\caption{The eigenvalues for each component are plotted for the eight subsets of statements selected as robust dimensions of the COMPASS. For this plot, Georgia Tech E\&M pre-instruction data was used. The first eigenvalue for each represents the fraction of the total variance that can be attributed to the first eigenvector (i.e., the first component). In each subset, the linear drop-off of the variance that can be attributed the other eigenvectors (i.e., the scree) in the subset is representative of a strong single factor (i.e., a single dimension). We have titled the dimensions (a) Perceived Ability, (b) Perceived Utility, (c) Real-world Connections, (d) Sense-making, (e) Expert behaviors, (f) Novice behaviors, (g) Personal Interest and (h) Avoiding Rote. \label{fig:compass-comp-em}}
\end{figure}

\begin{figure}[t]
\begin{center}
\includegraphics[clip, trim = 15mm 20mm 15mm 21mm, width=0.95\linewidth]{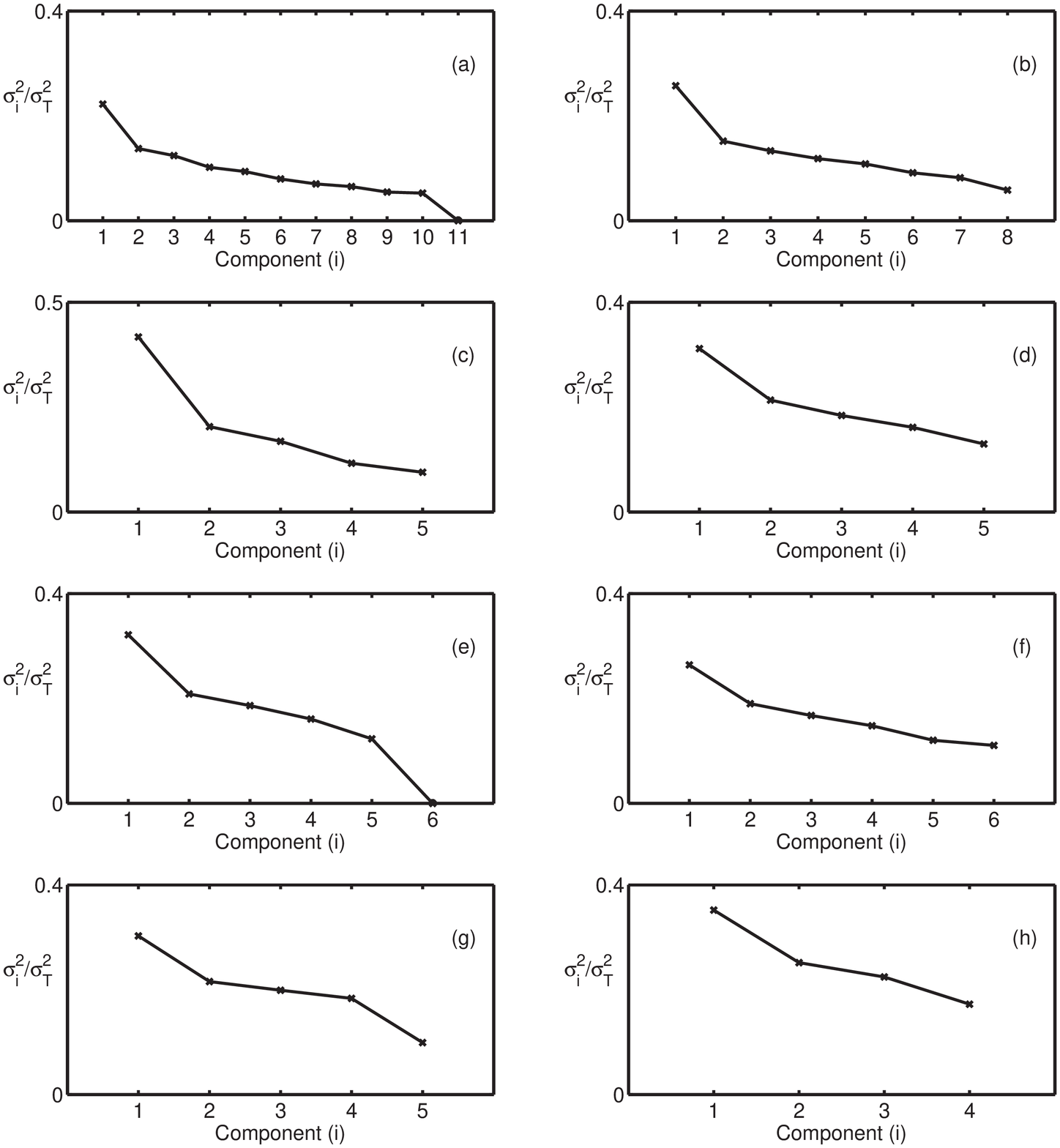}
\end{center}
\caption{The eigenvalues for each component are plotted for the eight subsets of statements selected as robust dimensions of the COMPASS. For this plot, NCSU mechanics pre-instruction data was used. The first eigenvalue for each represents the fraction of the total variance that can be attributed to the first eigenvector (i.e., the first component). In each subset, the linear drop-off of the variance that can be attributed the other eigenvectors (i.e., the scree) in the subset is representative of a strong single factor (i.e., a single dimension). We have titled the dimensions (a) Perceived Ability, (b) Perceived Utility, (c) Real-world Connections, (d) Sense-making, (e) Expert behaviors, (f) Novice behaviors, (g) Personal Interest and (h) Avoiding Rote. \label{fig:compass-comp-ncsu}}
\end{figure}

\begin{figure}[t]
\begin{center}
\includegraphics[clip, trim = 15mm 20mm 15mm 21mm, width=0.95\linewidth]{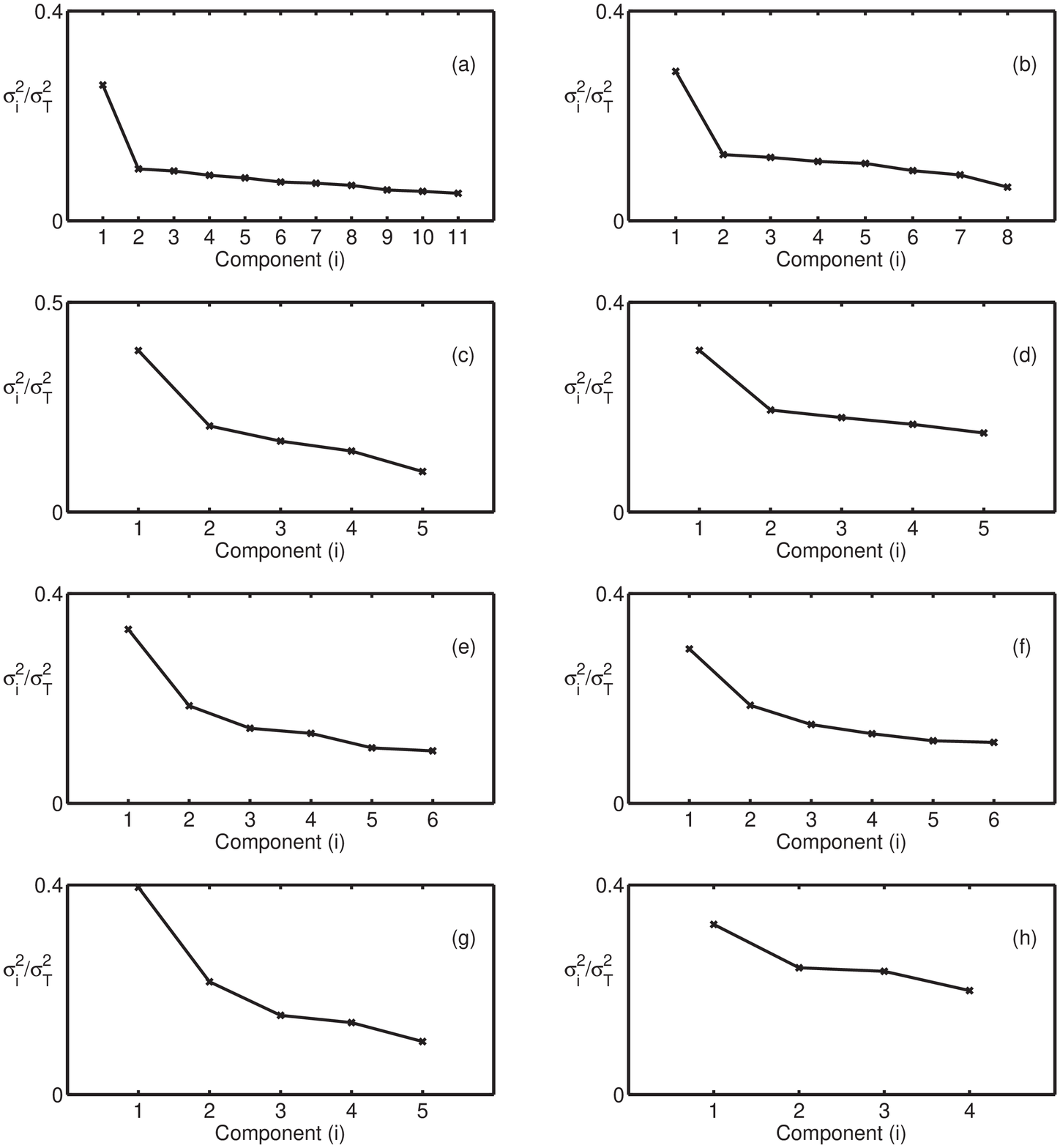}
\end{center}
\caption{The eigenvalues for each component are plotted for the eight subsets of statements selected as robust dimensions of the COMPASS. For this plot, all (Georgia Tech and NCSU) pre-instruction data was used. The first eigenvalue for each represents the fraction of the total variance that can be attributed to the first eigenvector (i.e., the first component). In each subset, the linear drop-off of the variance that can be attributed the other eigenvectors (i.e., the scree) in the subset is representative of a strong single factor (i.e., a single dimension). We have titled the dimensions (a) Perceived Ability, (b) Perceived Utility, (c) Real-world Connections, (d) Sense-making, (e) Expert behaviors, (f) Novice behaviors, (g) Personal Interest and (h) Avoiding Rote. \label{fig:compass-comp-all}}
\end{figure}

\begin{figure}[htp]
  \begin{center}
  \subfigure[Shift in COMPASS scores (Overall GPA)]{\label{fig:embygpa}\includegraphics[width=0.45\linewidth]{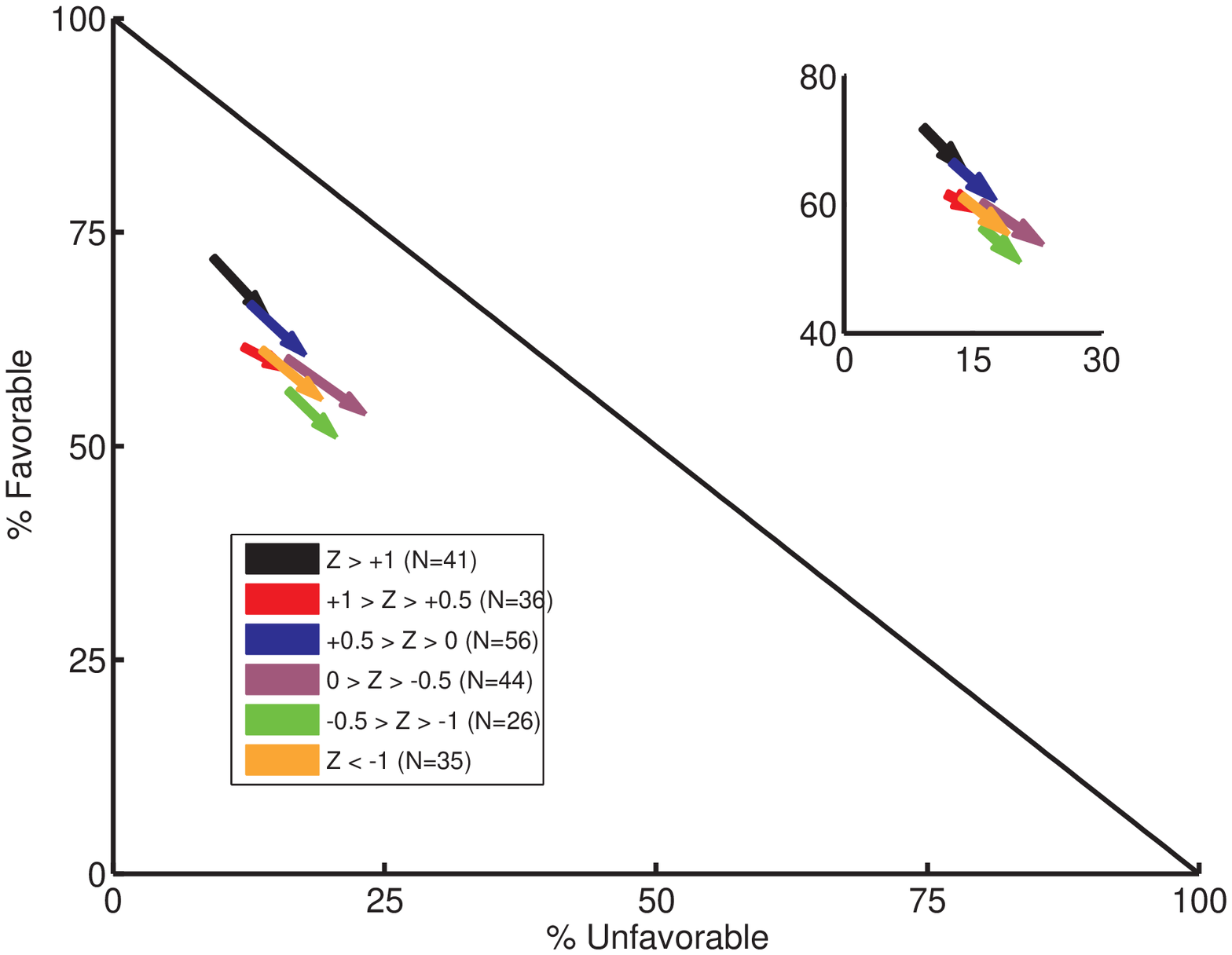}}                
  \subfigure[Shift in COMPASS scores (Course Grade)]{\label{fig:embygrade}\includegraphics[width=0.45\linewidth]{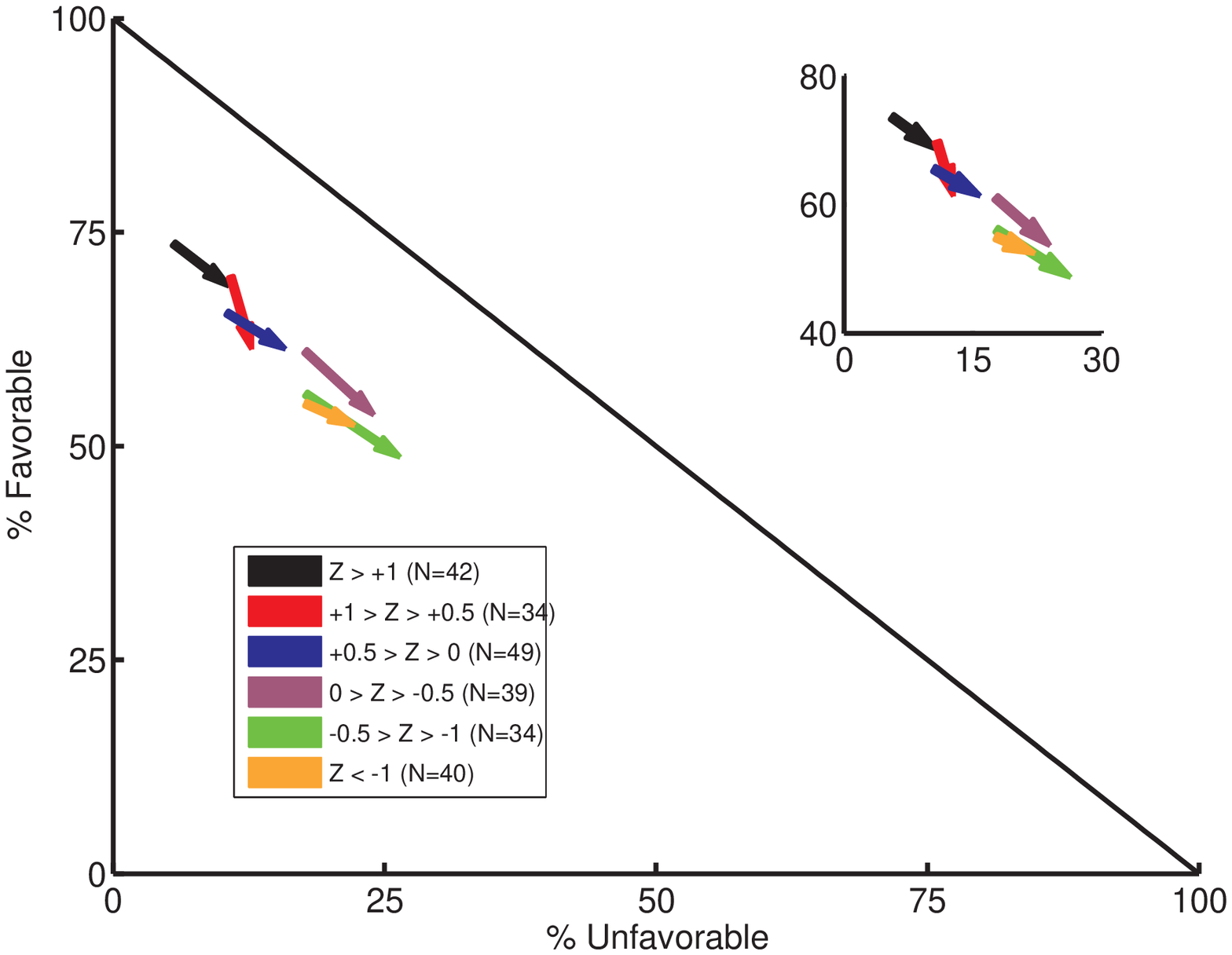}}\\
  \subfigure[Shift in COMPASS scores (Classification)]{\label{fig:embyclass}\includegraphics[width=0.45\linewidth]{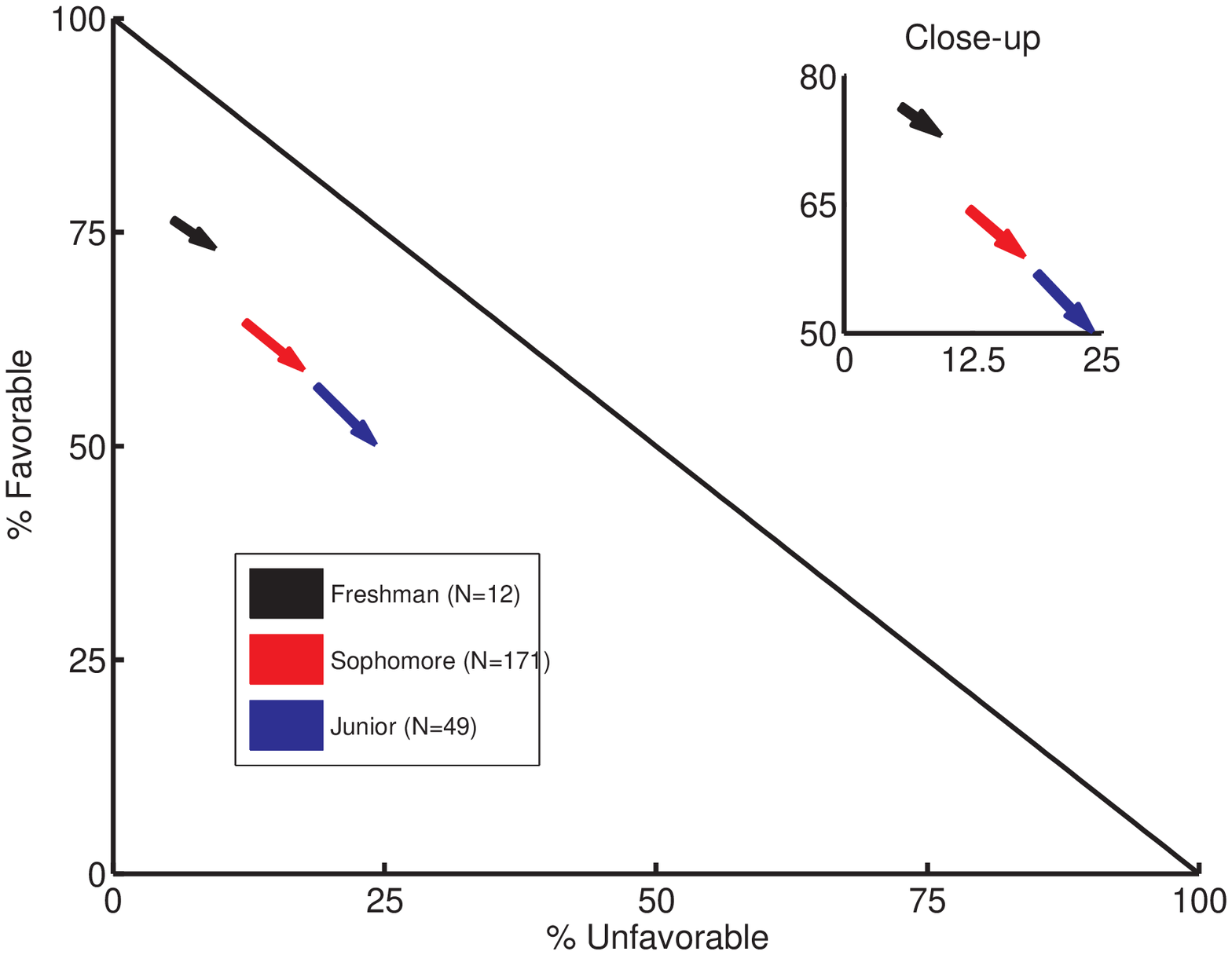}}
    \subfigure[Shift in COMPASS scores (College)]{\label{fig:embycollege}\includegraphics[width=0.45\linewidth]{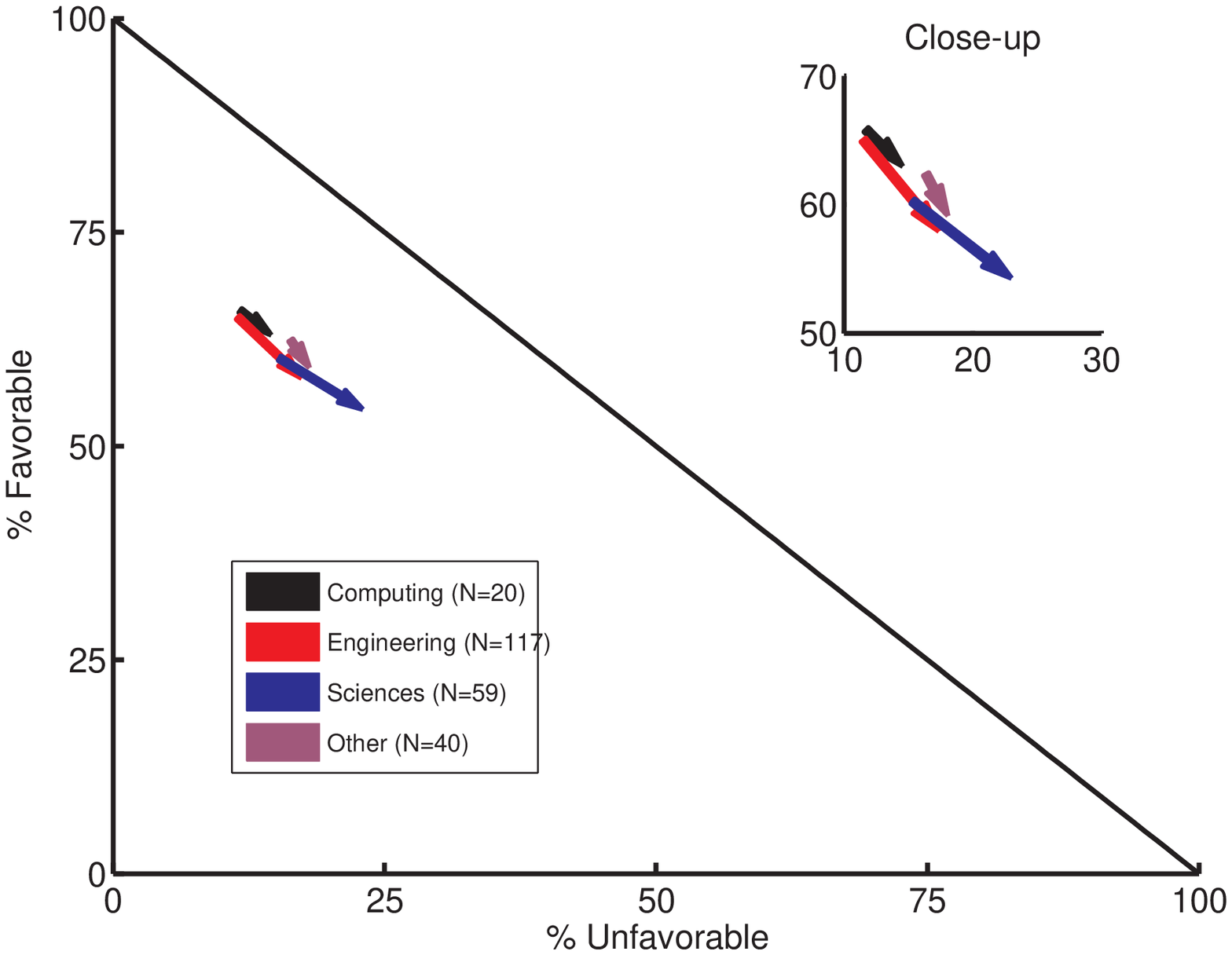}}
    \end{center}
  \caption{[Color] - The shift in the E\&M students' mean COMPASS scores are shown. Colored arrows indicate the magnitude and direction of the shift from pre- to post-instruction. Mean scores are shown for students based on: (a) their normalized Overall GPA, (b) their normalized score earned in an introductory mechanics course, (c) their official classification upon starting the E\&M course (typically, a sophomore level course) and (d) their declared major college upon entering the mechanics course. Architecture, Liberal Arts and Management compose the ``Other'' category, but the majority of these students were Architecture majors.}
  \label{fig:compass_dem-em}
\end{figure} 

\begin{figure}[t]
  \begin{center}
 \subfigure[Shift in COMPASS scores (Classification)]{\label{fig:ncsubyclass}\includegraphics[width=0.45\linewidth]{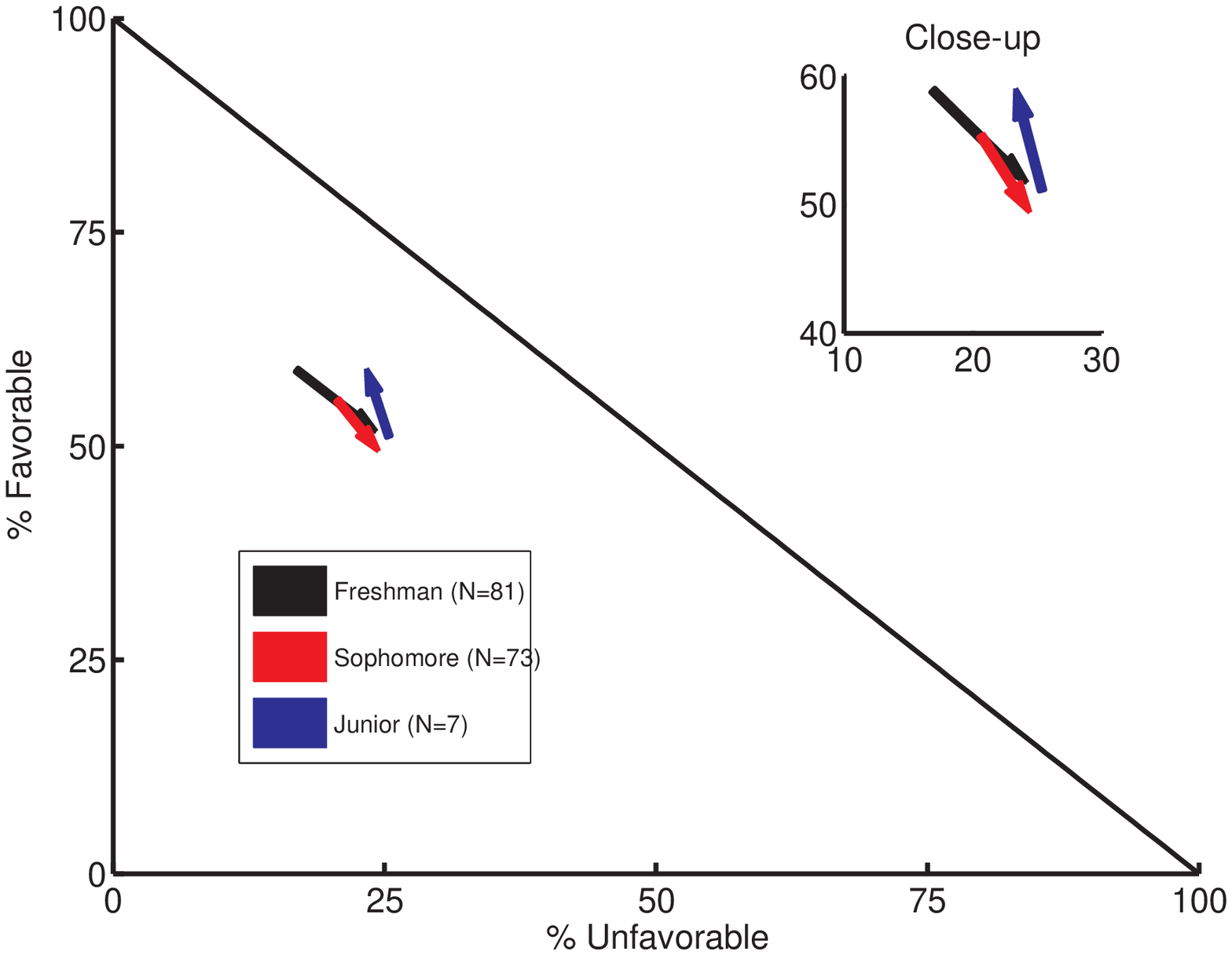}}
    \subfigure[Shift in COMPASS scores (College)]{\label{fig:ncsubycollege}\includegraphics[width=0.45\linewidth]{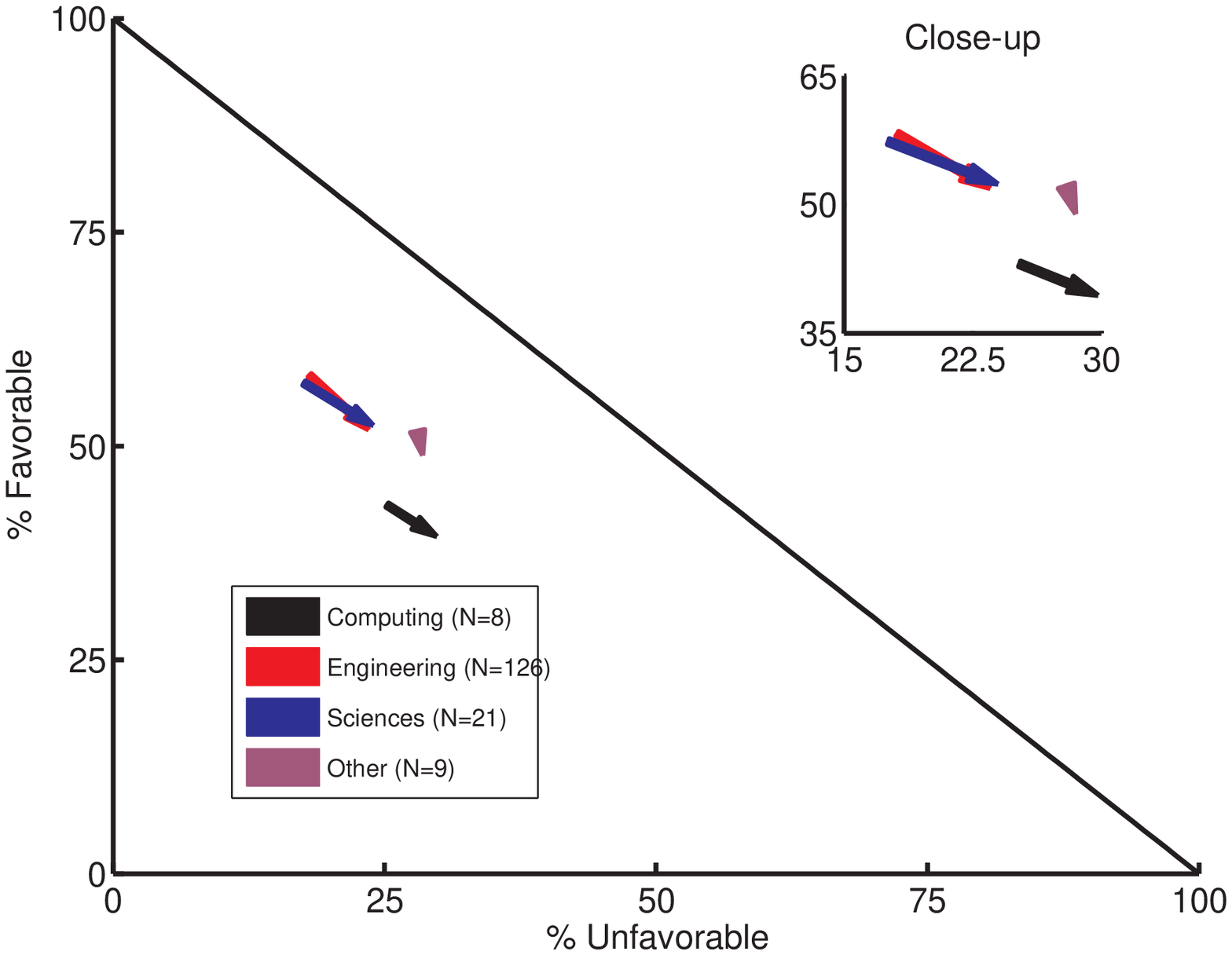}}\\
    \subfigure[Shift in COMPASS scores (Course Grade)]{\label{fig:ncsubygrade}\includegraphics[width=0.45\linewidth]{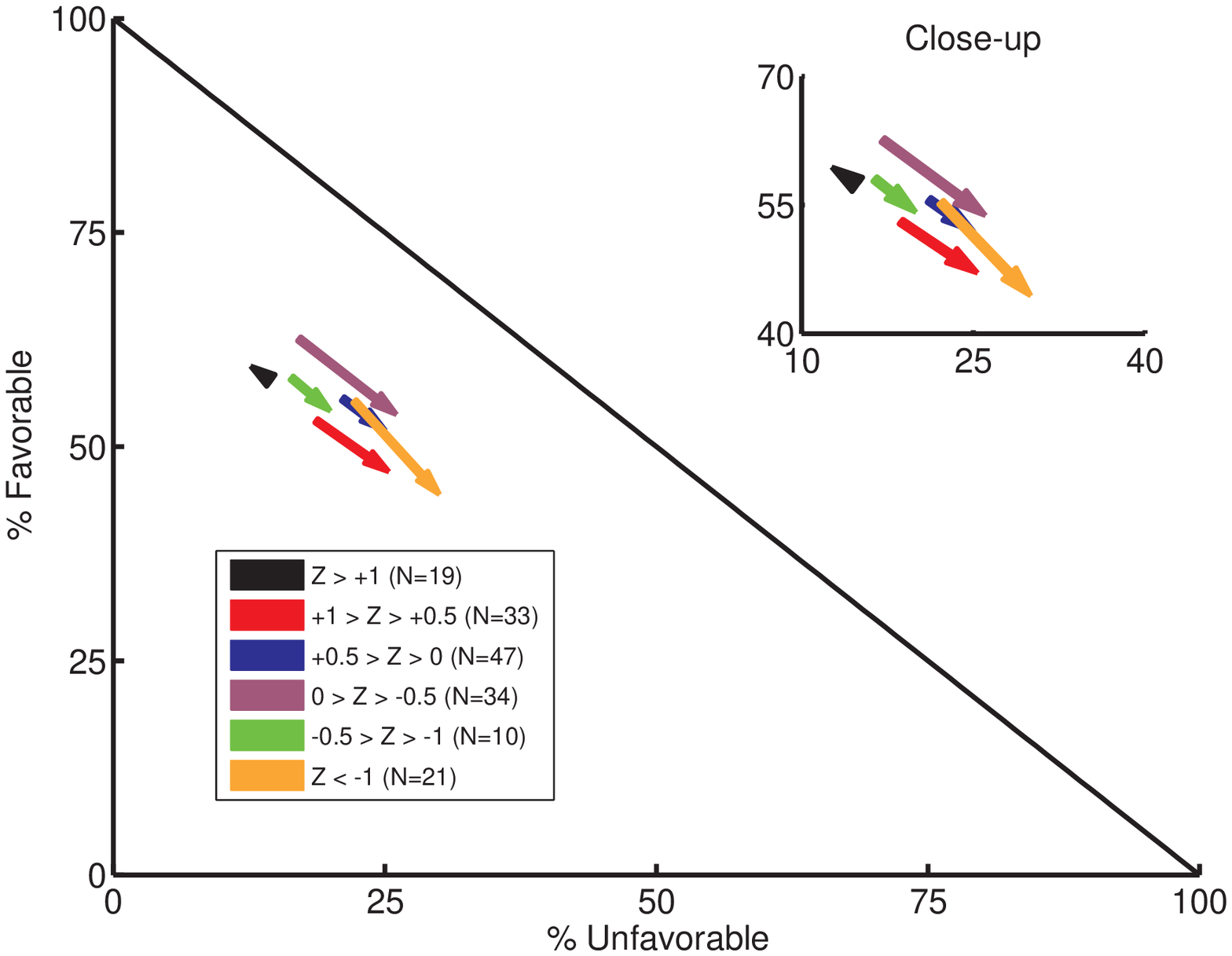}}
    \end{center}
  \caption{[Color] - The shift in the NCSU mechanics students' mean COMPASS scores are shown. Colored arrows indicate the magnitude and direction of the shift from pre- to post-instruction. Mean scores are shown for students based on: (a) their official classification upon starting the mechanics course, (b)  their declared major college upon entering the mechanics course and (c) their normalized score earned in an introductory mechanics course. All non-engineering and non-science majors were included for completeness but these students were roughly 5\% of the total population.}
  \label{fig:compass_dem-ncsu}
\end{figure}

%% file: appendix/app-stats.tex
\chapter{Statistical Techniques}\label{app-stats}

This appendix presents a bit more detail on some of the more esoteric statistical techniques used in this thesis, namely, hierarchical cluster analysis and contingency table analysis.

\section{Hierarchical Cluster Analysis\label{sec:clusteranalysis}}

Cluster analysis aims to organize observations into subsets that are similar in some fashion. 
In our work, we are attempted to uncover subsets of students who made similar errors as measured through an empirically developed set of codes (Table \ref{tab:vpcodes}). 
Using hierarchical cluster analysis, we determined which groups of students had similar binary code patterns. 
Furthermore, we can evaluate if these subsets of students' programs are characterized by some common underlying error or set of errors. 
In performing cluster analysis, one must prepare the data for analysis, choose metric with which to measure proximity or similarity of patterns and choose a method of linking groups of identical data into clusters. 

Cluster analysis is a robust and diverse data classification technique. 
This section is meant to give the reader a sense for how we used cluster analysis in our work not as a full introduction to it.
The interested reader is directed to the texts by Everitt \cite{clustereveritt}, Kaufmann \cite{clusterkaufman}, and Tan \cite{tan2006introduction}. We outline the procedures of cluster analysis first, discuss a few terms, and then proceed to illustrate with an example using our own data set.

\subsection{A Rough Outline of Things to Come}

Cluster analysis begins with a data set that the researcher believes has some underlying patterns to it.
This data set has some set of observations that correspond to a single event. 
Each event might be connected to another event through similarities in this observations. 
In our work, these events were individual students' programs and the observations were the codes used to classify the programs.
The patterns in the observations are compared across events using a metric. 
This metric dictates how the initial distance or similarity measurements are made between all events using their observations.
In our case, the patterns of students' affirmative (1) or negative (0) codes were compared and similarity measures computed.
These initial measurements are then scanned for the lowest value between pairs of events.
This value sets the threshold for the first cluster.
All events that are separated by this same value are fused into the first sets of clusters.
The measurements are then scanned for the next highest value; however, this is a bit different than the initial measurement because clusters have been formed.
Linkage functions are used to compute the distance between clusters and other clusters or clusters and events.
These linkage functions might simple use the closest distance or average them in some way.
The next lowest value is the next threshold used to form the next set of clusters.
This procedure continues until all events are in a single cluster.
Cluster analysis demands that eventually all events appear in a cluster, but this does not mean that these clusters are useful or meaningful.
Clusters are reviewed after they are formed for their coherence or utility.

\subsection{Preparing the Data Set}

Hierarchical cluster analysis works on the premise that similarities between observations can be used to connect observations into groups.
Hence, the data set must be organized to facilitate a similarity comparison. 
Most major cluster analysis software \cite{MATLAB:2010,eads-hcluster-software} requires that input data be numeric.
Hence, categorical variables should be converted in some manner to numeric values.
One can use binary codes where each categorical variable is represented by a single code which contains either an affirmative (1) or negative (0) value.
It might be tempting to use some sort of numeric rating system (e.g., Category 1 = 1, Category 2 = 2, etc.).
However, binary codes are more appropriate because the distances measured between categorical variables is meaningless.
In our work, the variables were already binary; no conversion was necessary.

An $n \times m$ binary matrix $\mathbf{F}$ represents the affirmative nature of $m$ codes for $n$ students.

\begin{equation}
\mathbf{F} = \overbrace{\left.\begin{bmatrix} F_{11} & \cdots & F_{1j} & \cdots & F_{1m}  \\ \vdots & \ddots & \vdots &\ddots & \vdots \\ F_{i1} & \cdots & F_{ij} & \cdots & F_{im}  \\ \vdots & \ddots & \vdots &\ddots & \vdots \\ F_{n1} & \cdots & F_{nj} & \cdots & F_{nm} \end{bmatrix}\right\} \begin{rotate}{270}\text{Students}\end{rotate}}^\text{Codes}
\end{equation}

The $j^{th}$ column of $\mathbf{F}$ is a binary column vector; the elements of which represent whether or not a student received an affirmative (1) or a negative (0) on the $j^{th}$ code.  
We call this vector $\vec{C}_j$, the code vector. 
Each of the $m$ codes has a corresponding $\vec{C}_j$.

\begin{equation}
\vec{C}_j = \begin{bmatrix} F_{1j} \\ \vdots \\ F_{ij} \\ \vdots \\ F_{nj} \end{bmatrix}
\end{equation}

The $i^{th}$ row of $\mathbf{F}$ is a binary row vector; the elements of which represent a code was marked as affirmative (1) or negative (0) for the $i^{th}$ student. 
We call this vector $\vec{S}_i$, the student vector. 
Each of the $n$ students has a corresponding $\vec{S}_i$.

\begin{equation}
\vec{S}_i = \begin{bmatrix} F_{i1} \hdots F_{ij} \hdots F_{im} \end{bmatrix}
\end{equation}

Patterns can be compared between each code's $\vec{C}_j$ to determine the similarity between codes. 
That is, which codes were most often applied together.
However, for our analysis, we chose to compare the patterns of each student's $\vec{S}_i$ to determine which students had similar errors that were picked up by the codes.
We next compare the binary patterns between pairs of student vectors to identify which ones are most similar.

\subsection{Choosing a Distance Metric}\label{sec:bdm}

For cluster analyzing data, there are several metrics available. 
The literature recommends trying a few to ensure that the resulting clusters are either invariant to the metric chosen or easily interpretable \cite{clustereveritt}. 
Some metrics that have been proposed are distance measures others are similarity measures. 
In this section, we justify our use of the Jaccard metric to determine the clusters of student errors presented in Sec. \ref{sec:cluster}. 
For now, we fix the linkage function (i.e., average linkage) which will be described later (Sec. \ref{sec:lf})

\subsubsection{Common Metrics for Continuous Data}

Metrics that are measures of distance are most often used in continuous data sets. 
However, such metrics may be used with binary data if each cell in the binary pattern is an independent code.
Distance metrics rely on the orthogonality of the dimensions.
For binary data without such properties, codes maybe collapsed to independent cell blocks thereby coarse graining the data set.
More notable distance measures for continuous data include the Manhattan, Euclidean and Minkoswki distances. 

The Manhattan (city-block) distance between two student vectors is distance between two vectors if the distance traversed was confined to a mesh.

\begin{equation}\label{eqn:cb}
D^{CB}_{kl} = \sum_{j=1}^m \left(S_{kj} - S_{lj}\right) 
\end{equation}

This distance is simply how far apart two vectors are ``as the taxi-cab drives''.

The Euclidean distance between two student vectors is simply the magnitude of their separation.

\begin{equation}\label{eqn:eu}
D^E_{kl}=\left|\vec{S}_k - \vec{S}_l\right|
\end{equation}

This distance is simply how far apart two vectors are ``as the crow flies''.

The Minkowski distance measures the distance between two student vectors using a generalization of both the Euclidean and Manhattan distances, the $p$-norm.  

\begin{equation}\label{eqn:mk}
D^M_{kl} = \left\{\sum_{j=1}^m \left(S_{kj} - S_{lj}\right)^p\right\}^{1/p}
\end{equation}

The Minkowski metric has an extra degree of freedom, $p$. 
For $p=1$ in Eq. \ref{eqn:mk}, we get Eq. \ref{eqn:cb} and for $p=2$, we get Eq. \ref{eqn:eu}.

\begin{figure}[t]
 \begin{center}
 \subfigure[Clusters formed using $D^{CB}$]{\includegraphics[width=0.45\linewidth, clip, trim=40mm 0mm 40mm 0mm]{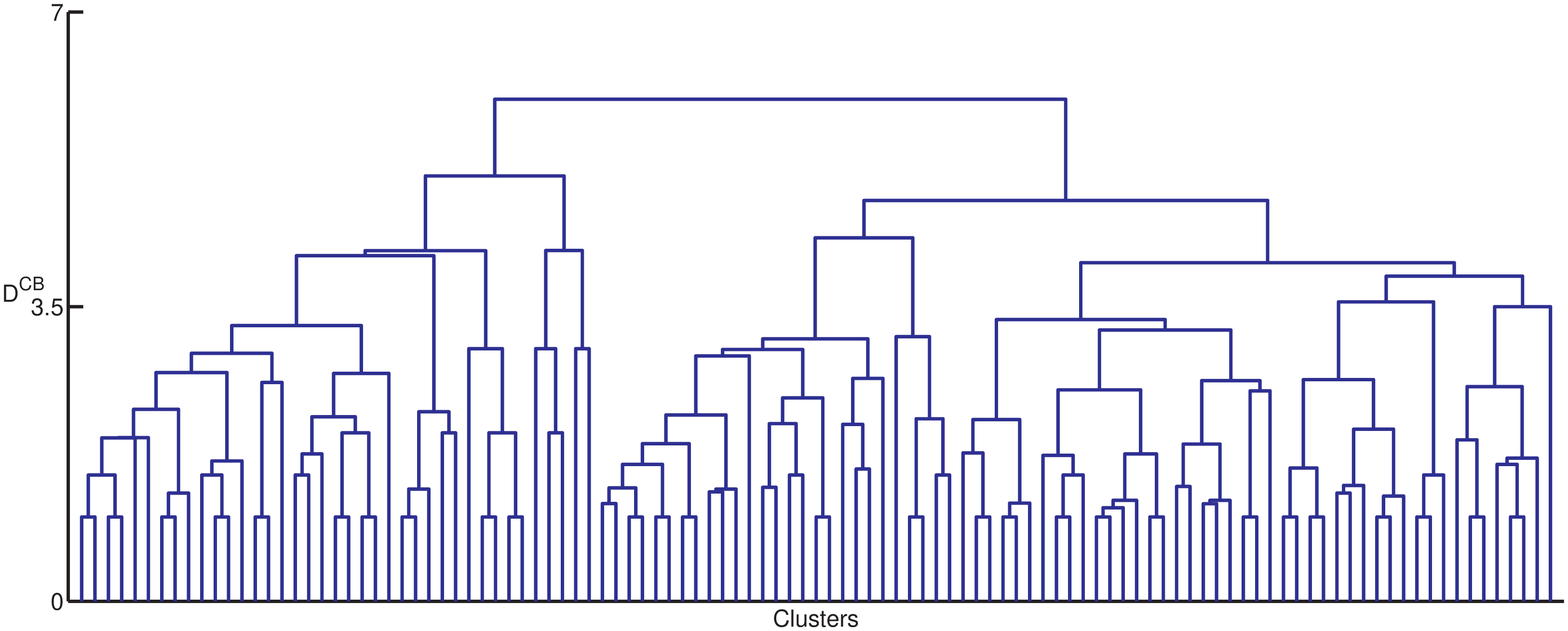}}
 \subfigure[Clusters formed using $D^{E}$]
 {\includegraphics[width=0.45\linewidth, clip, trim=40mm 0mm 40mm 0mm]{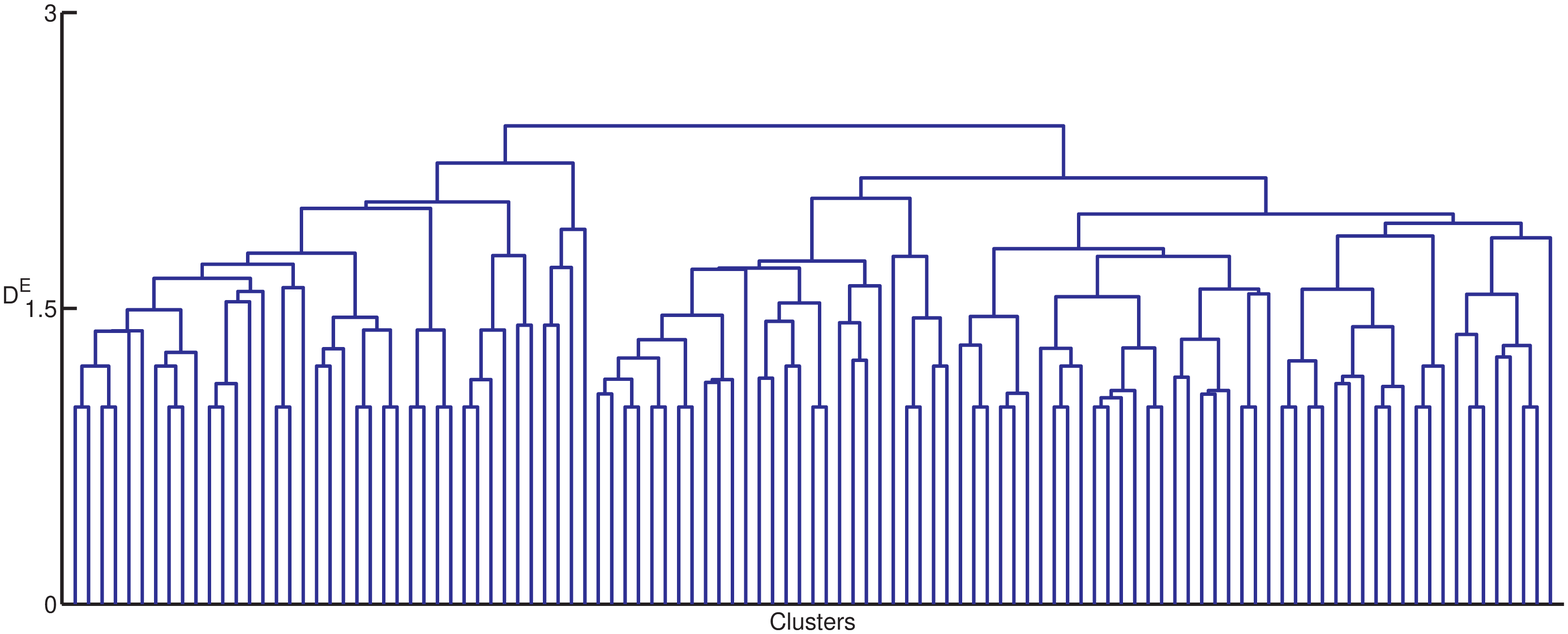}}
 \subfigure[Clusters formed using $D^{M3}$]
 {\includegraphics[width=0.45\linewidth, clip, trim=40mm 0mm 40mm 0mm]{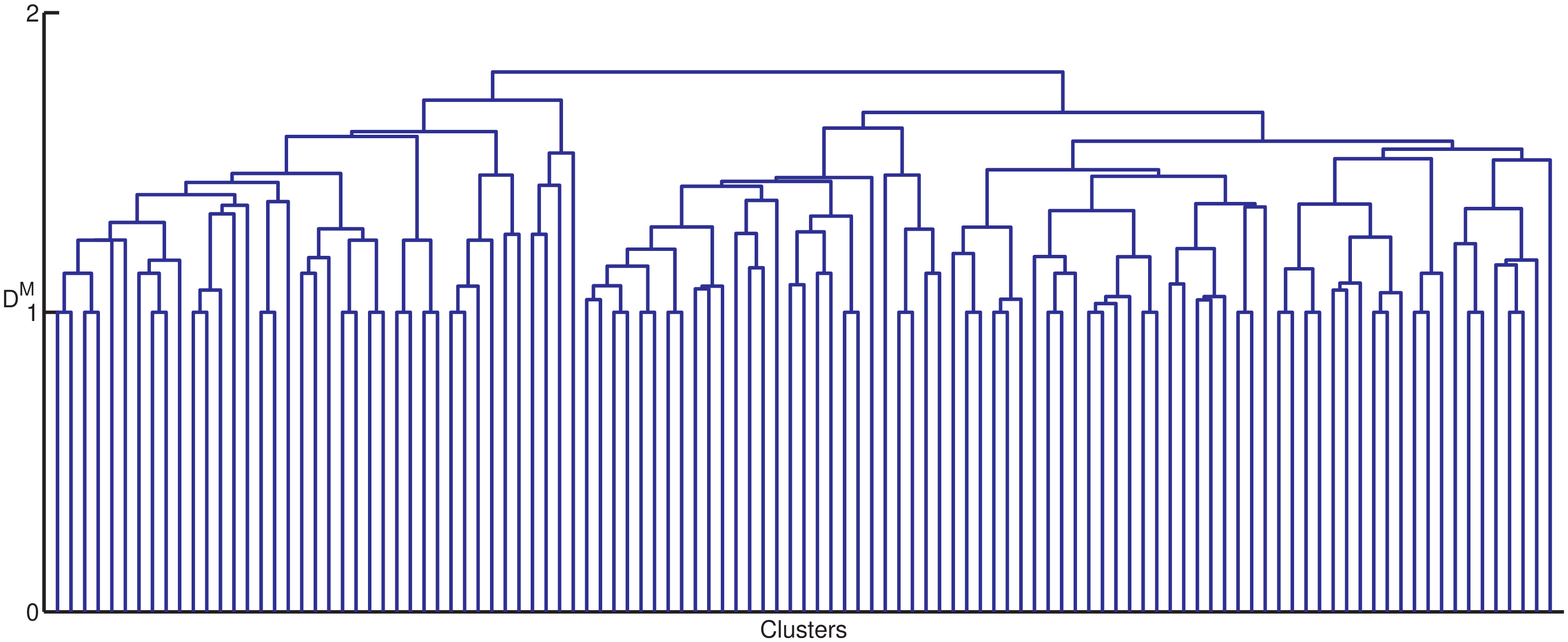}}
  \subfigure[Clusters formed using $D^{M4}$]
  {\includegraphics[width=0.45\linewidth, clip, trim=40mm 0mm 40mm 0mm]{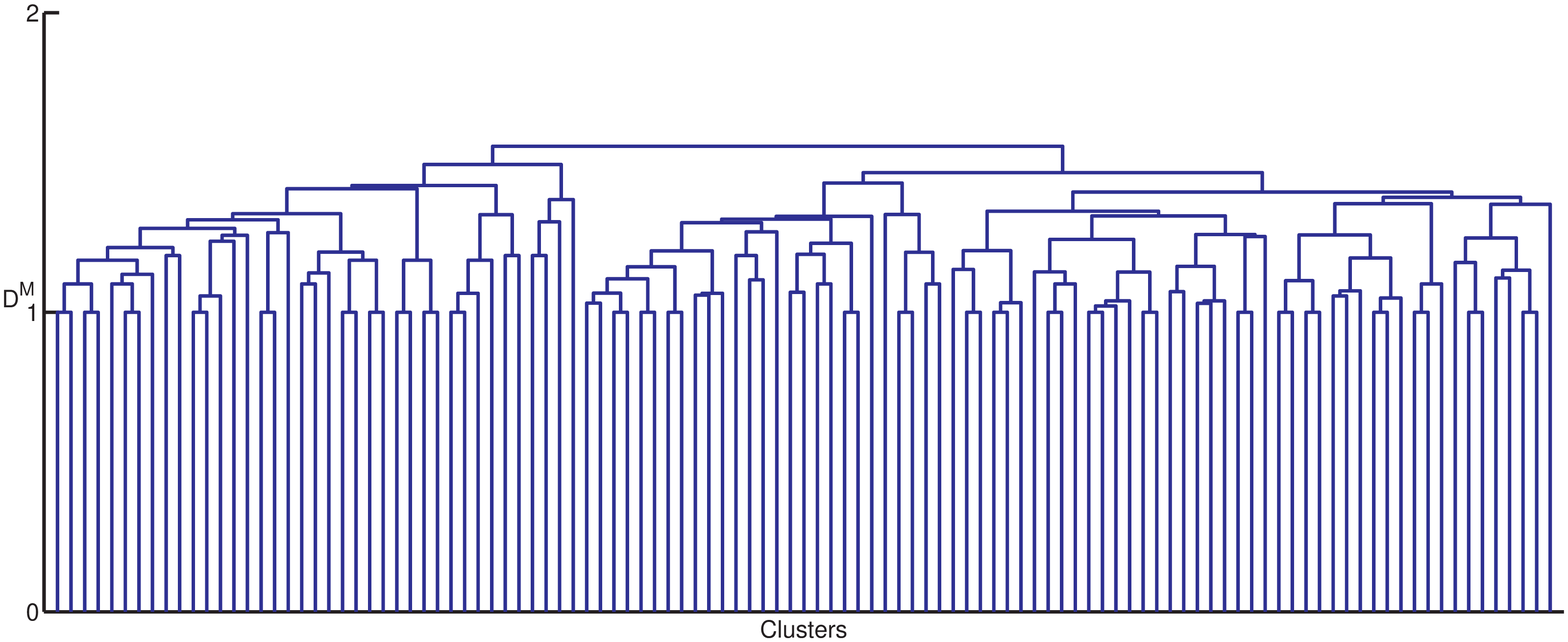}}
 \end{center}
 \caption{[Color] - Clusters formed using different distance metrics are shown: (a) the Manhattan distance, (b) the Euclidean distance, (c) the Minkowski distance with $p=3$ and (d) the Minkowski distance with $p=4$. Each cluster is structurally identical; the students in each cluster are the same.   \label{fig:clusterswithdist}}
\end{figure}

In Fig. \ref{fig:clusterswithdist}, we have plotted cluster diagrams (dendrograms) in which each of the previous metrics were used.
Dendrograms are a visual representation of the cluster procedure. 
In each figure, 111 distinct binary patterns are compared (the vertical lines at the bottom of each figure). 
Distances between each student vector are computed and compared.
The closest student vectors are grouped or ``fused''. 
Binary vectors are usually joined at their lowest possible value of the metric (e.g., for binary data in any Minkoswki space, this is 1). 
The joining of vectors or clusters of vectors is indicated by horizontal lines in the dendrogram. 
After the initial clusters are formed, the properties of the students in the cluster are compared to all other clusters using a linkage function (Sec. \ref{sec:lf}). 
Again the lowest possible outcomes are fused in a pairwise fashion.
This procedure is continued until no lone student vector exists.
Followed from the bottom to the top, the cluster analysis procedure unfolds for any choice of metric and linkage function.

Fig. \ref{fig:clusterswithdist} is simply meant to illustrate the similarity of clusters produced using Minkowski distance metrics. 
Each cluster shown in Fig. \ref{fig:clusterswithdist} is structurally identical; the students in each cluster are the same. 
However, we can easily see how the choice of $p$ dictates the spacing between leaves.
Ultimately, clusters produced by these metrics were not useful because each element of the student vector $\vec{S}_k$ was not independent. 
Hence, the components of the student vector were not orthogonal.
The resulting clusters had student vectors which had little (besides mathematically) to do with each other.

\subsubsection{Common Metrics for Binary Data}

Metrics that are measures of similarity are used most often used in binary data sets. 
These metrics generally do not rely on the orthogonality of the dimensions.
The distances that measured with similarity metrics are typically some scaling fraction of the elements that are similar between vectors.
More notable similarity measures for binary data are the Hamming and Jaccard distances.

The Hamming metric is the simplest choice for binary data. It defines the distance between student vectors $\vec{S}_{k}$ and $\vec{S}_{l}$ as the proportion of codes (elements of the vector) for which student $K$ and $L$ are inconsistent.

\begin{equation}\label{eqn:hamming}
D_{kl}^H = \frac{C_{kl}^{10} + C_{kl}^{01}}{C_{kl}^{11} + C_{kl}^{10} + C_{kl}^{01} + C_{kl}^{00}} = \frac{C_{kl}^{10} + C_{kl}^{01}}{n}
\end{equation}

Here $C_{kl}^{xy}$ represents the number of codes for which the $k^{th}$ and $l^{th}$ student received some mark $x$ and $y$ respectively. The superscripts ($xy$) indicate whether the code was marked affirmative (1) or negative (0). In the superscript, the first digit refers to  the $k^{th}$ student and the second digit refers to $l^{th}$ student. 

As an example, consider two student vectors $\vec{a} = [1 0 0 1 0 1]$ and $\vec{b} = [0 0 1 1 0 1]$. For these vectors, the Hamming distance is 1/3. Using Eq. \ref{eqn:hamming},

\begin{equation}\label{eqn:hammingex}
D_{ab}^H = \frac{C_{ab}^{10} + C_{ab}^{01}}{n} = \frac{1+1}{6} = \frac{1}{3}
\end{equation}

Computing this pairwise distance between all pairs of students produces a $n \times n$ symmetric distance matrix, $\mathbf{D}^H$, with 0's along the diagonal. The elements of this matrix, $D^H_{kl}$, give the proportion of codes for which students $k$ and $l$ agree. The extrema of any one element are 0 (completely disagree) and 1 (completely agree).

The Jaccard metric is also a valid choice for binary data. It is somewhat similar to the Hamming metric expect that it neglects codes for which both students received negatives (0). The Jaccard metric defines the distance between student vectors $\vec{S}_{k}$ and $\vec{S}_{l}$ as the proportion of codes for which student $K$ and $L$ are disagree compared to the total number of codes minus those that are both negative.

\begin{equation}\label{eqn:jaccard}
D_{kl}^J = \frac{C_{kl}^{10} + C_{kl}^{01}}{C_{kl}^{11} + C_{kl}^{10} + C_{kl}^{01}} = \frac{C_{kl}^{10} + C_{kl}^{01}}{n- C_{kl}^{00}}
\end{equation}

Using the same example vectors, $\vec{a} = [1 0 0 1 0 1]$ and $\vec{b} = [0 0 1 1 0 1]$, we find the inter-cluster distance is 1/2. Generally speaking, inter-cluster distances are larger using the Jaccard metric. Using Eq. \ref{eqn:jaccard},

\begin{equation}\label{eqn:jaccardex}
D_{kl}^J = \frac{C_{kl}^{10} + C_{kl}^{01}}{n - C_{kl}^{00}} = \frac{1 + 1}{6 - 2} = \frac{1}{2}
\end{equation}

An $n \times n$ symmetric distance matrix with 0's along the diagonal, $\mathbf{D}^J$, is formed by computing this pairwise distance between all pairs of students. The elements of this matrix, $D^J_{kl}$, give Jaccard distance between $k^{th}$ and $l^{th}$ students The extrema of any one element are 0 (identical) and 1 (completely opposite).

\begin{figure}[t]
\begin{center}
\subfigure[Clusters formed using the Hamming metric]{\includegraphics[width=0.45\linewidth, clip, trim=38mm 0mm 40mm 0mm]{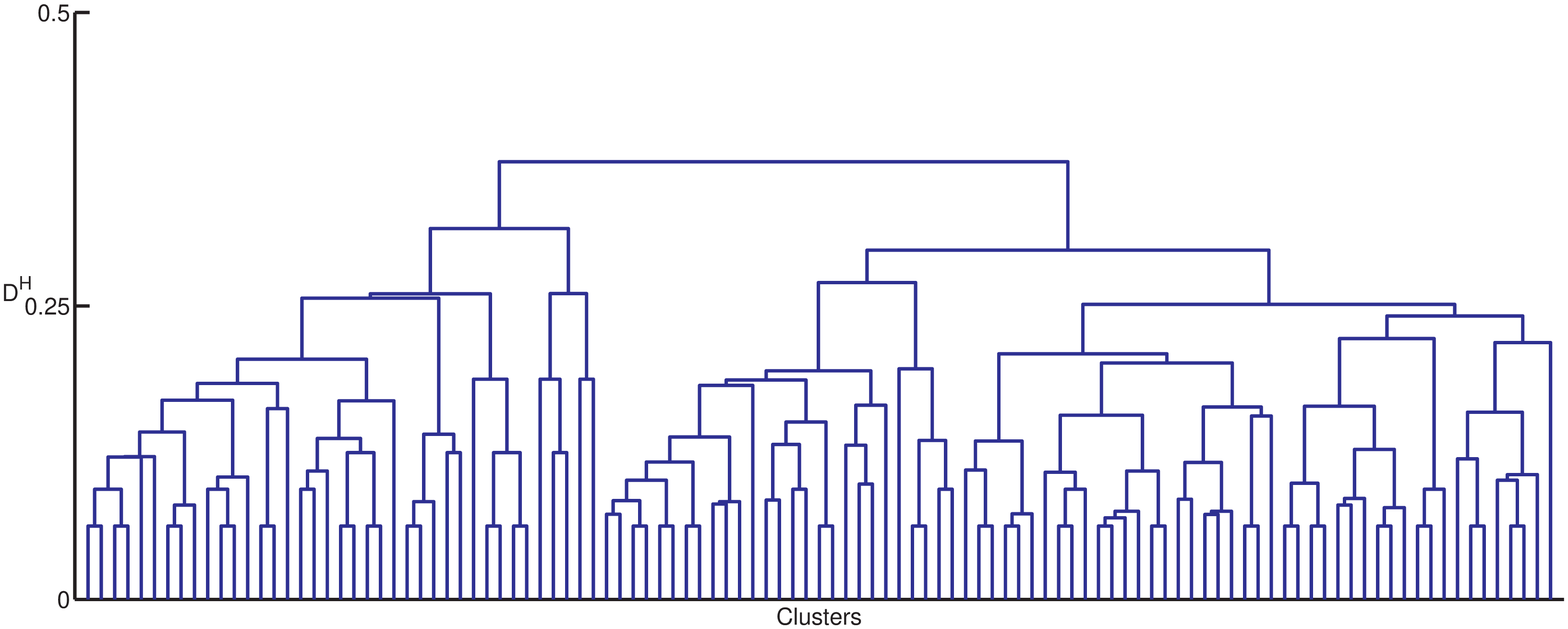}}
\subfigure[Clusters formed using the Jaccard metric]{\includegraphics[width=0.45\linewidth, clip, trim=40mm 0mm 40mm 0mm]{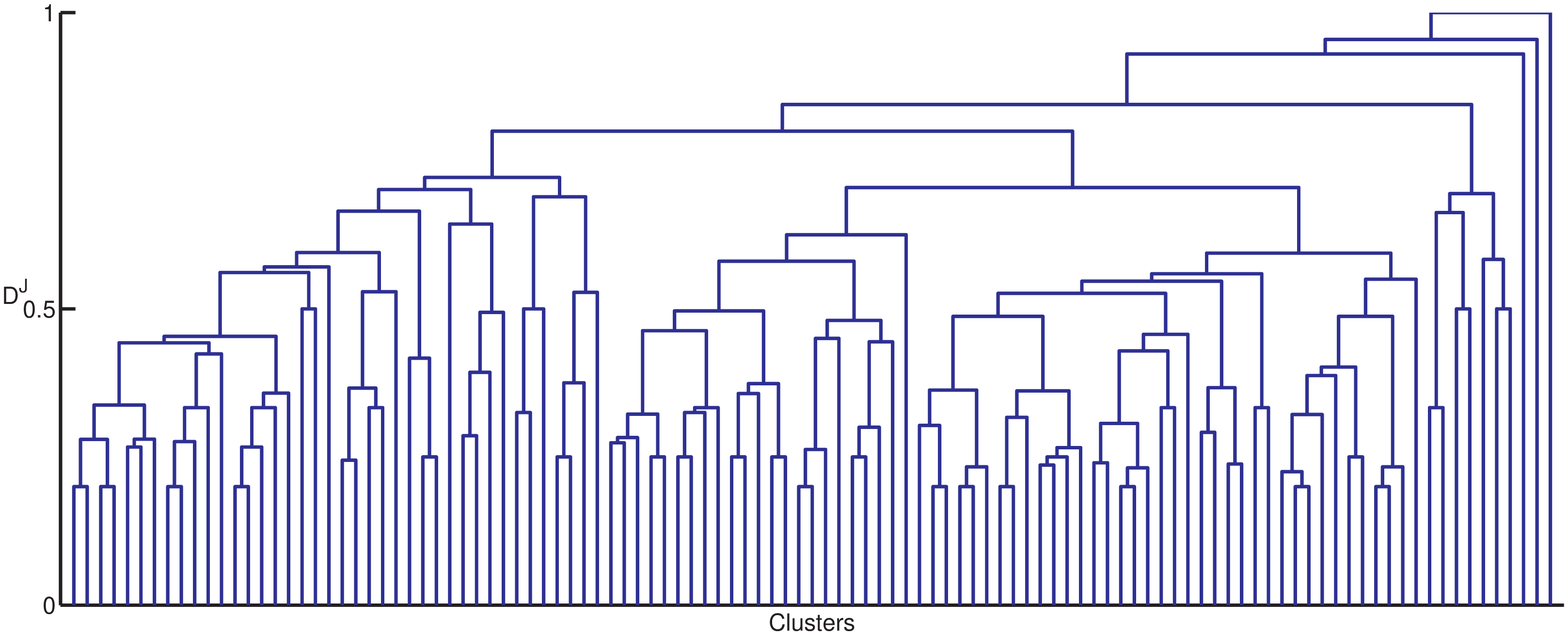}}
\end{center}
\caption{[Color] - Clusters formed using two different binary metrics are shown: (a) the Hamming metric and (b) the Jaccard metric. The Hamming metric treats elements of the student vectors as independent. The Jaccard takes into account the identity of the patterns (i.e., vectors are identical vs. vectors have the completely opposite pattern).\label{fig:clustersbin}}
\end{figure}

In Fig. \ref{fig:clustersbin}, we have plotted dendrograms using our data and each of the aforementioned binary metrics.
The structural differences of the clustering events are quite apparent. 
A review of the clusters that were linked in each case found that the Hamming metric was undesirable for our work.
The Hamming metric uses all possible similarities; it treats elements of the vector as if they are independent.
Hence, the resulting clusters were not informative.
The Jaccard metric neglects elements in which both are marked negative (0). 
This is a more useful feature because comparisons now take into account the identity of the whole pattern (i.e., vectors are identical vs. vectors have the completely opposite pattern).

The choice of metric is made for a variety of reasons: continuous vs binary data, presence vs. lack of independence between elements, et cetera.
However, clusters should be reviewed for coherence and utility no matter the choice of metric. 
We chose to use the Jaccard metric for our cluster analysis not only because it is a valid choice for binary data, but because of the coherence of the resulting clusters.

\subsection{Forming Clusters}\label{sec:lf}

Determining which student vectors (or clusters of student vectors) are fused together to form new clusters is an iterative pair-wise process. 
The inter-cluster distance is used to determine at what level clusters are fused.
The distance between clusters are computing using linkage functions; they
describe how ``close'' clusters are to each other. 
For the present example, the Jaccard distance, $D^J_{kl}$, is initially computed for each pair of student vectors those values form a $n \times n$ symmetric matrix, $\mathbf{D}^J$.
The matrix is searched for the lowest pair or pairs of values. 
The $s_1$ student vectors separated by this distance are fused at this level.
The distance matrix is then reduced to a $n-(s_1) \times n-(s_1)$ matrix where distances have now been computed using the linkage function. 
This new matrix is then searched again for the smallest element (corresponding to the closest pair of student vectors, newly formed cluster and student vector or pair of clusters).
This value is then used to fuse the next set of student vectors or clusters.
Throughout the procedure, we keep track of which student vectors are fused at what inter-cluster distance.

\begin{figure}[t]
 \begin{center}
 \subfigure[Simple Linkage]{\label{fig:simplelinkage}\includegraphics[clip, trim = 8.7mm 7mm 7.3mm 7mm, width=0.3\linewidth]{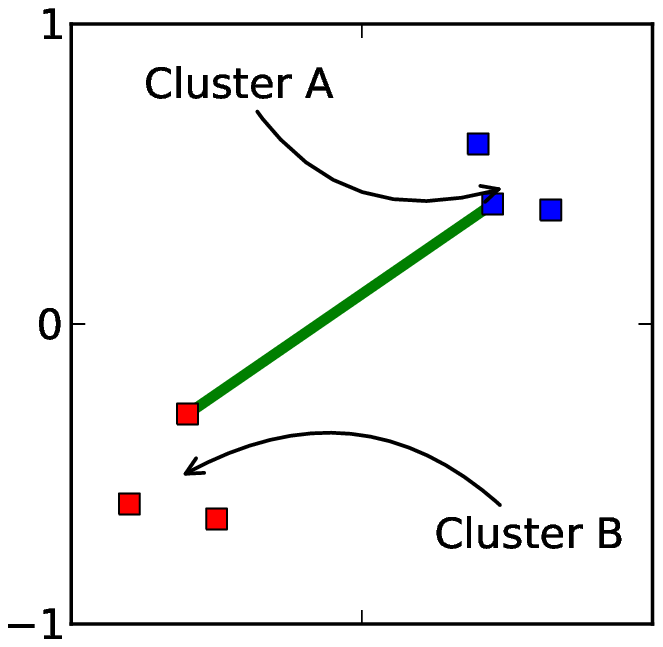}}
 \subfigure[Complete Linkage]
 {\label{fig:completelinkage}\includegraphics[clip, trim = 8.7mm 7mm 7.3mm 7mm, width=0.3\linewidth]{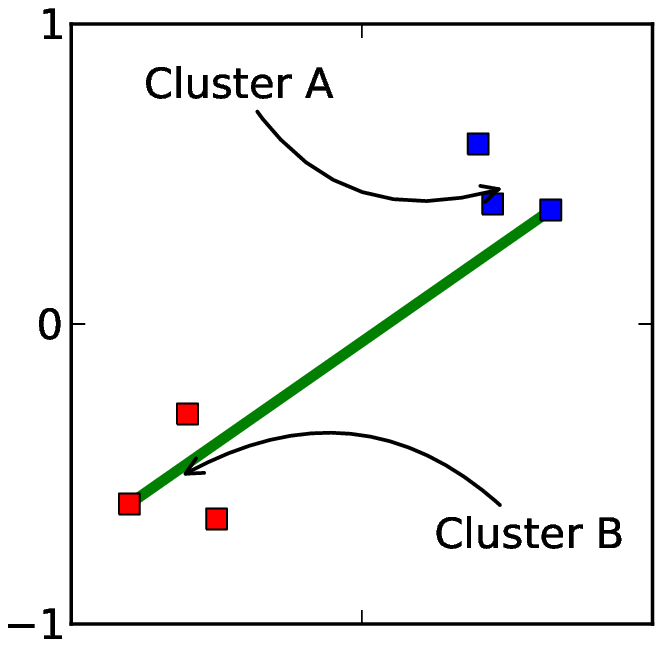}}
 \subfigure[Average Linkage]
 {\label{fig:meanlinkage}\includegraphics[clip, trim = 8.7mm 7mm 7.3mm 7mm, width=0.3\linewidth]{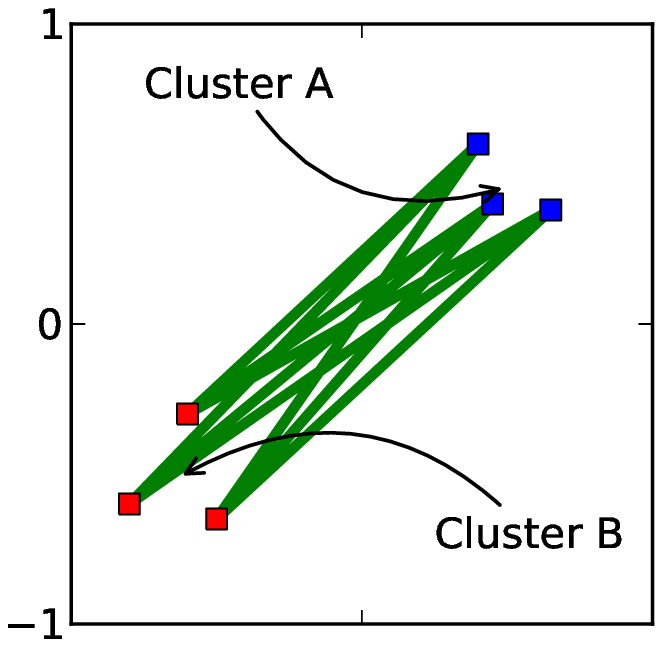}}
 \end{center}
 \caption{[Color] - Three common linkage functions used to compute distances between clusters are illustrated graphically: (a) simple or nearest-neighbor linkage, (b) complete or farthest-neighbor linkage and (c) average linkage. \label{fig:lfs}}
\end{figure}


To performing this iterative procedure, we must have chosen a method for recomputing the distances between student vectors (and clusters).
Linkage functions determine how the inter-cluster distances are computed. 
Several linkage functions exist: single linkage \cite{sneath1957application}, complete linkage \cite{sorenson1948method}, and average linkage \cite{sokal1975statistical}. 
A visual representation of the linkage functions appears in Fig. \ref{fig:lfs}.
Single linkage (Fig. \ref{fig:simplelinkage}) compares the shortest distance between two clusters and tends to be affected largely by ``chaining''; clustering events are somewhat sequential and give little useful structure.
Complete linkage (Fig. \ref{fig:completelinkage}) tends to form equal sized clusters .
Average linkage takes into account the structure of clusters (Fig. \ref{fig:meanlinkage}) and it is relatively robust \cite{clustereveritt}.

\begin{figure}
\begin{center}
\subfigure[Clusters formed using simple linkage]{\label{fig:simple}\includegraphics[width=0.45\linewidth, clip, trim=40mm 0mm 40mm 0mm]{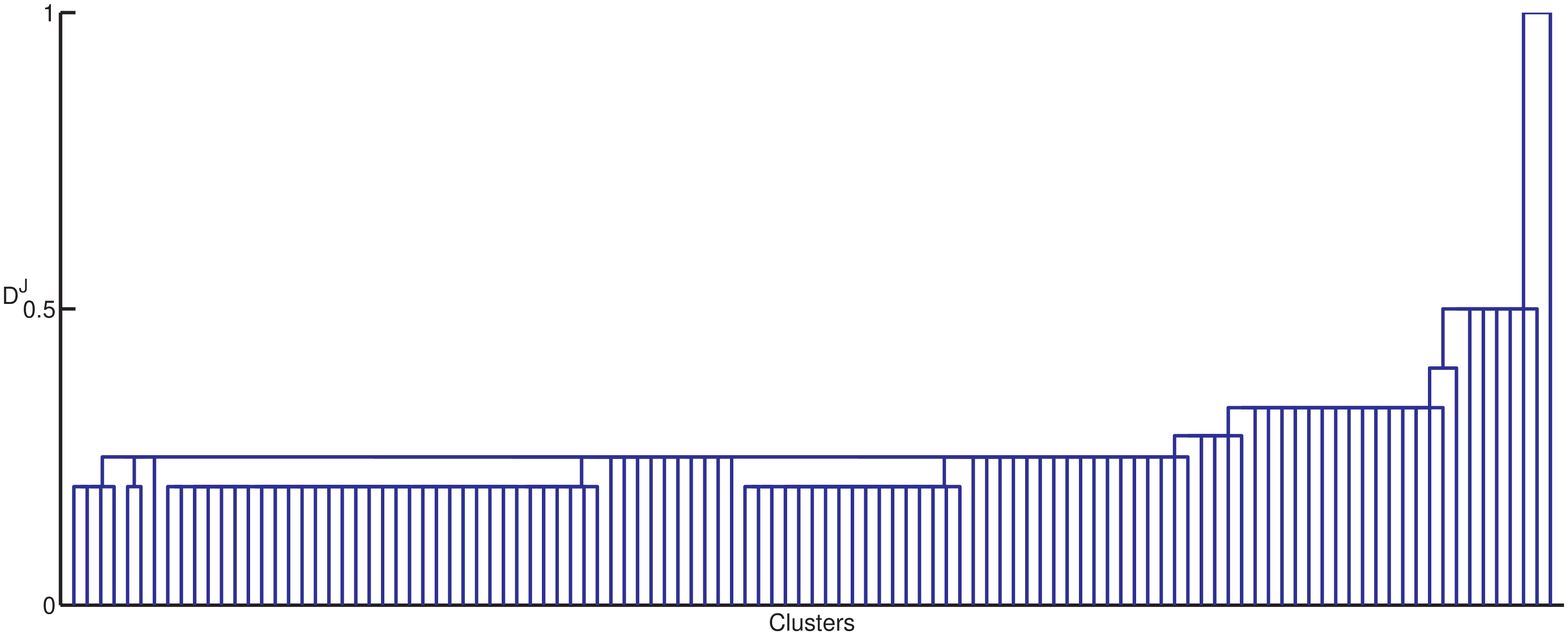}}
\subfigure[Clusters formed using complete linkage]
{\label{fig:complete}\includegraphics[width=0.45\linewidth, clip, trim=40mm 0mm 40mm 0mm]{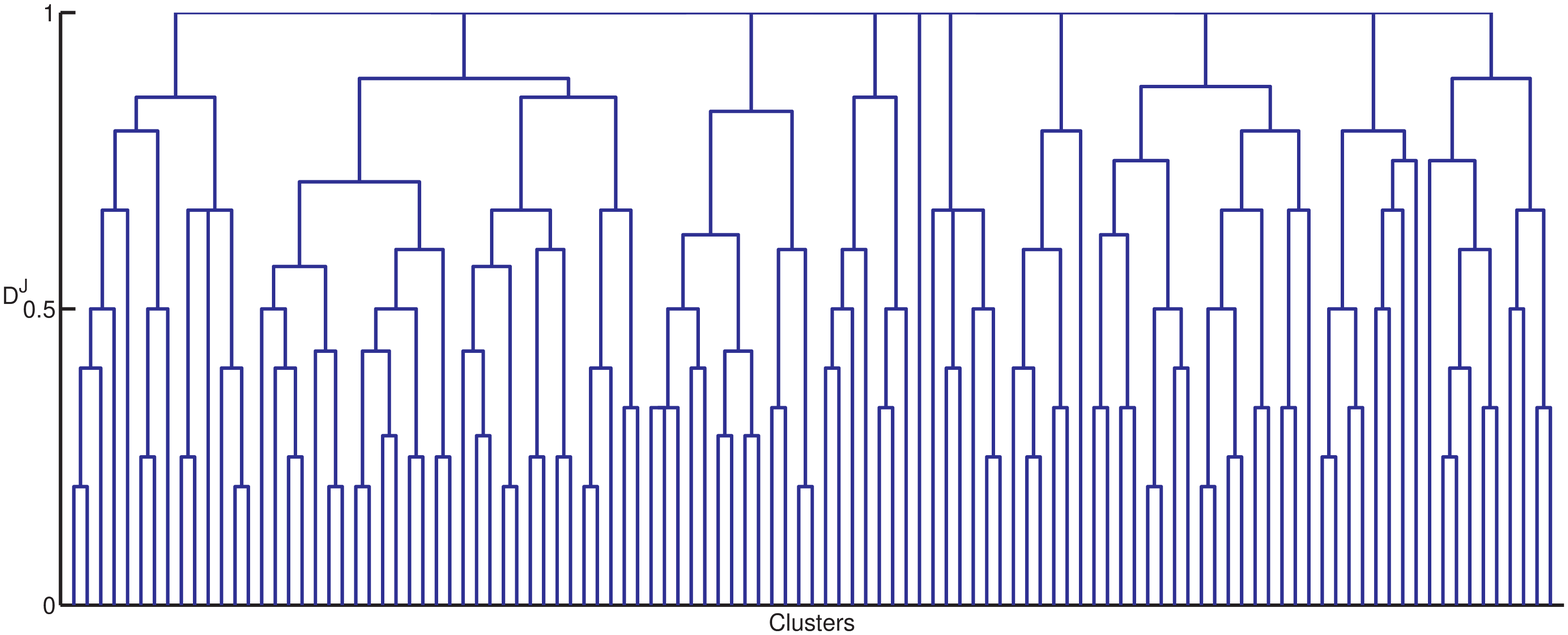}}
\end{center}
\caption{[Color] - Clusters formed using the Jaccard metric and either (a) single linkage or (b) complete linkage. Single linkage is prone to long chaining events typically render the analysis useless. Complete linkage tends to form equal size clusters.\label{fig:linkcompare}}
\end{figure}

\begin{figure}[t]
\begin{center}
\subfigure[Clusters formed using average linkage]{\label{fig:averagecolor}\includegraphics[width=0.45\linewidth, clip, trim=40mm 0mm 40mm 0mm]{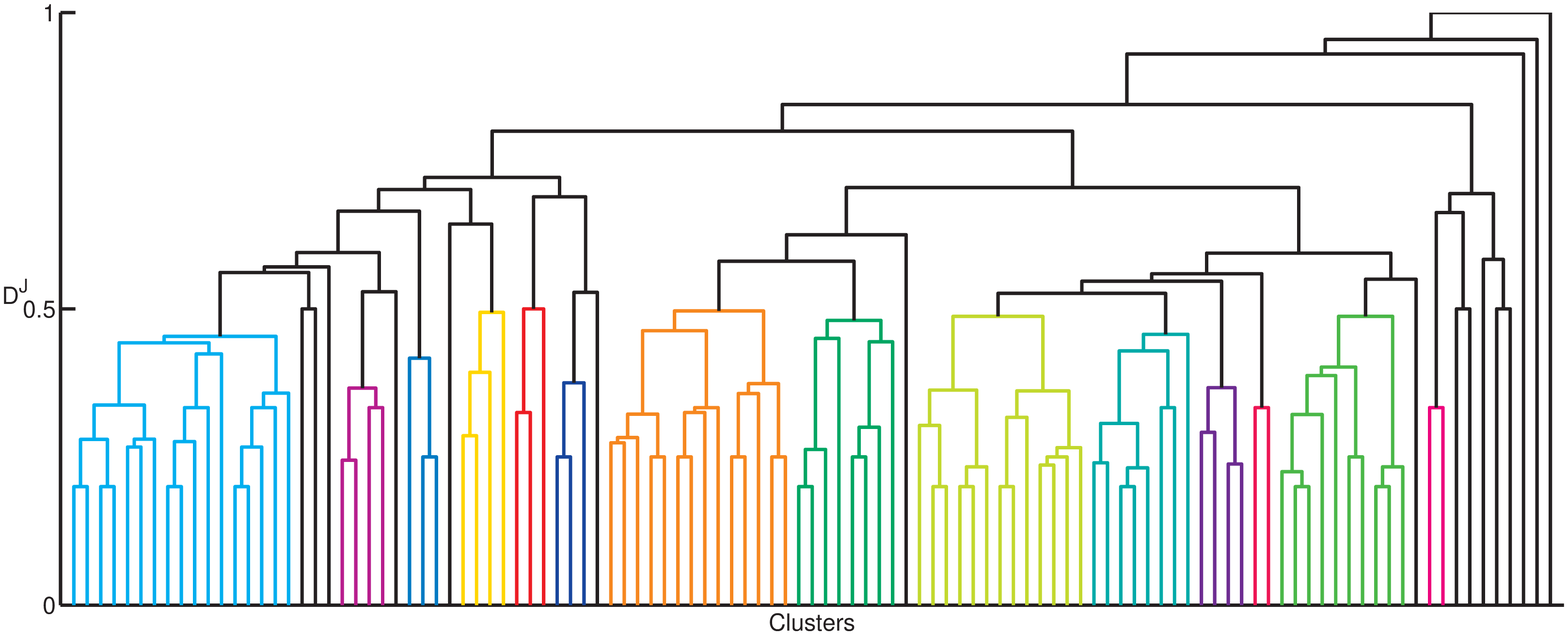}}
\subfigure[Same as (a) but collapsed]
{\label{fig:collapse}\includegraphics[width=0.45\linewidth, clip, trim=40mm 0mm 40mm 0mm]{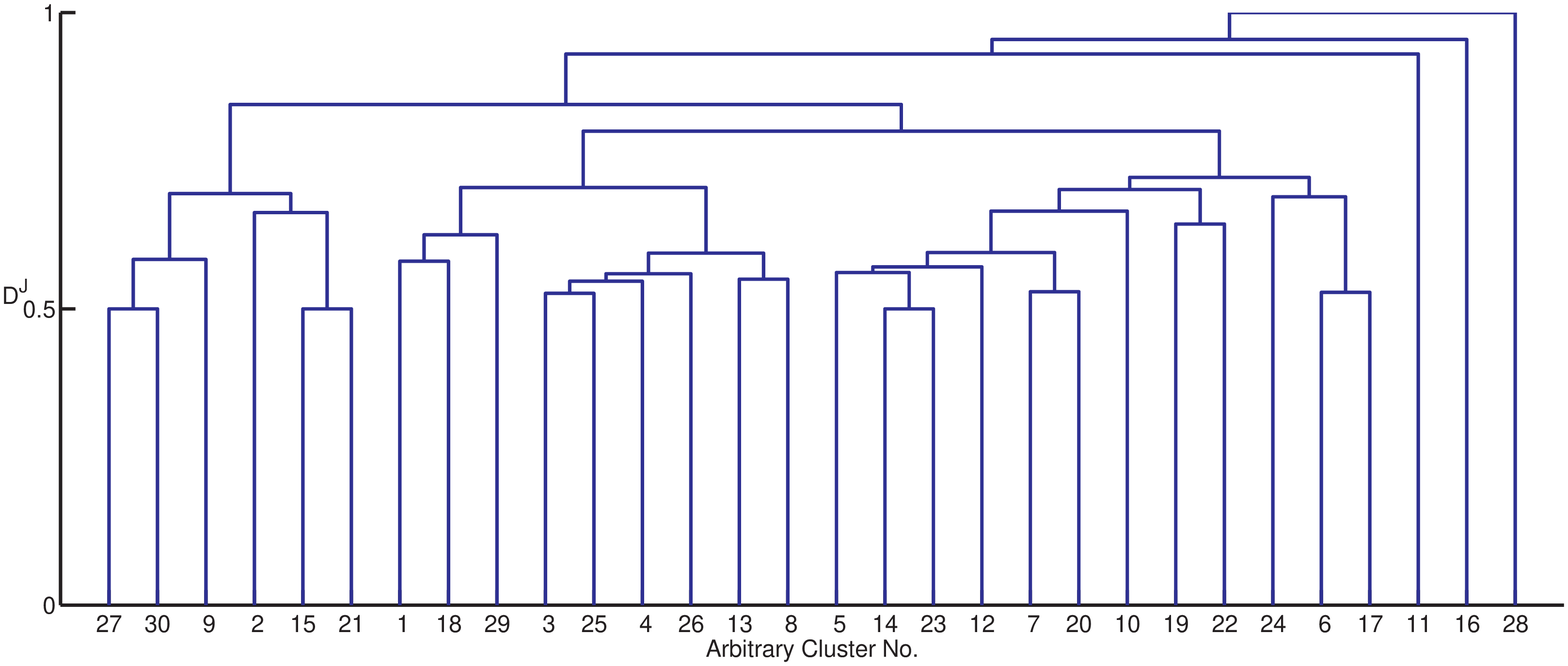}}
\end{center}
\caption{[Color] - Clusters formed using the Jaccard metric and the average linkage function. In (a), we highlight the fourteen clusters with more than one unqiue code. In (b), we collapse the details for the additional leaves to highlight the thirty clusters that were reviewed in detail (Sec. \ref{sec:cluster}).\label{fig:clusj}}
\end{figure}

We used each of these linkage functions and investigated which returned coherent and useful clusters.
Coherence and utility are both determined after the analysis is completed.
Clusters formed using the simple linkage function experienced severe chaining (Fig. \ref{fig:simple}). 
The coherence of this clusters was not apparent. 
Student vectors in clusters had nothing but mathematical connections.
Complete linkage produced roughly equally populated clusters with some coherence (Fig. \ref{fig:complete}), although many of the useful connections that we believed should have appeared from this analysis were lacking.
Neither the simple nor the complete linkage functions produced clusters that were as coherent as those formed using average linkage. 
The fourteen clusters which contained more than one unique pattern are highlighted in color in Fig. \ref{fig:averagecolor}.
We reviewed each of the unique 111 student vectors contained within each cluster.
These fourteen clusters were reduced, ultimately, to seven which had clear connections between student vectors they contained (Sec. \ref{sec:cluster}).

\section{Contingency Tables \label{sec:ctables}}

One can ask the question: is a group populated by a different set of students than another? 
Or more directly, can we find an association between a demographic feature and membership. 
Contingency table analysis can describe whether an association between some demographic feature (e.g., major, classification) and membership (e.g., class taken, honors status) exists and the confidence level of that association. 
When using contingency table analysis, one understands that the $p$-values obtained are conservative as compared to its parametric analogs \cite{numericalrecipes}.

The approach is to form a table of events. An event can be any number of countable items. 
An example might be the number of students with different majors taking two different classes. 
By separating the students into their given class, Class 1 vs. Class 2, and counting each students with a given major in each class, one has proposed a valid contingency table. 
The requirement being that no student is counted twice for a given set of events, hence one could not ``double major'' or be in both classes at once. 

After counting the events, labeled $N_{ij}$, the column and row sums for table are computed. Summing down the column,

\begin{displaymath}
 N_{i.} = \sum_j N_{ij}
\end{displaymath}

is equivalent to counting the number of responders in each treatment. While summing across the rows,

\begin{displaymath}
 N_{.j} = \sum_i N_{ij}
\end{displaymath}

is equivalent to counting the total number of responders with a given score regardless of treatment. One can determine the total number of responders by summing all rows and columns,

\begin{displaymath}
 N = \sum_{i,j} N_{ij} = \sum_i N_{i.} =\sum_j N_{.j}.
\end{displaymath}

We are able to compute an expected value for the number of events, $n_{ij}$, and compare that expectation value to the actual count. If there is no difference in the fraction of students with a particular major in the courses, we expect that the fraction of events in a given row are the same regardless of course. We can propose the null hypothesis,

\begin{displaymath}
 H_0:\:\frac{n_{ij}}{N_{.j}}=\frac{N_{i.}}{N}\:\mathrm{or}\:n_{ij}=\frac{N_{i.}N_{.j}}{N}
\end{displaymath}
with the alternative hypothesis,
\begin{displaymath}
 H_1:\:\frac{n_{ij}}{N_{.j}}\neq\frac{N_{i.}}{N}\:\mathrm{or}\:n_{ij}\neq\frac{N_{i.}N_{.j}}{N}.
\end{displaymath}

A chi-square analysis is performed with this expectation value, $n_{ij}$, where we sum over all events,

\begin{displaymath}
 \chi^2 = \sum_{i,j}\frac{\left(N_{ij} - n_{ij}\right)^2}{n_ij},
\end{displaymath}
\begin{displaymath}
 \nu = IJ-I-J+1
\end{displaymath}
where $\nu$ is the number of degrees of freedom in the chi-square analysis (number of rows, $I$, number of columns, $J$). One can compare the reduced form of this statstic, $\chi^2/\nu$, at a given confidence level, $\alpha$, to computed values given in relevant texts \cite{bevington} or using any statistical package \cite{MATLAB:2010,R2011}. 

After performing this analysis, we found no association between choice of major  and honor status for mechanics students at Georgia Tech (Sec. \ref{sec:compass-dimscores}). We found differences in the population (major and classification) of Georgia Tech mechanics and E\&M courses (Sec. \ref{sec:compass-em}), but no difference in the population (major and classification) between Georgia Tech and NCSU mechanics courses (Sec. \ref{sec:compass-ncsu}).

%% file: caballero-vita.tex
\begin{vita}
Marcos Daniel Caballero was born on May 1, 1982 in the balmy flats of McAllen, Texas. 
He moved to the Hill Country in the summer of 1991, where he completed elementary, middle and high school in San Antonio, Texas. 
He graduated with honors from John Marshall High School in 2000.

Marcos entered the University of Texas at Austin in the fall of 2000. 
While there, he worked overnight at a local copy center and performed undergraduate research.
Throughout his undergraduate tenure, Marcos tutored friends in physics and mathematics; he wanted his career to involve teaching physics.
After completing a BS in physics in 2004, Marcos spent a year working in Austin.

Eventually, he entered the physics PhD program at the Georgia Institute of Technology.
A traditional PhD program did not interest him; he preferred to focus on understanding how students learn physics and learning the craft of teaching. 
To this end, he helped found, with Drs. Michael Schatz, Richard Catrambone, and M. Jackson Marr, the Georgia Tech Physics Education Research group in 2007. 
Marcos helped to develop a group focused on scientific inquiry into the teaching and learning of physics. 

Marcos completed his Ph.D. work under the supervision of Dr. Michael Schatz in July, 2011 and subsequently began post-doctoral research with the Physics Education Research group at the University of Colorado at Boulder. 
In January, 2007, he married Jamie Su Hadden in Atlanta, Georgia. 
Their first child, Juniper Jane Caballero, was born on January 22, 2011.
\end{vita}